\documentclass[]{iopart}
\expandafter\let\csname equation*\endcsname\relax
\expandafter\let\csname endequation*\endcsname\relax
\usepackage{iopams,amsmath,amssymb,amsfonts,stmaryrd,wasysym,graphicx,multirow,color,bm}
\usepackage{url}\usepackage{diagrams}
\usepackage[colorlinks=true,citecolor=blue,urlcolor=blue]{hyperref}
\usepackage[square,sort&compress,comma,numbers]{natbib}

\begin{document}

\title{Globally symmetric topological phase: from anyonic symmetry to twist defect}

\author{Jeffrey C. Y. Teo}\ead{jteo@virginia.edu}\address{Department of Physics, University of Virginia, VA22904, USA}\date{\today}

\begin{abstract}
Topological phases in two dimensions support anyonic quasiparticle excitations that obey neither bosonic nor fermionic statistics. These anyon structures often carry global symmetries that relate distinct anyons with similar fusion and statistical properties. Anyonic symmetries associate topological defects or fluxes in topological phases. As the symmetries are global and static, these extrinsic defects are semiclassical objects that behave disparately from conventional quantum anyons. Remarkably, even when the topological states supporting them are Abelian, they are generically non-Abelian and powerful enough for topological quantum computation. In this article, I review the most recent theoretical developments on symmetries and defects in topological phases.
\end{abstract}

\maketitle 


\section{Introduction}

Topologically ordered states in two dimensions are exotic phases of matter that support fractional quasiparticle excitations with anyonic statistics (a review of these ideas can be found in Ref.~\cite{Wilczekbook,Fradkinbook,Wenbook}). For example the fundamental excitations in a fractional quantum Hall state are quasiparticles that carry fractions of an electric charge and obey neither bosonic nor fermionic statistics. These unconventional properties stem from the long range entanglement of a ground state, which is separated from all excitation states by an energy gap. Unlike ordinary phases of matter, such as a magnet or superconductor, which are characterized by Landau order parameters and broken symmetries, topological phases can be featureless so that there are no apparent local order parameters, and their {\em topological order}, i.e.~the anyonic quasiparticle properties, does not rely on a spontaneous symmetry breaking.

On the other hand, a topological phase {\em can} carry a symmetry or coexist with a symmetry breaking phase. The broadening of the concept of topological phases by symmetries lead to finer classes of symmetry protected/enriched topological phases (SPT/SET)~\cite{LevinGu12, ChenGuLiuWen12, LuVishwanathE8, MesarosRan12, EssinHermele13, WangPotterSenthil13, BiRasmussenSlagleXu14, Kapustin14}. For instance two phases with identical topological order can be separated by symmetries if there are no adiabatic path connecting the two phases without closing the bulk gap or breaking the symmetries. The most notable examples are fermionic topological insulators protected by time-reversal symmetry\cite{HasanKane10, QiZhangreview11, KaneMele2D1, Molenkamp07, MooreBalents07, Roy07, FuKaneMele3D, QiHughesZhang08, Hasan08}. These are short range entangled states that do not support fractional excitations with anyonic statistics. However the symmetry still enables a non-trivial bulk topology and anomalous boundary states. 

In this review article, instead of focusing on the classification of topological phases with symmetries, we concentrate on the consequence of global symmetries. In particular we pay attention to the concept and realization of {\em twist defects} in these systems. They are semiclassical objects relying on the winding of some global textures, and are {\em not} quantum excitations. The prototype of such defects is the zero energy Majorana bound state~\cite{Wilczek09, HasanKane10, QiZhangreview11, Beenakker11, Alicea12}. They were first proposed to exist at flux vortices of the Moore-Read fractional quantum Hall state~\cite{MooreRead, GreiterWenWilczek91, NayakWilczek96} and topological $p$-wave superfluids/superconductors~\cite{Kitaevchain, Volovik99, ReadGreen, Ivanov}. More recently, with the discovery of topological insulators and the application of strong spin-orbit coupled materials, tremendous effort has been put into superconducting-ferromagnet heterostructure with quantum spin Hall insulators~\cite{FuKane08, AkhmerovNilssonBeenakker09, FuKanechargetransport09, LawLeeNg09, GoldhaberGordon12}, nanowires~\cite{Kitaevchain,SauLutchynTewariDasSarma,OregRefaelvonOppen10,Sato2009,Kouwenhoven12,DengYuHuangLarssonCaroffXu12,Shtrikman12,RokhinsonLiuFurdyna12,ChangManucharyanJespersenNygardMarcus12,FinckHarlingenMohseniJungLi13} and atomic chains~\cite{NadjDrozdovLiChenJeonSeoMacDonaldBernevigYazdani14}. Majorana zero modes are {\em non-Abelian} objects. They carry non-trivial quantum dimensions so that multiple Majorana's give rise to a muti-dimensional ground state Hilbert space. The degeneracy is preserved in the thermodynamic limit where quasiparticle tunnelings between these defects are exponentially suppressed by spatial separation. Quantum information is therefore non-locally stored and is robust against local perturbation. Moreover, non-Abelian unitary operations on the degenerate Hilbert space can be obtained by defect braiding and pair measurement. These form the basis of a topological quantum computer~\cite{Kitaev97, OgburnPreskill99, Preskilllecturenotes, FreedmanKitaevLarsenWang01, ChetanSimonSternFreedmanDasSarma, Wangbook, SternLindner13}. The even more powerful parafermion or fractional Majorana bound state, which carry richer quantum degeneracies and braiding characteristics, were predicted at the SC-FM edge~\cite{LindnerBergRefaelStern, ClarkeAliceaKirill, MChen, Vaezi, mongg2} of fractional topological insulators~\cite{LevinStern09, LevinStern12, MaciejkoQiKarchZhang10, SwingleBarkeshliMcGreevySenthil11, LevinBurnellKochStern11}, helical 1D Luttinger liquids~\cite{OregSelaStern13}, and fractional quantum Hall states~\cite{BarkeshliQi, teo2013braiding, BarkeshliQi13, BarkeshliJianQi, khan2014, BarkeshliOregQi14}. All these can be unified in the framework of twist defects in a globally symmetric topological phase. 

\subsection{Outline and Objectives}
In this review article, we aim at providing the basic concepts and model realizations of globally symmetric topological phases and the twist defects they support. We focus more on individual examples using exactly solvable models rather than on the most general mathematical framework, although certain universal aspects will be highlighted. There has been numerous theoretical proposals on experimental implementations. It is not the main objective of this article to provide a completely summary and comparison. Instead we will present two major setups, the superconductor - ferromagnet - (fractional) quantum spin Hall insulator heterostructure and the bilayer fractional quantum Hall states, for illustration purpose. The review article will be organized according to the following.

A review on topological phases in two dimensions will be provided in section~\ref{sec:discretegaugetheory}. It will begin with the Kitaev's toric code model~\cite{Kitaev97} of a $\mathbb{Z}_2$ gauge theory in the deconfined limit. It is a lattice spin model, where the exact ground states and excitations can be written down explicitly. It exhibits many essential properties of a globally symmetric Abelian topological state. For instance the {\em anyonic symmetry} is realized as a lattice translation. Next in section~\ref{sec:fluxcharge}, we will generalize to discrete $G$-gauge theories~\cite{BaisDrielPropitius92,Propitius-1995,PropitiusBais96,Preskilllecturenotes,Freedman-2004,Mochon04}, where the gauge group $G$ can be non-Abelian and lead to a non-Abelian topological phase. We will see the appearance of non-Abelian anyonic excitations, which carry internal degeneracies and non-trivial quantum dimensions, and explain their non-Abelian braiding statistics. Lastly in section~\ref{sec:MTC} we will review the relevant mathematical language we will use to describe a general topological state. The topological information is encoded by a modular tensor category~\cite{Kitaev06, Turaevbook, BakalovKirillovlecturenotes, Wangbook, Walkernotes91, FreedmanLarsenWang00, Bondersonthesis}, which includes the fusion and splitting properties of anyonic excitations, and the exchange and monodromy braiding operations among anyons. 

The concept of {\em anyonic symmetry}~\cite{Kitaev06,EtingofNikshychOstrik10,Bombin,YouWen,BarkeshliJianQi,TeoRoyXiao13long,khan2014,BarkeshliBondersonChengWang14,TeoHughesFradkin15,TarantinoLindnerFidkowski15} will be presented in section~\ref{sec:anyonicsymmetry}. For a given topological order there
is a corresponding set of anyonic quasiparticles, and the anyonic symmetry group acts on the set of quasiparticles to permute the anyon labels. This is similar to, for example, the permutation of a discrete set of ground states in a conventional symmetry broken phase by the discrete symmetry operations. However, there is generically no local order paramter distinguishing the anyons and therefore the global symmetry is only {\em weakly} broken~\cite{Kitaev06}. We will begin the description of anyonic symmetries in Abelian topological phases in section~\ref{sec:globalsymmetryab} using effective Chern-Simons theory~\cite{WenZee92,Wenedgereview}. This will be followed by exactly solvable lattice models, bilayer and conjugation symmetric phases, as well as the $S_3$-symmetric $SO(8)_1$-state~\cite{khan2014, BarkeshliBondersonChengWang14, TeoHughesFradkin15}, which is intimately related to the surface of a bosonic topological insulator~\cite{BurnellChenFidkowskiVishwanath13, VishwanathSenthil12, WangPotterSenthil13, Senthil2014}. In section~\ref{sec:globalsymmetrynab}, we will demonstrate anyonic symmetries in non-Abelian topological states. We will focus on one example, the $S_3$-symmetric chiral ``4-Potts" state~\cite{TeoHughesFradkin15}, a non-Abelian topological state whose gapless $(1+1)$D boundary is described in low energy by a conformal field theory of a chiral sector of the 4-state Potts model at criticality~\cite{DijkgraafVafaVerlindeVerlinde99, CappelliAppollonio02}. Lastly we will provide an outlook in section~\ref{sec:classificationsymmetry} to the general classification and obstruction of quantum symmetries and reference therein for interested readers.

Section~\ref{sec:defect} will concentrate on twist defects~\cite{Kitaev06, EtingofNikshychOstrik10, barkeshli2010, Bombin, Bombin11, KitaevKong12, kong2012A, YouWen, YouJianWen, PetrovaMelladoTchernyshyov14, BarkeshliQi, BarkeshliQi13, BarkeshliJianQi, MesarosKimRan13, TeoRoyXiao13long, teo2013braiding, khan2014, BarkeshliBondersonChengWang14, TeoHughesFradkin15, TarantinoLindnerFidkowski15}. These are extrinsic topological defects defined by their relabeling action on orbiting anyons according to an anyonic symmetry. In other words, they are static fluxes of anyonic symmetries. These defects generically carry non-trivial quantum dimensions, which make them non-Abelian, even in an Abelian medium. In section~\ref{sec:twistdefectA}, we will demonstrate three main examples. First we will realize twist defects as dislocations and disclinations in exactly solvable lattice spin or rotor models in section~\ref{sec:defectlatticemodel}. This includes the Kitaev's toric code~\cite{Kitaev06, Bombin, KitaevKong12}, the Wen's plaquette $\mathbb{Z}_k$-rotor model~\cite{YouWen, YouJianWen} and the tri-color code model~\cite{Bombin11,TeoRoyXiao13long}. Next in section~\ref{sec:genons}, we move on to dislocations or {\em genons} in bilayer systems~\cite{BarkeshliQi, BarkeshliQi13, LevinGu12, Levin13, BarkeshliJianQi13, BarkeshliJianQi13long} and an experimental proposal by Barkeshli {\em et.al.}~\cite{BarkeshliQi13,BarkeshliOregQi14} on gated trenches in bilayer fractional quantum Hall states. In section~\ref{sec:parafermion}, we will explain how parafermions or fractional Majorana bound states in superconductor - ferromagnet - fractional topological insulator heterostructures~\cite{FuKane08, LindnerBergRefaelStern, ClarkeAliceaKirill, MChen, Vaezi, mongg2} can be categorized as twist defects of some extended globally symmetric topological phases. After the three examples, we will highlight the main features and ingredients of a general defect theory in section~\ref{sec:twistdefectB}. This will include the defect fusion rules and basis transformations, known as defect $F$-symbols. Explicitly derivation will be demonstrated for the Ising-like defects in the Kitaev's toric code. We will elaborate the unconventional defect fusion structures for the $S_3$ symmetric $SO(8)_1$-state as well as the non-Abelian chiral ``4-Potts" state, where fusion rules can contain degeneracies and even be non-commutative. Lastly we will present the projective braiding operations of twist defects in section~\ref{sec:defectbraiding}. We will illustrate using parafermionic defects in a $\mathbb{Z}_n$ gauge theory. 

This review will be concluded in section~\ref{sec:conclusion}, where we will provide a more elaborate summary of contents covered in this review article as well as some prospects beyond twist defects in globally symmetric topological phases.

\section{Review on \texorpdfstring{$(2+1)$}{(2+1)}D discrete gauge theories and topological phases}\label{sec:discretegaugetheory}
Discrete gauge theories~\cite{BaisDrielPropitius92,Propitius-1995,PropitiusBais96,Preskilllecturenotes,Freedman-2004,Mochon04} in two dimensions are the prototypes of topological phases. The quantum system carries a gauge symmetry of a discrete gauge group $G$. Anyonic excitations of the topological state consist of charges and fluxes that exhibit non-trivial mutual braiding statistics. When $G$ is an Abelian group so that group elements commute, $g_1g_2=g_2g_1$, the topological state is also Abelian so that braiding operations commute and does not change the quantum state up to a unitary phase. The simplest example is a $\mathbb{Z}_2$ gauge theory and can be realized by the exactly solvable Kiteav toric code model~\cite{Kitaev97} and the Wen plaquette model~\cite{Wenplaquettemodel} on a lattice. This will be reviewed in section~\ref{sec:toriccode}. When the gauge group $G$ is non-Abelian, the topological state carries non-Abelian anyons. They are excitations that support non-commuting braiding operations and multichannel fusion rules. The fusion, exchange and braiding structure of anyons is summarized by a mathematical framework, called a modular tensor category~\cite{Kitaev06, Turaevbook, BakalovKirillovlecturenotes, Wangbook, Walkernotes91, FreedmanLarsenWang00, Bondersonthesis}. This generalizes the notion of topological phases outside of discrete gauge theories and will be reviewed in section~\ref{sec:MTC}.

\subsection{The Kitaev toric code: a \texorpdfstring{$\mathbb{Z}_2$}{Z2} gauge theory}\label{sec:toriccode}

\begin{figure}[htbp]
\centering
\includegraphics[width=0.5\textwidth]{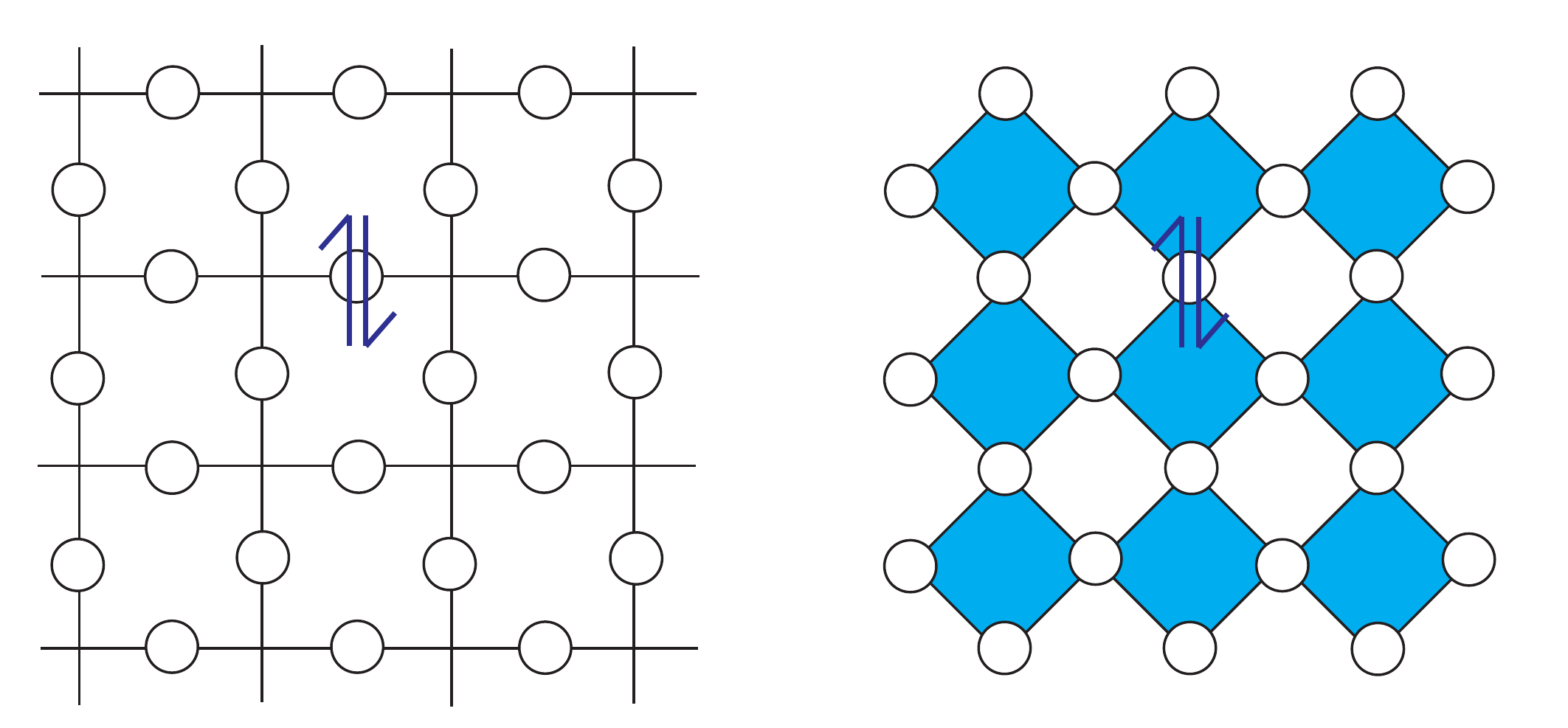}
\caption{Lattice spin model where a spin-$1/2$ degree of freedom is located at each link (left) or vertex (right).}\label{fig:toriccode0}
\end{figure}

The Kitaev's toric code~\cite{Kitaev97} is an exactly solvable lattice model that describes a $\mathbb{Z}_2$ gauge theory. It is also a prototype topological state with an anyonic symmetry. This toy model is intimately related to a two dimensional superconductor~\cite{HanssonOganesyanSondhi04,BondersonNayak13}, where its electric-magnetic symmetry is linked to fermion parity. The exactly solved spin model can be defined on any planar graph, and we will consider the simplest case when it lives on a square lattice. There is a spin-$1/2$ degree of freedom on each link (see the left diagram in figure~\ref{fig:toriccode0}). The Hamiltonian consists of mutually commuting local operators, known as stabilizers. There are two types of stabilizers, a vertex type $V$ and a plaquette type $P$, each constructed by a product of spin operators over the four adjacent links. \begin{align}\hat{V}=\prod_{i\in V}\sigma^z_i,\quad\hat{P}=\prod_{i\in P}\sigma^x_i\label{toriccodeVP}\end{align} where $\sigma_x=\left(\begin{smallmatrix}0&1\\1&0\end{smallmatrix}\right)$ and $\sigma_z=\left(\begin{smallmatrix}1&0\\0&-1\end{smallmatrix}\right)$ are spin operators acting on a spin $|\uparrow\rangle=(1,0)$, $|\downarrow\rangle=(0,1)$. The Hamiltonian is a sum of the stabilizers \begin{align}H=-\sum_V\hat{V}-\sum_P\hat{P}.\label{toriccodeH}.\end{align} This Hamiltonian is exactly solvable because the vertex and plaquette operators mutually commute and share simultaneous eigenstates. 

Equivalently, the $\mathbb{Z}_2$ topological state can also be realized by the Wen plaquette model~\cite{Wenplaquettemodel}. It is constructed on a checkerboard lattice where spins live on vertices instead of links (see the right diagram of figure~\ref{fig:toriccode0}). There are two types of plaquettes, one white and the other blue. They take the role of the previous vertex and plaquette respectively. In this case, by a basis transformation $U=e^{i\pi\sigma_y/4}$ that flips the spins $\sigma_x\to\sigma_z$, $\sigma_z\to-\sigma_x$ along the rows in the square lattice or the even rows in the checkerboard lattice, the stabilizers of the Wen plaquette model uniformly take the expression of \begin{align}\hat{P}_\Diamond=\sigma^x_l\sigma^x_r\sigma^z_t\sigma^z_b\label{WenP}\end{align} where $l,r,t,b$ are the left, right, top and bottom vertices of the white or blue plaquette $P_\Diamond$. The Hamiltonian \eqref{toriccodeH} now takes the form of \begin{align}H=-\sum_{P_\Diamond}\hat{P}_\Diamond\end{align} which is invariant under a half-translation $(1/2,1/2)$ that interchanges between $V$ and $P$ in the Kitaev toric code or between $\hat{P}_\Diamond$ of different color in the Wen plaquette model. We will see later that this corresponds to the electric-magnetic anyonic symmetry in a $\mathbb{Z}_2$ gauge theory.

Ground states of this model are simultaneous eigenstates of each $\hat{P}$ and $\hat{V}$ (or $\hat{P}_\Diamond$) with $\hat{P}=\hat{V}=+1$ (resp.~$\hat{P}_\Diamond=+1$) for all stabilizers so that the energy is minimized. Quasiparticles are excitations localized at vertex or squares with $\hat{V}=-1$ or $\hat{P}=-1$. The excitations can only be transported along the rows and columns from vertex to vertex or squares to squares. Or on the checkerboard, excitations live on a white or blue plaquette and can only hop to adjacent plaquettes of the same color. This is executed by acting on the quantum state with a spin operator at the connecting vertex. For example a vertex excitation $\hat{V}_1=-1$ can be transported to a neighbor vertex $V_2$ by acting the spin operator $\sigma^x_j$ on the quantum state, where $j$ is the link connecting $V_1$ and $V_2$. The new state now has eigenvalue $\hat{V}_1=+1$ since $\hat{V}_1$ and $\sigma^x_j$ anticommute. However, the value of the second vertex is flipped. As a result, the vertex excitation is effectively moved from $V_1$ to $V_2$. 

Repeating this motion will leave a string of spin operators along the quasiparticle trajectory. Since the lattice is bi-partite into vertices $V$ and squares $P$ (or plaquettes $P_\Diamond$ of opposite colors), it is easy to see there are two fundamental non-trivial quasiparticles, which we will label by the charge $e$ and flux $m$. They are bosons but obey mutual semionic statistics, i.e.~dragging one completely around the other will result in a $-1$ braiding phase. For example when moving a vertex excitation around a square, it leaves behind a $\sigma^x$ loop around the square. This exactly coincides with the plaquette operator $\hat{P}$ in \eqref{toriccodeVP}. If the square is occupied with a plaquette excitation, the $\sigma^x$ loop will register a $-1$ value on the quantum state. In general a vertex excitation can circle around a loop that encloses multiple squares, and the $\sigma^x$ {\em Wilson loop} along the quasiparticle trajectory is exactly the product of the square operators enclosed by the loop. This is because all interior spins cancel, $(\sigma^x)^2=1$, by the products adjacent square operators, leaving behind spins along the boundary. This $e$-Wilson loop operator then measures the number of enclosed $m$-plaquette excitations.

The flux $m$ and charge $e$ can also combine to form a composite quasiparticle $\psi=e\times m$ which, on the lattice, is equivalent to the excitation of adjacent vertex and square in the Kitaev's model (or plaquettes with opposite color in the Wen's model). The quasiparticle $\psi$ is a fermion due to the $-1$ twist phase upon a $360^\circ$ rotation of its internal structure. The anyon structure of quasiparticle excitations are summarized by the fusion rules \begin{align}e\times e=m\times m=\psi\times\psi=1,\quad e\times m=\psi,\quad e\times\psi=m,\quad m\times\psi=e\label{TCfusion}\end{align} where $1$ denotes the vacuum or ground state, as well as the exchange and braiding rules  \begin{gather}\vcenter{\hbox{\includegraphics[width=0.5in]{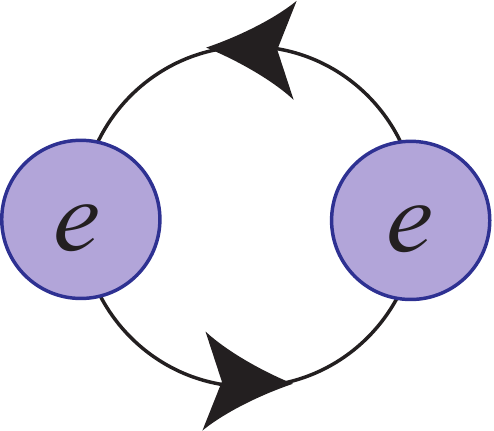}}}=\vcenter{\hbox{\includegraphics[width=0.5in]{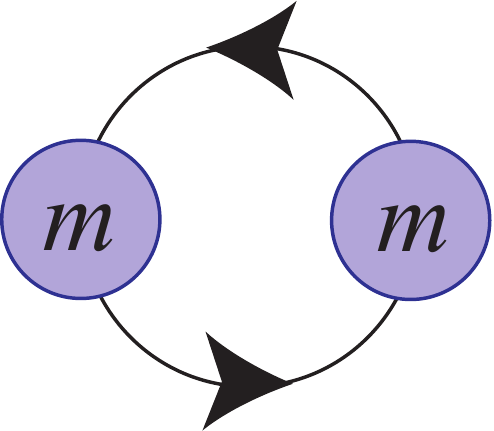}}}=1,\quad\vcenter{\hbox{\includegraphics[width=0.5in]{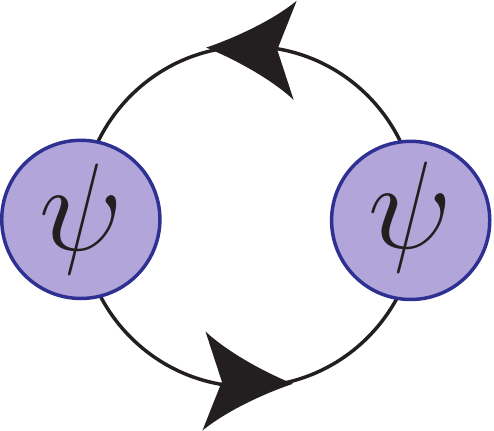}}}=-1,\quad\vcenter{\hbox{\includegraphics[width=0.7in]{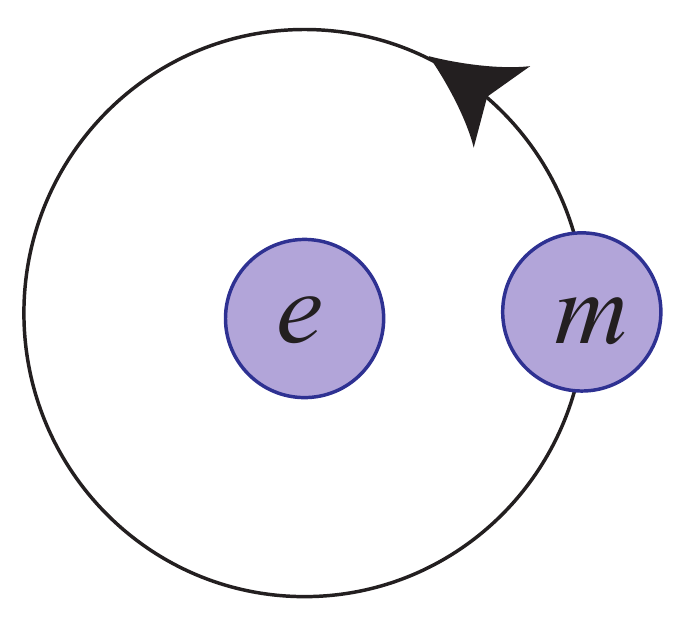}}}=\vcenter{\hbox{\includegraphics[width=0.7in]{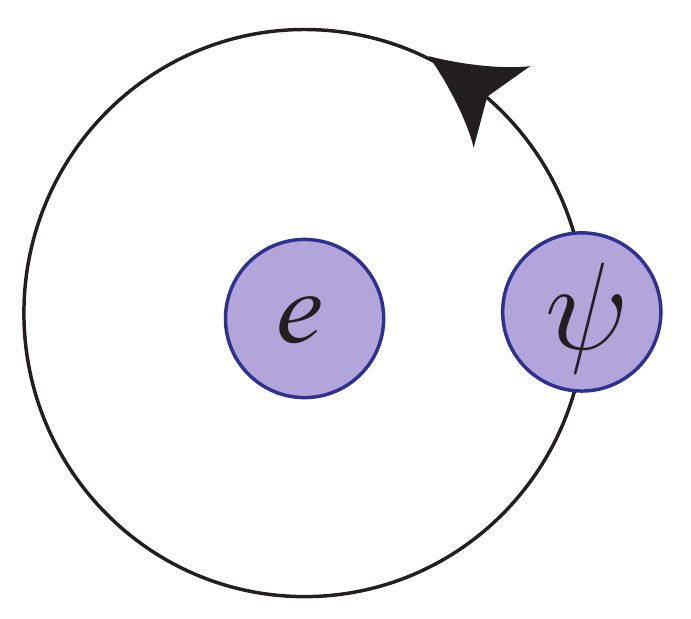}}}=\vcenter{\hbox{\includegraphics[width=0.7in]{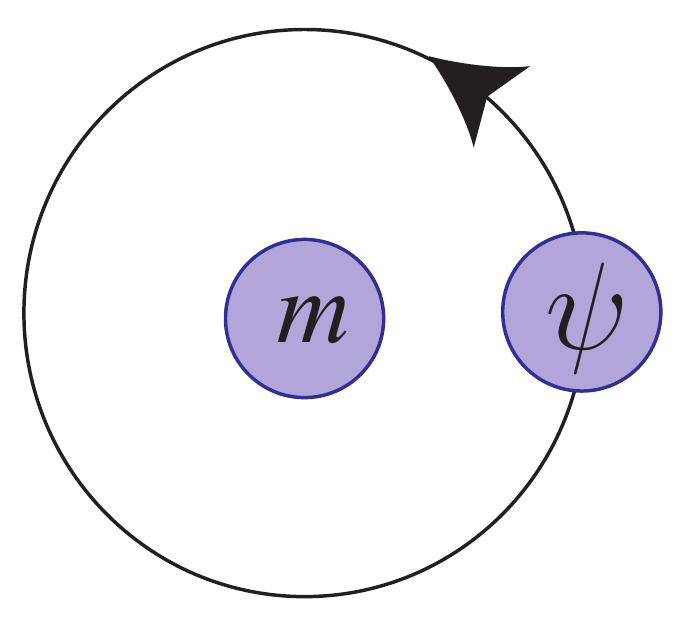}}}=-1.\end{gather} At this point we can already notice that there is an electric-magnetic symmetry in the anyon structure. The fusion, exchange and braiding information is invariant under the relabelling of $e\leftrightarrow m$.

To make a connection with more general Abelian topological states, we note that this topological phase can also be described by an two-component $U(1)$ Chern-Simons theory via the $K$-matrix formalism\cite{WenZee92,Wenedgereview} with the Lagrangian density \begin{align}\mathcal{L}=\frac{1}{4\pi} K_{IJ}\alpha_Id\alpha_J+\alpha_I\mathcal{J}^I\label{toriccodeCSaction}\end{align} where we have used a 2-component $U(1)$ gauge field $\alpha_I$ with the $2\times2$ $K$-matrix $K=2\sigma_x$. Here $\mathcal{J}^1$ and $\mathcal{J}^2$ correspond to the currents of the charge $e$ and flux $m$ quasiparticles respectively. The action is invariant under the electric-magnetic symmetry, which swaps the two gauge fields, $\alpha_1\leftrightarrow\alpha_2,$ and is represented by the matrix $M=\sigma_x$. The $K$ matrix is unchanged under conjugation by this operation, $MKM^T=K.$

\subsection{Flux and charge excitations}\label{sec:fluxcharge} 
The toric code model is a realization of a $\mathbb{Z}_2$ gauge theory. The plaquette operators $\hat{P}$ measure the $\mathbb{Z}_2$ flux $e^{i\iint b}=e^{i\oint{\bf a}\cdot d{\bf l}}$ across the square and can take $\pm1$ value. The spin operator $\sigma^x_j$ along the link $j$ represents the parallel transport $e^{i\int{\bf a}\cdot d{\bf l}}$ between neighboring vertices. A vertex operator is a $\mathbb{Z}_2$ gauge transformation that flips $\sigma^x_j\to-\sigma^x_j$ on adjacent links but leaves all $\mathbb{Z}_2$ fluxes unchanged. A vertex excitation is therefore naturally identified with the $\mathbb{Z}_2$ charge as a gauge transformation alter its quantum phase. Lastly a flux-charge composite, also known as a dyon, gives an emergent fermionic excitation.

This notion can be generalized to an arbitrary discrete gauge group $G$~\cite{BaisDrielPropitius92,Propitius-1995,PropitiusBais96,Preskilllecturenotes,Freedman-2004,Mochon04}. The resulting topological state is known as a quantum double and is denoted by $D(G)$. 
In a two dimensional gauge theory with a finite, discrete gauge group $G$, the anyon excitations are labeled by the 2-tuple $\chi=\left([M],\rho\right)$. The flux component is characterized by a {\em conjugacy class} \begin{align}[M]=\left\{M'\in G:M'=NMN^{-1}\mbox{ for some }N\in G\right\}\label{conjugacyclass}\end{align} of the gauge group. For $\mathbb{Z}_2=\{1,-1\}$, the vacuum is the trivial flux $1=[1]$ while a plaquette excitation is the non-trivial flux $m=[-1]$. Given a particular conjugacy class for the flux, the possible charge components are characterized by an irreducible representation $\rho:Z_M\to U(\mathcal{N}_\rho)$ of the {\em centralizer} of $M$ (or any representative of $[M]$) defined by \begin{align}Z_M=\left\{N\in G:NM=MN\right\}.\label{centralizer}\end{align} For the $\mathbb{Z}_2$ gauge group, the centers are always the group itself as it is Abelian. The are only two irreducible representations, a trivial one $\rho_+(\pm1)=1$ and a non-trivial one $\rho_-(\pm1)=\pm1$. The vertex excitation thus corresponds to a pure charge, $e=\rho_-$. 

\begin{figure}[htbp]
\centering\includegraphics[width=0.4\textwidth]{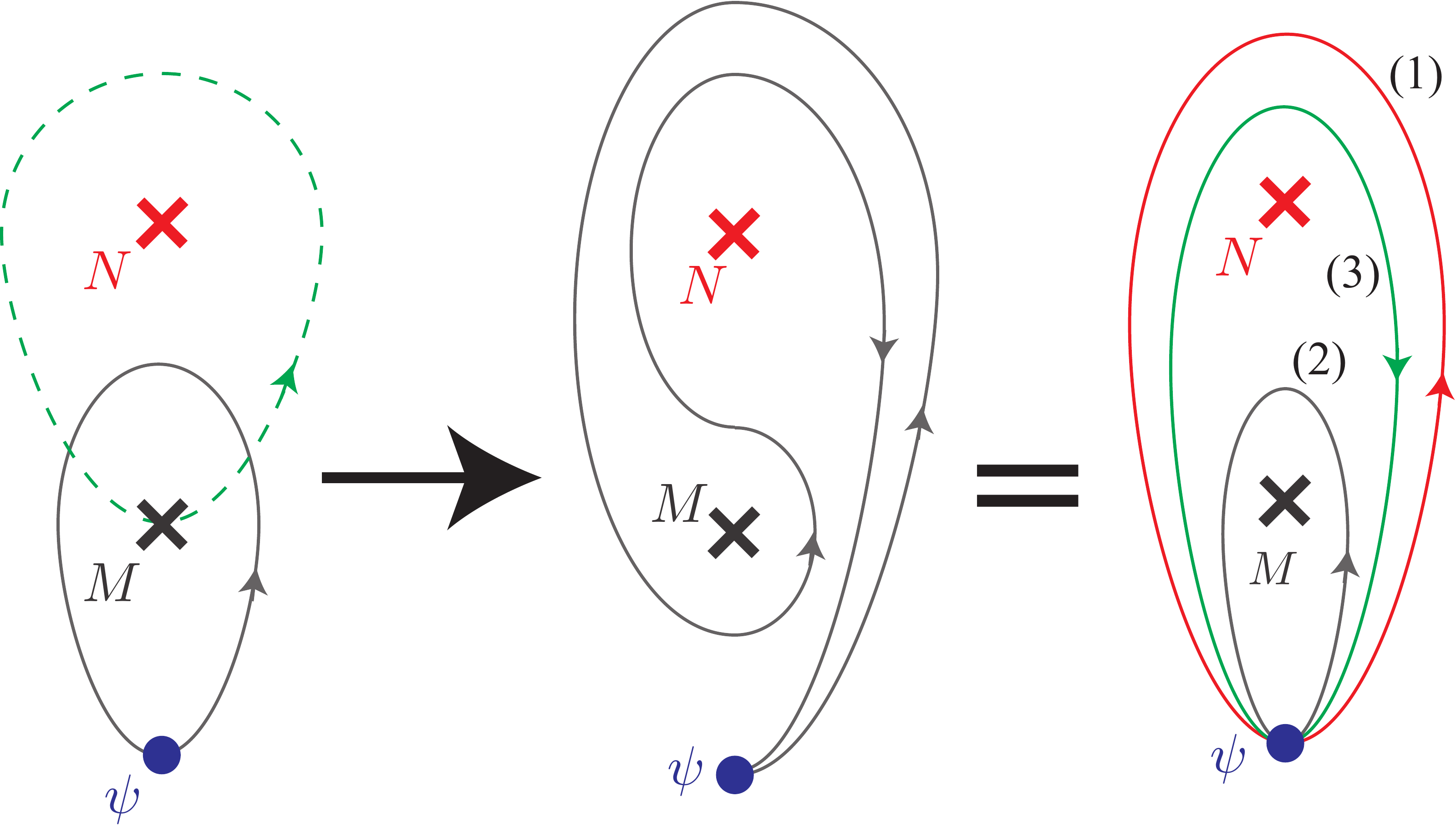}
\caption{Conjugation of holonomy $M\to N^{-1}MN$.}\label{fig:fluxconjugation}
\end{figure}
To understand the general rules for anyon construction, we begin with pure fluxes (i.e.~the charge components are represented by the trivial representation) and pure charges (the flux components are represented by the trivial conjugacy class). A charge $\psi^a$ will acquire a non-Abelian holonomy $\psi^a\to\rho(M)_b^a\psi^b$ when encircling a flux $M$ according to the representation $\rho$ which characterizes the charge. The holonomy, however, would change to $\rho(N^{-1}MN)$ if the flux $M$ is also braided around another flux $N$ (see Fig.~\ref{fig:fluxconjugation}). This means the holonomy measurement is defined only up to conjugacy. In fact, a projective measurement can only read off the character of the representation $\mbox{Tr}[\rho(M)]$, and therefore fluxes are naturally characterized by conjugacy classes rather than the individual elements of the class. Additionally, a fundamental charge should be described by an irreducible representation because if $\rho=\rho_1\oplus\rho_2$ is decomposable, then the charge can be split-off into simpler components $\psi=\psi_1+\psi_2$. Equivalent representations describe identical charges because they are merely related by basis transformation $\psi^a\to U_b^a\psi^b$ and are not topologically distinguishable.

Now that we understand why inequivalent fluxes are described by conjugacy classes, let us motivate why charges only take the representations of the centralizer group of a given class. When a charge $\psi^a$ is combined with a flux $M$ to form a dyon (flux-charge composite), the holonomy it acquires $\psi^a\to\rho(N)_b^a\psi^b$ when encircling another flux $N$ is only well-defined when $M$ and $N$ commute. If this were not the case, then the flux would change by conjugation $M\to M'=NMN^{-1}.$ Although this operation respects the conjugacy class it can permute the individual flux elements within the class. If this occurs then the evolution would not be cyclic, since the initial and final dyon Hilbert spaces could be different, and the holonomy could be gauged away by distinct basis redefinitions for $\psi^a_M\to U^a_b\psi^a_M$ and $\psi^a_{M'}\to U'^a_b\psi^a_{M'}$. As a result, the charge component of a dyon is characterized by the representation of the centralizer \eqref{centralizer} rather than the whole group. For pure charges the flux conjugacy class is that of the identity element, and thus, in that case, the centralizer is the entire group. This is expected since pure charges are not dyonic and therefore do not run into the same consistency issue. 

Using the theory of finite groups we can develop a systematic procedure to count the number of inequivalent anyon excitations. The quantum dimension of the dyon $\chi=\left([M],\rho\right)$ is the product \begin{align}d_\chi=\left|[M]\right|\dim(\rho)\end{align} where $|[M]|$ is the number group elements in the conjugacy class $[M]$ and $\dim(\rho)=\mathcal{N}_\rho$ is the dimension of the representation. $d_\chi$ counts the dimension of the Hilbert space associate to the dyon $\chi$ that is spanned by $\psi^a_{M_\mu}$ for $a=1,\ldots,\mathcal{N}_\rho$ and $M_\mu$ are group element in the same conjugacy class $[M]$. By choosing a set of conjugating representatives $N_\mu\in G$ ($\mu=1,\ldots,n=|[M]|$) such that $[M]=\{M_1=N_1MN_1^{-1},\ldots,M_n=N_nMN_n^{-1}\}$, it is straightforward to see that \begin{align}|G|=\left|[M]\right||Z_M|\label{cosetdecomposition1}\end{align} for any choice of $M\in G,$ since each group element can be uniquely be expressed as $N_\mu z$ for $z\in Z_M$ and $1\leq\mu\leq |[M]|$. A useful theorem in the representation theory of finite groups~\cite{grouptheorybook} relates the order of any finite group $H$ to the dimensions of its irreducible representations by \begin{align}|H|=\sum_{\mbox{\tiny irred. rep.}}\dim(\rho)^2.\label{repsumrule}\end{align} Eq.~\eqref{cosetdecomposition1} and \eqref{repsumrule} lead to the identification of the total quantum dimension \begin{align}\mathcal{D}=\sqrt{\sum_\chi d^s_\chi}\end{align} that characterizes the topological entanglement entropy~\cite{KitaevPreskill06,BrownBartlettDohertyBarrett13} $S_T=-\log(\mathcal{D})$ of the discrete gauge theory and the order of the gauge group $|G|$: \begin{align}\mathcal{D}^2&=\sum_{[M]}\left[|[M]|^2\mathop{\sum_{\mbox{\tiny irred. rep.}}}_{\mbox{\tiny of $Z_M$}}\dim(\rho)^2\right]=\sum_{[M]}|[M]|^2|Z_M|=\sum_{[M]}|[M]||G|=|G|^2.\label{DGTdimension}\end{align}

To illustrate this general structure we can use the simple example of $G=\mathbb{Z}_2=\{1,-1\}.$ This group is Abelian and has two conjugacy classes $[1], [-1],$ and the full group has two representations $\rho_{+}, \rho_{-},$ both of which are one-dimensional. Since the group is Abelian there is a nice simplification because the centralizer of every conjugacy class is just the entire group. Thus we can see that we should have four anyon excitations $\chi_1=([1],\rho_{+}), \chi_{e}=([1],\rho_{-}), \chi_{m}=([-1],\rho_{+}),$ and $\chi_{\psi=em}([-1],\rho_{-}).$ Each of the anyons has quantum dimension $d_\chi=1$ and the total quantum dimension is $\mathcal{D}=\sqrt{1^2+1^2+1^2+1^2}=2=\vert G\vert.$

The $360^\circ$ twist phase of a single dyon is equivalent to the $180^\circ$ exchange phase of a pair of identical dyons. The spin-exchange statistics of a dyon $\chi=([M],\rho)$ is therefore determined by the monodromy of its internal charge and flux components. For instance the fermion statistics of $\psi=e\times m$ in the toric code is a result of the $-1$ monodromy phase between the pure charge $e$ and flux $m$. In general the statistical phase of a dyon is given by \begin{align}\theta_\chi=e^{2\pi ih_\chi}=\rho(M)\end{align} which is always a $U(1)$ scalar because of the Schur's lemma as $M$ commutes with all elements in the centralizer $Z_M$ irreducibly represented by $\rho:Z_M\to U(\mathcal{N}_\rho)$.

The $360^\circ$-braiding of two dyons, $\chi=([M],\rho)$ and $\chi'=([M',\rho'])$, is controlled by the action of the first flux $[M]$ on the second charge $\rho'$ as well as the second flux $[M']$ on the first charge $\rho$. However as the flux conjugacy class of the first (or second) dyon may not lie insider the centralizer represeted by the charge of the second (resp.~first), we need to extend the representations $\rho:Z_M\to U(\mathcal{N}_\rho)$ to cover the entire gauge group $G$. Given a general dyon $\chi=([M],\rho)$ with $[M]=\{M_1=N_1MN_1^{-1},\ldots,M_n=N_nMN_n^{-1}\}$ and $\rho:Z_M\to U(\mathcal{N}_\rho)$, its associating Hilbert space \begin{align}\mathcal{H}_\chi=\mathrm{span}\left\{\psi^a_\mu:a=1,\ldots,\mathcal{N}_\rho,\;\mu=1,\ldots,n\right\}\end{align} transforms when braided around a general flux $L\in G$ according to an {\em induced} representation $\tilde\rho:G\to U(n\mathcal{N})$. \begin{align}\tilde\rho(L):\psi^a_\mu\longrightarrow\rho\left(N_\nu^{-1}LN_\mu\right)_b^a\psi^b_\nu\label{inducedrep}\end{align} where the indices change $\mu\to\nu$ according to $LM_\mu L^{-1}=M_\nu$. Notice that the operator $L_0=N_\nu^{-1}LN_\mu$ in \eqref{inducedrep} commutes with the group element $M$ and therefore lives in the centralizer $Z_M$ so that $\rho(L_0)$ is well-defined. In other words, the conjugation $L\to N_\nu^{-1}LN_\mu$ is required to compare the basis between states $\psi^a_\mu$ and $\psi^a_\nu$ in different representatives of the same flux class. However, since the induced representation depends on the arbitrary choice of the representatives $N_\mu,N_\nu$, it is gauge {\em dependent} when $\mu\neq\nu$. For instance the trace $\mathrm{Tr}[\tilde\rho(L)]$ only picks up the diagonal pieces when $LM_\mu L^{-1}=M_\mu$.

When two dyons $\chi=(M,\rho)$ and $\chi'=(M',\rho')$ braid around one another, the total associated Hilbert space is the tensor product $\mathcal{H}_\chi\otimes\mathcal{H}_{\chi'}$, and a projective measurement gives the character $\mathrm{Tr}\left[\tilde\rho(M')\otimes\tilde\rho'(M)\right]$. This defines the modular $S$-matrix \begin{align}S_{D(G)}^{\chi\chi'}=\frac{1}{|G|}\mathrm{Tr}\left[\tilde\rho(M')\otimes\tilde\rho'(M)\right]^\ast\label{SDG}\end{align} of the discrete $G$-gauge theory, where the normalization by $\mathcal{D}=|G|$ is to ensures the matrix $S=(S^{\chi\chi'})$ is unitary. The modular structure of fusion, exchange and braiding will be reviewed in the following subsection \ref{sec:MTC} in a more general context that applies to any topological phases.

The topological order of a discrete gauge theory can be destroyed by condensing gauge charges~\cite{BaisDrielPropitius92,Propitius-1995,BaisSlingerlandCondensation,Kong14}. Firstly the pure charges are mutually local bosons in the sense that they all have individual bosonic exchange statistics and trivial monodromy around one another. They can therefore Bose condense simultaneously. However in doing so, the condensate confines every remaining dyon that carries a non-trivial flux component and is non-local with respect to some pure charges. This is because the non-trivial holonomy phase of a charge around a flux requires a branch cut in the condensate wavefunction and cost an energy that grows with the length of the cut. For example, the vertex excitations $e$ that corresponds to the $\mathbb{Z}_2$ charge can be condensed by adding a strong enough string tension $h\sum_j\sigma^x_j$ over all links. The plaquette excitation $m$ that corresponds to the $\mathbb{Z}_2$ flux is now confined because the $\sigma^z$-Wilson string that separates a pair of $m$-excitations now costs an energy $hL$, for $L$ the length of the string. Therefore by condensing all pure gauge charges, the quantum system goes through a gap closing transition and drives the topological phase into a trivial phase. As fluxes are now confined defects, the $G$ symmetry becomes global. On the other hand, given a quantum phase with a global symmetry, one can construct defects associating to the group elements. The quantization of these defects then promotes them into dynamical anyonic excitations of a topological phase where the symmetry becomes local. Hence, one expect the condensation process can be reversed by a {\em gauging} process. \begin{align}\begin{diagram}
\stackrel{\mbox{trivial Boson condensate}}{\mbox{with global symmetry}}&\pile{\rTo^{\mbox{\small gauging}}\\\lTo_{\mbox{\small condensation}}}&\stackrel{\mbox{discrete gauge theory}}{\mbox{with local symmetry}}\end{diagram}.\label{D(G)gauging}\end{align} We will discuss this in a more general context with examples in later sections.

\subsection{Summary on \texorpdfstring{$(2+1)$}{(2+1)}D topological field theories}\label{sec:MTC}
Not all topological phases in two dimensions are discrete gauge theories. For instance, a gauge theory can only support anyons with integral quantum dimensions, but the Moore-Read fractional quantum Hall state and the Kitaev honeycomb spin model support Ising anyon excitation~\cite{MooreRead,Kitaev06} which has an irrational dimension $d_\sigma=\sqrt{2}$. Moreover a topological state can be chiral and carry low energy boundary modes that propagate only in a single direction. Any such state cannot be realized as a discrete gauge theory alone. Here we review the notion of a modular tensor category~\cite{Kitaev06, Turaevbook, BakalovKirillovlecturenotes, Wangbook, Walkernotes91, FreedmanLarsenWang00, Bondersonthesis} that describes the anyonic excitations of a general topological state.

First we begin in the fusion structure, called a fusion category, of anyons that encodes the fusion and splitting rules as well as a consistent set of basis transformations of quantum states. This structure does not include the exchange and braiding information but can be applied to describe non-local defect objects that are {\em not} anyonic excitations. A fusion category $\mathcal{F}=\langle x,y,z,\ldots\rangle$ consists of objects that are finite combinations of simple ones $x+y+\ldots$. In a topological state, a simple object is an anyon type. In a twist defect theory to be discuss later, a simple object is a defect-quasiparticle composite. In a discrete gauge theory where there is a Hilbert space associated to each dyon, the combination of dyons can be regarded as a direct sum of spaces $x\oplus y\oplus\ldots$. In a general topological state where the dimension of an anyon can be irrational, the combination is taken in an abstract sense but should be clearer when we discuss quantum states. A simple object is an object that cannot be decomposed into simpler ones.

Fusion and splitting of simple objects are described by the equation \begin{align}x\times y=\sum_zN_{xy}^zz\label{appfusionrule}\end{align} where the fusion matrix $N_x=(N_{xy}^z)$ has non-negative integer entries. $N_{xy}^z$ counts the multiplicity of distinguishable ways the ordered pair $(x,y)$ can be identified together (i.e.~fused) as the object $z$.  Equivalently, $N_{xy}^z$ also counts the splitting degeneracy -- the number of ways for $z$ to split into $x$ and $y$. We always assume there is an identity fusion element 1, the vacuum or ground state, so that $1\times x=x\times 1=x$ for any $x$. Moreover, given any simple object $x$, there must be a unique antipartner $\bar{x}$ so that ${x}\times\bar{x}=\bar{x}\times{x}=1+\ldots$, i.e.~$N_{x\bar{x}}^1=N_{\bar{x}x}^1=1$ and $N_{xy}^1=0$ whenever $y\neq\bar{x}$. Fusion rules are commutative $x\times y=y\times x$, i.e.~$N_{xy}^z=N_{yx}^z$, if the objects are anyons in a topological phase. This is because the ordering can be exchange by a $180^\circ$ exchange operation $R^{xy}$ to be discussed later. However fusions and splittings can be non-commutative for defects.

For example, we have seen the fusion rules for the Kitaev toric code in eq.\eqref{TCfusion} in section~\ref{sec:toriccode}. The emergent fermion is a composite of flux and charge, $\psi=e\times m$. All quasiparticle is self-conjugate in the sense that they are their own antipartner, $e\times e=m\times m=\psi\times\psi=1$. The fusion rules are said to be Abelian not because they are commutative but because they are single-channeled so that given any $x$ and $y$, there is a unique $z$ so that $x\times y=z$, i.e.~$N_{xy}^z=1$ and $N_{xy}^w=0$ if $w\neq z$. Non-Abelian topological states have multichannel fusion rules. For example Ising anyons satisfies $\sigma\times\sigma=1+\psi$~\cite{MooreRead} because there is Majorana zero mode at each Ising anyon $\sigma$, and a pair of them can have an even or odd fermion parity. Fibonacci anyons obey $\tau\times\tau=1+\tau$~\cite{ReadRezayi}. These are all non-Abelian anyons.

\begin{figure}[htbp]
\centering\includegraphics[width=0.45\textwidth]{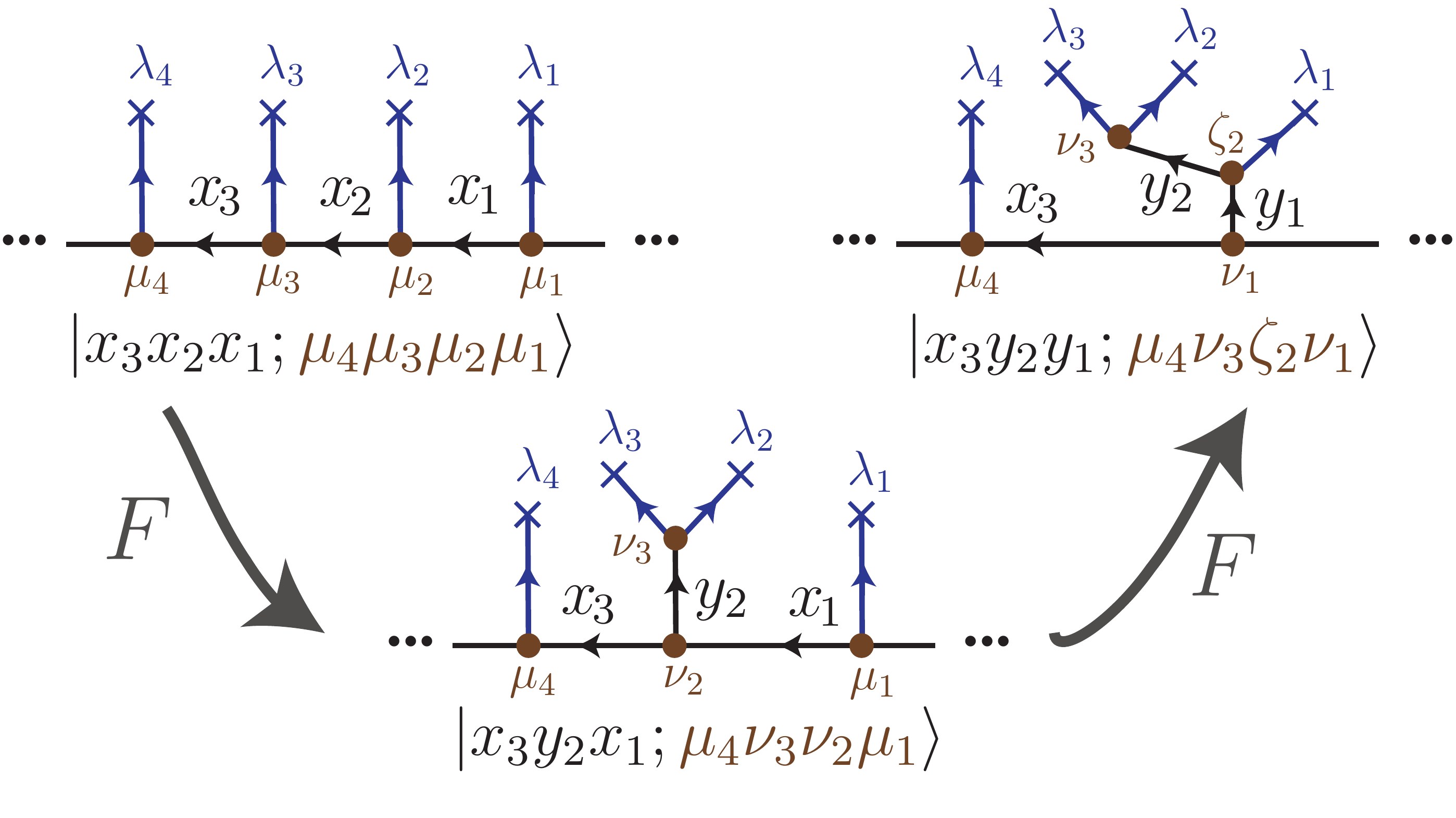}
\caption{Basis transformation of defect states by composition of fundamental $F$-moves. $\lambda_i$ are defect objects, $x_i,y_i$ are admissible internal fusion channels, and $\mu_i,\nu_i,\zeta_i$ label vertex degeneracies.}\label{fig:Fmoves}
\end{figure}

\begin{figure}[htbp]
\centering\includegraphics[width=0.45\textwidth]{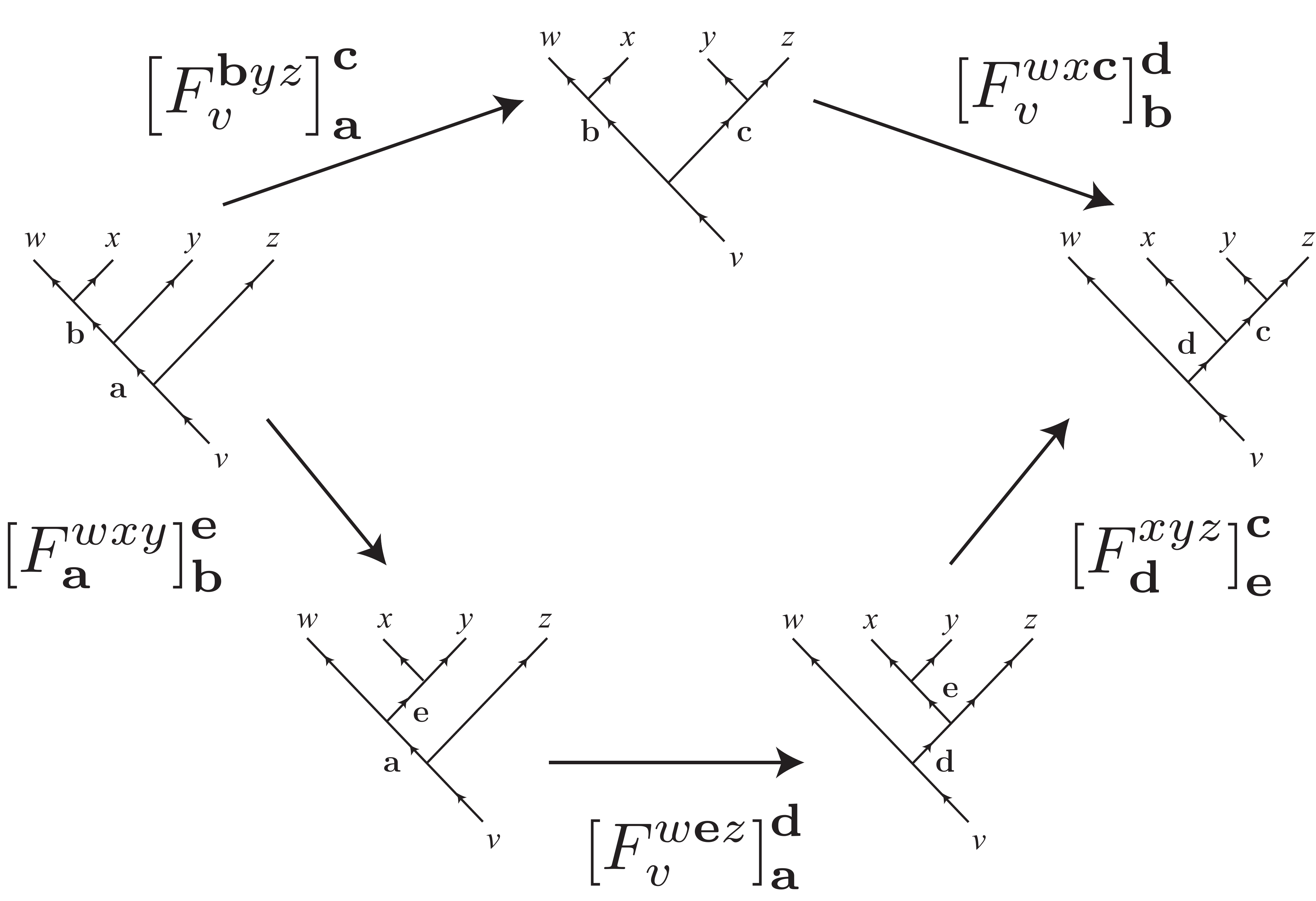}
\caption{The pentagon equation~\cite{Kitaev06}. Summation is taken over the internal anyon label ${\bf e}$. Vertex labels are suppressed.}\label{fig:Fpentagon}
\end{figure}

Eq.~\eqref{appfusionrule} is associative so that $x\times(y\times z)=(x\times y)\times z$, i.e.~$\sum_wN_{xw}^qN_{yz}^w=\sum_wN_{xy}^wN_{wz}^q$. The equality signs in fusion rules however only signify equivalences in the quantum level. Let us fix a finite number of simple objects $\{\lambda_1,\lambda_2,\ldots\}$ in a topological phase on a closed sphere, a quantum state can be specified by a {\em splitting tree} (or {\em fusion tree}) with known internal branches and vertices (see Fig.~\ref{fig:Fmoves} and Eq.~\eqref{splittingtreeapp} below). A splitting tree is a directed tree diagram, whose internal branches are labeled by anyons $x_v,$ and vertices -- which must be trivalent -- are labeled by {\em splitting states} $\mu_v$, to be explained below. The external branches are fixed by the known excitations $\lambda_i$. Each vertex describes the splitting of an anyon. It has one incoming branch $z,$ and two outgoing branches $x,y$ such that $N_{xy}^z>0$, i.e.~the fusion $x\times y\to z$ is admissible. The splitting degeneracy corresponds to a $N_{xy}^z$-dimensional space, whose orthonormal basis is labeled by $\mu_v=1,\ldots N_{xy}^z$. The set of admissible $|\{x_l\};\{\mu_v\}\rangle$ forms an orthonormal basis of the degenerate Hilbert space of quantum states with the fixed anyonic excitations $\{\lambda_1,\lambda_2,\ldots\}$.

Essentially, a splitting tree corresponds to a particular maximal set of mutually commuting statistical observables, i.e.~Wilson loops; and the internal branch and vertex labels correspond to simultaneous eigenvalues. For example, the fusion channel $x\times y\to z$ at a vertex determines the eigenvalue of the Wilson loop $\hat{\mathcal{W}}_w$ encirling $x$ and $y$ by the braiding between $w$ and $z$. Alternatively, a splitting tree describes a particular sequence of splittings that ultimately results in the creation of the excitations $\lambda_i$ on the external branches. Starting from the ground state $|GS\rangle$ with no excitations, anyons can be created by open Wilson string operators. For example, in the toric code (see Section~\ref{sec:toriccode}) the vacuum $1$ can be split into a pair of plaquette excitations, say $m\times m$, by a string of $\sigma_{x,z}$ operators $\hat{L}^{mm}_1$ connecting them. One of the plaquette excitations can subsequently be divided into $m\to e\times\psi$ by a string operator $\hat{L}^{e\psi}_m$ that branches from $m$ to $e$ and $\psi$. The quantum state with excitations $e,\psi,m$ can then be created by $\hat{L}^{e\psi}_m\hat{L}^{mm}_1|GS\rangle$. Notice that there are multiple ways an $m$-string can branch into $e$ and $\psi$. For instance, the $e$ and $\psi$ strings can braid and give a minus sign. The particular splitting operator $\hat{L}^{e\psi}_m$ specifies the phase of the quantum state. On the other hand, the string operators can be deformed by plaquette operators and still act identically on the ground state. These form equivalence classes of splitting string operators, such as $[\hat{L}^{e\psi}_m]$, with fixed open ends.

In more exotic cases, splitting can be degenerate and different branching operators can give orthogonal states. For example the $SU(3)_3$-state supports an anyon associated to the eight dimensional adjoint representation of $SU(3)$ which obeys the degenerate fusion rule ${\bf 8}\times{\bf 8}=1+{\bf 10}+\overline{\bf 10}+{\bf 8}+{\bf 8},$ such that $N_{\bf 8,8}^{\bf 8}=2$. There are two linearly independent local operators that split ${\bf 8}$ into ${\bf 8}\times{\bf 8}$. In general, the splitting $z\to x\times y$ at a vertex with degeneracy $N_{xy}^z$ is associated to a {\em splitting space} $V^{xy}_z$. It contains equivalence classes of local operators $[(\hat{L}^{xy}_z)_\mu]$ -- referred to as {\em splitting states} -- that connect an incoming $z$ to the outgoing $x$ and $y$. The collection of linearly independent splitting operators $\{(\hat{L}^{xy}_z)_\mu:\mu=1,\ldots,N_{xy}^z\}$ spans the splitting space $V^{xy}_z$, which forms an irreducible representation of the algebra of Wilson operators around $x$ and $y$: \begin{align}\hat{\mathcal{W}}(\hat{L}^{xy}_z)_\mu\hat{\mathcal{W}}^{-1}=\sum_{\nu}\mathcal{W}^\nu_\mu(\hat{L}^{xy}_z)_\nu.\end{align} A quantum state with known excitations is in general constructed by piecing the splitting string operators together with matching boundary conditions: \begin{align}\left|\vcenter{\hbox{\includegraphics[width=0.1\textwidth]{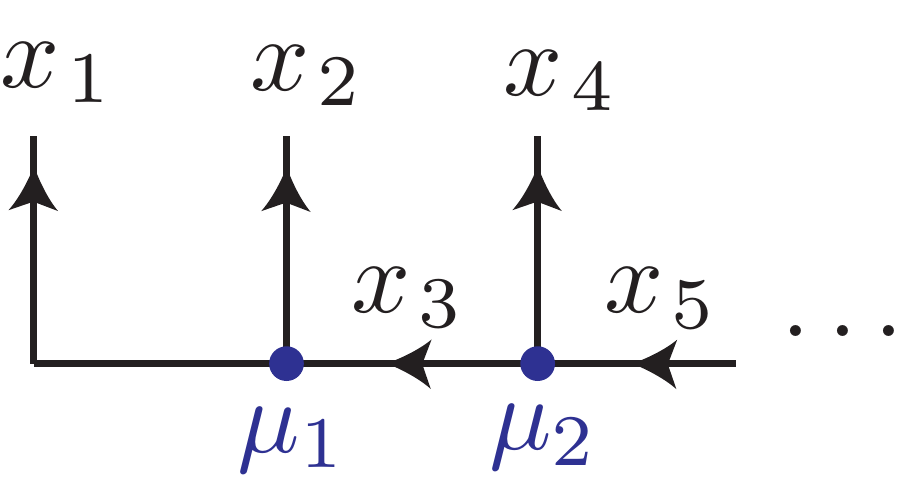}}}\right\rangle\propto\mathop{\sum_{boundary}}_{conditions}\left[(L^{x_1x_2}_{x_3})_{\mu_1}\right]\otimes\left[(L^{x_3x_4}_{x_5})_{\mu_2}\right]\otimes\ldots|GS\rangle.\label{splittingtreeapp}\end{align}

In an Abelian theory where fusion is single-channel, the quantum state \eqref{splittingtreeapp} is completely determined (up to a $U(1)$-phase) by its external branch labels, as all internal branch labels are fixed by the fusion rules. This means that the quantum state is completely determined by the anyon types of the excitations (objects on the external branches). This is not true for non-Abelian theories where fusion rules can be multi-channeled. These non-Abelian anyons give rise to degeneracies of quantum states since the internal branches and vertices of \eqref{splittingtreeapp} can carry different labels.

The quantum dimension $d_x$ of an anyon $x$ is defined to be the positive number that counts the quantum state degeneracy, which is proportional to $d_x^N$, of a closed system with $N$ type $x$ quasiparticle excitations in the thermodynamic limit when $N\to\infty$. Using the Perron-Frobenius theorem~\cite{Kitaev06}, the quantum dimension $d_x$ is given by the largest (absolute) eigenvalue of the fusion matrix $N_x=\left(N_{xy}^z\right)$. For example, an anyon is Abelian if and only if it has unit quantum dimension. The quantum dimension of an Ising anyon is $d_\sigma=\sqrt{2},$ and that of a Fibonacci anyon is $d_\tau=(1+\sqrt{5})/2$. In general, the fusion rules \eqref{appfusionrule} require \begin{align}d_xd_y=\sum_{z}N_{xy}^zd_z.\end{align} The total quantum dimension of the anyon theory is defined by $\mathcal{D}=\sqrt{\sum_xd_x^2}$.

Fusion associativity ${\bf a}\times({\bf b}\times{\bf c})=({\bf a}\times{\bf b})\times{\bf c}$ in the quantum state level is realized as basis transformations between different splitting trees (see Fig.~\ref{fig:Fmoves}). Primitive basis transformations are known as $F$-symbols. They relate  \begin{align}\left|\vcenter{\hbox{\includegraphics[width=0.5in]{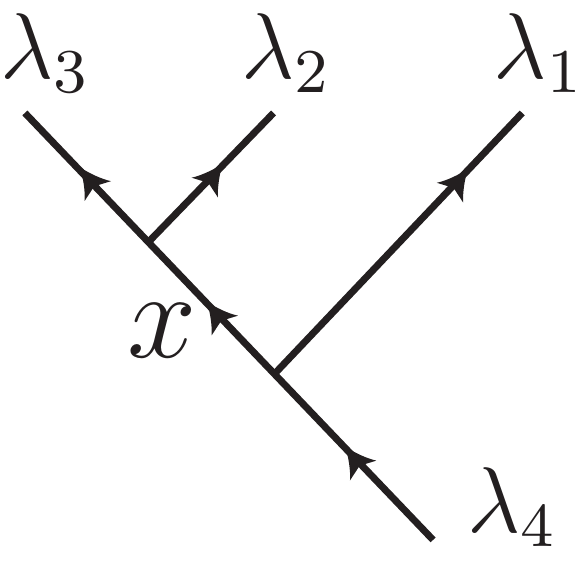}}}\right\rangle=\sum_{y}\left[F^{\lambda_3\lambda_2\lambda_1}_{\lambda_4}\right]_{x}^{y}\left|\vcenter{\hbox{\includegraphics[width=0.5in]{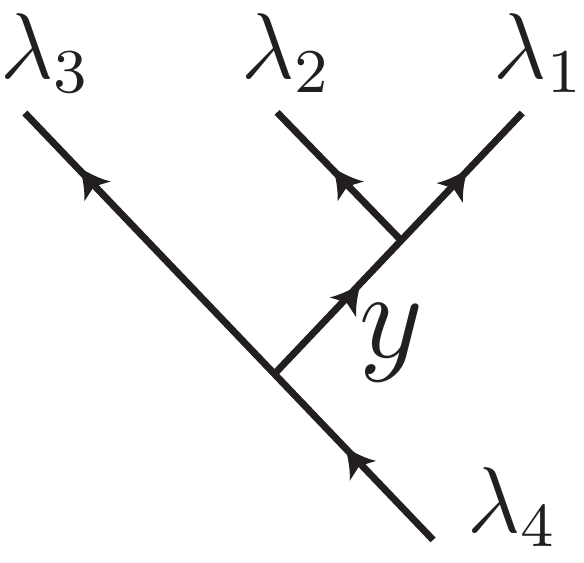}}}\right\rangle.\label{Fsymboldef}\end{align} where $F$-matrix entries are given by the inner product \begin{align}\left[F^{\lambda_3\lambda_2\lambda_1}_{\lambda_4}\right]_{x}^{y}=\left\langle\vcenter{\hbox{\includegraphics[width=0.5in]{F2}}}\right.\left|\vcenter{\hbox{\includegraphics[width=0.5in]{F1}}}\right\rangle .\label{Fdefinition}\end{align} Here the symbols $\mu_j$ for vertex degeneracies are suppressed. There are consistency relations, referred to as the pentagon equations, that the $F$-symbols have to obey (see Fig.~\ref{fig:Fpentagon})~\cite{Kitaev06}. They are required to ensure that any sequence of $F$-moves that connects two fixed initial and final splitting trees will give the identical overall basis transformation~\cite{MacLanebook}.

There are particle-antiparticle duality relationships in a fusion theory. For instance $x$ and $\bar{x}$ must have identical quantum dimensions. At the quantum state level, reversing the worldline of an anyon $x$ and replacing it by its conjugate $\bar{x}$ may result in an overall phase. This is determined by the bending diagram \begin{align}\vcenter{\hbox{\includegraphics[height=0.8in]{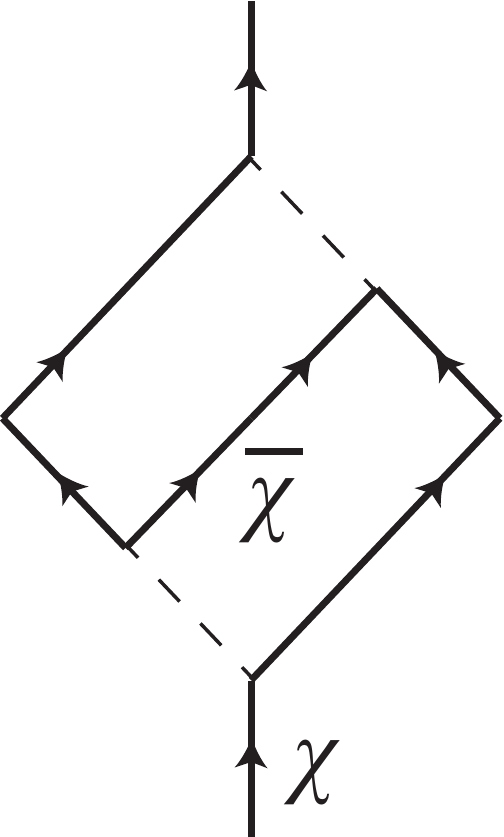}}}=d_\chi \left[F^{\chi\overline\chi\chi}_\chi\right]_1^1\;\vcenter{\hbox{\includegraphics[height=0.8in]{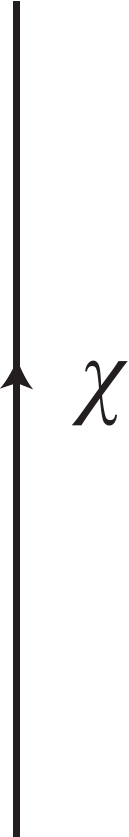}}}=\varkappa_\chi\;\vcenter{\hbox{\includegraphics[height=0.8in]{bending2}}}\label{bendingdef}\end{align}  where $d_\chi=\left|[F^{\sigma\sigma\sigma}_\sigma]_1^1\right|^{-1}$, or precisely the Frobenius-Schur (FS) indicator~\cite{FredenhagenRehrenSchroer92,Kitaev06} \begin{align}\varkappa_x=\frac{[F^{x\overline{x}x}_x]_1^1}{|[F^{x\overline{x}x}_x]_1^1|}\in U(1).\label{FSindicator}\end{align}

As a simple example, we consider four Ising anyons on a closed sphere, each is associated to a Majorana zero mode $\gamma_j$, for $j=1,2,3,4$. The twofold ground state degeneracy can be labeled by the fermion parity of a pair of Ising anyons, say $(-1)^{F_{12}}=i\gamma_1\gamma_2$. The total parity in a closed system is fixed, and is taken to be even $(-1)^{F_{12}+F_{34}}=(-1)^{F_{14}+F_{23}}=-\gamma_1\gamma_2\gamma_3\gamma_4=+1$. One can also pick another fermion parity, say $(-1)^{F_{23}}=i\gamma_2\gamma_3$, to label the states. As $(-1)^{F_{12}}$ and $(-1)^{F_{23}}$ anticommute, they do not share simultaneous eigenvalues, and the even and odd parity states with respect to these two operators are related by the  non-diagonal transformation: \begin{align}\left|\vcenter{\hbox{\includegraphics[width=0.08\textwidth]{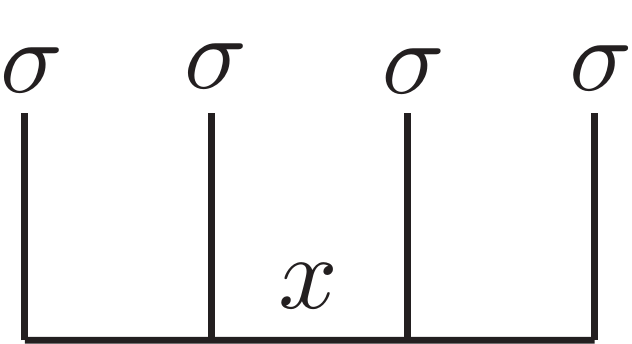}}}\right\rangle=\sum_{y=1,\psi}\left[F^{\sigma\sigma\sigma}_\sigma\right]_x^y\left|\vcenter{\hbox{\includegraphics[width=0.08\textwidth]{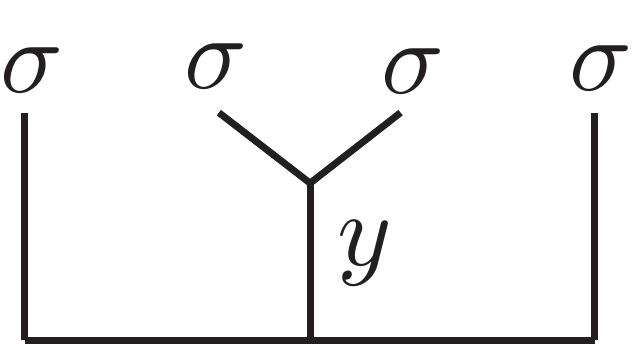}}}\right\rangle\end{align} where the $F$-matrix is  \begin{align}F^{\sigma\sigma\sigma}_\sigma=\frac{1}{\sqrt{2}}\left(\begin{array}{*{20}c}1&1\\1&-1\end{array}\right)\end{align} with its rows and columns arranged according to $x,y=1,\psi$ which label the parities $(-1)^{F_{12}}$ and $(-1)^{F_{23}}$ respectively. 

A fusion category is a theory that encodes associative fusion rules \eqref{appfusionrule} and a consistent collection of basis transformations \eqref{Fsymboldef}. In addition to this structure, an anyon theory also contains information about exchange and braiding. Fusion commutativity $x\times y=y\times x$ at the quantum state level translates to another kind of basis transformation. They are generated by the $R$-symbols \begin{gather}\vcenter{\hbox{\includegraphics[width=0.05\textwidth]{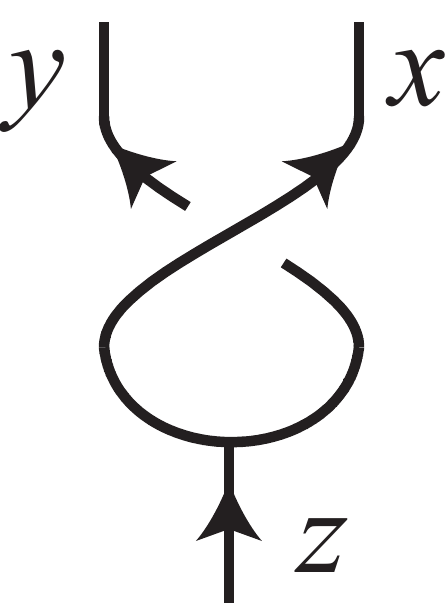}}}=R^{xy}_z\vcenter{\hbox{\includegraphics[width=0.05\textwidth]{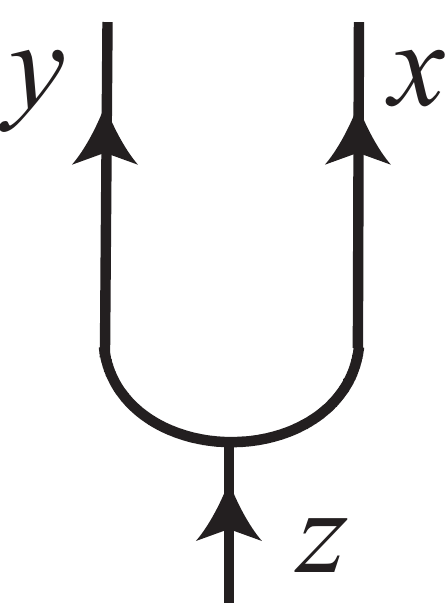}}}\label{Rsymboldefapp}\\\vcenter{\hbox{\includegraphics[width=0.05\textwidth]{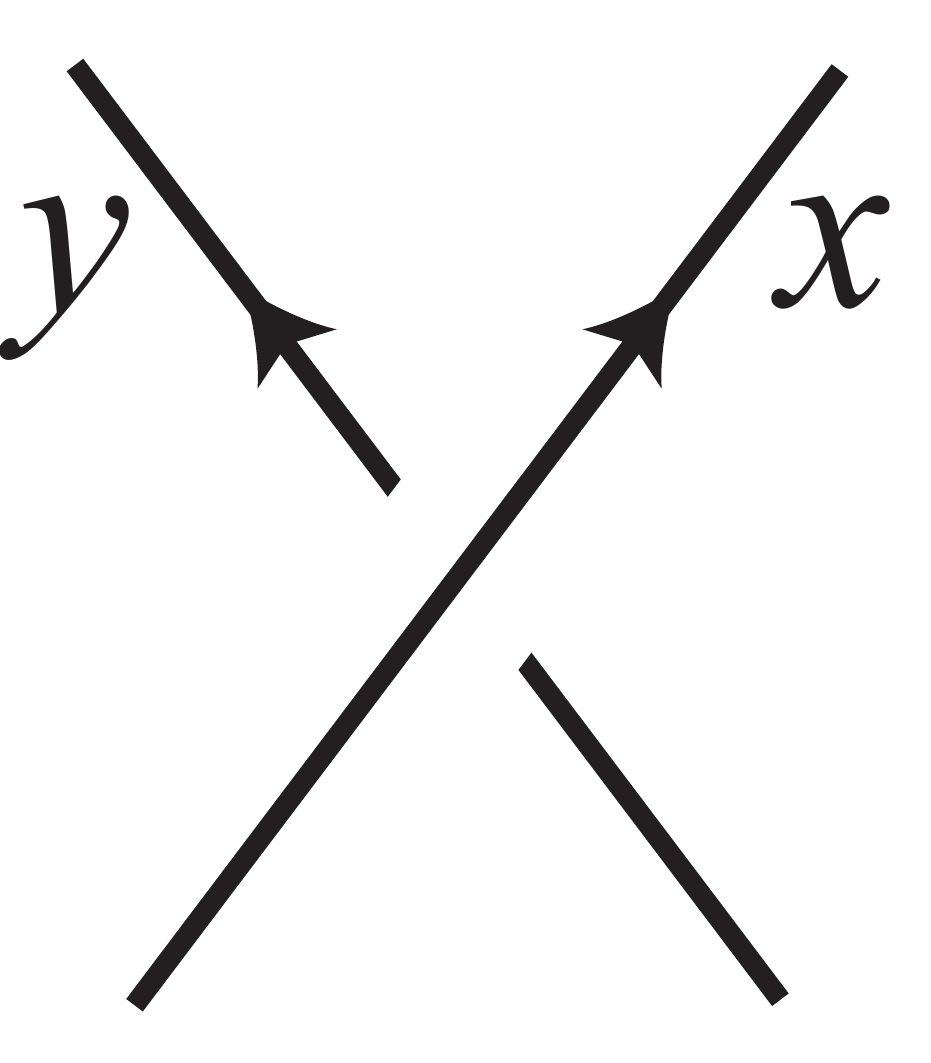}}}=\sum_{z}\sqrt{\frac{d_z}{d_xd_y}}R^{xy}_z\vcenter{\hbox{\includegraphics[width=0.05\textwidth]{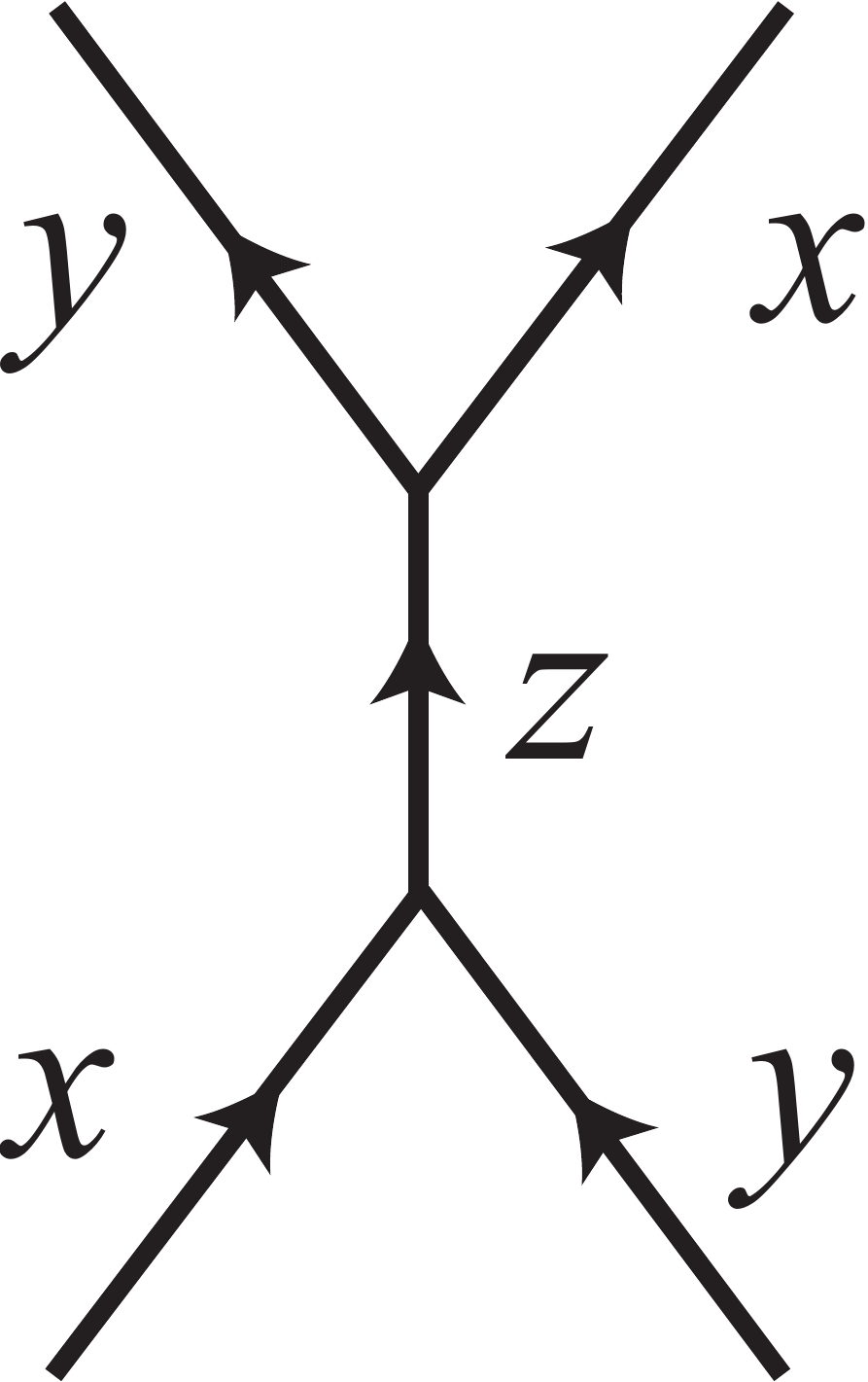}}}\label{Rsymboldefapp2}\end{gather} when the splitting $z\to x\times y$ is admissible. When splitting is degenerate, vertices in \eqref{Rsymboldefapp} and \eqref{Rsymboldefapp2} should be labeled, and $R^{xy}_z$ would be a $(N_{xy}^z)\times(N_{xy}^z)$ unitary matrix. 

\begin{figure}[ht]
\centering\includegraphics[width=0.4\textwidth]{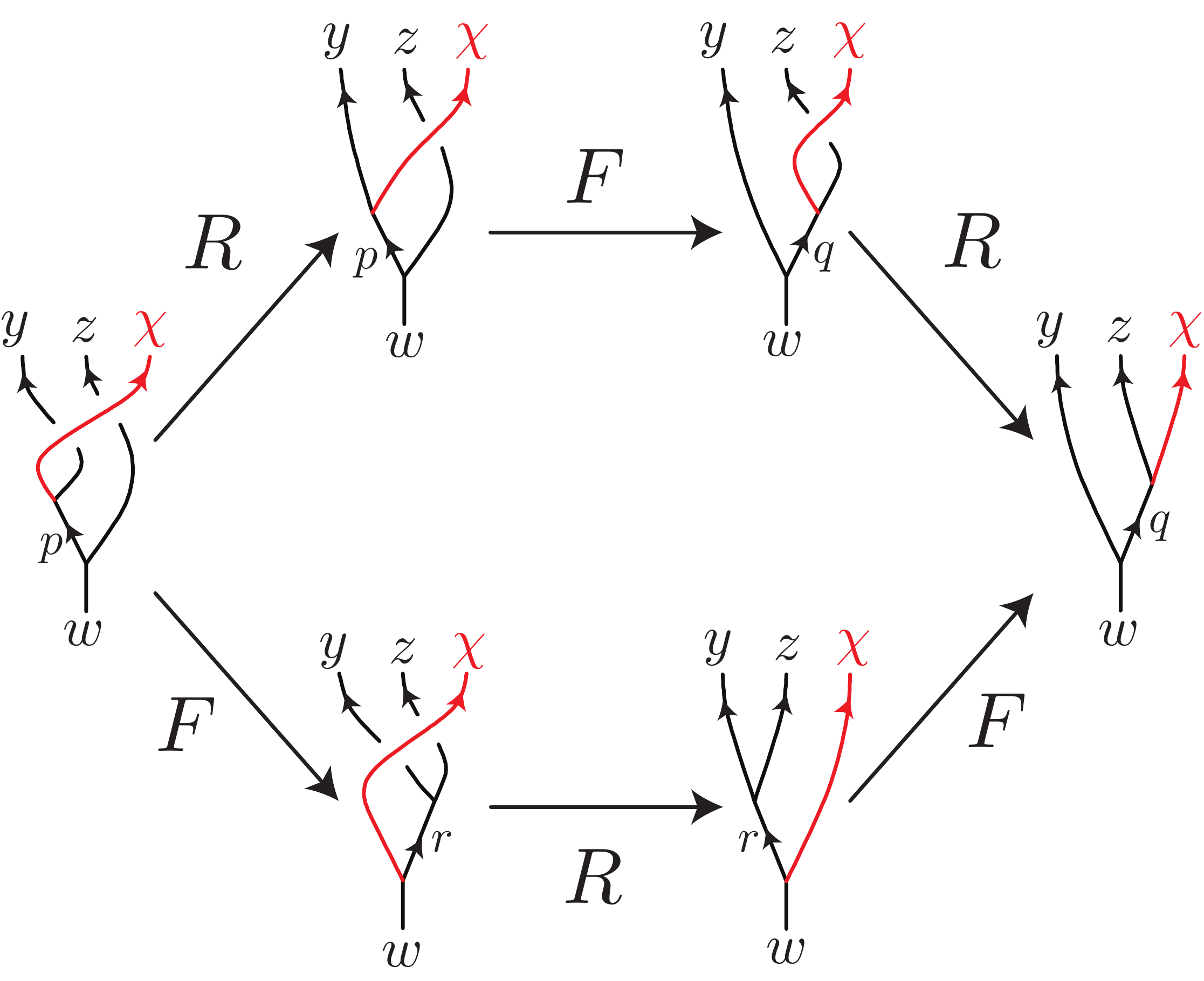}
\caption{Hexagon equation.}\label{fig:hexagon1}
\end{figure}

These exchange $R$-symbols follow the consistency relations called the {\em hexagon identity}~\cite{Kitaev06} \begin{align}R^{\chi y}_p\left[F^{y\chi z}_w\right]_p^qR^{\chi z}_q=\sum_r\left[F^{\chi yz}_w\right]_p^rR^{\chi r}_w\left[F^{yz\chi}_w\right]^q_r .\label{hexagoneq}\end{align} (see also Fig.~\ref{fig:hexagon1}). Essentially they guarantee fusion and exchange are compatible by requiring that the successive exchanges between $x,y$ and between $x,z$ are overall equivalent to the exchange between $x$ and $w=y\times z$. An anyon theory is therefore a {\em braided} fusion category that is equipped with a consistent exchange structure.

The spin statistics theorem equates the $\pi$-exchange phase \begin{align}\theta_x=\frac{1}{d_x}\sum_{y}d_y\mbox{Tr}\left(R^{xx}_y\right)\end{align} with the $2\pi$ twist $\theta_x=e^{2\pi ih_x}=\varkappa_x(R^{x\overline{x}}_1)^\ast$, where $h_x$ is referred to as the spin. Particle $x$ has identical spin to its conjugate $\bar{x}$. Exhange phases are related to $2\pi$ braiding phases by the {\em ribbon identity}
\begin{align}\theta_z\mathbb{I}=\vcenter{\hbox{\includegraphics[width=0.5in]{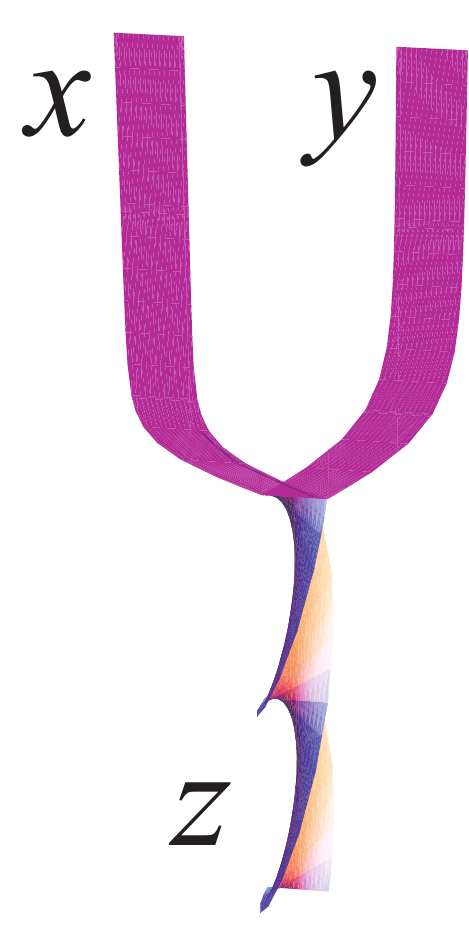}}}=\vcenter{\hbox{\includegraphics[width=0.5in]{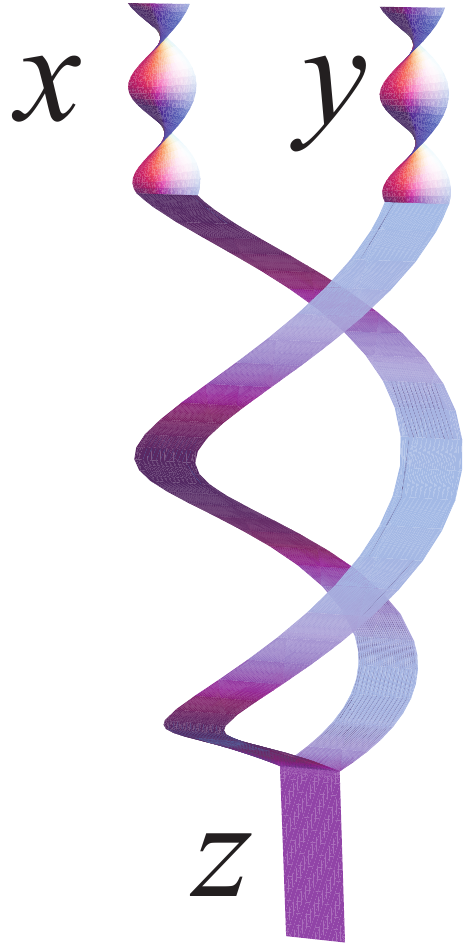}}}=R^{xy}_zR^{yx}_z\theta_x\theta_y\label{ribbonapp}\end{align} where $R^{{\bf a}{\bf b}}_{\bf c}R^{{\bf b}{\bf a}}_{\bf c}$ is the gauge independent ($2\pi$) braiding phase between ${\bf a}$ and ${\bf b}$ with a fixed overall fusion channel ${\bf c}$, and $\mathbb{I}$ is the $(N_{xy}^z)\times(N_{xy}^z)$ identity matrix. Heuristically Eq.~\eqref{ribbonapp} holds because twisting the overall quasiparticle ${\bf c}$ involves twisting its constituents as well as rotating its internal structure.

The braiding between anyons can be summarized by the average \begin{align}S_{xy}=\frac{1}{\mathcal{D}}\sum_{z}d_{z}\mbox{Tr}\left(R^{xy}_{z}R^{yx}_{z}\right)=\frac{1}{\mathcal{D}}\sum_{z}d_{z}N^{z}_{xy}\frac{\theta_{z}}{\theta_{x}\theta_{y}}\label{braidingSapp}\end{align} which are the matrix elements of the modular $S$-matrix. The $2\pi$ twists give the modular $T$-matrix $T_{xy}=\theta_{x}\delta_{xy}$. They (projectively) represent the modular group $SL(2;\mathbb{Z})$, i.e. the group of automorphisms of the torus, and obey the group relation \begin{align}(ST)^3=e^{2\pi i\frac{c_-}{8}}S^2\label{ST3=S2}\end{align} where $C=S^2$ is the conjugation matrix that relates $x\leftrightarrow\bar{x}$, squares to identity, and commutes with both $S$ and $T$. The projective phase is associated to the chiral central charge $c_-=c_R-c_L$ (mod 8) of the CFT along the $(1+1)$D boundary of the topological phase. Eq.~\eqref{ST3=S2} is equivalent to the Gauss-Milgram formula \begin{align}\exp\left(2\pi i\frac{c_-}{8}\right)=\frac{1}{\mathcal{D}}\sum_{x}d_{x}^2\theta_{x}.\label{GaussMilgram}\end{align} Moreover, the fusion matrices $N_{xy}^{z}$ that characterize fusion rules ${x}\times{y}=\sum_{z}N_{xy}^{z}{z}$ can in turned be determined by $S$-matrix throught the Verlinde formula~\cite{Verlinde88} \begin{align}N_{xy}^z=\sum_{w}\frac{S_{xw}S_{yw}S^\ast_{zw}}{S_{1w}}.\label{Verlindeformulaapp}\end{align} For example the $S$-matrix \eqref{SDG} determines the fusion rules of a discrete gauge theory.


\section{Anyonic symmetries}\label{sec:anyonicsymmetry}
Topological phases in $(2+1)$ dimensions can have extra global symmetries~\cite{Kitaev06,EtingofNikshychOstrik10,Bombin,YouWen,BarkeshliJianQi,TeoRoyXiao13long,khan2014,BarkeshliBondersonChengWang14,TeoHughesFradkin15,TarantinoLindnerFidkowski15}. 
For example in the Kitaev toric code in section~\ref{sec:toriccode}, we have seen that there is an electric-magnetic symmetry that relabels the charge and flux $e\leftrightarrow m$ without altering the fusion, exchange and braiding structure of the topological phase. This kind of anyon relabeling symmetry is inheritly non-local as it needs to be applied on all anyons at all positions. In this section, we will illustrate the symmetries in some known topological phases. We will mostly be focusing on Abelian topological phases. Section~\ref{sec:globalsymmetryab} will review the description of global symmetries with the help of an Abelian Chern-Simons theory. This includes the Kitaev toric code, bilayer fractional quantum Hall states, and the $SO(8)_1$ state that lives on the surface of a bosonic topological insulator~\cite{VishwanathSenthil12,BurnellChenFidkowskiVishwanath13,WangSenthil14,WangPotterSenthil13}. Anyonic symmetry also arises in non-Abelian topological state and we will demonstrate this in the 4-Potts state in section~\ref{sec:globalsymmetrynab}. Lastly we will give a brief discussion on the general classification of global symmetries and their obstructions in section~\ref{sec:classificationsymmetry}. A list of anyonic symmetries appears in this section can be found in table~\ref{tab:ASexamples}.

\begin{table}[htbp]
\centering
\begin{tabular}{lll}
Topological & Anyonic & Relabeling\\
Phase & Symmetries & Action\\\hline
All phases & $\mathbb{Z}_2$ conjugation & ${\bf a}\leftrightarrow\overline{\bf a}$\\\noalign{\smallskip}
Bi-layer systems & $\mathbb{Z}_2$ bilayer & ${\bf a}_\uparrow\leftrightarrow{\bf a}_\downarrow$\\
$(K=K_\uparrow\oplus K_\downarrow)$ & symmetry & \\\noalign{\smallskip}
$\mathbb{Z}_k$ gauge theory & $\mathbb{Z}_2$ e-m symmetry & $e^am^b\leftrightarrow m^ae^b$\\
$(K=k\sigma_x)$ & & \\\noalign{\smallskip}
$so(8)_1$ & Triality & permutation of\\
 & $S_3$-symmetry & fermions $\psi_1,\psi_2,\psi_3$\\\noalign{\smallskip}
Bi-layer toric code & $S_3\times\mathbb{Z}_2$ & $S_3$-permutation of\\
($so(8)_1^L\times so(8)_1^R$) & & $\psi_1^{L/R},\psi_2^{L/R},\psi_3^{L/R}$\\
 & & Bi-layer $\psi_i^L\leftrightarrow\psi_i^R$\\
$4$-state Potts & $S_3$ & permutation of\\
 & & bosons $\{j_1,j_2,j_3\}$,\\
 & & twist fields $\{\sigma_1,\sigma_2,\sigma_3\}$\\
 & & and $\{\tau_1,\tau_2,\tau_3\}$
\end{tabular}
\caption{Examples of anyonic symmetries in some topological phases and their matrix representations. Details of the $so(8)_1$ state, bilayer toric code and the 4-state Potts model can be found in Section~\ref{sec:globalsymmetryab} and \ref{sec:globalsymmetrynab}.}\label{tab:ASexamples}
\end{table}

\subsection{Global symmetries in Abelian topological phases}\label{sec:globalsymmetryab}
Abelian topological phases can be described by an effective field theory~\cite{WenZee92,Wenedgereview} $Z[\mathcal{J}]=\int[D\alpha({\bf r})]\exp(iS[\mathcal{J}])$. The Chern-Simons action is defined by \begin{align}S[\mathcal{J}]=\frac{1}{4\pi}\int d^2xdt\varepsilon^{\mu\nu\lambda}K_{IJ}\alpha^I_\mu\partial_\nu \alpha^J_\lambda+\alpha_I\mathcal{J}^I\label{CSaction}\end{align} where $a=(\alpha^1,\ldots,\alpha^N)$ is a $N$-component $U(1)$-gauge field. The integral $K$-matrix is symmetric non-dengenerate and encodes the fusion, spin and braiding properties of quasiparticle excitations. 

Quasiparticle excitations of the theory are labeled as $N$-component vectors ${\bf a}$ in an integer (anyon) lattice $\Gamma^\ast=\mathbb{Z}^r$. Quasiparticles $\psi^{\bf a}$ are sources for the currents $~\mathcal{J}_1^{a_1},\ldots\mathcal{J}_N^{a_N}$. At long distances, nearby quasiparticles combine to form single entity and have a fusion structure. The quasiparticle fusion rules coincide with lattice vector addition \begin{align}\psi^{\bf a}\times\psi^{\bf b}=\vcenter{\hbox{\includegraphics[width=0.5in]{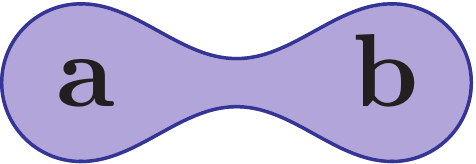}}}=\psi^{{\bf a}+{\bf b}}.\end{align}

The exchange statistics of a quasiparticle type is given by the Abelian phase factor \begin{align}\theta_{\bf a}=e^{2\pi ih_{\bf a}}=\vcenter{\hbox{\includegraphics[width=0.5in]{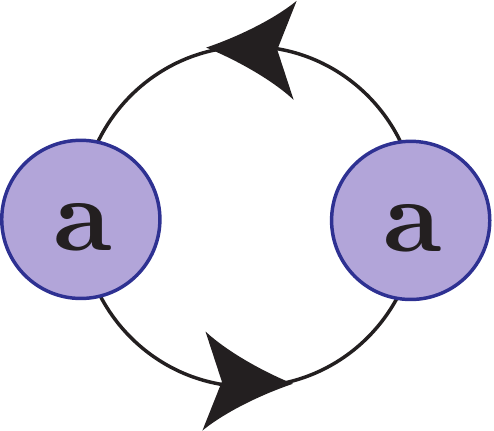}}}=e^{\pi i{\bf a}^TK^{-1}{\bf a}}\label{AQPexchange}\end{align} where the spin $h_{\bf{a}}$ of a quasiparticle $\psi^{\bf a}$ is given by ${\bf a}^tK^{-1}{\bf a}/2$. From the spin-statistics theorem, the exchange phase $\theta_{\bf a}$ is equivalent to the twist phase gained when a single quasiparticle is rotated by $2\pi$. The exchange phases of the quasiparticles is often summarized in terms of a $T$-matrix, $T_{\mathbf{a}\mathbf{b}}=\delta_{\mathbf{a},\mathbf{b}}\theta_{\mathbf{a}}$. 

The $2\pi$ braiding or monodromy phase when dragging $\psi^{\bf a}$ once around $\psi^{\bf b}$ is given by \begin{align}\mathcal{D}S_{{\bf a}{\bf b}}=\vcenter{\hbox{\includegraphics[width=0.7in]{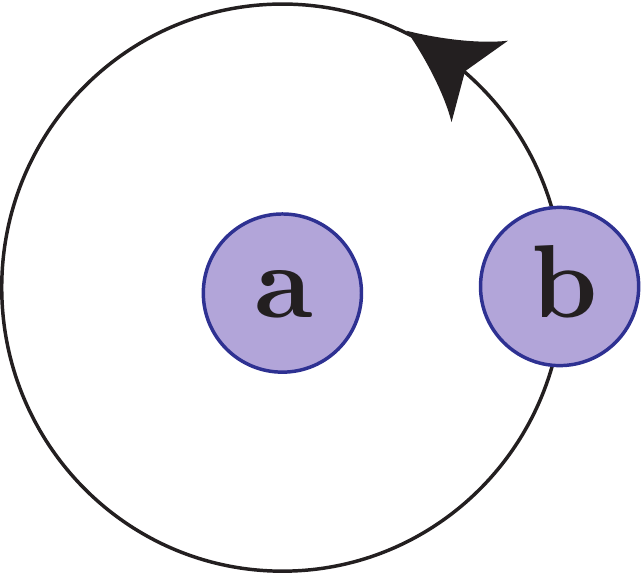}}}=e^{2\pi i{\bf a}^TK^{-1}{\bf b}}\label{AQPbraiding}\end{align} where $\mathcal{D}=\sqrt{|\det(K)|}$ is the total quantum dimension. $\mathcal{D}$ is defined this way so that the $\mathcal{D}^2$-dimensional $S$-matrix in eq.~\eqref{AQPbraiding}, which also agrees with eq.\eqref{braidingSapp}, that characterizes anyon braiding is normalized and unitary. The braiding phase is insensitive to the exact paths of the deconfined quasiparticle pair as long as the linking number of their world-lines is unchanged. 

The quasiparticles that occupy the sublattice $\Gamma=K\mathbb{Z}^r\subseteq\Gamma^\ast$ are called \emph{local} and only contribute trivial monodromy phases with all other quasiparticles. Intuitively they are the fundamental building blocks that are ``fractionalized" to form the topological state. When the diagonal entries of the $K$-matrix are all even, all local particles are bosonic and the topological state is said to be bosonic. Otherwise, the theory contains fermionic local particles, for instance, like electrons. In this review, we will assume bosonic topological states for the most of the time. This can be justified by extending a fermionic theory to include $\pi$-fluxes ($hc/2e$ fluxes for electrons) that are non-local with respect to the fermions. This extension is a special type of gauging, where the $\mathbb{Z}_2$ fermion parity symmetry is promoted to a local symmetry. 

Topological information encoded in the nonlocal braiding and exchange statistics of fractionalized quasiparticles is left invariant upon the addition of local bosonic particles ${\bf a}\to{\bf a}+K{\bf b}$, for $K{\bf b}\in\Gamma$. They therefore represent the same anyon type $[{\bf a}]=\{{\bf a}+K{\bf b}:K{\bf b}\in\Gamma=K\mathbb{Z}^N\}$. Distinct anyon types thus occupy the finite anyon quotient lattice $\mathcal{A}=\Gamma^\ast/\Gamma=\mathbb{Z}^r/K\mathbb{Z}^r$, which is an Abelian group with fusion product $[{\bf a}]\times[{\bf b}]=[{\bf a}+{\bf b}]$ and contains $\mathcal{D}^2=|\det(K)|$ elements.

For example in the Kitaev toric code in section~\ref{sec:toriccode}, the $K$-matrix is $K=2\sigma_x$ and the four anyon types are represented by $1=[0]$, $e=[{\bf e}_1]$, $m=[{\bf e}_2]$ and $\psi=[{\bf e}_1+{\bf e}_2]$, where the ground state is a condensate of local bosons $1=\gamma=\{a_1{\bf e}_1+a_2{\bf e}_2:a_1,a_2\in2\mathbb{Z}\}$. The quasiparticles form an anyon lattice of size 4, $\mathcal{A}=\mathbb{Z}_2\times\mathbb{Z}_2$. Vector addition recovers the fusion rules $\psi=e\times m$, $e^2=m^2=\psi^2=1$, and the quadratic form $K^{-1}$ reproduces the monodromy rules $\mathcal{D}S_{em}=\mathcal{D}S_{e\psi}=\mathcal{D}S_{m\psi}=-1$. 

A Chern-Simons description of a topological phase is not unique. A $K$-matrix would encode the identical fusion and braiding structure even after undergoing a basis transformation: $K\to WKW^T$ where $W$ is an invertible integer-valued matrix in $GL(N,\mathbb{Z})$. There are special transformations of this type, known as {\em automorphisms}, that leave the $K$-matrix invariant \begin{align}MKM^T=K.\label{automorphism}\end{align} Each such transformation $M$ corresponds to an {\em anyonic symmetry operation}~\cite{khan2014} that permutes the anyons that have the same fusion properties and braiding statistics: \begin{align}[M{\bf a}]\times[M{\bf b}]=[M({\bf a}+{\bf b})],\quad\theta_{M{\bf a}}=\theta_{\bf a},\quad S_{M{\bf a}M{\bf b}}=S_{{\bf a}{\bf b}},\quad T_{M{\bf a}M{\bf
b}}=T_{{\bf a}{\bf b}}.\label{ASdef2}\end{align} 

The collection of automorphisms forms a group \begin{align}\mbox{Aut}(K)=\left\{M\in GL(N;\mathbb{Z}):MKM^T=K\right\}\label{Autom}\end{align} that classifies the global symmetries of the topological quantum field theory \eqref{CSaction}. Within $\mbox{Aut}(K)$, there lies a sub-collection of trivial basis transformations, called {\em inner automorphisms}, that only rotate quasiparticles up to local particles, $N_0{\bf a}={\bf a}+K{\bf b}$. They act trivially on the anyon labels in $\mathcal{A},$ and form a normal subgroup \begin{align}\mbox{Inner}(K)=\left\{N_0\in\mbox{Aut(K)}:[N_0{\bf a}]=[{\bf a}]\right\}.\end{align} We are interested in non-trivial anyon relabeling actions, known as {\em outer automorphisms}. These are classified by ``taking out" the trivial inner automorphisms. Mathematically they live in the quotient group \begin{align}\mbox{Outer}(K)=\frac{\mbox{Aut}(K)}{\mbox{Inner}(K)}\label{outer}\end{align} so that the basis transformations $M$ and $N$ correspond to the same anyon relabeling action if they differ by an inner automorphism, i.e.~if $N=MN_0$ where $N_0$ is an inner automorphism, then we say $M$ is equivalent to $N$.

Most anyonic symmetries considered in this article can be described within this simple framework. Additionally, global anyonic symmetries are easy to identify as they are common features of many topological phases. For example, all topological phases have a conjugation symmetry $C:{\bf a}\to\overline{\bf a}=-{\bf a}$, and all bilayer systems support a layer-flipping symmetry that switches anyons living on opposite layers. These both represent global anyonic symmetries that can be represented by this framework. 

However there are exceptions. For instance the $K$-matrix formalism only applies to Abelian topological states. A more abstract description is required for non-Abelian ones. This will be highlighted in section~\ref{sec:globalsymmetrynab}. Secondly, the outer automorphism group \eqref{outer} may not contain all anyonic symmetry. For example, the theory with $K=12$, which is bosonic version of the Laughlin $\nu=1/3$ fractional quantum Hall state with gauged fermion parity, carries not only the conjugation symmetry $C:\psi^a\to\overline{\psi^a}=\psi^{-a}$ but also two other parity flip symmetries $P_5:\psi^a\to\psi^{5a}$ and $P_7:\psi^a\to\psi^{7a}$. These extra symmetries relabels the anyon types without changing their fusion rules or braiding statistics, but they cannot be represented by automorphisms unless the $K$ matrix is extended. \begin{align}K'=\left(\begin{array}{*{20}c}12&0&0\\0&0&1\\0&1&0\end{array}\right),\quad C=-\mathbb{I}_3,\quad P_7=\left(\begin{array}{*{20}c}7&-24&12\\2&-8&3\\-1&3&-2\end{array}\right),\quad P_5=-P_7.\end{align} Here the $3\times 3$ $K'$-matrix is {\em stably equivalent}~\cite{cano2013bulk,BarkeshliJianQi13long} to the original $K=12$ by a trivial bosonic state with $K_0=\sigma_x$. 

By extending the rank of the $K$-matrix, the group of outer automorphisms also extends. In general, one can consider a stable equivalent class $[K]$ of $K$-matrices, within which are matrices integral symmetric non-dengenerate matrices $K'$ stably equivalent to $K$, i.e.~\begin{align}K'\oplus T_1=W(K\oplus T_2)W^T,\end{align} where $W\in GL(N;\mathbb{Z})$ and $T_1$, $T_2$ are integral unimodular quadratic forms (i.e.~$T^T=T$ and $|\det(T)|=1$) that represents trivial topological states. For bosonic states, we requires $T_1$, $T_2$ to also be bosonic so that all diagonal entries are even. If one further requires equivalent $K$-matrices to have identical signature $c_-=c_R-c_L$, the difference between the number of positive and negative eigenvalues, then $T_i=\sigma_x\oplus\sigma_x\oplus\ldots$ up to unimodular transformations. If otherwise, the signatures of bosonic $K$ and $K'$ can be off by an integral multiple of 8, and $T_i$ can contain the quadratic forms of positive definite even unimodular lattices, like $E_8$ with dimension 8 and the Leech lattice with dimension 24~\cite{MilnorHusemollerbook}. The equivalent class $[K]$ thus consists of all $K$-matrices that describe the same topological phase.

We notice $[K]$ is a {\em direct system} in the sense that given any $K$-matrix, one can increase its dimension by $K\oplus T$, for $T$ an (even) unimodular quadratic form. Consequently the collection of outer automorphism groups $\mathrm{Outer}(K)$ is also directed in the sense that $\mathrm{Outer}(K)\subseteq\mathrm{Outer}(K\oplus T)$. We believe the entire global symmetry group of an Abelian topological phase is identical to the direct limit \begin{align}\mathrm{Outer}[K]=\lim_\to\mathrm{Outer}(K)\end{align} so that all anyonic symmetries have matrix representations given a big enough $K$ matrix. For example, the symmetry group of the theory with $K=12$ contains 4 elements and has a $\mathbb{Z}_2\times\mathbb{Z}_2$ structure.

\subsubsection{Exactly solved lattice models}\label{sec:symmlatticemodels}
The Kitaev toric code~\cite{Kitaev97} or equivalently the Wen plaquette model~\cite{Wenplaquettemodel}, the color code~\cite{BombinMartin06} and its $k$-rotor generalization~\cite{TeoRoyXiao13long} are exact solvable lattice models that describes abelian topological states. They are all equipped with global anyonic symmetries inherited from lattice translations and rotations. 

We start with the Wen's plaquette model in section~\ref{sec:toriccode} where the spin-$1/2$ degrees of freedom are put on the vertices of a rectangular checkerboard lattice. The bi-color structure of the plaquettes (see the right diagram in figure~\ref{fig:toriccode0}) corresponds to the two bosonic quasiparticles, the charge $e$ and the flux $m$, in the $\mathbb{Z}_2$ gauge theory. The $e$ excitations live on the white plaquettes while the $m$ excitations live on the blue ones.

In the Chern-Simons description with $K$-matrix $K=2\sigma_x$, the electric-magnetic symmetry that interchanges $e\leftrightarrow m$ is represented by $M=\sigma_x$. On the lattice level, it can be realized as a half-translation by $(1/2,1/2)$ so that the colors of the plaquettes are reversed and the anyon types are relabeled. Equivalently the anyonic symmetry can also be generated by a $\pi$-rotation of the rectangular lattice centered at the mid-point of an edge. These are clearly non-local operations as they alter the entire lattice. Identifying global symmetries of a topological phases by space group operations have the advantage of realizing twist defects as lattice defects. This will be discussed in the later section.

There are many exactly solvable lattice models that carries anyonic symmetries. For example the plaquette model can be generalized to describe a $\mathbb{Z}_k$ gauge theory. This can be done by replacing the spin-$1/2$ degrees of freedom by $k$-dimensional rotors \begin{align}\sigma_x=\left(\begin{array}{*{20}c}&1&&&\\&&1&&\\&&&\ddots&\\&&&&1\\1&&&&\end{array}\right),\quad\sigma_z=\left(\begin{array}{*{20}c}1&&&\\&w&&\\&&\ddots&\\&&&w^{k-1}\end{array}\right),\quad w=e^{2\pi i/k}\label{rotors}\end{align} that satisfies the commutation relations $\sigma_x\sigma_z=w\sigma_z\sigma_x$. The plaquette stabilizers $\hat{P}_\Diamond$ can be defined in the same way as in \eqref{WenP}. They are however not hermitian, and thus the Hamiltonian need to be modified to include the hermitian conjugates, $H=-\sum_{P_\Diamond}(\hat{P}_\Diamond+\hat{P}_\Diamond^\dagger)$. The theory carries $k^2$ quasiparticle types generally labeled by $e^{a_1}m^{a_2}$, where ${\bf a}=(a_1,a_2)$ are anyon lattice vectors under the $K$-matrix $K=k\sigma_x$. The electric-magnetic symmetry acts the same way by switching $e\leftrightarrow m$, and are represented by the same matrix $M=\sigma_x$ and half-translation or $\pi$-rotation on the lattice level.

\begin{figure}[htbp] \centering
  \includegraphics[width=0.65\textwidth]{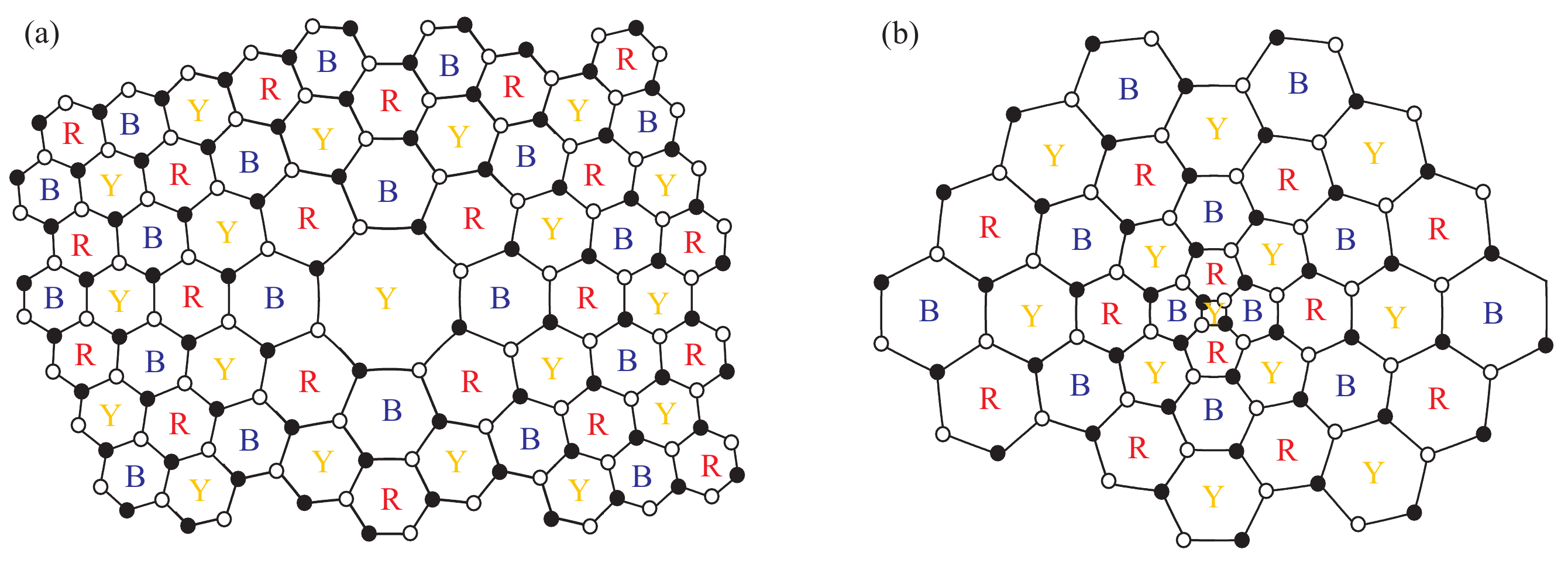} \caption{Trivial {\em twistless} defects that preserve local tri-coloring (YRB) bipartite ($\bullet,\circ$) order. (a) Negatively curved disclination with Frank angle $-120^\circ$; (b) Positively curved disclination with Frank angle $+120^\circ$.}\label{fig:squareoctagon} 
\end{figure}

Next we move on to another lattice model, the $k$-rotor tri-color model~\cite{BombinMartin06,TeoRoyXiao13long}, that possess a richer non-Abelian global symmetry group $S_3$, the permutation group of 3 elements. It can be defined on any bipartite tri-colored graph so that each vertex has three nearest neighbors with opposite $\bullet,\circ$-type and the plaquettes are colored by yellow (Y), red (R) or blue (B) so that no adjacent plaquettes share the same color, i.e.~the two types of vertices $\bullet,\circ$ are color vortices with opposite orientations (see figure~\ref{fig:squareoctagon}). The simplest regular lattice of this type is a honeycomb lattice.

Each plaquette $P$ carries two stablilizer operators \begin{align}\hat{P}_\bullet=\vcenter{\hbox{\includegraphics[width=0.7in]{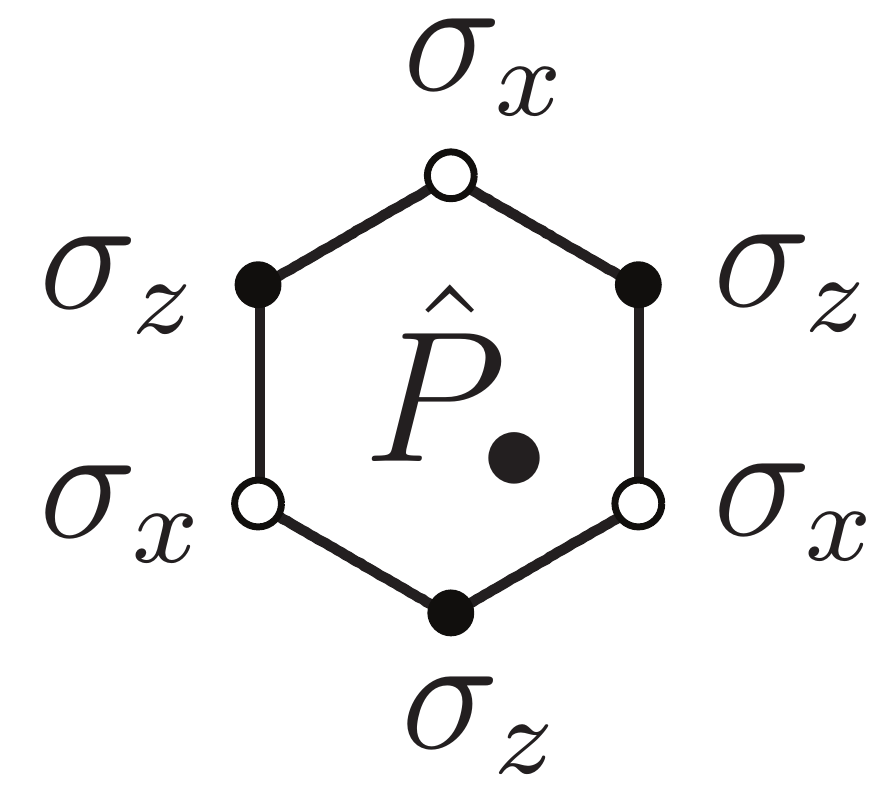}}}=\prod_{v_\bullet\in P}\sigma_z(v_\bullet)\prod_{v_\circ\in P}\sigma_x(v_\circ),\quad\hat{P}_\circ=\vcenter{\hbox{\includegraphics[width=0.7in]{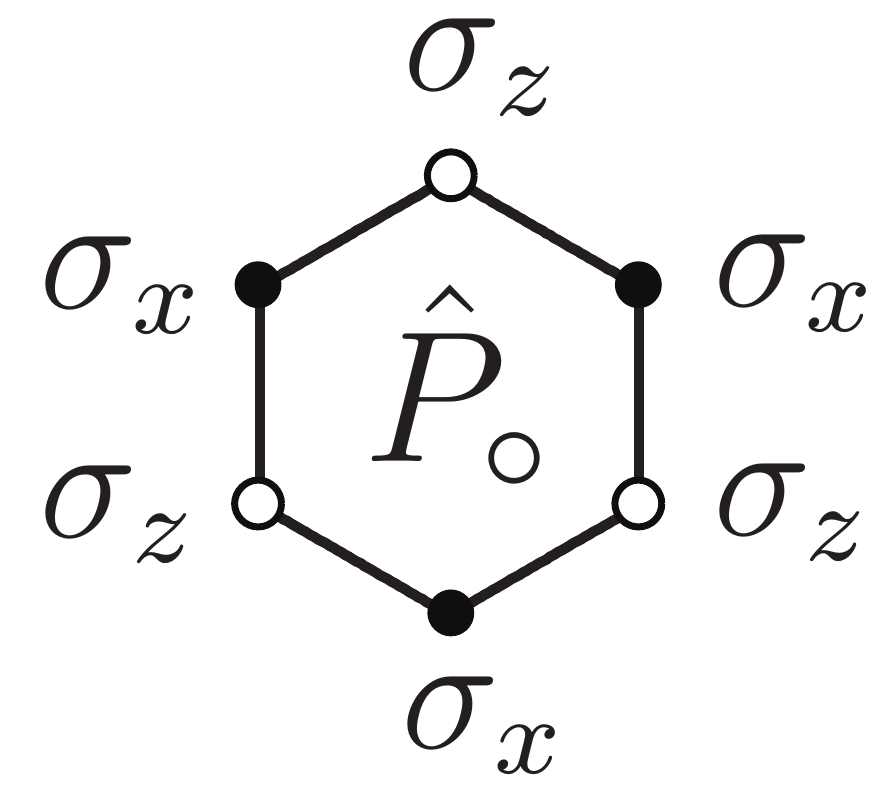}}}=\prod_{v_\bullet\in P}\sigma_x(v_\bullet)\prod_{v_\circ\in P}\sigma_z(v_\circ)\label{stabilizers1}\end{align} where there is a $k$-dimensional rotor degree of freedom, $\sigma_x,\sigma_z$ as defined in \eqref{rotors}, at each vertex. These plaquettes operators form a set of mutually commuting stabilizers and share simultaneous eigenvectors. The Hamiltonian is a sum of all the plaquette stabilizers \begin{align}H=-\sum_P(\hat{P}_\bullet+\hat{P}_\bullet^\dagger)-\sum_P(\hat{P}_\circ+\hat{P}_\circ^\dagger)\label{ham1}\end{align} The model is exactly solvable and the ground states are trivial flux configurations where $\hat{P}_\bullet=\hat{P}_\circ=1$ for all stabilizers. 

\begin{figure}[ht]
	\centering
	\includegraphics[width=0.55\textwidth]{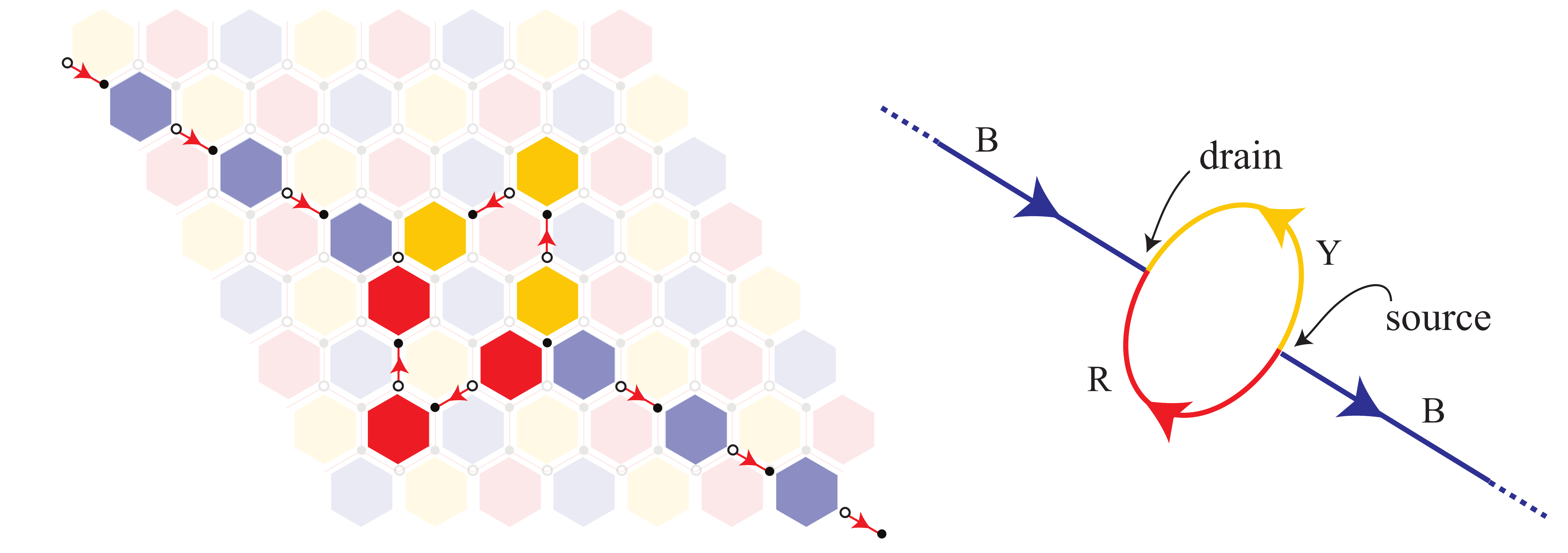}
 \caption{Color splitting of Wilson string through tricolor, trivalent, source and drain. Product of rotor operators are taken over highlighted $\bullet$ and $\circ$ vertices.}\label{fig:colorspliting}
\end{figure}

As plaquettes are tri-colored, one expect the theory to carry the six fundamental excitations $Y_{\bullet/\circ},R_{\bullet/\circ},B_{\bullet/\circ}$ that corresponds to $\hat{P}_{\bullet/\circ}=-1$ at a $Y$, $R$ or $B$ plaquette. In fact, one can construct colored Wilson strings by a product of rotors along links that connects plaquettes of the same color (see figure~\ref{fig:colorspliting}). They commute with all plaquette stabilizers except at the end of the strings. However, the six excitations are not independent. Three strings of distinct colors can come together at a vertex and does not cost extra energy. This means the excitations obey the fusion rules \begin{align}Y_\bullet\times R_\bullet\times B_\bullet=Y_\circ\times R_\circ\times B_\circ=1.\end{align} In general a quasiparticle excitation is a composite labeled by \begin{align}\psi^{\bf a}=(Y_\bullet)^{a_\bullet^1}(R_\bullet)^{a_\bullet^2}(Y_\circ)^{a_\circ^1}(R_\circ)^{a_\circ^2},\quad{\bf a}=({\bf a}_\bullet,{\bf a}_\circ)=(a_\bullet^1,a_\bullet^2,a_\circ^1,a_\circ^1)\in\mathbb{Z}^4\end{align} because $B_{\bullet/\circ}=Y_{\bullet/\circ}^{-1}R_{\bullet/\circ}^{-1}$ is redundant.

The topological state is Abelian and has a single-channeled fusion structure $\psi^{\bf a}\times\psi^{\bf b}=\psi^{{\bf a}+{\bf b}}$. The anyons have the exchange statistics \begin{align}\theta_{\bf a}=\vcenter{\hbox{\includegraphics[width=0.7in]{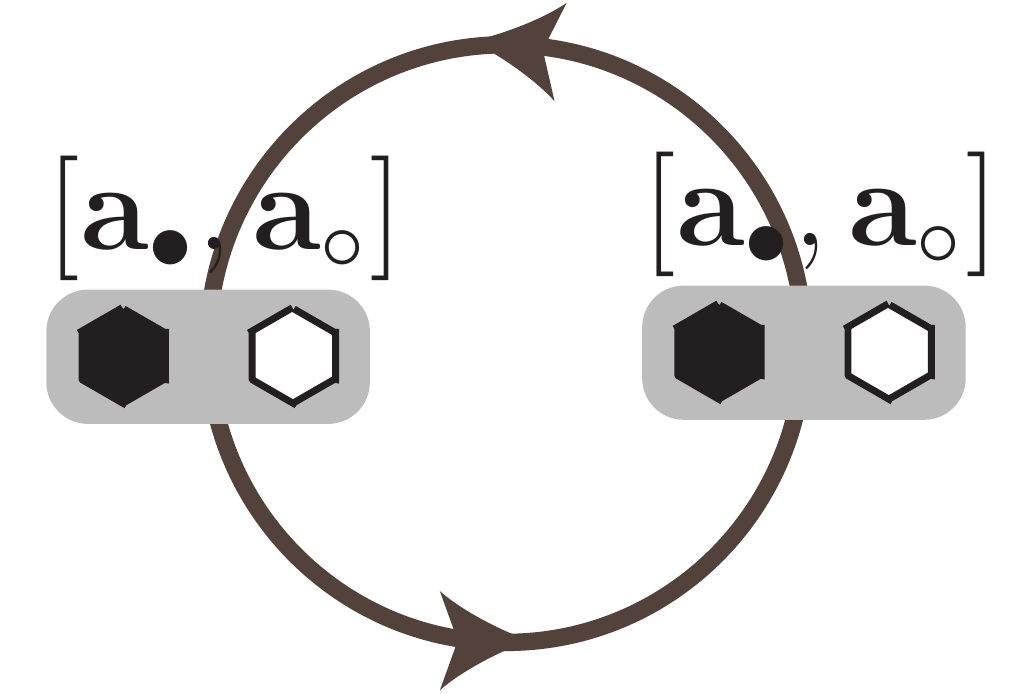}}}=e^{i\frac{2\pi}{k}{\bf a}_\circ^Ti\sigma_y{\bf a}_\bullet}\label{anyonexchangespin}\end{align} and follow the monodromy rules \begin{align}\mathcal{D}S_{\bf ab}=\vcenter{\hbox{\includegraphics[width=0.5in]{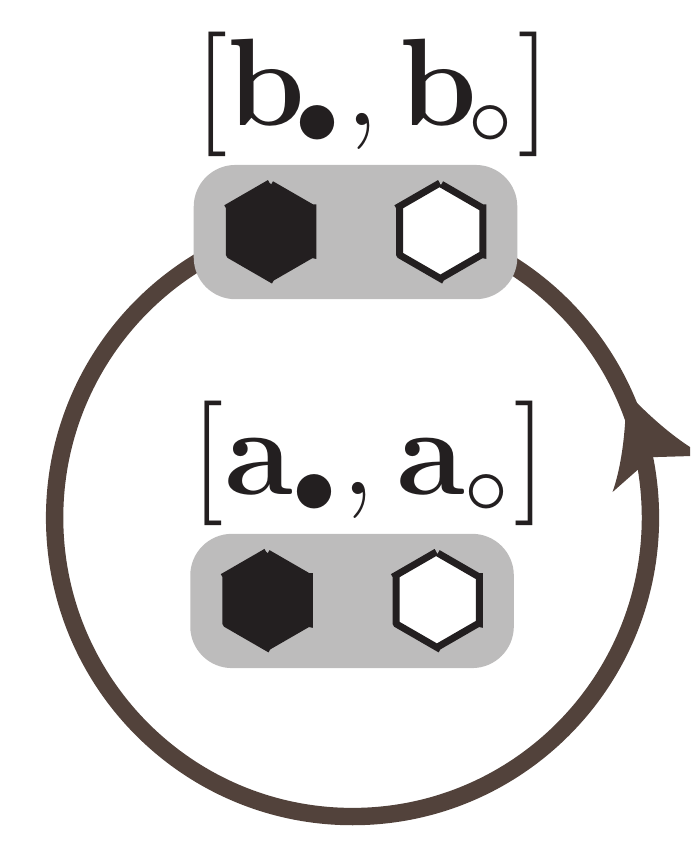}}}=e^{i\frac{2\pi}{k}({\bf a}_\circ^Ti\sigma_y{\bf b}_\bullet+{\bf b}_\circ^Ti\sigma_y{\bf a}_\bullet)}.\label{anyonfullbraiding}\end{align} These can be summarized by a Chern-Simons theory with the $4\times4$ $K$-matrix \begin{align}K=k\left(\begin{array}{*{20}c}0&0&0&-1\\0&0&1&0\\0&1&0&0\\-1&0&0&0\end{array}\right).\label{tricolorK}\end{align} The theory therefore has $\mathcal{D}^2=k^4$ quasiparticle types up to local bosons. The tri-color model has the same topological order as a bilayer $\mathbb{Z}_k$ gauge theory with $K$-matrix $K_{D(\mathbb{Z}_2^2)}=k\sigma_x\oplus k\sigma_x$. When $k=2$, the model is also topologically identical to the time reversal doublet $SO(8)_1\times\overline{SO(8)_1}$, which is equivalent to a quasi-2D slab of a bosonic topological insulator~\cite{WangPotterSenthil13,Senthil2014,TeoHughesFradkin15} that host the $SO(8)_1$ (or $\overline{SO(8)_1}$) state on the top (resp.~bottom) surface. The $K$-matrix now has a bigger dimension than that in \eqref{tricolorK} and decomposes into opposite chiral sectors $K_{SO(8)_1}\otimes\sigma_z=K_{SO(8)_1}\oplus K_{\overline{SO(8)_1}}$, where $K_{SO(8)_1}$ is a given by the Cartan matrix of $SO(8)$ (see eq.\eqref{so(8)Kmatrix} in section~\ref{sec:symmSO(8)}) and $K_{\overline{SO(8)_1}}=-K_{SO(8)_1}$. This theory is still stably equivalent to \eqref{tricolorK} but possess a larger symmetry group $(S_3\times S_3)\rtimes\mathbb{Z}_2$. However we will not be focusing on the most general symmetries but instead on those generated by the lattice $S_3$ symmetries. 

The $S_3$-anyonic symmetries are generated by threefold cyclic color permutation \begin{align}\Lambda_3:\left(\begin{array}{*{20}c}Y_\bullet&R_\bullet&B_\bullet\\Y_\circ&R_\circ&B_\circ\end{array}\right)\to\left(\begin{array}{*{20}c}R_\bullet&B_\bullet&Y_\bullet\\R_\circ&B_\circ&Y_\circ\end{array}\right)\label{Lambda3}\end{align} and twofold transposition of color and rotor types \begin{align}\Lambda_{B}:\left(\begin{array}{*{20}c}Y_\bullet&R_\bullet&B_\bullet\\Y_\circ&R_\circ&B_\circ\end{array}\right)\to\left(\begin{array}{*{20}c}R_\circ&Y_\circ&B_\circ\\R_\bullet&Y_\bullet&B_\bullet\end{array}\right).\label{LambdaB}\end{align} Under the $K$-matrix representation \eqref{tricolorK}, they take the matrix form \begin{align}\Lambda_3=\left(\begin{array}{*{20}c}0&-1&0&0\\1&-1&0&0\\0&0&0&-1\\0&0&1&-1\end{array}\right),\quad\Lambda_{B}=\left(\begin{array}{*{20}c}0&0&0&1\\0&0&1&0\\0&1&0&0\\1&0&0&0\end{array}\right)\end{align} and leave the $K$ matrix invariant, $\Lambda K\Lambda^T=K$. Other symmetries can be generated by compositions $\Lambda_3^{-1}=\Lambda_3^2$, $\Lambda_Y=\Lambda_3\Lambda_B\Lambda_3^{-1}$ and $\Lambda_R=\Lambda_3^{-1}\Lambda_B\Lambda_3$.

Suppose the model is defined on a honeycomb lattice. The threefold cyclic color permutation $\Lambda_3$ can be generated by primitive lattice translations or threefold rotations about a vertex. The twofold transpositions can be generated by a sixfold rotation centered at a hexagon. Notice the space group contatining rotations and translations is $\mathrm{Aut}=C_6\ltimes\mathcal{L}$, where $\mathcal{L}\cong\mathbb{Z}^2$ is the translation lattice. It contains a subgroup $\mathrm{Inner}=C_3\ltimes\mathcal{L}'$ that consists of threefold rotations about a hexagon and second nearest neighbor translations $\mathcal{L}'$. This subgroup leaves the bipartite vertex and tricolor plaquette structure invariant. The quotient is exactly the anyonic symmetry group $S_3=\mathrm{Aut}/\mathrm{Inner}$. Symmetry twist defects can therefore be realized as lattice topological defects, such as dislocations and disclinations. This will be discussed in the next section.

\subsubsection{Bilayer and conjugation symmetries in Abelian topological states}\label{sec:bilayerstates}
Bilayer symmetries appear in $(mmn)$-fractional quantum Hall (FQH) states~\cite{Wentopologicalorder90, WenchiralLL90, WenZee92, Wenbook, Fradkinbook, BoebingerJiangPfeifferWest90, SuenJoSantosEngelHwangShayegan91, EisensteinBoebingerPfeifferWestHe92} and are recently explored in Ref.~\cite{BarkeshliQi13,teo2013braiding}. The Chern-Simons theory is characterized by a $2\times 2$ $K$-matrix in the form of \begin{align}K=\left(\begin{array}{*{20}c}m&n\\n&m\end{array}\right)\end{align} and there is always a $\mathbb{Z}_2$ bilayer symmetry $M=\sigma_x$ and a conjugation symmetry $C=-\mathbb{I}$ that leaves the $K$-matrix invariant, $MKM^T=CKC^T=K$.

In some cases, they corresponds to distinct symmetries. For example the bilayer bosonic Laughlin-$1/2$ FQH state (which is also the $SO(4)_1$ state) has the $K$-matrix $K=2\mathbb{I}_2$, where $\mathbb{I}_2$ is the $2\times2$ identity matrix. It has 4 quasiparticle types $1,s_1,s_2,s_1s_2$ represented by the lattice vectors $0,{\bf e}_1,{\bf e}_2,{\bf e}_1+{\bf e}_2$ respectively. The bilayer symmetries interchanges the semions $s_1\leftrightarrow s_2$ while the conjugation symmetry is actually a local symmetry (but may be projective) that does not relabel anyon types. The Kitaev toric code, which has $K=2\sigma_x$, is topologically identical to the $(002)$-FQH state where the electric-magnetic symmetry is also a bilayer symmetry.

There are cases when both the bilayer and conjugation symmetries are non-local. The $\mathbb{Z}_3$ gauge theory, which is identical to the $(003)$-FQH state and has $K=3\sigma_x$, carries 9 quasiparticle types and are generated by the charge $e$ and flux $m$ with $e^3=m^3=1$ and mutual monodromy phase $\mathcal{D}_{em}=e^{2\pi i/3}$. The bilayer symmetry flip $M:e\leftrightarrow m$ while conjugation sends $C:e^{a_1}m^{a_2}\to e^{-a_1}m^{-a_2}$. A $\mathbb{Z}_3$ gauge theory therefore has a $\mathbb{Z}_2\times\mathbb{Z}_2=\{1,M,C,MC\}$ global symmetry group.

\begin{figure}[htbp]
\centering\includegraphics[width=0.8\textwidth]{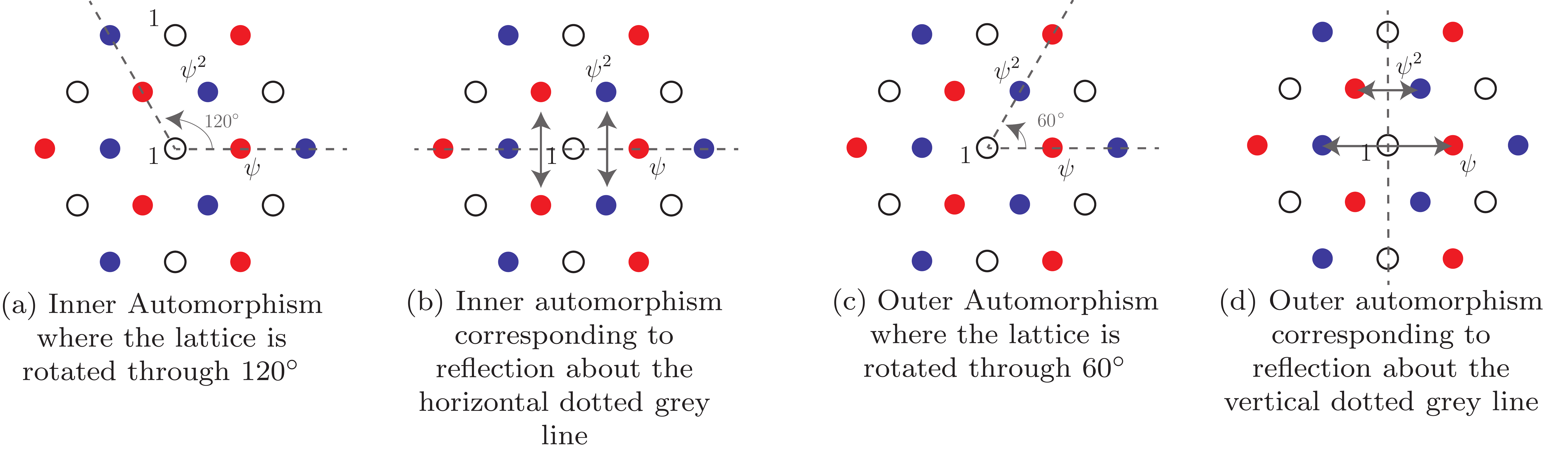}
\caption{Anyon lattice of $su(3)$ with inner and outer automorphisms. The white, red and blue circles refer to the distinct quasiparticles $1$, $\psi$ and $\psi^2$ respectively in the anyon quotient lattice $\mathbb{Z}^2/K\mathbb{Z}^2$. (a) and (b) are examples of inner automorphisms where anyon labels (colors of the circles) are preserved whereas (c) and (d) the outer automorphisms exchange $\psi$ and $\psi^2$ (blue and red circles)}
\label{fig:su3lattice}
\end{figure}

Sometimes the bilayer and conjugation symmetries can corresponds to the same anyon relabeling action. We take the $SU(3)_1$ state for example where the $K$-matrix is given by the Cartan matrix of the Lie group and is identical to that of the $(2,2,-1)$-FQH state. The anyon lattice can be represented by a triangular lattice (see figure~\ref{fig:su3lattice}). As $|\det(K)|=3$, there are three quasiparticle types $1,\psi,\psi^2$ represented by the vectors $0,{\bf e}_1,{\bf e}_2$ respectively. 

The group of automorphisms $\mbox{Aut}(K)$ in can be identified with the dihedral group $Dih_6$, the symmetry group of a hexagon. It is is generated by a sixfold ``rotation" $R_6$ and a twofold ``mirror" $M$ \begin{align}\mbox{Aut}\left(K_{su(3)}\right)=\left\langle\left. R_6=\left(
\begin{array}{cc} 0 & -1 \\ 1 & 1 \end{array}\right),M=\left(\begin{array}{cc} 0 & 1 \\ 1 & 0 \\\end{array}\right)\right|R_6^6=M^2=1,MR_6M^{-1}=R_6^{-1}\right\rangle.\end{align} We notice that these matrices are isometries with respect to the $K$-matrix for $su(3)$, $MKM^T=R_6KR_6^T=K$. Also, both $M$ and $R_6$ act on anyon labels by taking $e\leftrightarrow e^2$. They can be visualized  geometrically in Fig.\eqref{fig:su3lattice}(c) and (d) respectively. On the other hand the matrices $R_6M,R_6^2$ preserve anyon labels up to local particles, and therefore generate the group of inner automorphisms. $R_6^2$ and $RP^3$ are geometrically represented in Fig\ref{fig:su3lattice}(a) and (b) respectively. 
\begin{align}\mbox{Inner}\left(K_{su(3)}\right)=\left\langle R_6M,R_6^2\right\rangle=\mathbb{Z}_2\ltimes\mathbb{Z}_3=S_3.\end{align} The quotient \begin{align}\mbox{Outer}\left(K_{su(3)}\right)=\frac{\text{Aut}(K)}{\text{Inner}(K)}=\frac{Dih_6}{S_3}=\mathbb{Z}_2=\{1,\sigma=[M]=[R_6]\}\end{align} describes the twofold symmetry of the topological state, $\psi\leftrightarrow\psi^2$. For instance, the conjugation symmetry $C$ is the $\pi$-rotation of the anyon lattice and is given by $C=R_6^3$, which belongs to the same equivalent class $\sigma=[M]$.

It is also interesting to notice that the $\mathbb{Z}_3$ gauge theory with $K=3\sigma_x$ is identical to the time reversal doublet $SU(3)_1\otimes\overline{SU(3)_1}$ with $K=K_{SU(3)_1}\otimes\sigma_z=K_{SU(3)_1}\oplus K_{\overline{SU(3)_1}}$ with $K_{\overline{SU(3)_1}}=-K_{SU(3)_1}$. They both have nine quasiparticle types and are identified by equating $\psi=em$ and $\overline\psi=em^2$, where $e$ and $m$ are charge and flux of the $\mathbb{Z}_3$ gauge theory, and $\psi$ and $\overline\psi$ generate $SU(3)_1$ and $\overline{SU(3)_1}$ respectively. The electric-magnetic symmetry $e\leftrightarrow m$ in the gauge theory is equivalent to the bilayer symmetry in $\overline{SU(3)_1}$ while the conjugation symmetry $e\leftrightarrow e^2$, $m\leftrightarrow m^2$ is the combination of the bilayer symmetries in both $SU(3)_1$ and $\overline{SU(3)_1}$.

\begin{figure}[htbp]
	\centering\includegraphics[width=0.6\textwidth]{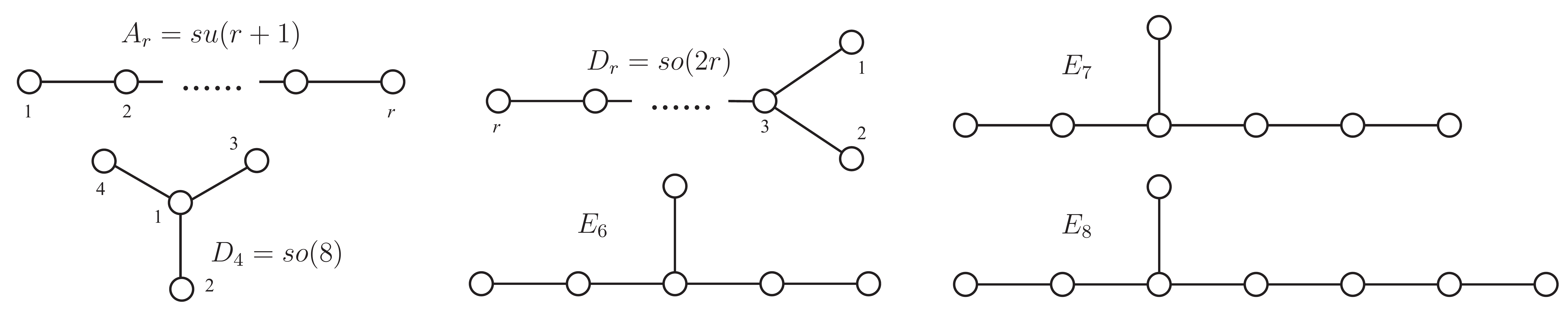}
	\caption{Dynkin Diagrams of simply-laced Lie algebras. The $A_r$, $D_r$ and $E_6$ diagrams are mirror symmetric. $D_4$ is $S_3=Dih_3$ symmetric.}\label{fig:Dynkin}
\end{figure}

There are series of topological states with Lie gauge groups. The simplest ones are given by simply-laced Lie groups, and at level 1, they form a class of $ADE$-states~\cite{mathieu1997conformal, FrohlichThiran94, FrohlichStuderThiran97, khan2014}. The $A$ and $D$-series give infinite sequences of states $A_r=SU(r+1)$ and $D_r=SO(2r)$. $E_6$, $E_7$ and $E_8$ are exception Lie groups. The $K$-matrices are identified with the Cartan matrix of the Lie algebras, which are summarized by the Dynkin diagrams (see figure~\ref{fig:Dynkin}). The nodes of the a Dynkin diagram are labeled by integers $I=1,2,\ldots,r$ for a rank $r$ Lie algebra. The Cartan matrix has dimension $r$ and diagonal entries 2. The off-diagonal entries $K_{IJ}$ are zero unless the nodes $I$ and $J$ are connected by a link, in which case $K_{IJ}=-1$. Simply-laced Lie algebras thus corresponds to series of Abelian topological states, whose boundary carries a chiral Kac-Moody current algebra (also known as a Wess-Zumino-Witten(WZW) conformal field theory (CFT)) of the Lie algebra at level 1~\cite{bigyellowbook}.

The symmetries of the Dynkin diagrams corresponds to anyonic symmetries of the Abelian topological states. In the $A_r$-series, there is a left-right mirror symmetry $I\to r-I$. As a consequence, $K_{SU(r+1)}$ is invariant under the symmetry $M_{IJ}=\delta_{I,r-J}$. This corresponds to the conjugation symmetry of the $SU(r+1)_1$-state. The $SU(3)_1$ case, for instance, was discussed above in the context of a bilayer FQH state. A $D_r$ topological state always carries four quasiparticle types, $1,\psi,s_+,s_-$, where $\psi$ is fermionic and $s_\pm$ are non-local $\pi$-fluxes with respect to the fermion. The mirror symmetry $M$ that flips node 1 and 2 in the $D_r$ Dynkin diagram corresponds to the $\mathbb{Z}_2$ symmetry that interchanges the fermion parities of $\pi$-fluxes, $M:s_+\leftrightarrow s_-$ where $s_+=s_-\times\psi$. For instance, $SO(16)_1$ has the same topological order as the Kitaev toric code (i.e.~$\mathbb{Z}_2$ gauge theory) except for the difference in chiral central charge $c_-=8$. The mirror symmetry is identical to the electric-magnetic symmetry in the toric code as $s_\pm$ are identified with the charge $e$ and flux $m$. The mirror symmetry of $E_6$ corresponds to the conjugation symmetry that interchange the two non-trivial anyon types $\psi\to\overline{\psi}$. $E_7$ and $E_8$ do not have non-trivial anyonic symmetry because $E_7$ has only two quasiparticle types, $1,\psi$, both self-conjugate, while $E_8$ has a trivial topological order with no non-trivial anyonic excitations. The $K$-matrix for $E_8$ is the smallest even unimodular quadratic form (up to equivalence) with non-vanishing signature. It can be combined with the $K$-matrix of any Abelian topological phase, $K\to K\oplus K_{E_8}$, without changing its topological order except increasing the chiral central charge $c_-$ by $8$.

\subsubsection{Triality symmetry in \texorpdfstring{$so(8)_1$}{so(8)}}\label{sec:symmSO(8)}
Now we move onto an Abelian topological state that carries a non-Abelian anyonic symmetry~\cite{khan2014, BarkeshliBondersonChengWang14, TeoHughesFradkin15}. The $SO(8)_1$ fractional quantum Hall state is described by a four-component Chern-Simons effective action \eqref{CSaction} with \begin{align}K_{SO(8)_1}=\left(\begin{array}{*{20}c}2&-1&-1&-1\\-1&2&0&0\\-1&0&2&0\\-1&0&0&2\end{array}\right).\label{so(8)Kmatrix}\end{align} The $K$-matrix is identical to the Cartan matrix of the Lie algebra $D_4=so(8)$ whose Dynkin diagram is shown in figure~\ref{fig:Dynkin}. As a result, the edge CFT carries a chiral $so(8)$ Kac-Moody structure at level 1~\cite{bigyellowbook, khan2014}. This topological phase has four quasiparticle types $1,\psi_1,\psi_2,\psi_3$, where $1$ is the local bosonic vacuum, and the $\psi_i$ are all fermions with mutual semionic statistics: $\mathcal{D}S_{\psi_i\psi_j}=-1$ for $i\neq j$. The fermions obey the fusion properties \begin{align}\psi_i^2=1,\quad\psi_1\times\psi_2\times\psi_3=1.\label{so(8)fusion}\end{align}

Since all non-trivial quasiparticles have the same spin, and the fusion rules are also invariant under permutation, the $SO(8)_1$ state has a triality anyonic symmetry $S_3$ that permutes the three fermions $\psi_i$. The triality symmetry is also apparent from the Dynkin diagram \ref{fig:Dynkin} where the $S_3$ symmetry permutes the $2^{\mathrm{nd}}$, $3^{\mathrm{rd}}$ and $4^{\mathrm{th}}$ external nodes.  The permutation group $S_3=\mathbb{Z}_2\ltimes\mathbb{Z}_3$ is generated by a twofold reflection $\sigma_1$ and a threefold rotation $\rho$ represented by \begin{align}\sigma_1=\left(\begin{array}{*{20}c}1&0&0&0\\0&1&0&0\\0&0&0&1\\0&0&1&0\end{array}\right),\quad \rho=\left(\begin{array}{*{20}c}1&0&0&0\\0&0&0&1\\0&1&0&0\\0&0&1&0\end{array}\right).\end{align} The reflection generator $\sigma_1$ interchanges $\psi_2\leftrightarrow\psi_3$ while fixing $\psi_1$, and $\rho$ (or $\rho^2$) cyclicly rotates $\psi_i\to\psi_{i+1}$ (resp.~$\psi_i\to\psi_{i-1}$). The other two reflections are defined as \begin{align}\sigma_2=\sigma_1\rho,\quad\sigma_3=\sigma_1\rho^2,\end{align} and they fix $\psi_2$ and $\psi_3$ respectively, while interchanging the remaining two fermions.

This topological state is proposed to appear on the gapped surface of a $(3+1)$D topological paramagnet (the type-II $efmf$ state)~\cite{WangPotterSenthil13,Senthil2014} and a bosonic topological insulator or symmetry protected topological state ~\cite{BurnellChenFidkowskiVishwanath13, VishwanathSenthil12}. The $SO(8)_1$-state belongs to a sixteenfold periodic class of topological states~\cite{Kitaev06, TeoHughesFradkin15} and can be modeled by eight copies of chiral Ising states (the Kitaev honeycomb model in its $B$-phase) after the condensation of fermion pairs. It can also be constructed by a coupled Majorana wire model~\cite{SahooZhangTeo15}.

Although the theory is chiral so that its boundary carries a chiral CFT with central charge $c_-=4$, it can be doubled and embedded in an exactly solvable lattice model. As eluded in section~\ref{sec:symmlatticemodels}, the tri-color spin model~\cite{BombinMartin06,TeoRoyXiao13long} (with $k=2$) and the bilayer toric code~\cite{MesarosKimRan13} have identical topological order with the $SO(8)_1\otimes\overline{SO(8)_1}$ state. For example the fermions $\psi_i$, $\overline{\psi}_i$ can be identified with the combinations of plaquette excitations in the tri-color model \begin{gather}\psi_1=Y_\bullet R_\circ,\quad\psi_2=R_\bullet B_\circ,\quad\psi_3=B_\bullet Y_\circ,\quad\overline{\psi}_1=R_\bullet Y_\circ,\quad\overline{\psi}_2=B_\bullet R_\circ,\quad\overline{\psi}_3=Y_\bullet B_\circ.\end{gather} The $S_3$-symmetries \eqref{Lambda3} and \eqref{LambdaB} of the tri-color model are identified with the triality symmetries of both $SO(8)_1$ and $\overline{SO(8)_1}$ sectors \begin{align}\Lambda_3=\rho\otimes\overline\rho,\quad\Lambda_B=\sigma_3\otimes\overline{\sigma_3}.\end{align} These are however not all the symmetries of the doubled state. For example the triality symmetries could act independently and differently on the two time reversal sectors. This gives the symmetry group $S_3\times S_3$ which contains the lattice symmetries of the tri-color model as its diagonal subgroup. Moreover, there is an additional flip symmetry the sends $\psi_i\leftrightarrow\overline{\psi}_i$ or equivalently $(Y_\bullet,R_\bullet,B_\bullet)\leftrightarrow(Y_\circ,R_\circ,B_\circ)$. This symmetry can only be present in a closed system with no boundary as it switches between $SO(8)_1\leftrightarrow\overline{SO(8)_1}$, which have opposite chiral central charge. It can only be represented in the $K$-matrix formalism if the $K$-matrix is extended by an $E_8$ state so that the state is stably equivalent to $SO(8)_1\otimes SO(8)_1$. If present, the global symmetry group of the tri-color model is extended to $(S_3\times S_3)\rtimes\mathbb{Z}_2$.

\subsection{Global symmetries in non-Abelian topological phases}\label{sec:globalsymmetrynab}
Non-abelian topological states do not have an effective field theory description similar to \eqref{CSaction}. Anyonic symmetries are abstract symmetries in modular tensor categories~\cite{EtingofNikshychOstrik10, BarkeshliBondersonChengWang14, TeoHughesFradkin15, TarantinoLindnerFidkowski15}. It is out of the scope of this review to cover this general description in detail. Loosely speaking anyonic symmetry in the non-Abelian setting is a global symmetry $M$ that leaves the fusion and exchange statistics unchanged, i.e.~\begin{align}N_{\bf ab}^{\bf c}=N_{(M{\bf a})(M{\bf b})}^{M{\bf c}},\quad\theta_{\bf a}=\theta_{M{\bf a}}\end{align} where $N_{\bf a}=(N_{\bf ab}^{\bf c})$ is the fusion matrix that counts fusion degeneracies ${\bf a}\times{\bf b}=\sum_{\bf c}N_{\bf ab}^{\bf c}{\bf c}$ and $\theta_{\bf a}=e^{2\pi ih_{\bf a}}$ is the exchange phase (or $2\pi$-twist) of anyon ${\bf a}$. The ribbon identity \eqref{ribbonapp} then ensures that the modular $S$-matrix \eqref{braidingSapp} is also left invariant \begin{align}S_{(M{\bf a})(M{\bf b})}=S_{\bf ab}.\end{align}

The simplest example of anyon symmetry in a non-Abelian state is the flip symmetry in a bilayer Ising theory. The topological state is a tensor product and contains anyons $\{1,\psi_1,\sigma_1\}\otimes\{1,\psi_2,\sigma_2\}$, where $\psi_i$ are Abelian fermions in each sector and $\sigma_i$ are the Ising anyons that satisfy the fusion rules $\sigma\times\sigma=1+\psi$ and $\psi\times\sigma=\sigma$, and has spin $h_\sigma=1/16$. The bilayer symmetry simply switches $\psi_1\leftrightarrow\psi_2$ and $\sigma_1\leftrightarrow\sigma_2$. Like the $SO(8)_1$ state, global symmetries can also be non-commutative in a non-Abelian topological state. An example will be given below.

\subsubsection{Triality symmetry in the chiral 4-state Potts model}\label{sec:trialitysymmetry}
The 4-state Potts model is a 2D classical statistical model. At the self-dual critical point, it can be mapped to a $(1+1)$D rational conformal field theory, whose chiral sector has a $Dih_2$-orbifold structure of $su(2)_1$~\cite{Ginsparg88,DijkgraafVafaVerlindeVerlinde99,CappelliAppollonio02}. By the bulk-boundary correspondence, this associates a $(2+1)$D non-Abelian topological phase, which we are going to denote by \begin{align}``4-\mathrm{Potts}"=SU(2)_1/Dih_2\end{align} where the quotation marks indicate that it is referring to the $(2+1)$D bulk.

\begin{table}[htbp]
\centering
\begin{tabular}{lll}
Anyons $\chi$&Quantum Dim $d_\chi$& Conformal Dim $h_\chi$\\\hline
1&1&0\\
$j_1,j_2,j_3$&1&1\\
$\Phi$&2&1/4\\
$\sigma_1,\sigma_2,\sigma_3$&2&1/16\\
$\tau_1,\tau_2,\tau_3$&2&9/16\\
\end{tabular}
\caption{The spins (encoded as conformal dimensions)  $h_\chi$ and quantum dimensions $d_\chi$ of anyons in the $S_3$-symmetric parent state $``SU(2)_1/Dih_2"=\mbox{``4-Potts"}$.}\label{tab:4statePottsanyons}
\end{table}
The topological state under consideration here has an anyon content that corresponds to the 11 primary fields in one of the chiral sectors of the 4-state Potts model~\cite{DijkgraafVafaVerlindeVerlinde99, CappelliAppollonio02}. The $(1+1)D$ boundary of the (2+1)D topological parent state we want to consider is characterized by the orbifold CFT $SU(2)_1/Dih_2$~\cite{Ginsparg88, Harris88}, where $Dih_2$ is the (double cover) group of $\pi$-rotations about $x,y,z,$ which are a subgroup in the continuous 3D rotation group $SU(2)$. The spins (encoded as conformal dimensions) and the quantum dimensions of the anyons in this theory are listed in Table~\ref{tab:4statePottsanyons}. 

The fusion rules are generated by \begin{gather}j_a\times j_a=1,\quad j_a\times j_{a\pm1}=j_{a\mp1}\nonumber\\\Phi\times j_a=\Phi,\quad\Phi\times\Phi=1+j_1+j_2+j_3\nonumber\\\sigma_a\times\sigma_a=1+j_a+\Phi\label{4statePottsfusion}\\\sigma_a\times\sigma_{a\pm1}=\sigma_{a\mp1}+\tau_{a\mp1}\nonumber\\\sigma_a\times j_a=\sigma_a,\quad\sigma_a\times j_{a\pm1}=\tau_a,\quad\sigma_a\times\Phi=\sigma_a+\tau_a\nonumber\end{gather} where $a=1,2,3$ modulo $3\mathbb{Z}$. The $F$-symbols that generate basis transformations can be evaluated (up to gauge transformations) by solving the hexagon and pentagon equation (Eq.!\eqref{hexagoneq} and Fig.~\ref{fig:Fpentagon}). In particular we choose \begin{gather}F^{j_\mu j_\nu j_\lambda}_{j_\mu\times j_\nu\times j_\lambda}=F^{j_a\Phi j_a}_\phi=F^{\Phi j_a\Phi}_{j_a}=1\nonumber\\F_\Phi^{\Phi j_aj_a}=F_\Phi^{j_aj_a\Phi}=F_{j_a}^{j_a\Phi\Phi}=F_{j_a}^{\Phi\Phi j_a}=1\nonumber\\F^{j_a\Phi j_{a\pm1}}_\Phi=F^{\Phi j_a\Phi}_{j_{a\pm1}}=-1\label{4PottsFsymbols}\\F_\Phi^{\Phi\Phi\Phi}=\frac{1}{2}\left(\begin{smallmatrix}1&1&1&1\\1&1&-1&-1\\1&-1&1&-1\\1&-1&-1&1\end{smallmatrix}\right),\quad F^{\sigma_a\sigma_a\sigma_a}_{\sigma_a}=\frac{1}{2}\left(\begin{smallmatrix}1&1&\sqrt{2}\\1&1&-\sqrt{2}\\\sqrt{2}&-\sqrt{2}&0\end{smallmatrix}\right)\nonumber\\F_\Phi^{\Phi j_aj_{a\pm1}}=\left(F_\Phi^{j_aj_{a\pm1}\Phi}\right)^{-1}=F_{j_a}^{j_{a\pm1}\Phi\Phi}=\left(F_{j_{a\pm1}}^{\Phi\Phi j_a}\right)^{-1}=\pm i\nonumber\end{gather} where the rows and columns for $F^{\Phi\Phi\Phi}_\Phi$ are arranged according to the internal fusion channels $\{1,j_1,j_2,j_3\}$ of $\Phi\times\Phi,$ and those for $F^{\sigma_a\sigma_a\sigma_a}_{\sigma_a}$ are arranged according to $\{1,j_a,\Phi\}$. These will be useful in understanding the defect fusion category later.

The modular $S$-matrix that characterizes braiding can be generated from the spin and fusion properties via Eq.~\eqref{braidingSapp} and is given by~\cite{DijkgraafVafaVerlindeVerlinde99, CappelliAppollonio02}:
\begin{align}S=\frac{1}{\mathcal{D}_0}\left(
\begin{smallmatrix}
 1 & 1 & 1 & 1 & 2 & 2 & 2 & 2 & 2 & 2 & 2\\
 1 & 1 & 1 & 1 & 2 & 2 & -2 & -2 & 2 & -2 & -2\\
 1 & 1 & 1 & 1 & 2 & -2 & 2 & -2 & -2 & 2 & -2\\
 1 & 1 & 1 & 1 & 2 & -2 & -2 & 2 & -2 & -2 & 2\\
 2 & 2 & 2 & 2 & -4 & 0 & 0 & 0 & 0 & 0 & 0\\
 2 & 2 & -2 & -2 & 0 & \sqrt{8} & 0 & 0 & -\sqrt{8} & 0 & 0\\
 2 & -2 & 2 & -2 & 0 & 0 & \sqrt{8} & 0 & 0 & -\sqrt{8} & 0\\
 2 & -2 & -2 & 2 & 0 & 0 & 0 & \sqrt{8} & 0 & 0 & -\sqrt{8}\\
 2 & 2 & -2 & -2 & 0 & -\sqrt{8} & 0 & 0 & \sqrt{8} & 0 & 0\\
 2 & -2 & 2 & -2 & 0 & 0 & -\sqrt{8} & 0 & 0 & \sqrt{8} & 0\\
 2 & -2 & -2 & 2 & 0 & 0 & 0 & -\sqrt{8} & 0 & 0 & \sqrt{8}\\
\end{smallmatrix}
\right)\label{4statePottsSmatrix}\end{align}
where the total quantum dimension is $\mathcal{D}_0=4\sqrt{2}$, and the entries are arranged to have the same order as the anyon listed in Table~\ref{tab:4statePottsanyons}.

The chiral ``4-state Potts" phase is $S_3$-symmetric because the fusion, spin, and braiding properties are invariant under the simultaneous permutation of $\{j_1,j_2,j_3\}$, $\{\sigma_1,\sigma_2,\sigma_3\},$ and $\{\tau_1,\tau_2,\tau_3\}$. The threefold ($\theta$) and twofold ($\alpha_a$) generators of the group respectively relabel \begin{align}\begin{array}{*{20}c}\theta:(j_a,\sigma_a,\tau_a)\to(j_{a+1},\sigma_{a+1},\tau_{a+1})\\\alpha_a:(j_{a\pm1},\sigma_{a\pm1},\tau_{a\pm1})\to(j_{a\mp1},\sigma_{a\mp1},\tau_{a\mp1})\end{array}\label{S3operation4Potts}\end{align} while fixing the other anyons.

\subsection{Classification and obstructions of quantum symmetries}\label{sec:classificationsymmetry}
Anyonic symmetries are quantum symmetries that act not only as permutations of anyon labels but also as operations on quantum states. These quantum operations should obey certain additional consistency requirements, and there could be multiple inequivalent sets of such quantum operations. The classification and obstruction to these quantum symmetries can be systematically characterized under a mathematical framework using group cohomologies~\cite{groupcohomologybook,EtingofNikshychOstrik10,BarkeshliBondersonChengWang14,TeoHughesFradkin15,TarantinoLindnerFidkowski15}. These relate to projective symmetry group~\cite{Wenspinliquid02} and non-symmorphic symmetries of the anyon lattice~\cite{TeoHughesFradkin15} as well as symmetry protected phases~\cite{ChenGuLiuWen11, GuWen12, ChenGuLiuWen12, LuVishwanath13, ChenBurnellVishwanathFidkowski14} and symmetry fractionalization~\cite{BarkeshliBondersonChengWang14,TarantinoLindnerFidkowski15}. The complete general description would require a review entirely dedicated to the task and will not be presented here. However, we will discuss one classical example and see how the same anyonic relabeling action can have multiple distinct quantum realizations.

We consider the bosonic Abelian $U(1)_n$ topological state with $K=2n$. (Here we adopt the level $n$ convention according to one commonly used in conformal field theory contexts~\cite{bigyellowbook}.) It is identical to the bosonic Laughlin FQH state with filling $\nu=1/2n$. However we will not require the topological state to preserve $U(1)$ charge conservation. In the case when $n=2m$ and $m$ is odd, the Abelian topological state also describes the fermion Laughlin FQH state with filling $\nu=1/m$ and coupled with a $\mathbb{Z}_2$ gauge theory so that $\pi$-fluxes with $\Phi=hc/2e$ are deconfined and fractionalizes the Laughlin $q=1/m$ quasiparticle, which is a $2\pi$-flux with $\Phi=hc/e$. The $K=2n$ state carries the anyon types $\psi^a$, for $a=-n+1,\ldots,-1,0,1,\ldots,n$, and each carries spin $h_{\psi^a}=a^2/4n$. They satisfy the fusion rule $\psi^a\times\psi^b=\psi^c$ for $c\equiv a+b$ mod $2n$.

When charge symmetry is broken, the $\mathbb{Z}_2$ conjugation symmetry flips $\sigma:\psi^a\leftrightarrow\psi^{-a}$ and fixes the self-conjugate $1=\psi^0$ and $\psi^n$. We are interested in the {\em quantum} action of the symmetry $\hat\sigma$ on anyon states $|a\rangle$. Loosely speaking, $|a\rangle$ can be thought of as a state with a local excitation $\psi^a$. However, the non-trivial anyonic statistics of $\psi^a$ requires the system to carry its antipartner $\psi^{-a}$ somewhere else. Nevertheless we are only interested in the {\em local} quantum action of $\hat\sigma$ on $\psi^a$ alone. There are ways to make this more precise. For example, instead of local anyonic excitation, one can take a closed torus geometry and from using the one-to-one correspondence between anyon types and the degenerate ground states, the state $|a\rangle$ labels the ground state the carries a $\psi^a$ flux across one of the cycle of the torus. The quantum action of $\hat\sigma$ can be made precise on anyonic excitations by realizing the quantum state of a particular excitation configuration can be represented by a splitting tree (see figure~\ref{fig:Fmoves}), which consists of anyon labeled branches as well as possibly degenerate trivalent splitting vertices. The overall quantum action $\hat\sigma$ then can be derived from the product of local actions on the components. Here we ocus only on the local action of $\hat\sigma$ on $|a\rangle$ and assume for simplicity that the action on splitting vertices is trivial.

The quantum operation \begin{align}\hat\sigma:|a\rangle\longrightarrow e^{i\varphi(a)}|-a\rangle\end{align} is generically gauge dependent when $\psi^a\neq\psi^{-a}$. One can consider the square operation \begin{align}\hat\sigma\hat\sigma:|a\rangle\longrightarrow e^{i\phi(a)}|a\rangle,\quad\phi(a)=\varphi(a)+\varphi(-a).\end{align} As we have assumed that $\hat\sigma$ acts trivially on the fusion operator of $\psi^a\times\psi^b\to\psi^c$, it must therefore preserves fusion \begin{align}e^{i\phi(a)}e^{i\phi(b)}=e^{i\phi(c)}.\end{align} In other words one can find an anyon $p=-n+1,\ldots,n$ such that $e^{i\phi(a)}=e^{2\pi ipK^{-1}a}$.

There are consistency requirements the phases $\phi(a)$ have to satisfy. Associativity $\hat\sigma(\hat\sigma\hat\sigma)=(\hat\sigma\hat\sigma)\hat\sigma$ requires the phases $e^{i\phi(a)}$ and $e^{i\phi(-a)}$ to be identical. Moreover, they must also be conjugate to each other as $\psi^a\times\psi^{-a}=1$. These force $e^{i\phi(a)}=\pm1$. There are therefore two distinct scenarios \begin{align}e^{i\phi(a)}=1\quad\mbox{or}\quad(-1)^a=e^{2\pi inK^{-1}a}.\end{align} These two cases represents two distinct quantum symmetries. For instance, when $K=2$, although the $\sigma$ symmtry is not relabeling anyons, the non-trivial phase corresponds to the projective symmetry~\cite{Wenspinliquid02} of the $SU(2)_1=U(1)_1$ state.

For a general global symmetry group $G$, there are quantum phase degrees of freedom $e^{i\phi_{M,N}({\bf a})}$ that tell the difference between the composite action $\hat{M}\hat{N}$ from $\widehat{MN}$ on the anyon state $|{\bf a}\rangle$. Gauge inequivalent choices of $e^{i\phi_{M,N}({\bf a})}$ consistent with associativity $L(MN)=(LM)N$ are classified by the second group cohomology $H^2(G,\mathcal{B}^\times)$ where $G$ acts on the group of Abelian anyons $\mathcal{B}^\times$ by the relabeling action. In the previous case, $G=\mathbb{Z}_2$ and $\mathcal{B}^\times=\mathbb{Z}_{2n}$. Quantum symmetries are therefore $H^2(\mathbb{Z}_2,\mathbb{Z}_{2n})=\mathbb{Z}_2$ classified. There are also cases when the anyon relabeling symmetry cannot be promoted to a true quantum symmetry. This obstruction originates from the non-trivial symmetry action of fusion operation ${\bf a}\times{\bf b}\to{\bf c}$. It is classified by the third group cohomology $H^3(G,\mathcal{B}^\times)$ and the symmetry is not obstructed only when the quantum information associates a trivial cohomology element. This is out of the scope of this review and we refer the readers to Ref.~\cite{EtingofNikshychOstrik10,BarkeshliBondersonChengWang14,TeoHughesFradkin15,TarantinoLindnerFidkowski15}.

\section{Twist defects}\label{sec:defect}
\begin{figure}[htbp]
\centering\includegraphics[width=0.3\textwidth]{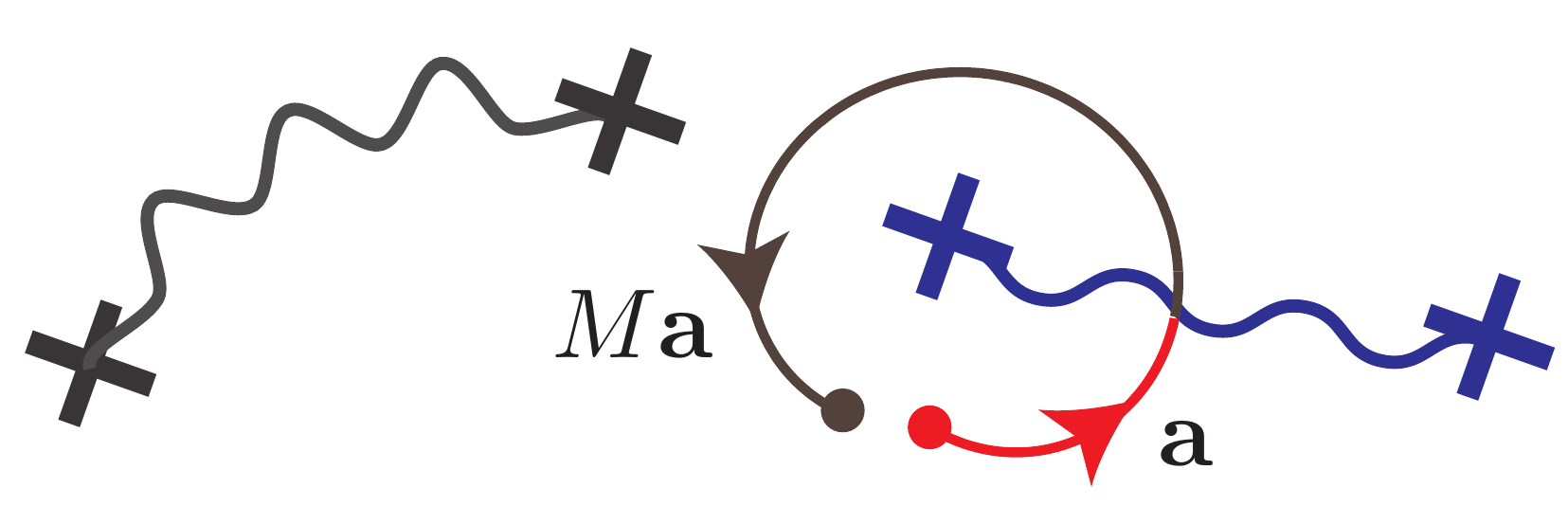}
\caption{Twist defects (crosses) connected by arbitrary branch cuts (curvy lines) where a passing anyon changes type ${\bf a}\to M{\bf a}$ according to an anyonic symmetry $M$.}\label{fig:defectanyoncuts}
\end{figure}

Now that we have discussed anyonic symmetries we will proceed to the construction of twist defects, which are semiclassical static fluxes of anyonic symmetries. There have been numerous recent developments in the theory of twist defects in topological phases~\cite{Kitaev06, EtingofNikshychOstrik10, barkeshli2010, Bombin, Bombin11, KitaevKong12, kong2012A, YouWen, YouJianWen, PetrovaMelladoTchernyshyov14, BarkeshliQi, BarkeshliQi13, BarkeshliJianQi, MesarosKimRan13, TeoRoyXiao13long, teo2013braiding, khan2014, BarkeshliBondersonChengWang14, TeoHughesFradkin15, TarantinoLindnerFidkowski15}. A twist defect is a semiclassical topological point defect (with an attached branch-cut) labeled by an anyonic symmetry $M$. It is characterized by its action on anyons that encircle the defect. A quasiparticle will change type according to the symmetry operation $M:{\bf a}\to M{\bf a}$ when it travels once, counter-clockwise, around the defect (see Fig.~\ref{fig:defectanyoncuts}). In a system with a finite number of defects, there exists a quasi-global definition of anyon labels that covers the system almost everywhere in space except along certain {\em branch cuts} between defects where the anyon label definition changes (see Fig.~\ref{fig:defectanyoncuts}).

Unlike anyonic quasiparticles, topological twist defects are not dynamical excitations of a quantum Hamiltonian. They are classical configurations or textures that vary slowly away from the defect points/cores. For example, in the absence of vortices the phase of an $s$-wave superconductor order parameter is locally uniform, but the phase winds by $2\pi$ around a flux vortex. A distortion in the defect texture in two dimensions usually generates a confining potential between defect partners that grows at least logarithmically in their separation. Therefore, one would not expect unbound defect pairs to appear spontaneously at long length scales. 

Moreover, as a twist defect permutes the anyon type of an orbiting quasiparticle, the Wilson string along the anyon trajectory around the defect does not close back to itself. Therefore unlike abelian anyons which can be locally detected by small Wilson loops, there are no local Wilson observables detecting a twist defect state. This non-locality is a central theme of many non-abelian anyons, such as vortex-bound Majorana fermions in chiral $p+ip$ superconductors~\cite{Ivanov,ReadGreen}, Ising anyon in the Kitaev's honeycomb model~\cite{Kitaev06} and Pfaffian fractional quantum Hall state~\cite{MooreRead}. The non-abelian object associated with twist defects however are not fundamental deconfined excitations of a true topological phase. They are qualitatively more similar to (fractional) Majorana excitations at SC-FM heterostructures with (fractional) topological insulators~\cite{FuKane08, LindnerBergRefaelStern, ClarkeAliceaKirill, MChen, Vaezi} or strongly spin-orbit coupled quantum wires~\cite{SauLutchynTewariDasSarma, OregSelaStern13}. Their existence rely on the topological winding of certain classical non-dynamical {\em order parameter field}, such as pairing and spin/charge gap~\cite{TeoKane}. 

In many example models, topological order and discrete spatial order are intertwined, especially in lattice spin or rotor models with topological order. In these cases, twist defects can manifest themselves as lattice defects such as dislocations and disclinations, where the order parameters also associate to translation or rotation symmetry breaking. For example, a dislocation in the toric code~\cite{Kitaev06, Bombin, KitaevKong12} (see Section~\ref{sec:twistdefectA}) switches the anyon type $e\leftrightarrow m$ of a quasiparticle after it travels once around the dislocation~\cite{Bombin}. Other examples include the defects in the $\mathbb{Z}_k$ plaquette model of Wen~\cite{YouWen, YouJianWen, teo2013braiding}, the Kitaev honeycomb model~\cite{Kitaev06, PetrovaMelladoTchernyshyov14} and the color code model~\cite{Bombin11, TeoRoyXiao13long}. Twist defects can also capture the fusion properties of parafermionic zero modes trapped at domain walls of fractional quantum spin Hall edges~\cite{ClarkeAliceaKirill,LindnerBergRefaelStern,MChen,mongg2,Vaezi}. 
Theoretical examples will be given in section~\ref{sec:twistdefectA}. A mathematical framework necessary to describe twist defects will be presented in section~\ref{sec:twistdefectB}.

\subsection{Non-abelian defects in abelian topological states}\label{sec:twistdefectA}

\subsubsection{Dislocations and disclinations in exactly solved lattic models}\label{sec:defectlatticemodel}
Twist defects were initially proposed as lattice defects in exact solvable models. They first appeared as dislocations in the Kitaev's toric code~\cite{Kitaev06, Bombin, KitaevKong12}. They were then generalized in the Wen's plaquette $\mathbb{Z}_k$-rotor model~\cite{YouWen, YouJianWen} and the color code model~\cite{Bombin11,TeoRoyXiao13long}. Certain topological quantities, like the topological entanglement entropy~\cite{KitaevPreskill06,BrownBartlettDohertyBarrett13}, can be evaluated exactly in these defect systems.

\begin{figure}[htbp]
\centering\includegraphics[width=0.6\textwidth]{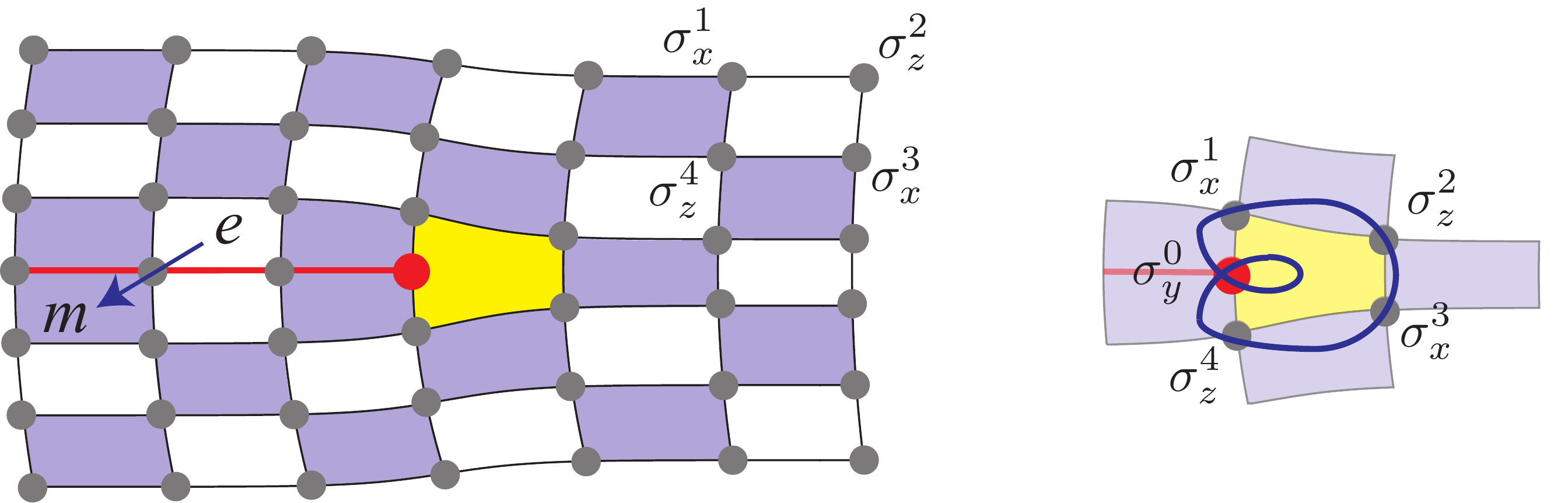}
\caption{Wen plaquette-model (for our purposes this is essentially equivalent to the toric code) with a dislocation (yellow pentagon). The pentagon operator $P_{\pentagon}$ is identical to the Wilson loop (blue) up to a $\pm i$ phase.}
\label{fig:toriccode}
\end{figure}

\begin{figure}[t]
\centering\includegraphics[width=0.8\textwidth]{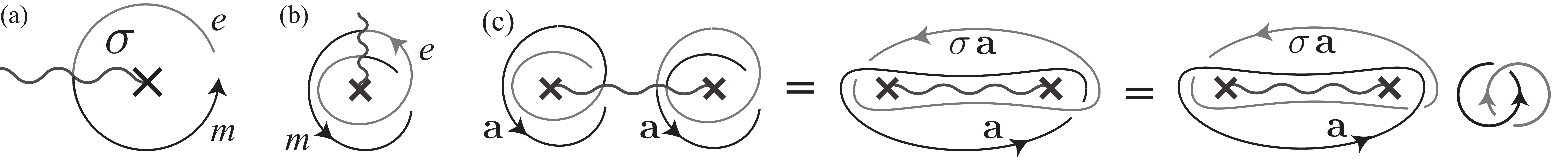}
\caption{(a) A quasiparticle changing type when traveling across a branch cut terminated at a twist defect $\sigma$. (b) The Wilson loop $\Theta$ that distinguishes defect species. (c) Joining two independent local $\Theta$ loops around two defects into two non-contractible Wilson operators around the pair.}\label{fig:defect1}
\end{figure}

We begin with defects in the Kitaev toric code, or equivalently the Wen plaquette model, which was reviewed previously section~\ref{sec:toriccode}. For convenience, we put the spin-$1/2$ degrees of freedom on the vertices of a rectangular checkerboard lattice (see figure~\ref{fig:toriccode}). A twist defect is given by a dislocation of the rectangular lattice. It is characterized by a topological quantity known as the Burgers vector. For the square lattice, a dislocation is a line of atoms (lattice sites) that end at a trivalent vertex adjacent to a pentagon plaquette. The plaquette operator at the pentagon in the Hamiltonian is modified to be $P_{\pentagon}^{\pm}=\pm\sigma_y^0\sigma_x^1\sigma_z^2\sigma_x^3\sigma_z^4$, where the additional $0^{th}$ site is the trivalent vertex, and the sign can be arbitrarily fixed locally at each defect. This operator commutes with all other plaquette operators and the model is still exactly solvable. However, the charge $e$ and flux $m$ quasiparticles can no longer be globally distinguished. This is because the bi-colored checkerboard pattern cannot be globally defined in the presence of a dislocation, and there is a branch cut -- represented by the red line in Fig.~\ref{fig:toriccode} -- originating from the defect where neighboring plaquettes share identical colors. As quasiparticles move diagonally from plaquette to plaquette, they change type across the branch cut according to the electric-magnetic anyon symmetry $e\leftrightarrow m$ (see Figs.~\ref{fig:toriccode} and \ref{fig:defect1}(a)). 

In general, a twist defect in the toric code is a topological defect of the lattice that switches the anyon labels of quasiparticles being dragged adiabatically around the twist defect according to the electric-magnetic symmetry $e\leftrightarrow m$ (see Fig.~\ref{fig:defect1}(a)). It can be constructed from any defect that violates the checkerboard lattice pattern, such as a dislocation or a disclination with an odd-coordinated vertex.  The sign of the defect plaquette operator $P_{\pentagon}^\pm$ corresponds to two distinct defect species, $\sigma_0$ and $\sigma_1$~\cite{TeoRoyXiao13long,teo2013braiding}. The species labels $\lambda=0,1$ can be more generically distinguished by the Wilson loop operator $\Theta$ (see Fig.~\ref{fig:defect1}(b)) formed by dragging either the $e$ or $m$ quasiparticle around the defect \emph{twice} to form a closed loop. For instance the pentagon operator $P_{\pentagon}^\pm$ acts identical to the ground state as the Wilson operator up to a $\pm i$ phase (see the right diagram in figure~\ref{fig:toriccode}). \begin{align}\Theta|GS\rangle=iP_{\pentagon}|GS\rangle=i(-1)^\lambda|GS\rangle.\end{align} This is because the Wilson string self-intersects at the trivalent vertex and gives $\sigma_z^0\sigma_x^0=i\sigma_y^0$.

This Wilson loop operator can therefore be absorbed into the ground state, but with a remaining eigenvalue phase $i$ or $-i$ depending on the defect species. Defect species can also mutate between $0$ and $1$ by absorbing or emitting an $e$ or $m$ quasiparticle. This is because the additional quasiparticle string emanating from a mutated defect will intersect the double loop operator $\Theta$ and contribute a minus sign. The species mutation can be facilitated by the pumping process $H(\varphi)=\cos(\varphi)P_{\pentagon}$, where the defect changes from $\sigma_\lambda\to\sigma_{\lambda+1}$ when the $\varphi$ phase increases by $\pi$. However, adding or subtracting a fermion $\psi$ to/from the defect does not change its species because the $\psi$-string would intersect $\Theta$ twice and give the trivial phase. These results are summarized by the following fusion (or equivalently splitting) rules of the twist defects and quasiparticles
\begin{align}
\sigma_\lambda\times\psi=\sigma_\lambda,\quad\sigma_\lambda\times e=\sigma_\lambda\times m=\sigma_{\lambda+1}.
\label{toriccodedefectfusion1}
\end{align} 
The defect species $\lambda=0,1$ (mod 2) therefore counts a topological charge that corresponds to the quasiparticle bounded at the defect. However as the Wilson loop $\Theta$ or equivalently the pentagon operator $P_{\pentagon}$ cannot distinguish $e$ from $m$ (or $1$ from $\psi$). $\sigma_0\times e$ and $\sigma_0\times m$ are topologically identical and belong to the same defect object $\sigma_1$. 

Next we move on to the ground state degeneracy of a pair of defects. A quasiparticle loop surrounding a pair of defects is non-contractible and therefore can have non-trivial eigenvalues. The available eigenvalues are restricted by the fusion rules. For instance two $e$-loops around the defect pair can join and annihilate themselves because $e\times e=1$. Moreover there are relationships between the $e$ and $m$-loop around a defect pair. They can be turned into each other by leaving behind two $\Theta$ loops, which condense and absorbed by the ground state (see figure~\ref{fig:defect1}(c)). There is no net phase in the process because the two $\pm i$ eigenvalues of $\Theta$ are canceled by the $-1$ linking phase between the $e$ and $m$ loop. This identifies the $e$ and $m$ loop around a defect pair and each can take eigenvalues $\pm1$. This gives the two degenerate ground states, $|1\rangle$ and $|\psi\rangle$, associated with a pair of $\sigma$ defects. In other words, we have the fusion rules \begin{align}\sigma_\lambda\times\sigma_\lambda=1+\psi\end{align} which resembles that of an Ising theory. If the two defects have opposite species labels, then the $\Theta$ loops will have opposite eigenvalues so that the values of the $e$ and $m$ loop around the defect pair will also be opposite. In this case we have the modified fusion rules \begin{align}\sigma_\lambda\times\sigma_{\lambda+1}=e+m\end{align} because the overall channels now have distinct statistical signatures measured by the $e$ and $m$ loop around the defect pair.

In a system with multiple defects, the ground state degeneracy scales as $\sqrt{2}^N$ where $N=2n$ is the total number of dislocations. This is bscause there is a twofold degeneracy associating to each defect pair. The fermion parity $1$ and $\psi$ (or $e$ and $m$ for opposite species) can be locally defined by the $\pm1$ eigenvalue of the $\mathcal{W}^{2i-1,2i}_e$ and $\mathcal{W}^{2i-1,2i}_m$ loop around the $(2i-1)^{\mathrm{th}}$ and $2i^{\mathrm{th}}$ defect. These Wilson loops do not overlap and therefore mutually commute and share simultaneous eigenvalues. The local fermion parities thus label the $2^n$ orthogonal ground states. The degeneracy is protected by another set of Wilson loops $\mathcal{W}^{2i,2i+1}_e$ and $\mathcal{W}^{2i,2i+1}_m$ around the $2i^{\mathrm{th}}$ and $(2i+1)^{\mathrm{th}}$ defect. They intersect once with their neighbors $\mathcal{W}^{2i-1,2i}_{e/m}$ and $\mathcal{W}^{2i+1,2i+2}_{e/m}$, hence flipping the fermion parities of the $(2i-1,2i)$ and $(2i+1,2i+2)$ defect pairs. The $2^n$ ground state manifold forms an irreducible representation of the non-commutative Wilson algebra \begin{align}{\mathcal{W}^{j,j+1}}^\dagger{\mathcal{W}^{j',j'+1}}^\dagger\mathcal{W}^{j,j+1}\mathcal{W}^{j',j'+1}=(-1)^{\delta_{|j-j'|,1}},\quad(\mathcal{W}^{j,j+1})^2=\Theta_j\Theta_{j+1}=\pm1.\label{Wilsonalgebratoriccode}\end{align} 

As all the Wilson operators commute with the Hamiltonian, all $2^n$ states must share the same ground state energy. The degeneracy can only be lifted if some of the Wilson operator are added to the Hamiltonian or by strong perturbation that closes the bulk excitation energy gap and forces a topological phase transition. In other words the degeneracy is protected against weak local perturbations that do not couple spatially separated defects. This property is immensely powerful for the robust storage of quantum information as a quantum state is now stable against local decoherence. 


The dislocation twist defect can be generalized to the $\mathbb{Z}_k$-version of the Kitaev toric code or Wen plaquette model that encodes a $\mathbb{Z}_k$ gauge theory. This can be done simply by replacing the spin operators $\sigma_{x/z}$ by rotors in \eqref{rotors}. The Abelian $D(\mathbb{Z}_k)$ topological state supports anyons of the form of $e^{a_1}m^{a_2}$ for $a_i=0,1,\ldots,k-1$ with the fusion information $e^k=m^k=1$ and the statistical properties $\theta_e=\theta_m=1$, $\mathcal{D}S_{em}=e^{2\pi i/k}$. Dislocations have the same relabeling action $e\leftrightarrow m$ on orbiting anyons. The species label of a defect now runs between $\lambda=0,1,\ldots,k$ as the Wilson operator $\Theta$, whose $k^{\mathrm{th}}$ power is the sign $(-1)^{k-1}$, now have $k$ eigenvalues $e^{\frac{2\pi i}{k}\left(\lambda+\frac{k-1}{2}\right)}$. They count the quasiparticle content bounded at the defect. \begin{align}\sigma_\lambda\times e^{a_1}m^{a_2}=\sigma_{\lambda+a_1+a_2}.\end{align} Similarly the $e$ (or $m$) loop around a defect pair has $k$ distinct eigenvalues and corresponds a $k$-fold ground state degeneracy. In other words the twist defects satisfy the {\em parafermionic} or {\em fractional Ising} fusion structure \begin{align}\sigma_\lambda\times\sigma_{\lambda'}=e^{\lambda+\lambda'}\times\left(1+\psi+\psi^2+\ldots+\psi^{k-1}\right)\end{align} for $\psi=em^{-1}$ the parafermion with $\theta_\psi=e^{-2\pi i/k}$ and $\psi^k=1$. The ground state degeneracy of a system with $N$ defects scale as $\sqrt{k}^N$ and therefore each defect carries a quantum dimension of $d_\sigma=\sqrt{k}$.

\begin{figure}[htbp]
\centering\includegraphics[width=0.8\textwidth]{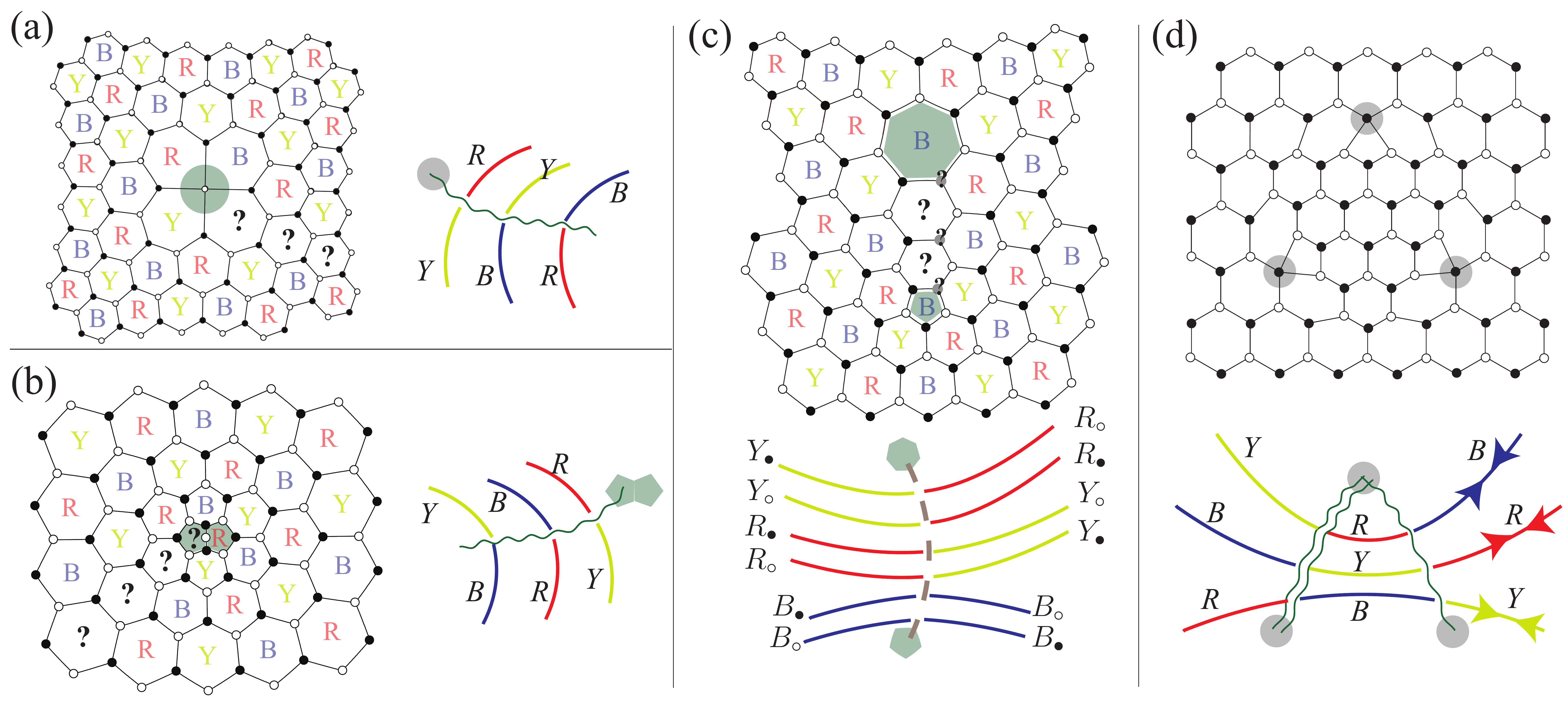}
\caption{Disclination and dislocation twist defects (shaded plaquettes and vertices). Color or sublattice frustrations (question marks) give rise to branch cuts (curly or dashed lines) that alter anyon types (colored lines) across. (a) A threefold twist defect generated by a $+120^\circ$-disclination centered at a tetravalent vertex. Sublattice types $\bullet,\circ$ are not affected by the defect. (b) An anti-threefold twist defect generated by a $-120^\circ$-disclination centered at a bivalent vertex and permutes colors in the opposite direction around. (c) A pair of twofold twist defects generated by a $\pm60^\circ$-disclination dipole. Each violates both tricoloring and bipartite structure. (d) A triplet of dislocation threefold twist defects connected by two color permutation branch cuts (curly lines).}\label{fig:twistdefect}
\end{figure}

The defect structure is more elaborate in the tricolor model described in section~\ref{sec:globalsymmetryab}. For simplicity, we consider the case when $k=2$ so that the lattice model is built out of spins and is topologically equivalent to the bilayer toric code $D(\mathbb{Z}_2\times\mathbb{Z}_2)$ and the time reversal double $SO(8)_1\otimes SO(8)_1$ (see section~\ref{sec:globalsymmetryab}). The general discussion for arbitrary $k$ can be found in Ref.~\cite{TeoRoyXiao13long}. When defined on a honeycomb lattice, the $S_3$ anyonic symmetry is linked with the rotation and translation symmetries of the lattice. The threefold symmetry $\Lambda_3$ in eq.\eqref{Lambda3} is realized as the $120^\circ$ rotation that rotate the plaquette colors about a vertex. It can also be realized as a primitive lattice translation. Twofold symmetry $\Lambda_B$ in eq.\eqref{LambdaB} is related to the sixfold symmetry that switches the $Y$ and $R$ hexagons as well as the $\bullet$ and $\circ$ vertices about a $B$-plaquette. Twist defects in the lattice model can therefore be generated by disclinations and dislocations that associate to the lattice symmetries (see figure~\ref{fig:twistdefect}).

A threefold defect, denoted by $[\Lambda_3]$, changes the color $Y\to R\to B\to Y$ of an anyon circling the defect in the counter-clockwise direction (see figure~\ref{fig:twistdefect}(a,b,d)). Its antipartner $[\overline\Lambda_3]$ swiches the colors in the opposite way. There are three distinct twofold defects $[\Lambda_\chi]$ for $\chi=Y,R,B=1,2,3$ mod 3, each switches the colors $(\chi-1)\leftrightarrow(\chi+1)$ as well as the vertex type $\bullet\leftrightarrow\circ$ of an orbiting anyon (see figure~\ref{fig:twistdefect}(c)). For concreteness, we take $\chi=B$, but the result applies to all colors by cyclic permutation. Similar to twist defects in the Kitaev toric code or the Wen plaquette model, the twofold defect $[\Lambda_B]$ comes with a species label ${\bf l}=(l_\bullet,l_\circ)$ for $l_\bullet,l_\circ=0,1$. The species can be statistically measured by dragging the anyon $R_\bullet$ or $R_\circ$ twice around the defect, which is the analogue of the Wilson operator $\Theta$ in the toric code case (c.f.~figure~\ref{fig:defect1}(b)). Defect species can be changed by emitting or absorbing an additional anyon, and thus ${\bf l}$ counts the quasiparticle charge bounded to the defect. It obey the fusion rule \begin{align}[\Lambda_\chi]_{\bf l}\times[{\bf a}]=[\Lambda_\chi]_{{\bf l}'},\quad\left\{\begin{array}{*{20}l}(l'_\bullet,l'_\circ)=(l_\bullet+1,l_\circ),&\mbox{for ${\bf a}=R_\bullet,Y_\circ$}\\(l'_\bullet,l'_\circ)=(l_\bullet,l_\circ+1),&\mbox{for ${\bf a}=R_\circ,Y_\bullet$}\end{array}\right.,\end{align} and the result for a general anyons ${\bf a}$ can be dervied from above together with associativity as well as symmetry \begin{align}\left([\Lambda_B]_{\bf l}\times[{\bf a}]\right)\times[{\bf b}]=[\Lambda_B]_{\bf l}\times[{\bf a}\times{\bf b}],\quad[\Lambda_B]_{\bf l}\times[{\bf a}]=[\Lambda_B]_{\bf l}\times[\Lambda_B{\bf a}].\end{align} As there are no non-trivial Wilson operators around a threefold defect, there are no species labels associate to $[\Lambda_3]$ and it does not bound quasiparticles. \begin{align}[\Lambda_3]\times[{\bf a}]=[\Lambda_3],\quad[\overline\Lambda_3]\times[{\bf a}]=[\overline\Lambda_3].\end{align}

\begin{figure}[ht]
	\centering
	\includegraphics[width=1\textwidth]{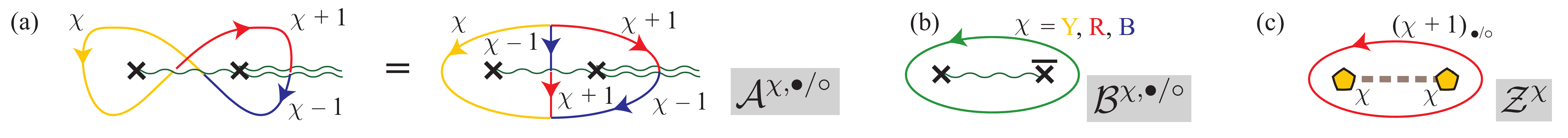}
	\caption{Non-local closed Wilson loops around twist defects. (a) A closed string $\mathcal{A}^{\chi,\bullet/\circ}$, either entirely $\bullet$-type or $\circ$-type, enclosing two threefold twist defects $\rho$ (black crosses). It can be equivalently represented by a loop (left) or a closed branched path with a tricolored source and drain (right). (b) A Wilson loop $\mathcal{B}^{\chi,\bullet/\circ}$ surrounding a threefold twist defect $\rho$ and its antipartner $\overline\rho$. (c) A Wilson loop $\mathcal{Z}^{\chi}$ containing two twofold twist defects of the same color $\sigma_\chi$, $\chi=Y,R,B$ (colored pentagons).}\label{fig:defectWilsonloop}
\end{figure}

In a system of twist defects, the non-local Wilson loops are generated by fundamental ones shown in figure~\ref{fig:defectWilsonloop}. They consist of non-contractible quasiparticle paths that go around a defect pair. The degenerate ground states in a defect system can be labeled by the eigenvalues of a maximal set of mutually commuting Wilson operators. Equivalently, the eigenvalues also represent the different fusion channels as well as fusion degeneracies. For example the eigenvalues of the $\mathcal{Z}^B_{\bullet/\circ}$-loops around a pair of bare $[\Lambda_B]_0$ defects determines the outcome of the fusion \begin{align}[\Lambda_B]_0\times[\Lambda_B]_0=1+Y_\bullet R_\circ+R_\bullet Y_\circ+B_\bullet B_\circ.\end{align} The non-commutative Wilson algebra  \begin{align}\left[\mathcal{A}^{\chi,\bullet},\mathcal{A}^{\chi',\bullet}\right]=\left[\mathcal{A}^{\chi,\circ},\mathcal{A}^{\chi',\circ}\right]=\left\{\mathcal{A}^{\chi,\bullet},\mathcal{A}^{\chi\pm1,\circ}\right\}=0\end{align} generated by loops around a pair of threefold defects associate a fourfold ground state degeneracy and corresponds to the degenerate fusion rule \begin{align}[\Lambda_3]\times[\Lambda_3]=4[\overline\Lambda_3],\quad[\overline\Lambda_3]\times[\overline\Lambda_3]=4[\Lambda_3].\end{align} The eigenvalues of the $\mathcal{B}$-loops around a $[\Lambda_3]$-$[\overline\Lambda_3]$ pair determines the fusion outcome \begin{align}[\Lambda_3]\times[\overline\Lambda_3]=\sum_{\bf a}[{\bf a}].\end{align} Equating the dimensions of the fusion equations, the threefold and twofold defects have quantum dimensions $d_{[\Lambda_3]}=4$ and $d_{[\Lambda_\chi]}=2$.

\begin{figure}[htbp]
\centering\includegraphics[width=0.15\textwidth]{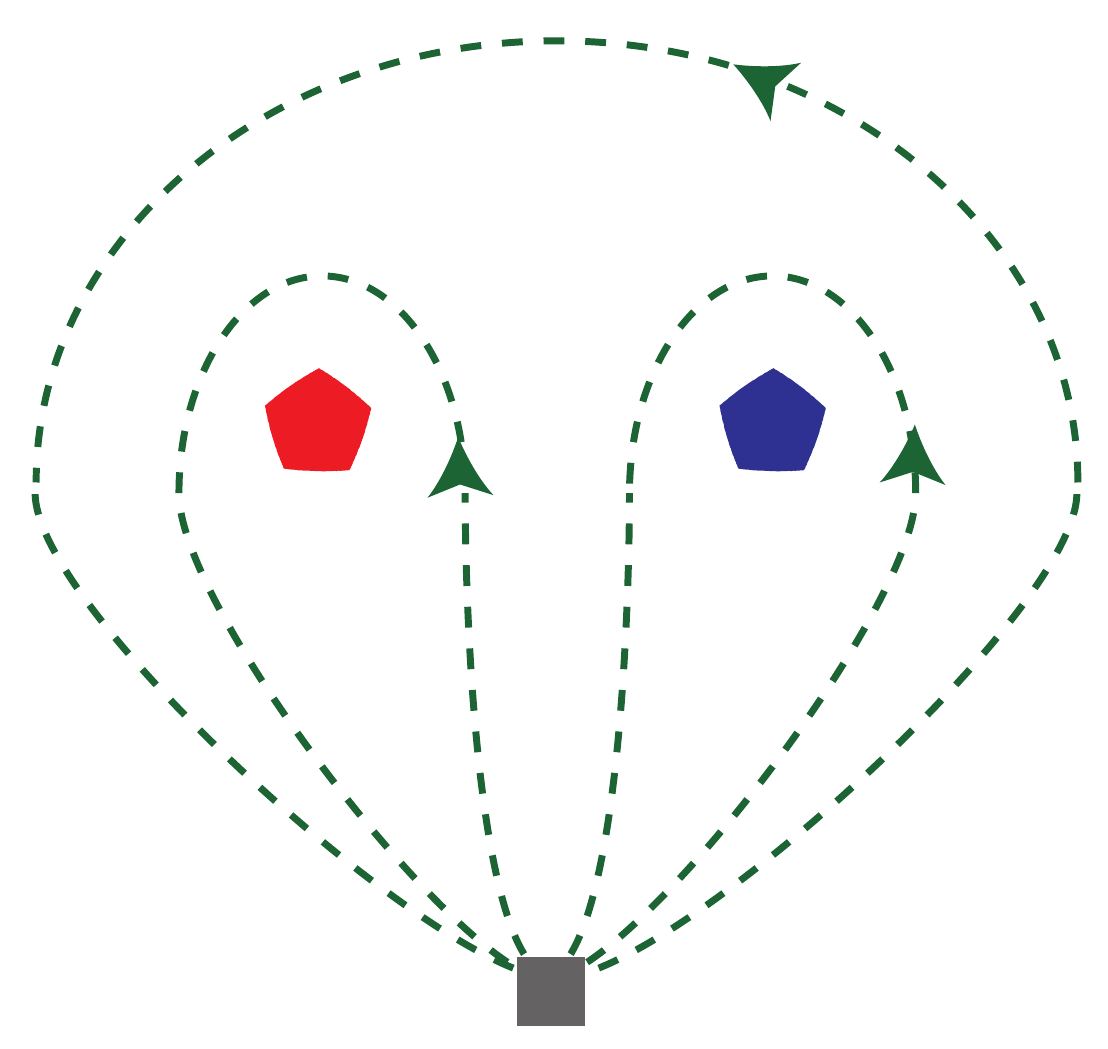}
\caption{Quasiparticle trajectory around a defect pair (colored pentagons).}\label{fig:defectloops}
\end{figure}

As the symmetry group $S_3$ is non-Abelian, defect fusion is non-commutative and order dependent. When the twofold defect $[\Lambda_\chi]$ is put on the left side of another one $[\Lambda_{\chi+1}]$ with a different color, a quasiparticle travelling in the anticlockwise direction around the pair will first circles $[\Lambda_{\chi+1}]$ before it orbits $[\Lambda_\chi]$ (see figure~\ref{fig:defectloops}). The overall relabeling action on the quasiparticle is therefore given by the product $\Lambda_\chi\circ\Lambda_{\chi+1}=\Lambda_3$. However if the order of the defects is inverted, then the relabeling action will be reversed because $\Lambda_{\chi+1}\circ\Lambda_\chi=\overline\Lambda_3$. Thus we have the order dependent fusion rules \begin{align}[\Lambda_\chi]_{\bf l}\times[\Lambda_{\chi+1}]_{{\bf l}'}=[\Lambda_3],\quad[\Lambda_\chi]_{\bf l}\times[\Lambda_{\chi-1}]_{{\bf l}'}=[\overline\Lambda_3]\end{align} and similarly for the fusion between a threefold and a twofold defect \begin{gather}[\Lambda_\chi]_{\bf l}\times[\Lambda_3]=\sum_{{\bf l}'}[\Lambda_{\chi+1}]_{{\bf l}'},\quad[\Lambda_3]\times[\Lambda_\chi]_{\bf l}=\sum_{{\bf l}'}[\Lambda_{\chi-1}]_{{\bf l}'},\\ [\Lambda_\chi]_{\bf l}\times[\overline\Lambda_3]=\sum_{{\bf l}'}[\Lambda_{\chi-1}]_{{\bf l}'},\quad[\overline\Lambda_3]\times[\Lambda_\chi]_{\bf l}=\sum_{{\bf l}'}[\Lambda_{\chi+1}]_{{\bf l}'}.\end{gather} Unlike anyonic excitations, defect fusion rules can be non-commutative because twist defects are objects equipped with an extensive texture that distinguishes the order of defect arrangement.

\subsubsection{Genons in bilayer systems}\label{sec:genons}
Non-abelian defects can be constructed as dislocations in abelian {\em topological nematic states}~\cite{BarkeshliQi, BarkeshliQi13} such as a multiple Chern band with symmetry~\cite{WangRan11, LuRan12}, described by {\em genons} in effective field theory~\cite{BarkeshliJianQi} and classified by Wilson structures of non-chiral gapped edges~\cite{LevinGu12, Levin13, BarkeshliJianQi13, BarkeshliJianQi13long}. Experimental proposals are already available in double layer FQH systems~\cite{BarkeshliQi13,BarkeshliOregQi14}. 

\begin{figure}
\centering\includegraphics[width=0.35\textwidth]{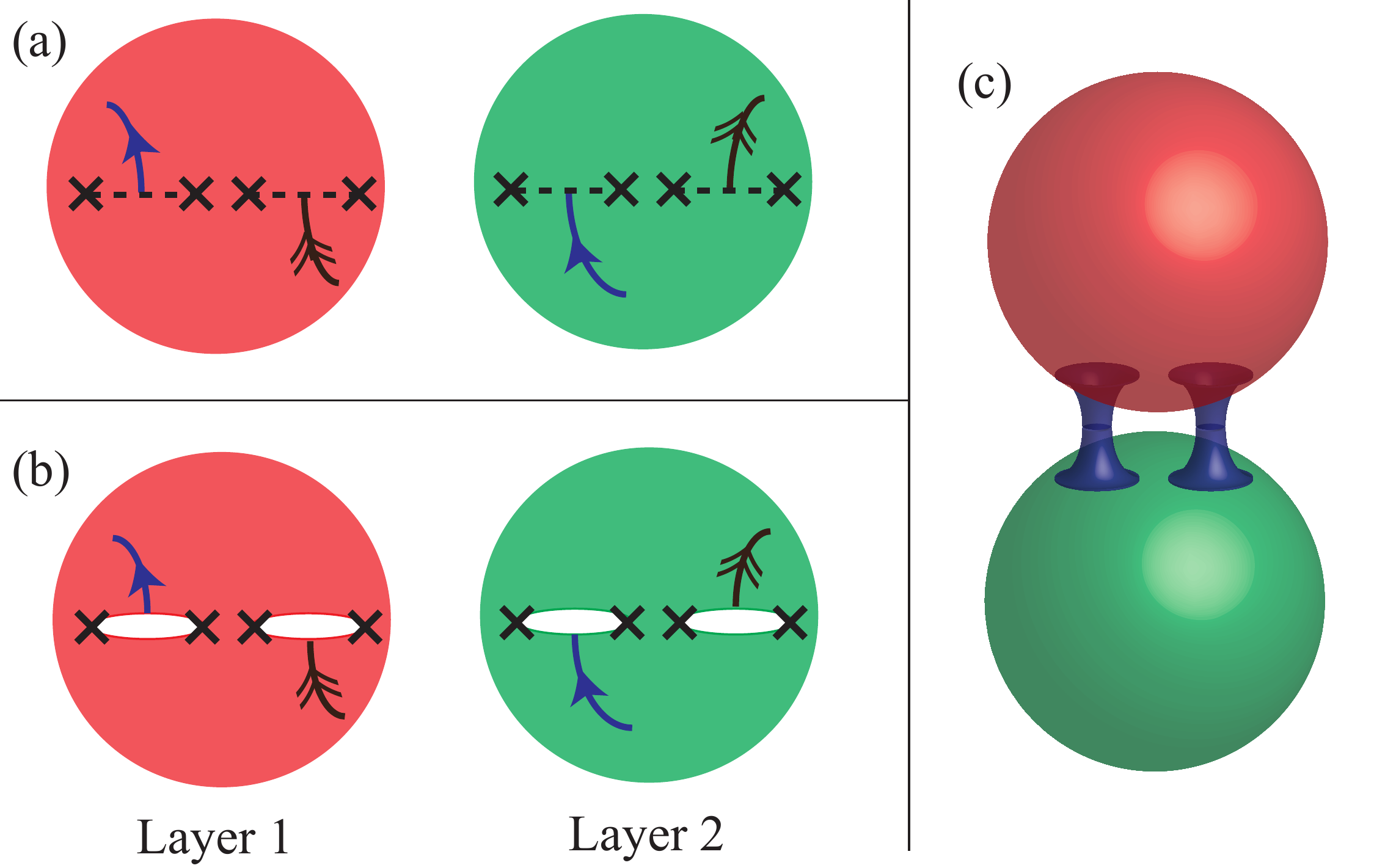}
\caption{Four twist defects (crosses) of a bilayer state in a closed system. (a) Interlayer branch cuts (dashed lines). (b) Enlarging branch cuts to holes. (c) Pasting holes between layers to form handles.}\label{fig:genon}
\end{figure}

Here we first look at a geometric interpretation of twist defects in in an Abelian bilayer system. The Chern-Simons action \eqref{CSaction} is characterized by the $K$-matrix of the block diagonal form \begin{align}K=\left(\begin{array}{*{20}c}K_1&0\\0&K_1\end{array}\right)\end{align} where $K_1$ is the $K$-matrix of a single layer. This bilayer state has a $\mathbb{Z}_2$ symmetry $\sigma=\sigma_x$ that interchanges the layers. When an anyon orbits around a $\sigma$ defect, it passes from one layer to another. This can be represented by a branch cut, where the layer switches, joining a pair of defects (see figure~\ref{fig:genon}).

From an analysis similar to that in the previous subsection~\ref{sec:defectlatticemodel} on lattice models, the fusion outcomes of a pair of defects are determined by the values of the Wilson loops around the pair. Suppose $\mathcal{A}_1$ is the collection of the different anyon types in a single layer. The fusion rule of a pair of bare defects is given by \begin{align}\sigma\times\sigma=\sum_{{\bf a}\in\mathcal{A}_1}{\bf a}\otimes\overline{\bf a}\label{bilayerfusion}\end{align} where the tensor product ${\bf a}\otimes\overline{\bf a}$ denotes the combination of an anyon ${\bf a}$ on the first layer and its antipartner $\overline{\bf a}$ on the second layer. Thus a defect pair corresponds to $|\mathcal{A}_1|=|\det(K_1)|$ ground states, or equivalently each defect has the quantum dimension $d_\sigma=|\det(K_1)|$.

However in a closed system, the overall topological charge is fixed and is identical to the vacuum 1. This extra constraint reduces to ground state degeneracy. For example, a closed sphere containing a pair of defects carries no ground state degeneracy. This is because Wilson loops around the defect pair are actually contractible by wrapping around the back of the sphere. The minimal system that supports non-trivial ground state degeneracy contains four defects (see figure~\ref{fig:genon}). By separating the two layers onto two spheres, the four twist defects can be represented by the ends of two branch cuts where the layers switch. Geometrically this is equivalently to pasting the two spherical layers back to each other along the branch cuts. The resulting object is a single-layer torus, where the defects are now smoothened into a pair of handles (see figure~\ref{fig:genon}(b,c)).

The ground state degeneracy is well-known~\cite{WenNiu90} and is identical to number of anyon types in a single-layer system, $|\mathcal{A}|=|\det(K_1)|$. It is not surprising that the degeneracy matches exactly with the number of fusion channels in \eqref{bilayerfusion}. In fact there is a one-to-one correspondence between the ground states of a single-layer torus and the fusion outcomes in \eqref{bilayerfusion} \begin{align}|{\bf a}\rangle=\left|\vcenter{\hbox{\includegraphics[width=0.7in]{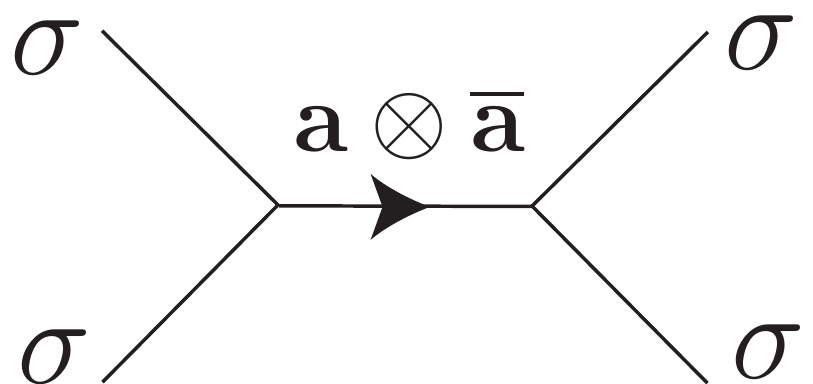}}}\right\rangle\end{align} where the intermediate branch of the fusion tree is labeled by the admissible fusion channel of $\sigma\times\sigma$.

The Wilson algebra are generated by two types of loops around the two cycles of the torus. The first kind $\mathcal{W}^{12}_{\bf a}$ surrounds the first two defects by an ${\bf a}$-string on the same layer. It corresponds to a loop around the first handle in figure~\ref{fig:genon}(c). The second kind $\mathcal{W}^{23}_{\bf a}$ surrounds the second and third defects by dragging ${\bf a}$ across a branch cut to another layer and closing back by crossing the other branch cut. It corresponds to a loop connecting the two spheres by going through the two handles in figure~\ref{fig:genon}(c). All other Wilson loops can be generated by these two. They satisfy the commutation relation \begin{align}\mathcal{W}^{12}_{\bf a}\mathcal{W}^{23}_{\bf b}=\mathcal{D}S_{\bf ab}\mathcal{W}^{23}_{\bf b}\mathcal{W}^{12}_{\bf a},\quad\mathcal{D}S_{\bf ab}=e^{2\pi i{\bf a}^TK_1^{-1}{\bf b}}\label{Wilsonalgebrabilayer}\end{align} due to the crossing phase $\mathcal{D}S_{\bf ab}$ between the ${\bf a}$ and ${\bf b}$ strings. The non-commutative Wilson algebra is irreducibly represented by, and therefore protecting, the degenerate ground states.

A bilayer system with $2n$ twist defects in a closed spherical geometry can be mapped by a similar way to a single layer system in a closed surface with $n-1$ handles, i.e.~genus $n-1$. The $|\det(K_1)|^{n-1}$ degenerate ground states form an irreducible representation of Wilson loops going around or across different handles. As the genus scales with the number of defect pairs, twist defects can be regarded genus-carrying objects, or {\em genons}. 

\begin{figure}[htbp]
\centering\includegraphics[width=0.4\textwidth]{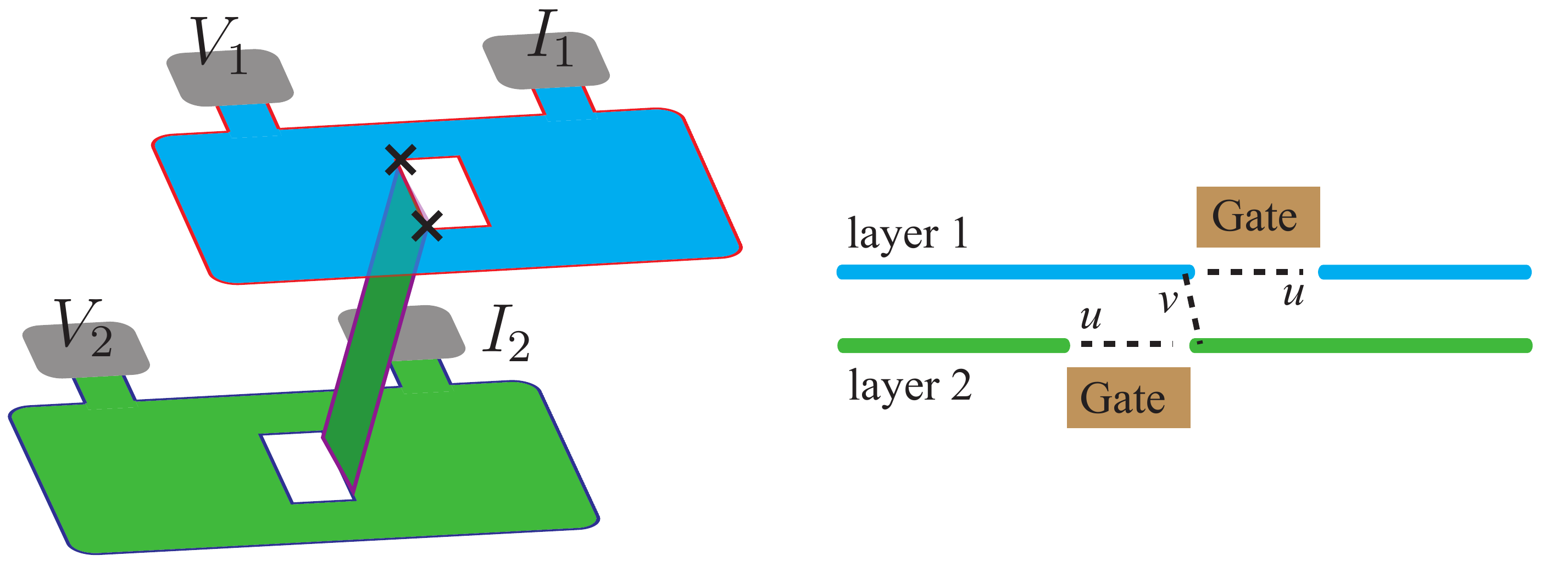}
\caption{A pair of twist defects (crosses) constructed by pasting edges of a bilayer FQH state.}\label{fig:FQHbilayer}
\end{figure}

Theoretical proposal has been made to realize genons in bilayer FQH state and we review a particular one proposed in Ref.~\cite{BarkeshliQi13,BarkeshliOregQi14} (see figure~\ref{fig:FQHbilayer}). A defect pair is constructed by locally bridging the two layers by a top and bottom gate. The gates deplete the quantum Hall state and create a trench on each layer. The edge modes along the trenches couple by quasiparticle tunneling, which can be tunned by the gating configuration. When the inter-trench tunneling $v$ is stronger than the intra-trench tunnling $u$, the top layer is effectively connected to the bottom layer. This facilitate a layer-flip for a passing quasiparticle and corresponds to a pair of defects located at the two ends of the trench.

The defect structure gives rise to non-trivial transport signatures. Take the bilayer Laughlin $\nu=1/3$ FQH state for example (see figure~\ref{fig:FQHbilayer}). The 4-terminal electrical current $I_1-I_2=\frac{1}{3}\frac{e^2}{h}(V_1-V_2)$ in the uncoupled bilayer limit will be modified in the presence of the interlayer bridge. When the trench is finite, the three ground states $|a\rangle$, $a=1,2,3$, associate to the defect pair split in energy $E_{ab}=E_a-E_b\sim e^{-l/l_{\mathrm{loc}}}$, where $l$ is the size of the trench and $l_{\mathrm{loc}}$ is some localization length scale. Quasiparticle tunneling between the defects and the outer edge modifies the steady state 4-terminal transport by continuously leaking charge into or out of the trench. This process requires the bias to overcome the energy barrier $E_{ab}$ for all states $a$ and $b$. If one measures the modification $G=dI/dV$ of conductance, there will be two resonant peaks at bias $e^\ast=\mathrm{max}(E_{ab})$ or $-\mathrm{min}(E_{ab})$, for $e^\ast=e/3$ the charge of the Laughlin quasiparticle. Unlike the case for Majorana zero modes, these two peaks are asymmetric about zero bias and are distinct signatures of twist defects in the Laughlin $\nu=1/3$ FQH state.

\subsubsection{Parafermions in superconducting heterostructures}\label{sec:parafermion} 
Twist defects can also be realized as parafermions or fractional Marjoana zero modes in superconducting heterostructures. The simplest construction was proposed by Fu and Kane~\cite{FuKane08}, where a zero energy Majorana bound state (MBS) is located at the junction between a quantum spin Hall insulator (QSHI), a superconductor (SC) and a ferromagnetic insulator (FMI). Later realization of the Kitaev $p$-wave wire~\cite{Kitaevchain}, which hosts boundary MBS, were proposed and tested in proximity induced superconducting strong spin-orbit coupled nanowires in magnetic fields~\cite{LutchynSauDasSarma10,OregRefaelvonOppen10,Sato2009,Kouwenhoven12,DengYuHuangLarssonCaroffXu12,Shtrikman12,RokhinsonLiuFurdyna12,ChangManucharyanJespersenNygardMarcus12,FinckHarlingenMohseniJungLi13} and ferromagnetic atomic chains on a superconductor~\cite{NadjDrozdovLiChenJeonSeoMacDonaldBernevigYazdani14}. Generalizing zero energy Majorana bound states at superconductor heterostructures, the more exotic fractional Majorana states -- which carry richer fusion and braiding characteristics -- are predicted at the SC-FM edge~\cite{LindnerBergRefaelStern, ClarkeAliceaKirill, MChen, Vaezi, mongg2} of fractional topological insulators~\cite{LevinStern09, LevinStern12, MaciejkoQiKarchZhang10, SwingleBarkeshliMcGreevySenthil11, LevinBurnellKochStern11} and helical 1D Luttinger liquids~\cite{OregSelaStern13}. 

\begin{figure}[htbp] 
\centering\includegraphics[width=0.85\textwidth]{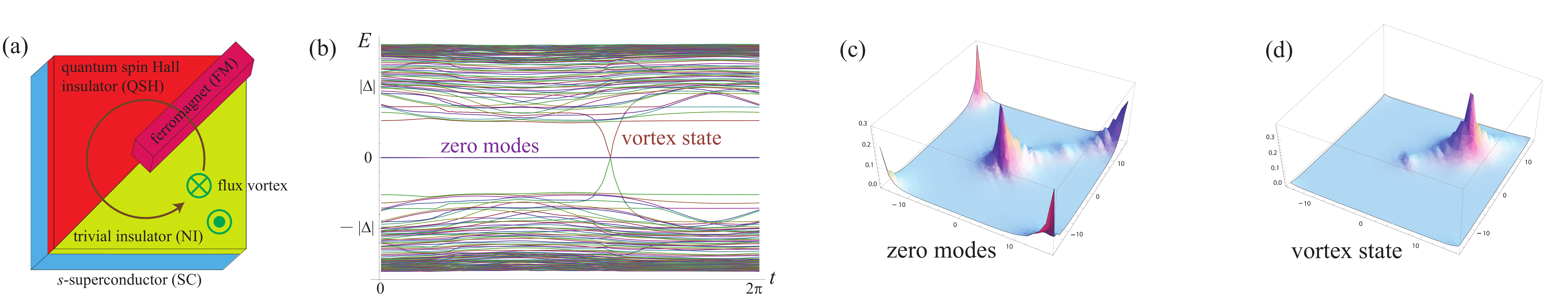}
\caption{When a vortex orbits the MBS in a QSH-SC-FM heterostructure (a), a single level-crossing occurs (b). (c) and (d) respectively show the wavefunction magnitudes of the zero energy MBS pair and the vortex state at the crossing.}\label{fig:QSHSCFM} 
\end{figure}
These heterostructures are smoothly connected to an isotropic superconducting medium where certain order parameter winds continuously around a point defect. We begin with the two-dimensional QSHI-SC-FMI heterostructure. The edge of a QSHI carries a helical Dirac mode, which aquires an energy gap when time reversal symmetry or charge conservation are broken when it sits next to a FMI or SC. \begin{align}\mathcal{H}_{\mathrm{edge}}=i\hbar v{\bf c}^\dagger\sigma_z\partial_x{\bf c}+h(x){\bf c}^\dagger\sigma_x{\bf c}+i\Delta(x){\bf c}^T\sigma_y{\bf c}+h.c.\label{FuKaneH}\end{align} where ${\bf c}=(c_R^\uparrow,c_L^\downarrow)$ are counter-propagating (complex) Dirac fermions. The (anti)ferromagnetic backscattering $h(x)=h\Theta(-x)$ and superconducting pairing $\Delta(x)=\Delta\Theta(x)$ are competing orders, and are present along the negative and positive $x$-axis respectively, for $\Theta(x)=0$ when $x<0$ or 1 when $x>0$. This Hamiltonian can be solved exactly and a zero energy Majorana bound state (MBS) is localized at the interface at $x=0$.

Eq.\eqref{FuKaneH} can be rewritten in bosonized form \begin{align}\mathcal{H}_{\mathrm{edge}}=\tilde{v}[(\partial_x\tilde\phi_R)^2+(\partial_x\tilde\phi_L)^2]+\tilde{h}(x)\cos(\tilde\phi_R-\tilde\phi_L)+\tilde\Delta(x)\cos(\tilde\phi_R+\tilde\phi_L)\label{FuKaneHb}\end{align} where $c_{R/L}\sim\exp(i\tilde\phi_{R/L})$ and the chiral bosons satisfy the equal time commutation relations \begin{align}\left[\tilde\phi_\mu(x),\tilde\phi_{\mu'}(x')\right]=i\pi\left[(-1)^\mu\delta_{\mu\mu'}\mathrm{sgn}(x'-x)+\mathrm{sgn}(\mu'-\mu)\right]\end{align} where $\mu=R,L=0,1$ and $\mathrm{sgn}(y)=y/|y|$ for $y\neq0$ or 0 for $y=0$. The superconducting and (anti)ferromagnetic backscatterings, $\tilde\Delta(x)$ and $\tilde{h}(x)$, pin the bosons $\langle\tilde\phi_R(x)\pm\tilde\phi_L(x)\rangle$ along the positive and negative $x$-axis, and associate an insulating and pairing gap respectively. 

\begin{figure}[htbp]
\centering\includegraphics[width=0.75\textwidth]{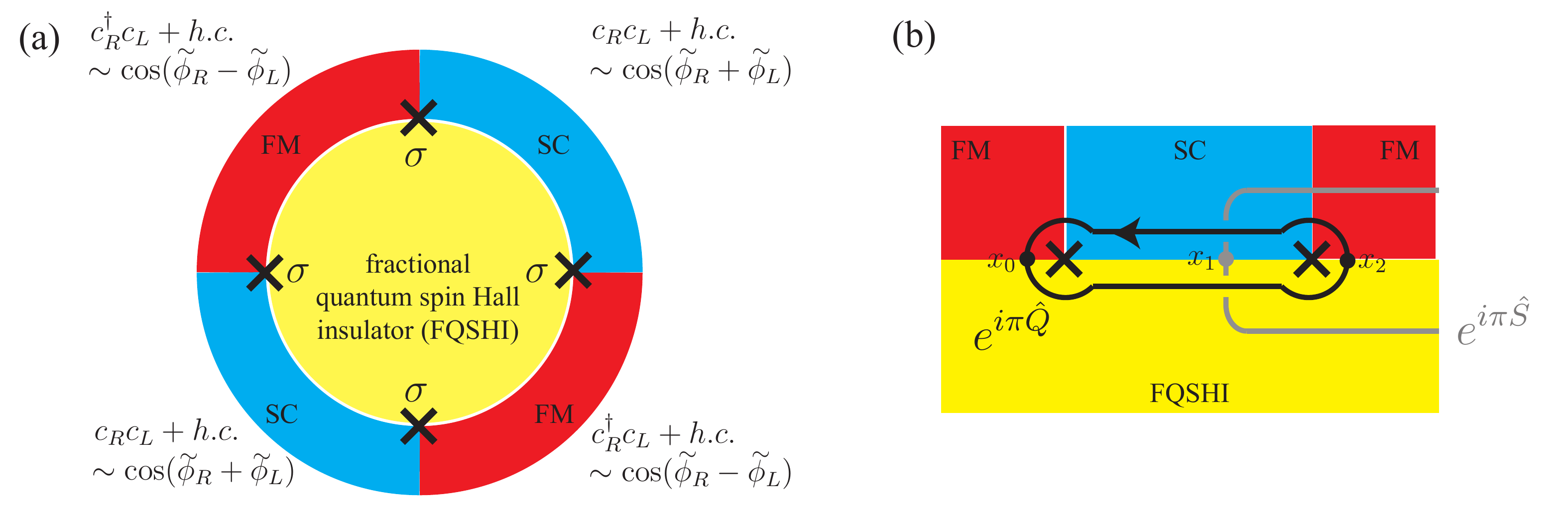}
\caption{(a) Parafermions or fractional Majorana's (crosses) at superconducting (blue) and ferromagnetic (red) interfaces of a fractional quantum spin Hall insulator. (b) Wilson observables.}\label{fig:FQSHISCFM} 
\end{figure}
This can be generalized directly to the fractional case when the QSHI is repaced with a fractional quantum spin Hall insulator (FQSHI). The $\nu=1/m$ FQSHI is identical to a time reversal double of the Laughlin $\nu=1/m$ FQH state~\cite{LevinStern09,LevinStern12}. It has the $K$-matrix of $K=m\sigma_z$ in the effective Chern-Simons description \eqref{CSaction}, and carries a helical boundary that hosts quasiparticles $\Psi^n_{R/L}\sim e^{in\phi_{R/L}}=e^{in\tilde\phi_{R/L}/m}$ with fractional electric charge $ne^\ast=en/m$, for $c_{R/L}=e^{i\tilde\phi_{R/L}}$ the electron operator with charge $e$. The boson variables now satisfy the commutation relation \begin{align}\left[\phi_\mu(x),\phi_{\mu'}(x')\right]=\frac{i\pi}{m}\left[(-1)^\mu\delta_{\mu\mu'}\mathrm{sgn}(x'-x)+\mathrm{sgn}(\mu'-\mu)\right]\label{paracomm}\end{align} for $\mu,\mu'=R,L=0,1$. The FQHSI edge can be gapped by superconducting (SC) or (anti)ferromagnetic (FM) backscattering, $c_Rc_L$ or $c_R^\dagger c_L$, shown in figure~\ref{fig:FQSHISCFM}. Generalizing \eqref{FuKaneHb}, the edge is described by the bosonized Hamiltonian \begin{align}\mathcal{H}_{\mathrm{edge}}=\tilde{v}[(\partial_x\phi_R)^2+(\partial_x\phi_L)^2]+h(x)\cos(m(\phi_R-\phi_L))+\Delta(x)\cos(m(\phi_R+\phi_L))\label{paraHb}\end{align} where the (anti)ferromagnetic and superconducting terms are given by electron backscattering and pairing, $c_R^\dagger c_L+h.c.\sim\cos(\tilde\phi_R-\tilde\phi_L)$ and $c_Rc_L+h.c.\sim\cos(\tilde\phi_R+\tilde\phi_L)$ respectively. A parafermion or fractional Majorana zero mode $\sigma$ is bounded at each SC-FM interface where $h(x)-\Delta(x)$ changes sign.

A pair of $\sigma$'s corresponds to $2m$ degenerate ground state, each associates a distinct good quantum number of an observable. We focus on a superconducting segment in between a $\sigma$ pair (see figure~\ref{fig:FQSHISCFM}(b)). Similar argument holds for the (anti)ferromagnetic segment by replacing $\phi_R+\phi_L$ by $\phi_R-\phi_L$. The electric charge of the SC segment in units of $e$ is given by the number \begin{align}Q=N_R+N_L=\frac{1}{2\pi}\int_{x_0}^{x_2}dx\left(\partial_x\phi_R(x)-\partial_x\phi_L(x)\right)\end{align} where $\rho_\mu=(-1)^\mu\partial_x\phi_\mu/2\pi=c_\mu^\dagger c_\mu$ are the electron number density for the $\mu=R/L$ sector, and the SC segment sits between $x_0$ and $x_2$. As number conservation is broken by condensation of Cooper pairs in a SC, $Q$ is well defined modulo 2 and we have the observable $e^{i\pi Q}$. This operator can also be viewed as a Wilson loop \begin{align}e^{i\pi Q}=e^{i(\phi_R(x_2)-\phi_L(x_2)-\phi_R(x_0)+\phi_L(x_0)/2)}\sim\left(\Psi_R^{1/2}(x_2)\Psi_L^{1/2}(x_2)^\dagger\right)\left(\Psi_L^{1/2}(x_0)\Psi_R^{1/2}(x_0)^\dagger\right)\end{align} where a further fractionalized quasiparticle $\Psi^{1/2}\sim e^{i\phi/2}$ is dragged around the $\sigma$ pair (see figure~\ref{fig:FQSHISCFM}(b)). As the Laughlin quasiparticle $\Psi\sim e^{i\phi}$ is corresponds to a $2\pi$-flux $\Phi=hc/e$ in the 2D bulk, the half quasiparticle $\Psi^{1/2}$ is therefore a $\pi$-flux with $\Phi=hc/2e$, which is the fundamental flux quantum in a superconducting medium.

We assume $h(x)$ and $\Delta(x)$ are heavy-side step functions and $x_0$, $x_2$ are displaced slightly to the two FM regions so that $\Delta(x_{0,2})=0$. In this case, the observable $e^{i\pi Q}$, which only involves quasiparticles at $x_0$ and $x_2$, commutes with the FM gapping term $\cos[m(\phi_R-\phi_L)]$ and thus is diagonalized by the ground states. It can take $2m$ eigenvalues $e^{i\pi Q}=e^{in\pi/m}$ because a Laughlin quasiparticle $\Psi^n$ carries $Q=n/m$. (The effect of the $\pi$-flux $\Psi^{1/2}$ is more subtle and will be described later.) This indicates a $2m$ ground state degeneracy.

These degenerate ground states irreducibly represents a non-commutative algebra of Wilson observables. If we conside the next parafermion pair that sandwich a FM segment, then we would have the new Wilson operator \begin{align}e^{i\pi S}=e^{i(\phi_R(x_3)+\phi_L(x_3)-\phi_R(x_1)-\phi_L(x_1)/2)}\sim\left(\Psi_R^{1/2}(x_3)\Psi_L^{1/2}(x_3)\right)\left(\Psi_R^{1/2}(x_1)^\dagger\Psi_L^{1/2}(x_1)^\dagger\right)\end{align} for $S$ the spin operator in units of $\hbar/2$, \begin{align}S=N_R-N_L=\frac{1}{2\pi}\int_{x_1}^{x_3}dx\left(\partial_x\phi_R(x)+\partial_x\phi_L(x)\right)\end{align} where the FM segment is bounded between points, $x_1$ and $x_3$, slightly inside the neighboring SC segments. The spin observable also commutes with the sine-Gordon potential in \eqref{paraHb}, but does not commute with the neighboring observable $e^{i\pi Q}$. By using the canonical commutation relation \eqref{paracomm} and the fact that $x_0<x_1<x_2$, \begin{align}\left[S,Q\right]=\frac{1}{(2\pi)^2}\int_{x_0}^{x_2}\left[\phi_R(x_1)+\phi_L(x_1),\partial_x\phi_R(x)-\partial_x\phi_L(x)\right]=\frac{i}{m\pi}.\end{align} This implies the commutation relation of adjacent Wilson observables \begin{align}e^{i\pi S}e^{i\pi Q}=e^{i\pi/m}e^{i\pi Q}e^{i\pi S}.\end{align} They can be irreducibly represented by the $2m\times2m$ rotor matrices \begin{align}\langle n|e^{i\pi Q}|n'\rangle=e^{i\pi n/m}\delta_{n,n'},\quad\langle n|e^{i\pi S}|n'\rangle=\delta_{n,n'+1}\end{align} where $n,n'$ label the degenerate ground states. 

This applies to a ring geometry (see figure~\ref{fig:FQSHISCFM}(a)) with four segments where the total charge and spin are fixed. In general for a system with $N$ SC and $N$ FM segements, there are $2N$ Wilson observables $\mathcal{W}_{2j-1}=e^{i\pi Q_j}$ and $\mathcal{W}_{2j}=e^{i\pi S_j}$ associating to the $j^{\mathrm{th}}$ SC and FM segment respectively. They satisfies the algebraic relation \begin{align}\mathcal{W}_j\mathcal{W}_{j'}=e^{i\pi(\delta_{j+1,j'}-\delta_{j-1,j'})/m}\mathcal{W}_{j'}\mathcal{W}_j\label{Wilsonalgebraparafermion}\end{align} for $j$ being defined modulo $2N$. Fixing the total charge and spin operators $\prod_je^{i\pi Q_j}$ and $\prod_je^{i\pi S_j}$, there is a $(2m)^{N-1}$ ground state degeneracy labeled by the eigenvalues of $e^{i\pi Q_j}$. As the system contains $2N$ parafermions, each carries a quantum dimension of $d_\sigma=\sqrt{2m}$.

As the ground state associating to a pair of parafermions is characterized by its charge $Qe$ modulo $2e$, we can write the fusion rules as \begin{align}\sigma\times\sigma=1+\Psi+\Psi^2+\ldots+\Psi^{2m-1}\label{parafermionfusion}\end{align} which generalizes the Ising fusion rules $\sigma\times\sigma=1+\psi$ of Majorana zero modes, for $\Psi^n$ now represents the Laughlin $Q=n/m$ quasiparticle. In fact, the fusion outcome is statistically indistinguishable from a true Laughlin quasiparticle by monodromy measurements when the parafermion pair is brought together. 

Parafermions or fractional Majorana zero modes can be understood in the context of twist defects. In the normal quantum spin Hall case, when the SC proximity effect is strong and the energy gap is large enough, the MBS is surrounded by a superconducting region. This is shown in figure~\ref{fig:QSHSCFM}, where a bulk SC is put beneath a QSHI and a trivial insulator. The SC quantizes fluxes in units of $hc/2e$, which gives a $\pi$ quantum phase to an orbiting electron. Although $\pi$-fluxes are always confined in a 3D SC, in a theoretical perspective it is useful to consider them as deconfined excitations by extending the topological state by {\em gauging} the $\mathbb{Z}_2$ fermion parity symmetry. A SC-QSHI heterostructure is now identical to a $\mathbb{Z}_2$ gauge theory, or the Kitaev toric code, where the $\pi$-flux corresponds to the $m$ excitation (see section~\ref{sec:toriccode}).

When the $\pi$-flux orbits around the MBS at a QSHI-SC-FMI junction, the Caroli-de Gennes-Matricon vortex states it carries undego a single level crossing (see figure~\ref{fig:QSHSCFM}(b))~\cite{KhanTeoVishveshwaraappearsoon}. A single BdG fermion $\psi$ is adiabatically pumped into or out of the flux vortex after one complete cycle. This changes the fermion parity and switches the quasiparticles label $m\leftrightarrow e=m\times\psi$ of an orbiting vortex. The MBS can therefore be regarded as a twist defect associating to the electric-magnetic symmetry in a $\mathbb{Z}_2$ gauge theory.

This can be generalized to the fractional quantum spin Hall scenario~\cite{khan2014, KhanTeoVishveshwaraappearsoon, KhanTeoHughesVishveshwarafuture}. We consider a FQSHI that consists of a $\nu=1/m$ Laughlin FQH state for spin $\uparrow$ with chirality $R$ and its time reversal copy occupying spin $\downarrow$ and opposite chirality $L$. Similar to the previous quantum spin Hall case, we assume there is a bulk SC beneath the system and quantizes $\pi$-fluxes with $\Phi=hc/2e$. When passing across each FQH copy, a $\pi$-flux associate a $q=1/2m$ electric charge (in units of $e$) due to the fractional Hall conductance $\sigma_{xy}=(1/m)e^2/h$~\cite{Laughlinargument}. We extend the fermionic topological state by coupling to a $\mathbb{Z}_2$ gauge theory to include $\pi$-fluxes. The resulting bosonic topological state can be described by an effective Chern-Simons action \eqref{CSaction} with a $K$ matrix $K=4m$. The extended theory now has $4m$ quasiparticle type $\Psi^n$, for $n=0,1/2,1,\ldots,2m-1,2m-1/2$ mod $2m$. $\Psi^{1/2}$ is the anyon associates to a $\pi$-flux. It carries spin $h_{\Psi^{1/2}}=1/8m$, and is half of the Laughlin quasiparticle $\Psi$. $\Psi^m$ was the local electronic operator, and is now become non-local with respect to the $\pi$-flux. 

This Abelian theory has a conjugation symmetry $C:\Psi^n\leftrightarrow\Psi^{-n}$. The parafermion or fractional Majorana at a FM-SC junction is now a twist defect with respect to this symmetry. This is because when a quasiparticle $\Psi^n_R=e^{in\phi_R}$ in the spin $\uparrow$ sector is brought to the FM edge, it is backscattered by the FM coupling $h(x)$ in \eqref{paraHb} into $\Psi^n_L=e^{in\phi_L}$ in the spin $\downarrow$ sector. However when it goes around the junction to the SC edge, it is backscattered by the SC coupling $\Delta(x)$ in \eqref{paraHb} that flips the sign of the boson operator $\phi_L\to-\phi_R$. The quasiparticle switches back to the spin $\uparrow$ sector but has a conjugated label $\Psi^{-n}_R=e^{-in\phi_R}$. As a result a full cycle around the parafermion changes the anyon type from $\Psi^n\to\Psi^{-n}$.



\subsection{Quantum field theory in the presence of static defects}\label{sec:twistdefectB}
Here we describe a framework that describes the quantum states of a topological state in the presence of twist defects as well as the projective braiding operations. This can be summarized mathematically as a ``$G$-crossed tensor category"~\cite{EtingofNikshychOstrik10,BarkeshliBondersonChengWang14,TeoHughesFradkin15,TarantinoLindnerFidkowski15}. However such general description is out of the scope of this review, which aims at introducing the basic ideas of twist defect. Instead we will demonstrate some salient features using some simple examples. This subsection will begin with a more in depth discussion on defect fusion rules and quantum state basis transformation. This extends the modular tensor category (see section~\ref{sec:MTC}) of the underlying topological phase into a $G$-graded fusion category. Next we will demonstrate the projective non-Abelian operations when defects are braided among each other. Lastly, we will look at the {\em restricted} modular transformations in a {\em twisted} torus geometry. We notice in passing that given a globally symmetric topological state, a complete set of defect braiding information, known as $G$-crossed braiding, can be defined consistently with defect fusion. This general description of a $G$-crossed tensor category can be found in Ref.~\cite{BarkeshliBondersonChengWang14}, but will be omitted in this review.

\subsubsection{Defect fusion rules}
Previously we have already seen the fusion rules for twist defects in lattice models, bilayer systems and parafermions in SC heterostructures. The simplest one comes from the Ising fusion rules of defects $\sigma_\lambda$ of the $\mathbb{Z}_2$ symmetric Kitaev's toric code or the Wen's plaquette model (see section~\ref{sec:defectlatticemodel}). \begin{align}\sigma_\lambda\times\psi=\sigma_\lambda,\quad\sigma_\lambda\times\sigma_\lambda=1+\psi.\end{align} Defects generically come with species labels that characterize the quasiparticles trapped at the defects. For example there are two dislocation species, $\sigma_0$ and $\sigma_1$, in the toric code, which differ from each other by \begin{align}\sigma_\lambda\times e=\sigma_\lambda\times m=\sigma_{\lambda+1}.\end{align} We have also seen defect fusion rules of the Bombin color code model in section~\ref{sec:defectlatticemodel}. Unlike those in a conventional anyon theory, fusion could be degenerate and even non-commutative because of the non-Abelian symmetry group $S_3=\{1,\Lambda_3,\overline\Lambda_3,\Lambda_Y,\Lambda_R,\Lambda_B\}$. For example, \begin{align}[\Lambda_Y]_{\bf l}\times[\Lambda_R]_{{\bf l}'}=[\Lambda_3],\quad[\Lambda_R]_{\bf l}\times[\Lambda_Y]_{{\bf l}'}=[\overline\Lambda_3],\quad[\Lambda_3]\times[\Lambda_3]=4[\overline\Lambda_3].\end{align}

These can be summarized by a general $G$-graded {\em defect fusion category}, where $G$ is the global anyonic symmetry group. The fundamental objects in this semiclassical description are defect-quasiparticle composites denoted by $M_\lambda$, where $M$ is an anyonic symmetry element in $G$ associated to the defect, and $\lambda$ is the species label representing the equivalence class of the anyon charge bound at the defect-quasiparticle composite. For example, $\lambda$ could specify the fractional electric charge carried by a twist defect in a FQH state, or more general anyonic charges. Additionally, a species label can change, or \emph{mutate}, when the defect is fused with a quasiparticle. For example, \begin{align}{\bf a}\times M_\lambda=M_\lambda\times{\bf a}=M_{\lambda'}.\label{generalspeciesmutation}\end{align} In general, the species label can be statistically distinguished by a Wilson loop measurement via dragging a quasiparticle, ${\bf a}$, $p$ times around the defect, where $M^p{\bf a}={\bf a}$. For instance, the two species of dislocations in the toric code give distinct phase factors under the double Wilson loop operator $\Theta$ in Fig.~\ref{fig:defect1}(b), which is well defined since the anyonic symmetry is twofold.

Defect species can be described more explicitly if the globally (anyonic) symmetric parent topological state is Abelian. In a $K$-matrix description \eqref{CSaction}, Abelian quasiparticles are labeled by an anyon lattice $\mathcal{A}=\mathbb{Z}^N/K\mathbb{Z}^N$ with lattice addition reflecting the fusion rule $\psi^{\bf a}\times\psi^{\bf b}=\psi^{{\bf a}+{\bf b}}$. Species of twist defects associated to an anyonic symmetry operation $M$ can be labeled by the quotient lattice~\cite{teo2013braiding, khan2014} \begin{align}\mathcal{A}_M=\frac{\mathcal{A}}{(1-M)\mathcal{A}}.\label{defectsectorquotient}\end{align} This is because combining a defect with quasiparticles that are related by the symmetry should give the identical defect-quasiparticle composite: \begin{align}M_\lambda=\psi^{\bf l}\times M_0=\psi^{{\bf l}+(M-1){\bf b}}\times M_0.\label{defectQPcompositeeq}\end{align} This can be diagrammatically explained by comparing topologically equivalent quasiparticle string patterns of the composite object: \begin{align}\vcenter{\hbox{\includegraphics[width=0.2\textwidth]{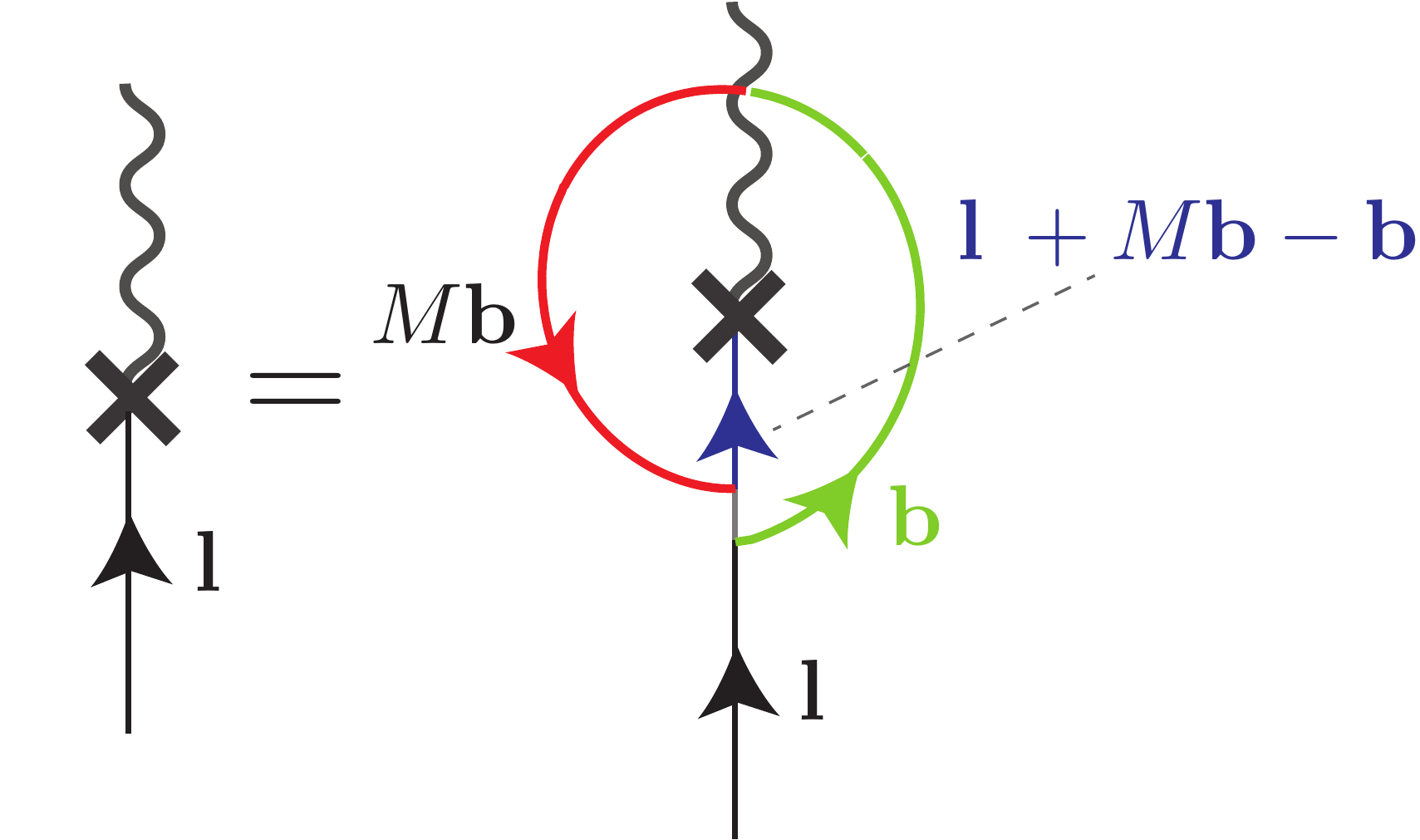}}}\label{defectQPcomposite}\end{align} where the ${\bf l}$ QP string attached to the defect $M_\lambda$ can be modified to ${\bf l}+M{\bf b}-{\bf b}$ by splitting off a ${\bf b}$ QP and letting it orbit once around the defect. This process cannot change the defect specices $\lambda$ as it cannot be detected by any Wilson measurements. For the toric code, $\mathcal{A}=\{1,e,m,\psi\}=\mathbb{Z}_2\oplus\mathbb{Z}_2$ and $(1-\sigma)\mathcal{A}=\{1,\psi\}=\mathbb{Z}_2$. The quotient is thus $\mathcal{C}_\sigma=\mathbb{Z}_2\oplus\mathbb{Z}_2/\mathbb{Z}_2=\mathbb{Z}_2$ and contains the two species of twist defects $\sigma_0,\sigma_1$.

The defect fusion category~\cite{EtingofNikshychOstrik10,TeoRoyXiao13long,teo2013braiding,BarkeshliBondersonChengWang14,TeoHughesFradkin15,TarantinoLindnerFidkowski15} is a {\em $G$-graded} extension of the modular tensor category (MTC)~\cite{Kitaev06, Walkernotes91, Turaevbook, FreedmanLarsenWang00, BakalovKirillovlecturenotes, Wangbook} that describes the underlying globally symmetric topological phase. \begin{align}\mathcal{C}=\bigoplus_{M\in G}\mathcal{C}_M.\label{Ggradedfusion}\end{align} $\mathcal{C}_1$, the subcategory corresponding to the identity element 1 in $G$, encodes the fusion data of the original MTC. Each other sector $\mathcal{C}_M$ is generated by twist defects associated to the anyonic symmetry $M$ with different species labels. A quasiparticle encircling two defects associated to symmetries $M$ and $N$ is relabeled by the combination $MN$ (see figure~\ref{fig:defectloops}). Group multiplication is therefore carried over to defect fusion \begin{align}\mathcal{C}_{M}\times\mathcal{C}_{N}\longrightarrow\mathcal{C}_{MN}\label{Ggradedfusionf}\end{align} and does not in general commute when $G$ is non-Abelian. 

As an example, the defect fusion category of the toric code has $\mathcal{C}_1=\langle1,e,m,\psi\rangle$ and $\mathcal{C}_\sigma=\langle\sigma_0,\sigma_1\rangle$, where $\mathbb{Z}_2=\{1,\sigma\}$ is the electric-magnetic anyonic symmetry group. Additionally, defect fusion respects group multiplication. For instance, the Ising fusion rule $\sigma_0\times\sigma_0=1+\psi$ is consistent with the requirement that $\mathcal{C}_\sigma\times\mathcal{C}_\sigma\to\mathcal{C}_1$ according to the twofold group structure $\sigma^2=1$. Moreover, the quasiparticle sector $\mathcal{C}_1$ acts transitively on the defect sectors $\mathcal{C}_\sigma$ and gives rise to distinct species labels $e\times\sigma_\lambda=m\times\sigma_\lambda=\sigma_{\lambda+1}$. In a general defect theory, the quasiparticle sector $\mathcal{C}_1$ is always closed under fusion, $\mathcal{C}_1\times\mathcal{C}_1\to\mathcal{C}_1,$ and it acts on individual defect sectors \begin{align}\mathcal{C}_1\times\mathcal{C}_M\longrightarrow\mathcal{C}_M,\quad\mathcal{C}_M\times\mathcal{C}_1\longrightarrow\mathcal{C}_M\end{align} by combining with the defects to form defect-quasiparticle composites. Mathematically, each defect sector $\mathcal{C}_M$ is known as a {\em $\mathcal{C}_1$-bimodule} (like a $\mathcal{C}_1$-vector space) and is equipped with associative ``vector addition" and ``scalar multiplication" operations such as $\sigma_0+\sigma_1=(1+e)\times\sigma_0$ and $e\times(m\times\sigma_0)=(e\times m)\times\sigma_0=\sigma_0$.

\begin{figure}[htbp]
\centering\includegraphics[width=0.5\textwidth]{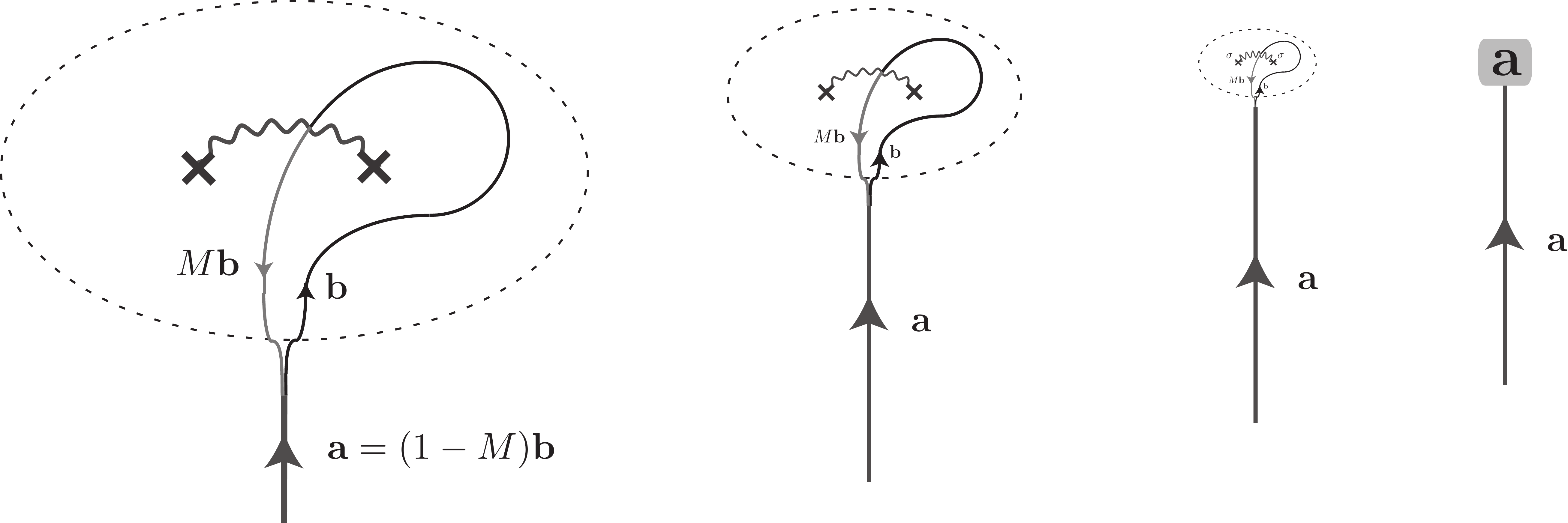}
\caption{Fusion of a pair of (bare) defects associated to opposite anyonic symmetries $M$ and $M^{-1}$. Wavy line represents branch cut where passing anyons change labels.}\label{fig:splittingstategeneraldefect1}
\end{figure}

Next we consider the fusion of a conjugate defect pair, but allowing for the possibility of different species labels on each defect. We assume the underlying globally symmetric phase is Abelian for simplicity. Conjugate pair fusion can be diagrammatically represented by two point defects with canceling branch cuts (see Fig.~\ref{fig:splittingstategeneraldefect1}). Their fusion outcome must be a trivial defect, i.e.~an Abelian quasiparticle (in $\mathcal{C}_1$), because of their trivial overall relabeling action to anyons encircling the pair of defects. The overall fusion channel depends on how quasiparticle strings are hung between the defects. Fig.~\ref{fig:splittingstategeneraldefect1} shows the general admissible string configurations that are irremovable when the defects fuse together. The overall open string contributes ${\bf a}=(1-M){\bf b}$ to the fusion channel of the defect pair. In other words, there is a one-to-one correspondence between the admissible fusion channels and the anyon sublattice $(1-M)\mathcal{A}.$ This leads to the fusion rule structure \begin{align}M_\lambda\times\overline{M}_{\lambda'}={\bf e}\times\sum_{{\bf a}\in(1-M)\mathcal{A}}{\bf a}\label{conjugatedefectfusion}\end{align} where $\overline{M}_{\lambda'}$ is a defect conjugate to $M_\lambda$ associated to the inverse symmetry $M^{-1}$, and ${\bf e}$ is some Abelian quasiparticle that depends on the species labels $\lambda,\lambda'$ as well as the quantization of the symmetry. Each defect, in general, is attached to a quasiparticle string ${\bf l}$ that reflects its species label $\lambda$ (see Eqs.~\eqref{defectQPcompositeeq} and \eqref{defectQPcomposite}). Hence, the quasiparticle strings ${\bf l}$ and ${\bf l}'$ of the defects $M_\lambda$ and $\overline{M}_{\lambda'}$ reflect their species labels. Both strings contribute to the fusion outcome in \eqref{conjugatedefectfusion}, and are encapsulated by the anyon label ${\bf e}$. Given a fixed species label, the choice of ${\bf l}$ and ${\bf l}'$ is not unique, however the choice does not affect the overall outcome since all anyons in $(1-M)\mathcal{A}$ are summed over in \eqref{conjugatedefectfusion}. 
Moreover ${\bf e}$ also depends on the projective action of symmetry on quantum states. For example the projective $\mathbb{Z}_2$ symmetry~\cite{Wenspinliquid02} on the Abelian semion phase $SU(2)_1=U(1)_1$ with $K=2$, leads to the defect fusion rule $\sigma\times\sigma=s$ for $s$ the semion, whereas a trivial $\mathbb{Z}_2$ symmetry would give $\sigma\times\sigma=1$ instead. For simplicity, this type of complications will be omitted in the examples we are going to discuss.

Since Abelian quasiparticles have unit quantum dimension, and defects with different species labels are related by absorbing or emitting Abelian quasiparticles \eqref{generalspeciesmutation}, all defects in the same sector $\mathcal{C}_M$ must share identical quantum dimension $d_M$. [This is {\em not} true if the globally symmetry parent state is non-Abelian. An example is the chiral ``4-Potts" state defined in Section~\ref{sec:trialitysymmetry}.] Moreover, since conjugate objects in a fusion theory must carry identical quantum dimension, we have $d_M=d_{\overline{M}}$. By equating the quantum dimensions on both sides of Eq.~\eqref{conjugatedefectfusion}, each defect carries a dimension of \begin{align}d_M=\sqrt{\left|(1-M)\mathcal{A}\right|}.\label{defectdimensionabelian}\end{align} This number determines the ground state degeneracy of a system of $\mathcal{N}$ defects in the thermodynamic limit, $G.S.D.\propto (d_M)^{\mathcal{N}}$ for $\mathcal{N}\to\infty$. For example, twist defects in the toric code satisfy $\sigma_\lambda\times\sigma_\lambda=1+\psi$ and have quantum dimension $d_\sigma=\sqrt{2}.$

The total quantum dimension of the defect fusion category is defined so that it squares to the sum of the squares of the dimensions of all the simple objects in $\mathcal{C}$: \begin{align}\mathcal{D}_{\mbox{\small defect}}\equiv\sqrt{\sum_{M\in G}\sum_{\lambda\in\mathcal{A}_M}d^2_{M}}=\mathcal{D}_0\sqrt{|G|}\label{defecttotaldimension}\end{align} where we note that sum also contains the the trivial identity element of $G,$ $\mathcal{D}_0=\sqrt{|\mathcal{A}|}$ is the total quantum dimension of the globally symmetric Abelian parent state without considering defects, and $|\ast|$ is the number of elements in $\ast$. Eq.~\eqref{defecttotaldimension} can be proven by seeing that all defect sectors $\mathcal{C}_M$ have identical total dimension. \begin{align}\sum_{\lambda\in\mathcal{A}_M}d^2_M=\left|\frac{\mathcal{A}}{(1-M)\mathcal{A}}\right|\times\left|(1-M)\mathcal{A}\right|=\left|\mathcal{A}\right|\end{align} where the quotient accounts for the number of defect species $\lambda$ (see Eq.~\eqref{defectsectorquotient}) and its denominator cancels $d_M^2$ (see Eq.~\eqref{defectdimensionabelian}). 
A proof of Eq.\eqref{defecttotaldimension} by a general argument that applies even to non-Abelian parent states can be found in Ref.~\cite{BarkeshliBondersonChengWang14}.

\subsubsection{Basis transformation}\label{sec:Fbasistransformation}
The quantum ground states of a system of defects can be labeled by the quantum numbers of a maximal set of commuting Wilson-line observables, whose eigenvalues are associated to the internal fusion channels and vertex degeneracies of a fusion tree (see Fig.~\ref{fig:Fmoves}). For example in the toric code case studied in section~\ref{sec:defectlatticemodel}, the $2^N$ ground state of a system with $2N$ dislocations can be specified by the simultaneous eigenvalues of the $e$-Wilson loops $\mathcal{W}^{2j-2,2j}$ surrounding the $(2j-1)^{\mathrm{th}}$ and $(2j)^{\mathrm{th}}$ defects. For instance, quantum states in a system of four Ising defects (or zero energy Majorana bound states) $\sigma_1,\sigma_2,\sigma_3,\sigma_4$ can be labeled by the commuting local fermion parities $\{\mathcal{W}^{12}=(-1)^{F_{12}},\mathcal{W}^{34}=(-1)^{F_{34}}\}$ of pairs of zero modes. They label the fusion channels $1$ or $\psi$ of $\sigma_1\times\sigma_2$ and $\sigma_3\times\sigma_4$.

Different fusion trees correspond to distinct sets of Wilson operators that may not commute. Each tree defines a complete basis for the degenerate ground states, and basis transformations between different fusion trees are generated by the $F$-symbols (see Fig.~\ref{fig:Fmoves}). They are defined in Eq.~\eqref{Fsymboldef} and \eqref{Fdefinition}.
Continuing with our example, quantum states of the four Ising defects in the toric code can also be labeled by a different set of local fermion parities $\{\mathcal{W}^{23}=(-1)^{F_{23}},\mathcal{W}^{41}=(-1)^{F_{41}}\}$, which does not commute with the original. Basis states with respect to these two sets of local observables are related by a unitary transformation $F^{\sigma\sigma\sigma}_\sigma$ to be discussed a bit later. 

\begin{figure}[htbp]
\centering\includegraphics[width=0.5\textwidth]{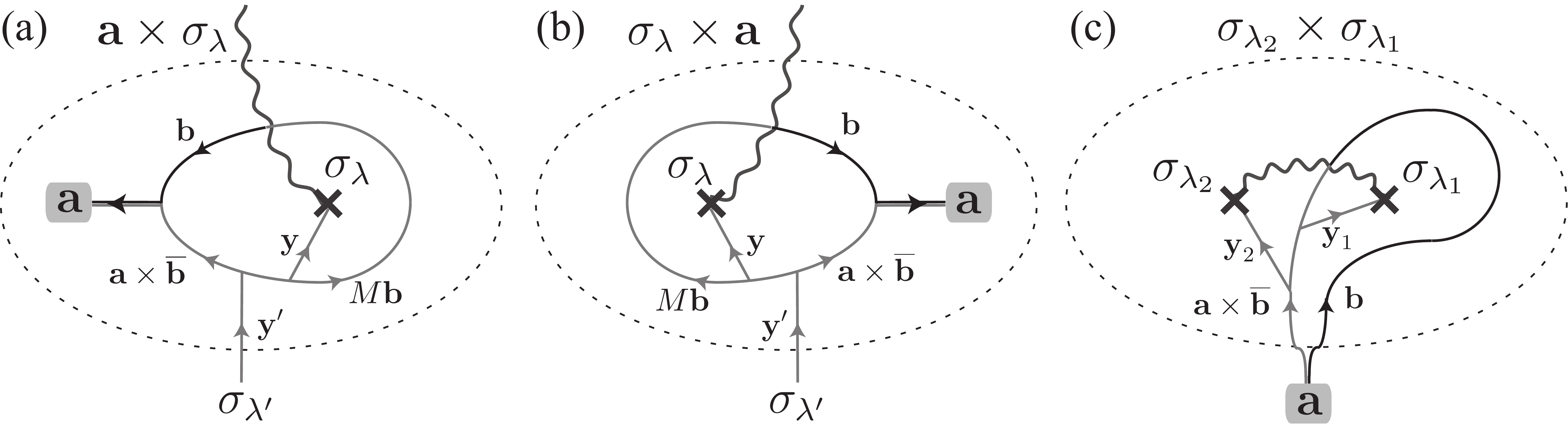}
\caption{Quasiparticle string configurations for splitting operator $L^{x_1x_2}_{x_3}$. Branch cuts (wavy lines) switch labels ${\bf b}\to M{\bf b}$ of passing quasiparticles according to the e-m symmetry $M:e\leftrightarrow m$. ${\bf b}=1$ when ${\bf a}=1,e$, or ${\bf b}=e$ when ${\bf a}=m,\psi$. String ${\bf y}$ determines species of the defect where it ends. ${\bf y}=1$ for $\sigma_0$ and ${\bf y}=e$ for $\sigma_1$.}\label{fig:splittingstatestoriccode}
\end{figure}

In general any two fusion trees, or maximally commuting sets of observables, can be connected by a seqence of $F$-transformations, like the one shown in figure~\ref{fig:Fmoves}. The overall basis transformation between any two fusion trees is path independent. This cocycle consistency condition is ensured by the pentagon equation ``$FF=FFF$" (see Fig.~\ref{fig:Fpentagon}) and MacLane's coherence theorem (see Refs.~\cite{Kitaev06, MacLanebook}). Instead of solving the algebraic pentagon equation, the $F$-matrices can be more efficiently computed directly from their definition~\cite{TeoRoyXiao13long,teo2013braiding,TeoHughesFradkin15}. This is done by choosing a particular set of {\em splitting operator} $[L^{xy}_z]$ for an admissible fusion process $x\times y\to z$. Each operator can be diagrammatically represented by quasiparticle Wilson strings around $x$ and $y$ with fixed boundary conditions associated to the fusion outcome $z$. An example of this is given for defects in the toric code (see Fig.~\ref{fig:splittingstatestoriccode}). In fact the string configuration in Fig.~\ref{fig:splittingstategeneraldefect1} can also be treated as a choice of splitting state for a conjugate pair of bare defects in a general Abelian system. 

We first demonstrate the splitting of a fermion into a pair of bare defects, $\psi\to\sigma_0\times\sigma_0$, in the toric code. $\psi$ can be created from the ground state by an open (semi-infinite) Wilson string of $\sigma_{x,z}$ operators that ends at $\psi$ (see left side of \eqref{splittingdefinition} below). Splitting is the procedure of cutting out a disc containing $\psi,$ and then replacing the interior by a new lattice with two dislocation defects $\sigma$. \begin{align}\vcenter{\hbox{\includegraphics[width=0.4\textwidth]{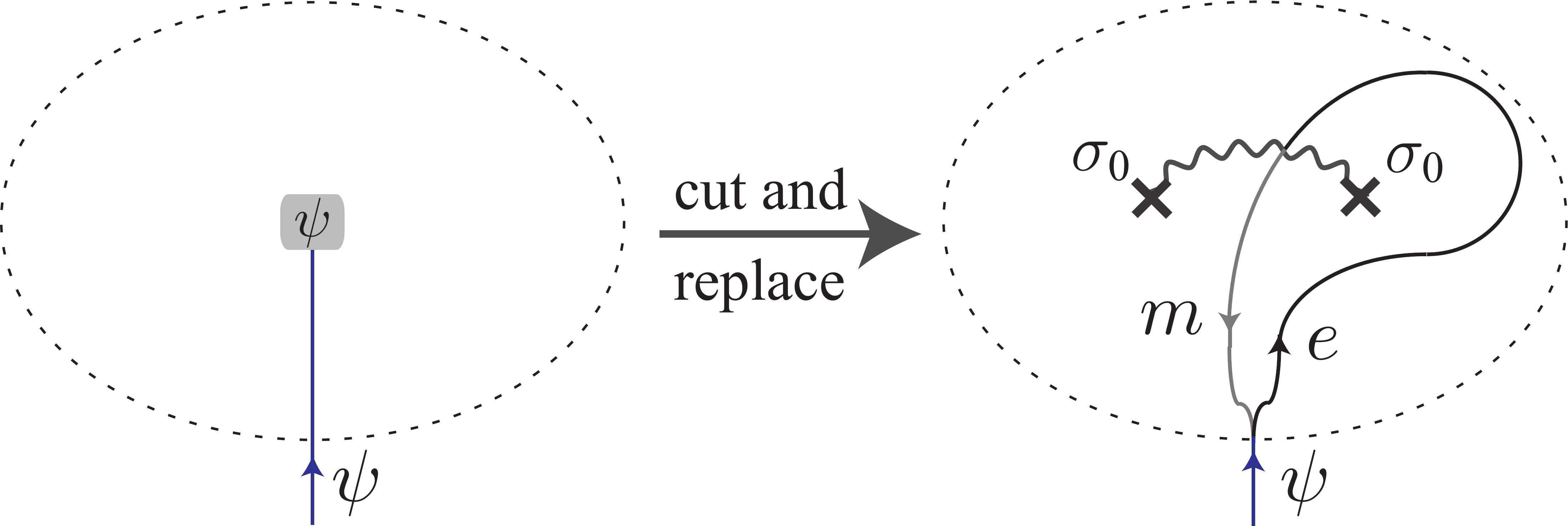}}}.\label{splittingdefinition}\end{align} The replacement must match the boundary condition (along the dashed line) so that the quasiparticle string continues into the domain. However, the string cannot terminate, otherwise there would be another fermion in the new domain, and splitting would be $\psi\to\sigma\times\sigma\times\psi$ instead of $\psi\times\sigma\times\sigma$. This means the Wilson string must wind non-trivially around the defects. One particular possibility is presented in \eqref{splittingdefinition}. A splitting state $[L^{\sigma_0,\sigma_0}_\psi]$ is the equivalence class of such string configurations. For example, the string could deform inside the region and would correspond to the same splitting state. However, the mirror image of \eqref{splittingdefinition} is an inequivalent string pattern, and differs from the original by an $e$-loop around the defect pair. The $e$-loop cuts the $\psi$-string and gives an additional minus sign from crossing. Fig.~\ref{fig:splittingstatestoriccode} picks the string patterns of particular splitting states for $\sigma_{\lambda'}\to{\bf a}\times\sigma_{\lambda}$, $\sigma_{\lambda'}\to\sigma_\lambda\times{\bf a}$ and ${\bf a}\to\sigma_{\lambda_1}\times\sigma_{\lambda_2}$.

Now let us describe the basis transformations that determine the set of $F$-symbols. Given a particular configuration of defects and quasiparticles, an eigenstate of a maximal set of commuting Wilson loop operators can be specified by the intermediate channels of a fusion tree. For example, the eigenvalues of the following three Wilson loops encodes the same information as the three Abelian anyons ${\bf a}_1,{\bf a}_2,{\bf a}_3$ in the fusion tree:
\begin{align}
\vcenter{\hbox{\includegraphics[width=0.23\textwidth]{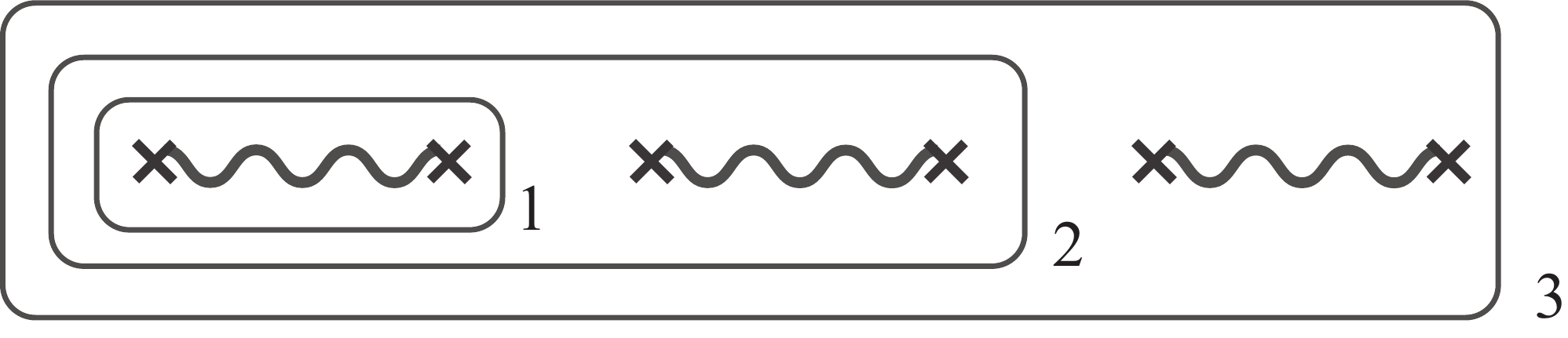}}}\leftrightarrow\left|\vcenter{\hbox{\includegraphics[width=0.16\textwidth]{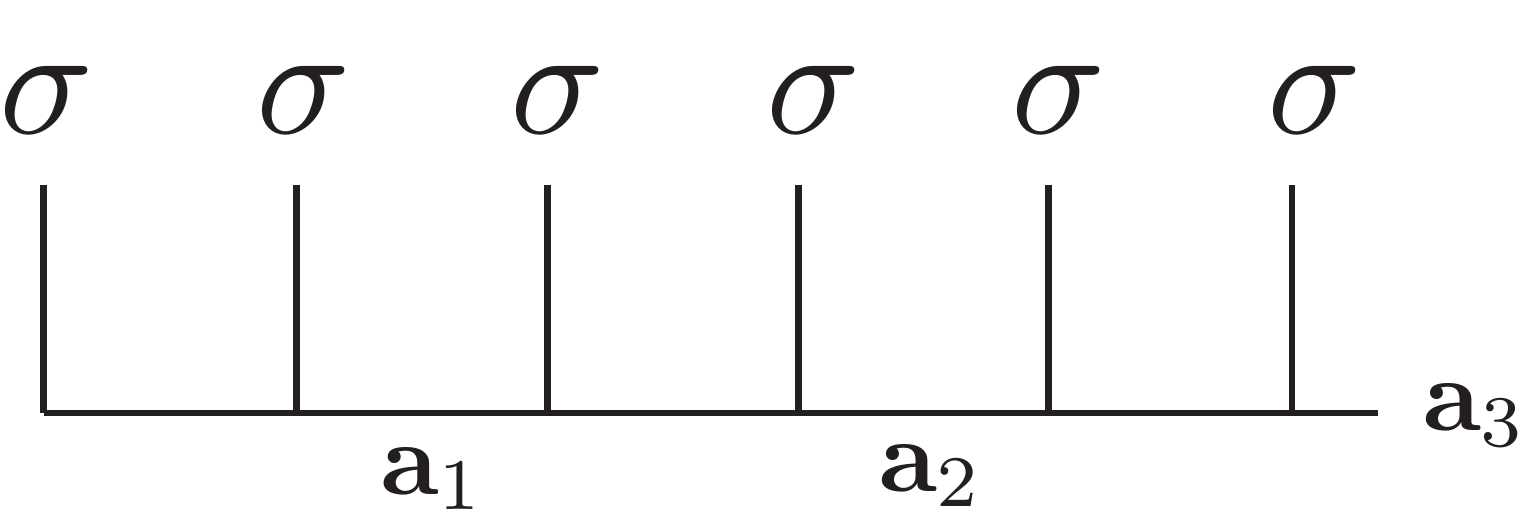}}}\right\rangle
\label{toriccodefusiontree}\end{align}
 Each leg of a fusion tree is labeled by a quasiparticle or defect. All vertices must be trivalent and admissible according to the fusion rules. Each vertex associated to a splitting state is diagrammatically represented by specific quasiparticle string patterns as shown in Fig.~\ref{fig:splittingstatestoriccode}. Piecing together the local string patterns of each vertex gives the global string structure. The strings must be attached continuously by matching boundary conditions between splitting states (c.f.~\eqref{splittingtreeapp}).
For example, by patching the splitting state string patterns of $\psi\to\sigma_0\times\sigma_0$ and $\sigma_0\to\psi\times\sigma_0$ from Fig.~\ref{fig:splittingstatestoriccode}(a) and (c), \begin{align}\left|\vcenter{\hbox{\includegraphics[width=0.07\textwidth]{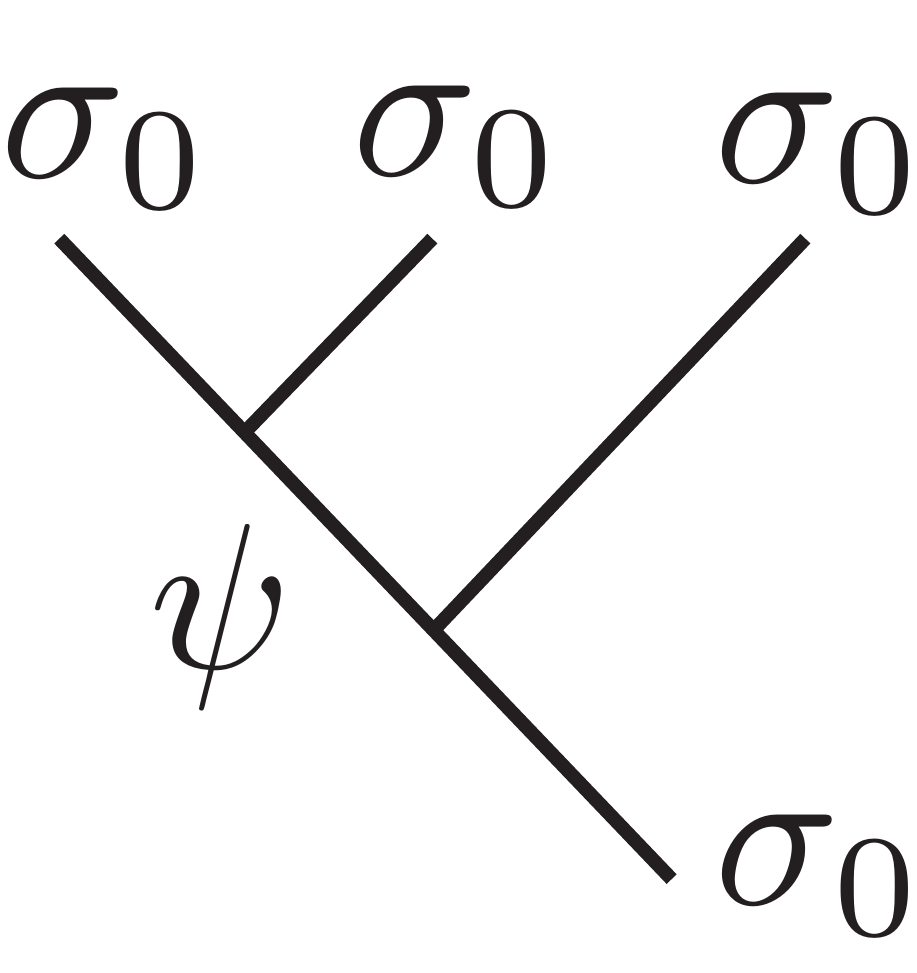}}}\right\rangle=\vcenter{\hbox{\includegraphics[width=0.25\textwidth]{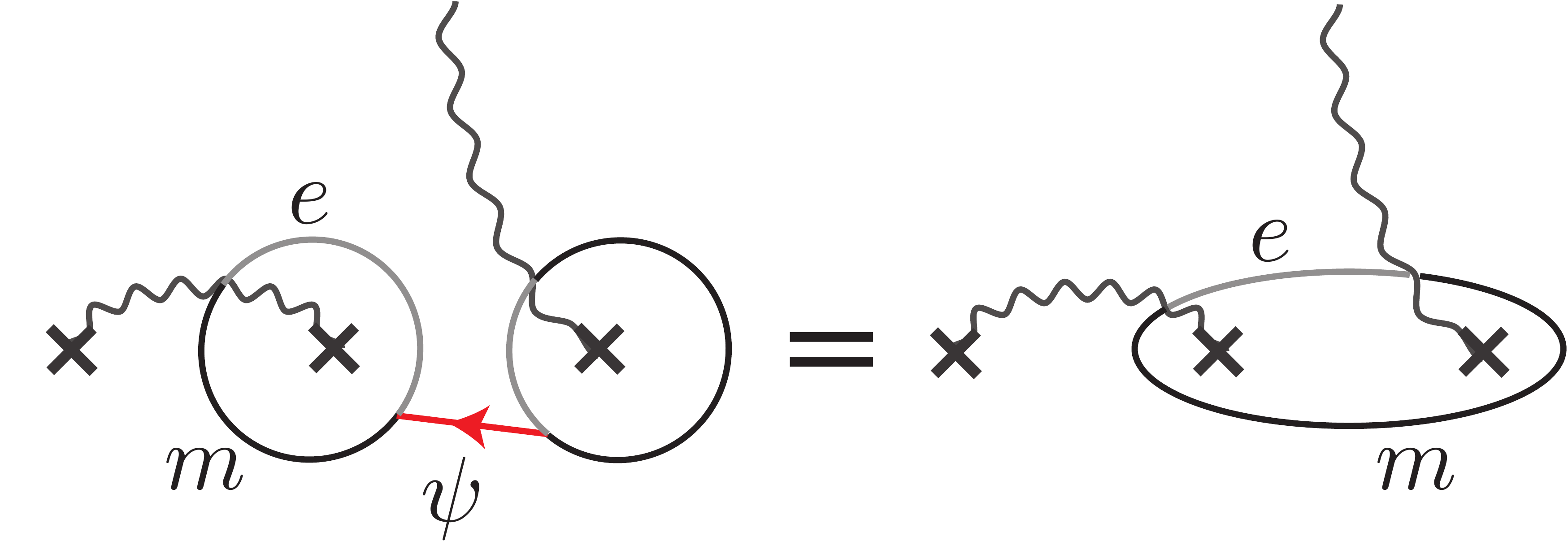}}}.\label{Fsss1toriccode}\end{align} 

The vacuum splitting $1\to\sigma_0\times\sigma_0$ is diagramatically represented by a pair of defects joined together by a branch cut with no quasiparticle strings. The original vacuum requires all Wilson loops $\mathcal{W}$ around the pair to condense to the ground state with a trivial phase, i.e. act trivially on the ground state. In the case of three defects, different configurations of branch cuts correspond to different ground states. For instance, \begin{align}\left|\vcenter{\hbox{\includegraphics[width=0.07\textwidth]{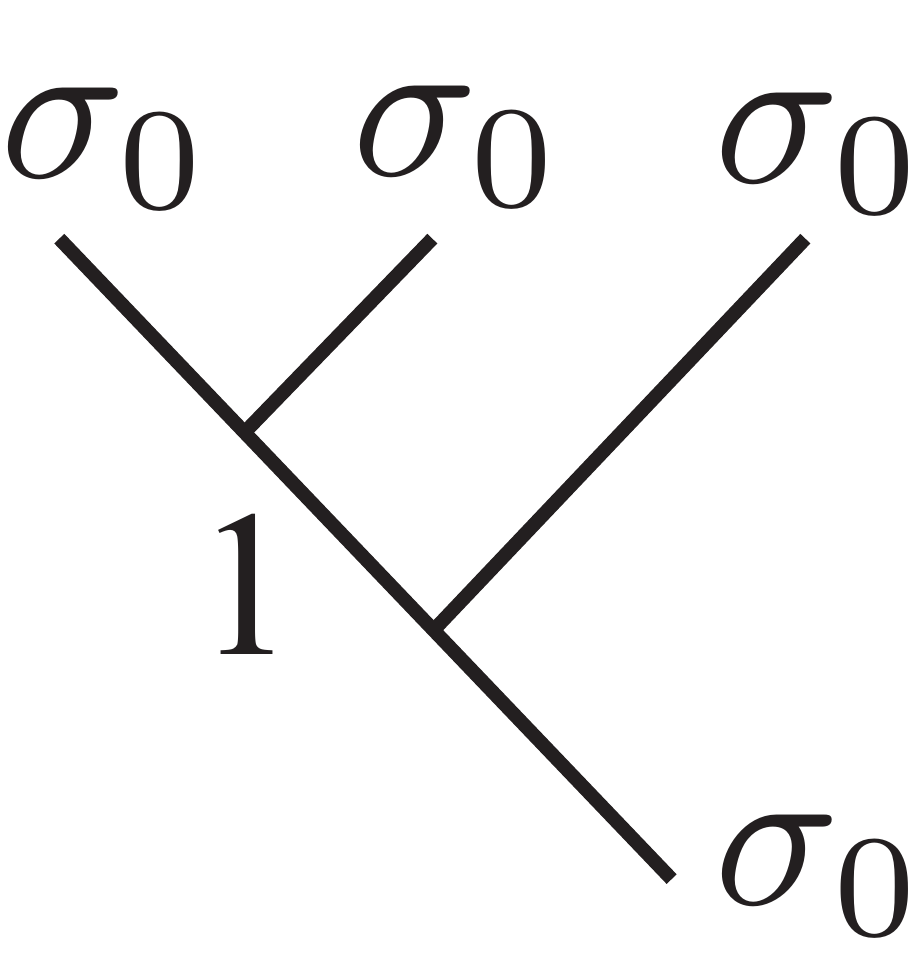}}}\right\rangle&=\vcenter{\hbox{\includegraphics[width=0.1\textwidth]{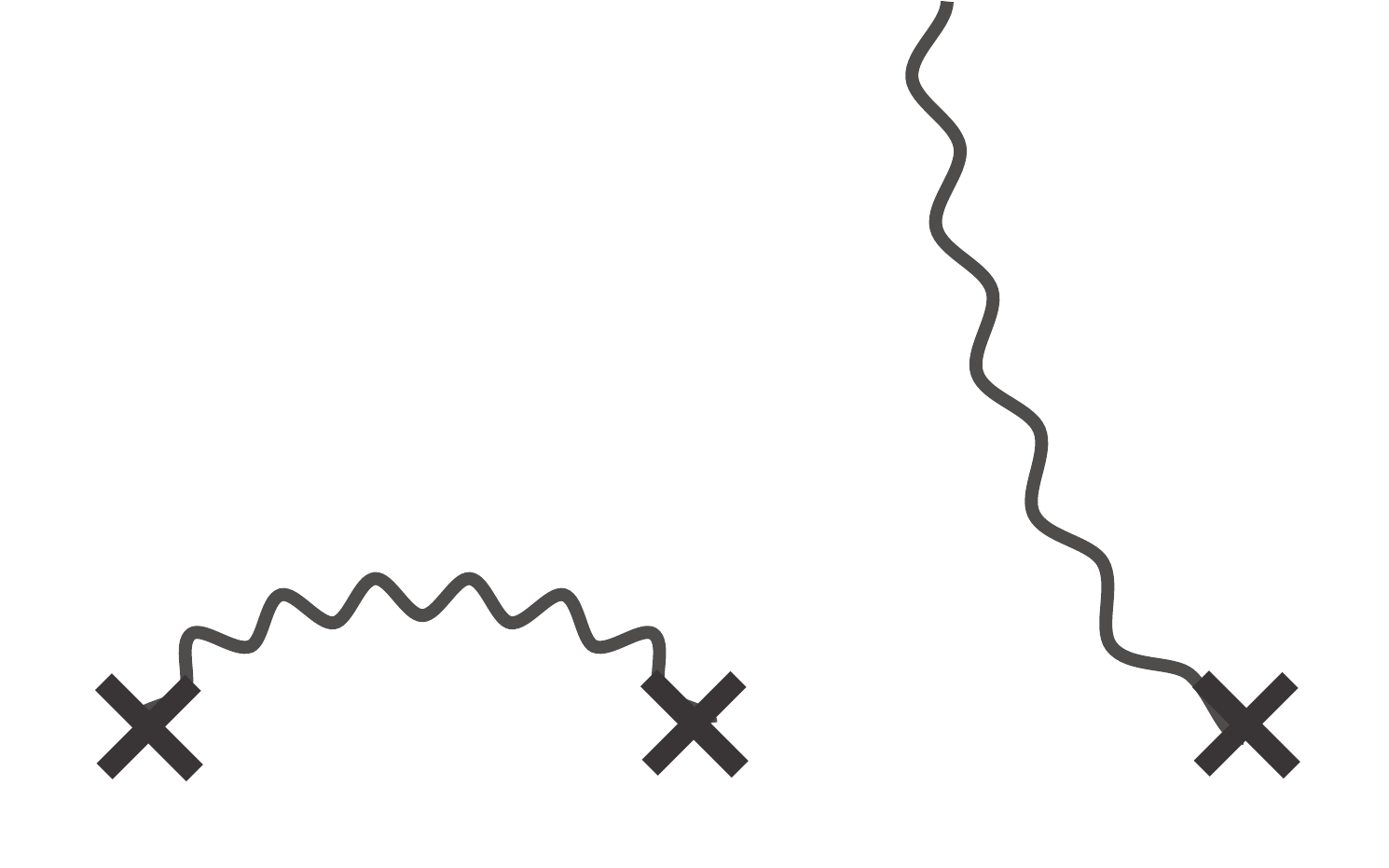}}}=\frac{(-1)^s}{\sqrt{2}}\sum_{{\bf a}=1,m}\vcenter{\hbox{\includegraphics[width=0.1\textwidth]{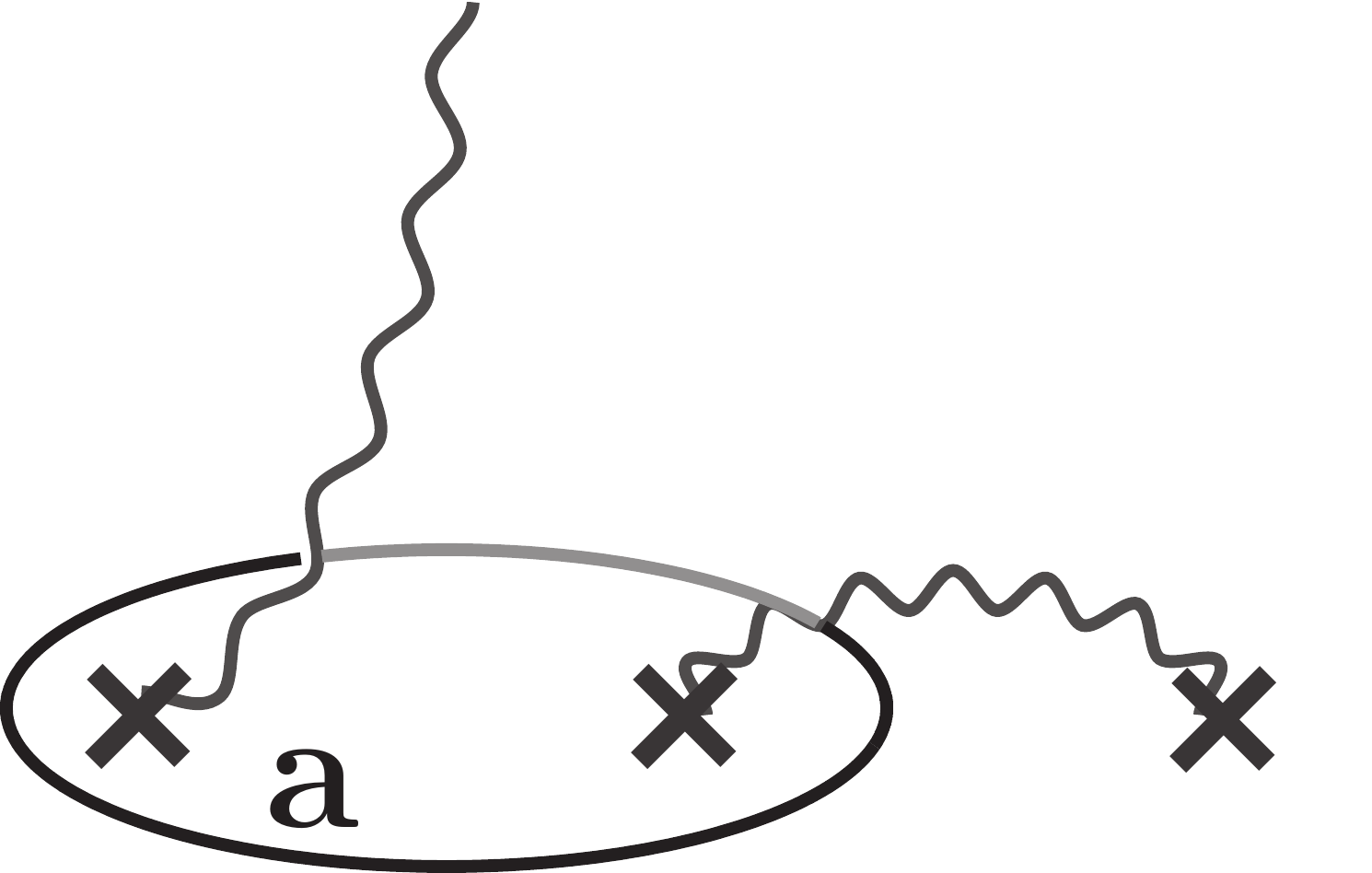}}}\nonumber\\&=\frac{(-1)^s}{\sqrt{2}}\left[\vcenter{\hbox{\includegraphics[width=0.1\textwidth]{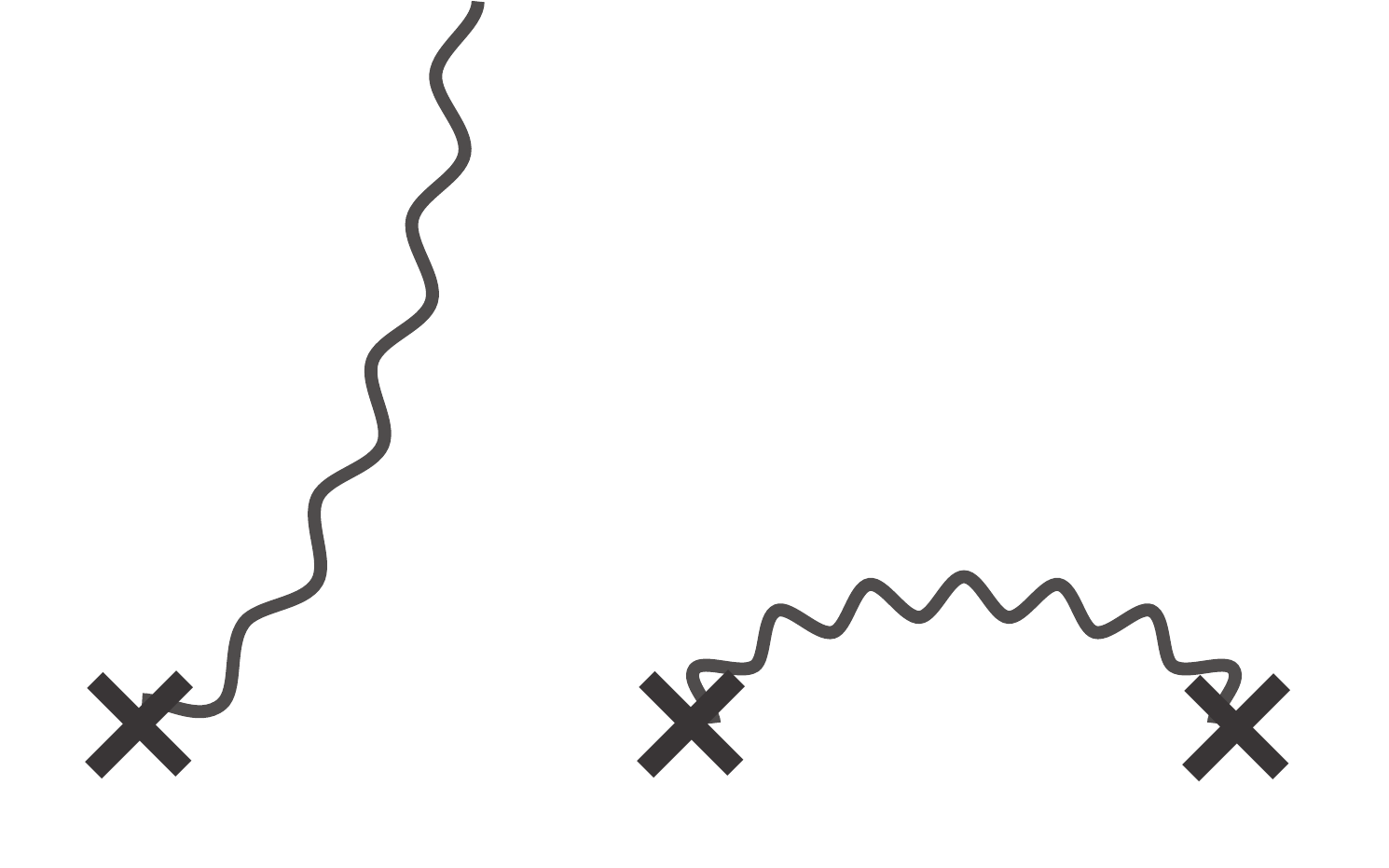}}}\right]+\frac{(-1)^s}{\sqrt{2}}\left[\vcenter{\hbox{\includegraphics[width=0.1\textwidth]{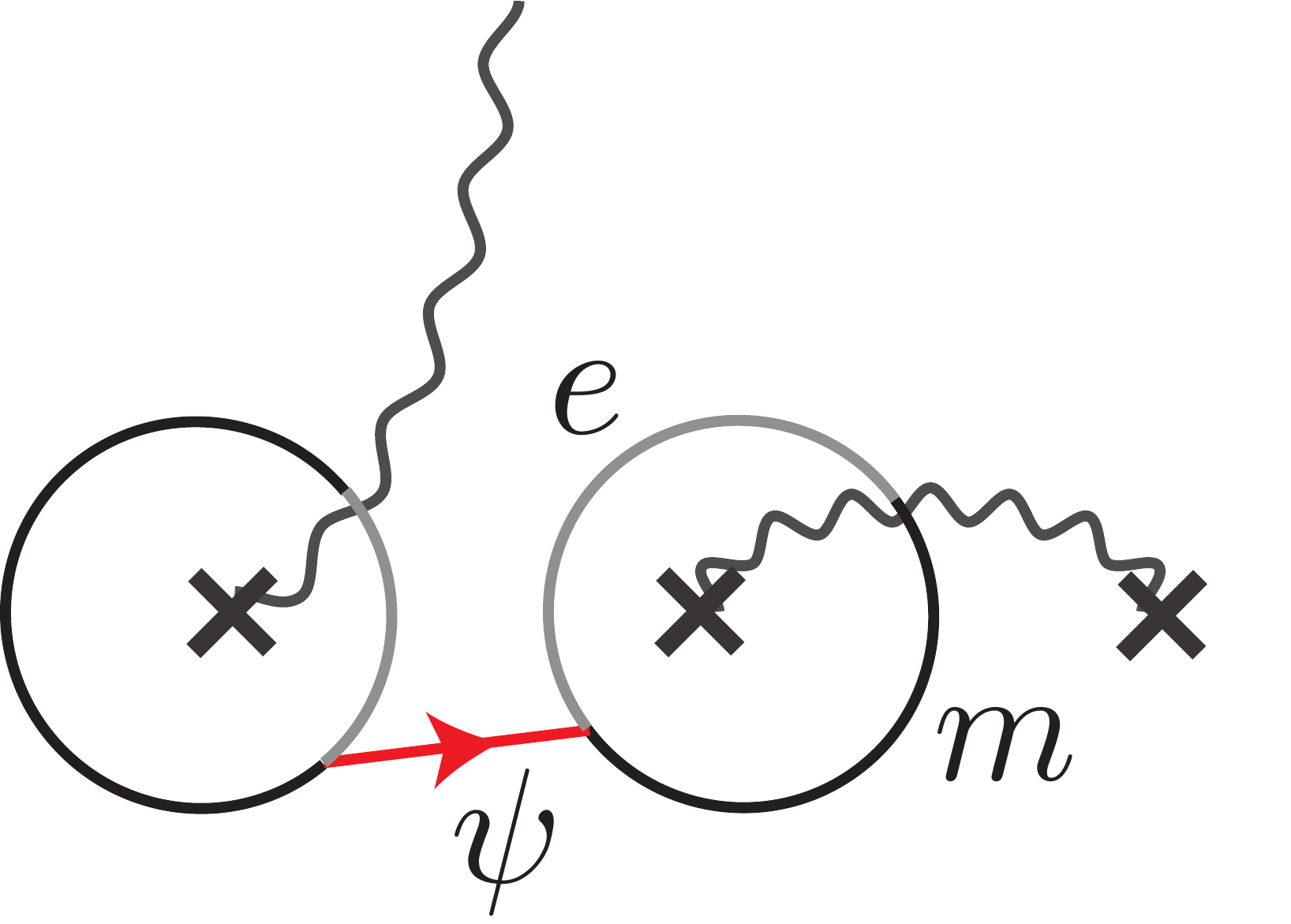}}}\right]\nonumber\\&=\frac{(-1)^s}{\sqrt{2}}\left|\vcenter{\hbox{\includegraphics[width=0.07\textwidth]{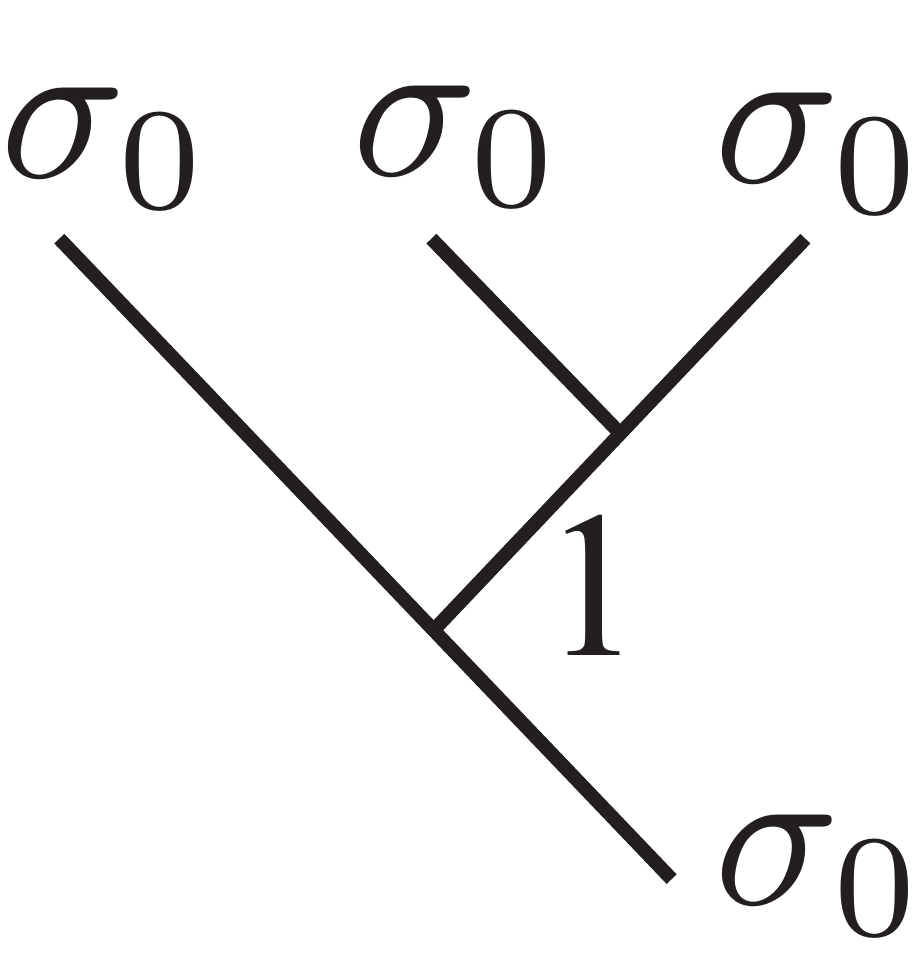}}}\right\rangle+\frac{(-1)^s}{\sqrt{2}}\left|\vcenter{\hbox{\includegraphics[width=0.07\textwidth]{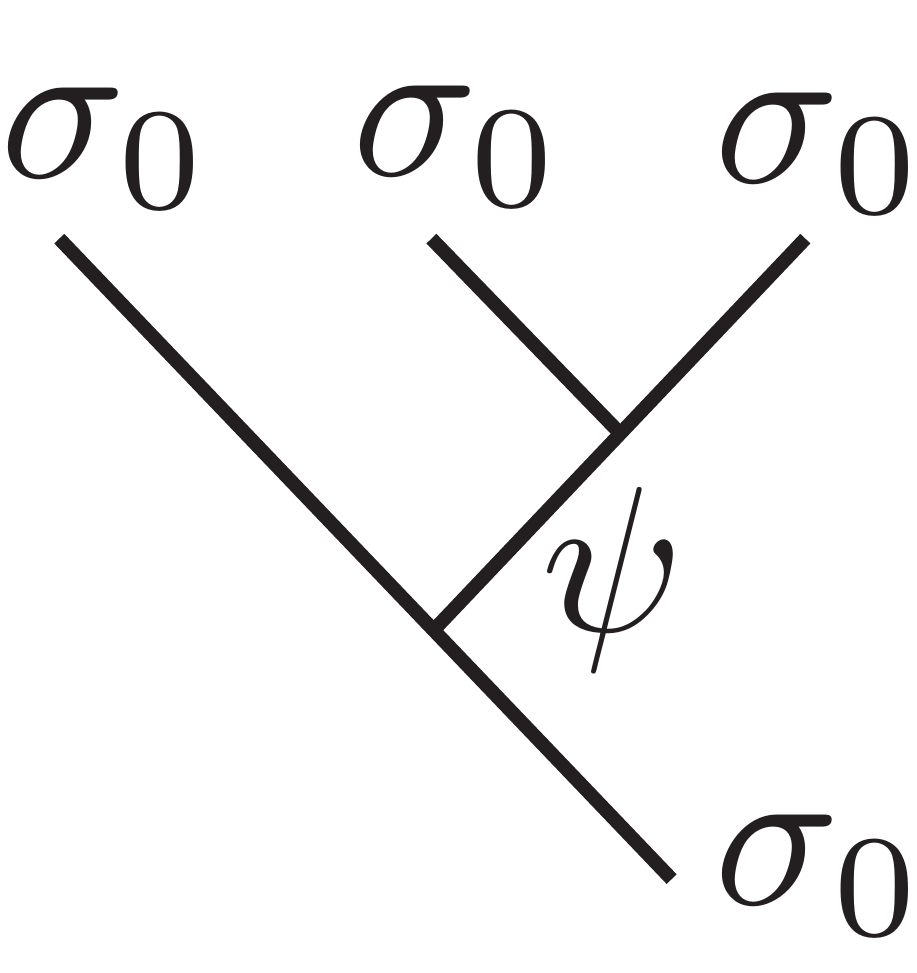}}}\right\rangle\label{Fsss2toriccode}\end{align} where the sum of the Wilson operators ${\bf a}=1,m$ around the first two defects forms a projection operator and forces the vacuum fusion channel. For the toric code, the sign is positive ($s=0$). However when the lattice model is modified by a $\mathbb{Z}_2$-symmetry protected topological state, the sign is flipped ($s=1$). This sign determines the Frobenius-Schur indicator $\varkappa_\sigma=(-1)^s$ (see Eq.~\eqref{FSindicator}), which is classified by $H^3(\mathbb{Z}_2,U(1))=\mathbb{Z}_2$.

Now combining \eqref{Fsss1toriccode} and \eqref{Fsss2toriccode}, and keeping track of the crossing phase of the Wilson loops, we have \begin{align}\left|\vcenter{\hbox{\includegraphics[width=0.07\textwidth]{Fsss0toriccode}}}\right\rangle&=\frac{(-1)^s}{\sqrt{2}}\sum_{{\bf a}=1,m}\vcenter{\hbox{\includegraphics[width=0.1\textwidth]{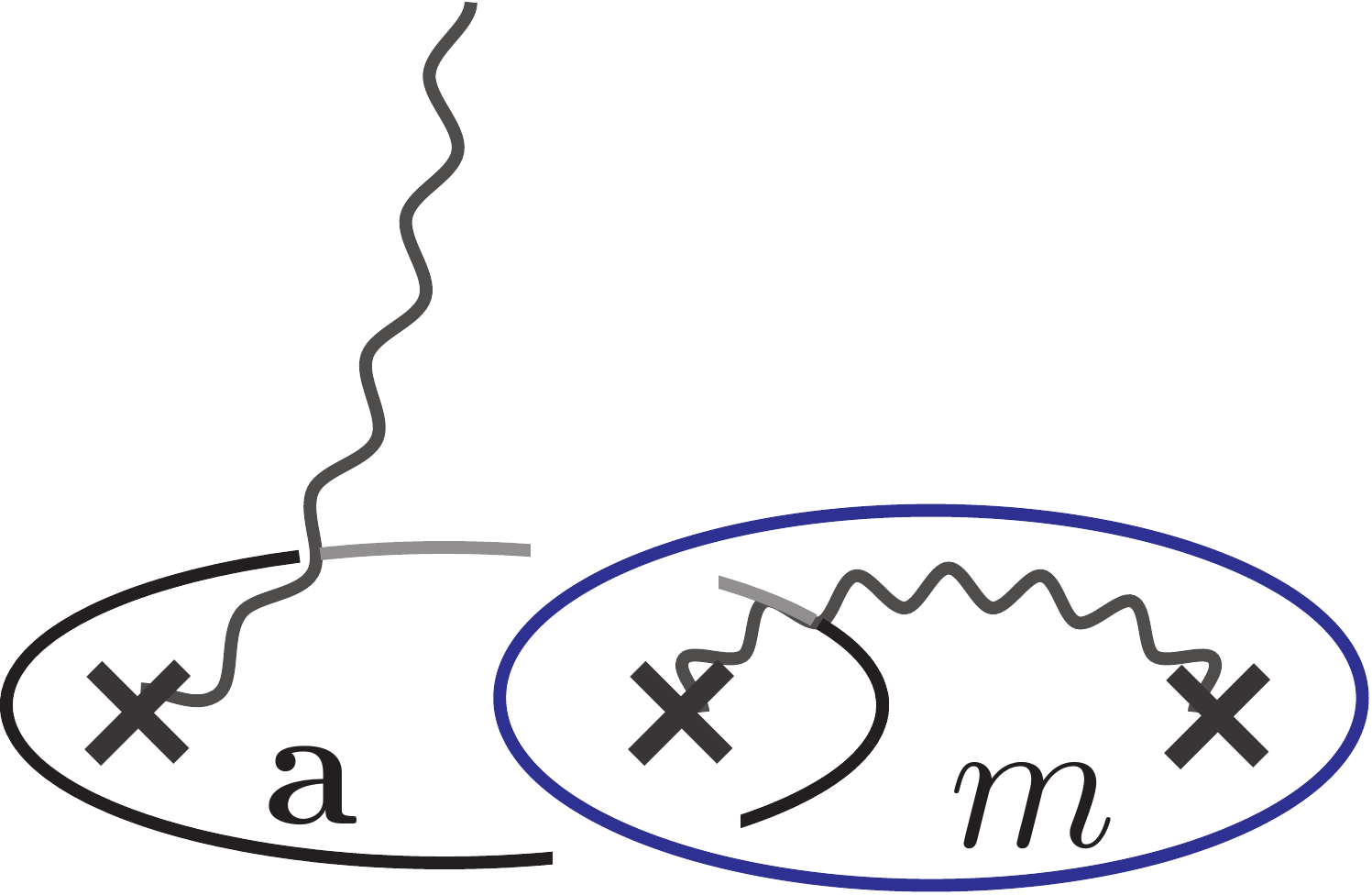}}}\nonumber\\&=\frac{(-1)^s}{\sqrt{2}}\left[\vcenter{\hbox{\includegraphics[width=0.1\textwidth]{Fsss8toriccode}}}\right]-\frac{(-1)^s}{\sqrt{2}}\left[\vcenter{\hbox{\includegraphics[width=0.1\textwidth]{Fsss5toriccode}}}\right]\nonumber\\&=\frac{(-1)^s}{\sqrt{2}}\left|\vcenter{\hbox{\includegraphics[width=0.07\textwidth]{Fsss6toriccode}}}\right\rangle-\frac{(-1)^s}{\sqrt{2}}\left|\vcenter{\hbox{\includegraphics[width=0.07\textwidth]{Fsss7toriccode}}}\right\rangle.\label{Fsss3toriccode}\end{align}
Eqs.~\eqref{Fsss2toriccode} and~\eqref{Fsss3toriccode} describe the basis transformations between different fusion trees. They transform between eigenstates of different sets of Wilson operators. In general, basis transformations are generated by $F$-symbols defined in \eqref{Fsymboldef}. 

Following a similar procedure to that of Eqs.~\eqref{Fsss1toriccode}, \eqref{Fsss2toriccode}, and \eqref{Fsss3toriccode}, we can obtain a consistent set of $F$-symbols with arbitrary admissible labels $\lambda_i=1,e,m,\psi,\sigma_0,\sigma_1$ along the external legs. The fusion trees \begin{align}\left|\vcenter{\hbox{\includegraphics[width=0.5in]{F1}}}\right\rangle&=\mathop{\sum_{boundary}}_{condition}\left[L^{\lambda_3\lambda_2}_{x}\right]\otimes\left[L^{x\lambda_1}_{\lambda_4}\right]|GS\rangle\nonumber\\\left|\vcenter{\hbox{\includegraphics[width=0.5in]{F2}}}\right\rangle&=\mathop{\sum_{boundary}}_{condition}\left[L^{\lambda_3y}_{\lambda_4}\right]\otimes\left[L^{\lambda_2\lambda_1}_{y}\right]|GS\rangle\end{align} can be diagrammatically represented by patching the splitting states defined in Fig.~\ref{fig:splittingstatestoriccode} together with matching boundary conditions. Their overlap can be derived by keeping track of how quasiparticle strings are deformed and intersect. The results for the F-symbols of the defect fusion category based on the toric code anyonic symmetry are summarized in Table~\ref{tab:Fsymbolstoriccode}.

\begin{table}[htbp]
\centering
\begin{tabular}{ll}
\multicolumn{2}{c}{Defect $F$-symbols for the (bare) toric code}\\\hline
$F^{{\bf a}{\bf b}{\bf c}}_{\bf d}$, $F^{{\bf a}{\bf b}\sigma}_{\sigma}$, $F^{\sigma{\bf a}{\bf b}}_{\sigma}$, $F^{{\bf a}\sigma\sigma}_{\bf b}$, $F^{\sigma\sigma{\bf a}}_{\bf b}$ & $1$ \\
$F^{{\bf a}\sigma{\bf b}}_{\sigma}$, $F^{\sigma{\bf a}\sigma}_{\bf b}$ & $(-1)^{a_2b_2}$ \\
$\left[F^{\sigma\sigma\sigma}_{\sigma}\right]_{\bf a}^{\bf b}$ & $\frac{1}{\sqrt{2}}(-1)^{a_2b_2}$
\end{tabular}
\caption{Admissible $F$-symbols for defects in the toric code with quasiparticle decomposition ${\bf a}=e^{a_1}m^{a_2}$, ${\bf b}=e^{b_1}m^{b_2}$. They do not depend on the defect species $\sigma=\sigma_0,\sigma_1$.}\label{tab:Fsymbolstoriccode}
\end{table}

\subsubsection{\texorpdfstring{$S_3$}{S3} defects in \texorpdfstring{$SO(8)_1$}{SO(8)}}\label{sec:defectSO(8)}
The toric code is an Abelian topological phase with an Abelian global symmetry. In general both the topological phase and the symmetry group can be non-Abelian. Here we illustrate the defect structure of two non-Abelian examples. We first begin with the chiral $SO(8)_1$ state, which is Abelian but carries a non-Abelian symmetry (see section~\ref{sec:symmSO(8)}). It has three fermionic but mutually semionic fermions $\psi_1,\psi_2,\psi_3$, which satisfy the fusion rules $\psi_i^2=1$ and $\psi_i\times\psi_{i+1}=\psi_{i-1}$, for $i=1,2,3$ mod 3. The topological state has a $S_3=\{1,\rho,\overline\rho,\sigma_1,\sigma_2,\sigma_3\}$ symmetry generated by the threefold cyclic rotation $\rho:\psi_i\to\psi_{i+1}$ and the twofold permutation $\sigma_i:\psi_{i-1}\leftrightarrow\psi_{i+1}$. There are therefore six sectors in the $S_3$-graded defect fusion category \begin{align}\mathcal{C}=\mathcal{C}_1\oplus\mathcal{C}_{\rho}\oplus\mathcal{C}_{\overline\rho}\oplus\mathcal{C}_{\sigma_1}\oplus\mathcal{C}_{\sigma_2}\oplus\mathcal{C}_{\sigma_3}\end{align} where $\mathcal{C}_1=\langle1,\psi_1,\psi_2,\psi_3\rangle$ is the fusion content of the underlying $SO(8)_1$ state. From \eqref{defectsectorquotient} we can write down each defect sectors explicitly \begin{align}\mathcal{C}_\rho=\langle\rho\rangle,\quad\mathcal{C}_{\overline\rho}=\langle\overline\rho\rangle,\quad\mathcal{C}_{\sigma_i}=\langle(\sigma_i)_0,(\sigma_i)_1\rangle\end{align} where $(\sigma_i)_\lambda\times\psi_{i\pm1}=(\sigma_i)_{\lambda+1}$, for $\lambda=0,1$ mod 2 are the species labels for the twofold defect.

The defects obey the fusion rules \begin{gather}(\sigma_i)_0\times(\sigma_i)_0=1+\psi_i,\quad(\sigma_i)_0\times(\sigma_i)_1=\psi_{i-1}+\psi_{i+1},\nonumber\\\rho\times\rho=2\overline{\rho},\quad\rho\times\overline{\rho}=1+\psi_1+\psi_2+\psi_3,\nonumber\\(\sigma_i)_\lambda\times(\sigma_{i+1})_{\lambda'}=\rho,\quad(\sigma_i)_\lambda\times\rho=\overline\rho\times(\sigma_i)_\lambda=(\sigma_{i+1})_0+(\sigma_{i+1})_1,\nonumber\\(\sigma_i)_\lambda\times(\sigma_{i-1})_{\lambda'}=\overline{\rho},\quad(\sigma_i)_\lambda\times\overline{\rho}=\rho\times(\sigma_i)_\lambda=(\sigma_{i-1})_0+(\sigma_{i-1})_1.\nonumber\end{gather} Notice that, similar to the tricolor model which is also $S_3$ symmetric described in section~\ref{sec:defectlatticemodel}, fusion rules here contain degeneracies and some are non-commutative.

\begin{table}[ht]
\centering
\begin{tabular}{ll}
\multicolumn{2}{c}{Threefold defect $F$-symbols for $SO(8)_1$}\\\hline\noalign{\smallskip}
$F^{{\bf a}{\bf b}{\bf c}}_{{\bf a}+{\bf b}+{\bf c}}$ & $1$\\
$F^{\rho{\bf a}{\bf b}}_\rho$ & $R^{{\bf a}(\Lambda_3{\bf b})}$\\
$F^{{\bf a}{\bf b}\rho}_\rho$ & $R^{{\bf b}(\Lambda_3^2{\bf a})}$\\
$F^{{\bf a}\rho{\bf b}}_\rho$ & $\mathcal{D}S_{(\Lambda_3{\bf a}){\bf b}}$\\
$F^{\rho\overline\rho{\bf a}}_{\bf b}$ & $R^{{\bf a}(\Lambda_3^2{\bf b})}$\\
$F^{{\bf a}\rho\overline\rho}_{\bf b}$ & $R^{({\bf a}+{\bf b})(\Lambda_3^2{\bf a})}$\\
$F^{\rho{\bf a}\overline\rho}_{\bf b}$ & $\mathcal{D}S_{{\bf a}(\Lambda_3{\bf b})}R^{{\bf a}(\Lambda_3{\bf a})}$\\
$F^{\rho\rho{\bf a}}_{\overline\rho}$ & $R^{{\bf a}{\bf a}}\mathcal{A}_{\bf a}$\\
$F^{{\bf a}\rho\rho}_{\overline\rho}$ & $\mathcal{A}_{\Lambda_3^2\bf a}$\\
$F^{\rho{\bf a}\rho}_{\overline\rho}$ & $R^{{\bf a}(\Lambda_3^2{\bf a})}\mathcal{A}_{\Lambda_3\bf a}$\\
$F^{\rho\rho\rho}_{\bf a}$ & $\mathcal{A}_{\bf a}\exp\left[\frac{\pi}{3}\left(\frac{\mathcal{A}_{\psi_1}+\mathcal{A}_{\psi_2}+\mathcal{A}_{\psi_3}}{\sqrt{3}}\right)\right]$\\
$\left[F^{\rho\overline\rho\rho}_{\rho}\right]^{\bf b}_{\bf a}$ & $\frac{1}{2}\mathcal{D}S_{(\Lambda_3{\bf a}){\bf b}}R^{{\bf b}(\Lambda_3{\bf b})}$\\
$\left[F^{\overline\rho\rho\rho}_\rho\right]_{\bf a}$ & $\frac{1}{\sqrt{2}}R^{{\bf a}{\bf a}}\mathcal{A}_{\Lambda_3{\bf a}}\mathcal{A}_{\psi_2}$\\
$\left[F^{\rho\rho\overline\rho}_\rho\right]^{\bf b}$ & $\frac{1}{\sqrt{2}}R^{{\bf b}{\bf b}}\exp\left[\frac{\pi}{3}\left(\frac{\mathcal{A}_{\psi_1}+\mathcal{A}_{\psi_2}+\mathcal{A}_{\psi_3}}{\sqrt{3}}\right)\right]\mathcal{A}_{\psi_1}\mathcal{A}_{\bf b}$\\
\end{tabular}
\caption{The $R$-symbols are defined in \eqref{Rso8abelianapp}. $\mathcal{D}S_{{\bf a}{\bf b}}=R^{{\bf a}{\bf b}}R^{{\bf b}{\bf a}}$ is the braiding phase of ${\bf a}$ around ${\bf b}$. The $2\times2$ matrices $\mathcal{A}_{\bf a}$ act on splitting degeneracy $\overline\rho\to\rho\times\rho$ or $\rho\to\overline\rho\times\overline\rho,$ and can be represented by Pauli matrices \eqref{so(8)Arepapp}. $F$-symbols with interchanged $\rho\leftrightarrow\overline\rho$ are not listed explicitly but can obtained by replacing $\Lambda_3\leftrightarrow\Lambda_3^2$.}\label{tab:so(8)Fsymbols}
\end{table}

Next we highlight the $F$-transformations. Instead of presentating the full computation, we will explain the $F$-symbols involving only threefold defects. The detail derivations can be found in Ref.~\cite{TeoRoyXiao13long, TeoHughesFradkin15} and the results are summarized in table~\ref{tab:so(8)Fsymbols}. To express the $F$-symbol basis transformations in a simple notation, it is convenient to represent the Abelian quasiparticles by two dimensional $\mathbb{Z}_2$-valued vectors ${\bf a}=(0,0)=1$, $(1,0)=\psi_1$, $(0,1)=\psi_2,$ and $(1,1)=\psi_3$. In this notation, the threefold symmetry, for example, is represented by the $\mathbb{Z}_2$-valued matrix \begin{align}\Lambda_3=\left(\begin{array}{*{20}c}0&-1\\1&-1\end{array}\right)\equiv\left(\begin{array}{*{20}c}0&1\\1&1\end{array}\right).\end{align} The exchange $R$-symbols \eqref{Rsymboldefapp} between Abelian anyons in $SO(8)_1$ can be chosen to be \begin{align}R^{{\bf a}{\bf b}}=(-1)^{{\bf a}^T\sigma_x\Lambda_3^2{\bf b}},\label{Rso8abelianapp}\end{align} so that the braiding phase, $\mathcal{D}S_{{\bf a}{\bf b}}=R^{{\bf a}{\bf b}}R^{{\bf b}{\bf a}}=(-1)^{{\bf a}^T\sigma_x{\bf b}},$ agrees with that from the conventional $K$-matrix description \eqref{so(8)Kmatrix}. 


First, the $R$-symbol in eq.~\ref{Rso8abelianapp} is bilinear in ${\bf a}$ and ${\bf b},$ and therefore the $F$-symbols involving only Abelian quasiparticles can be chosen to be trivial, $F^{{\bf a}{\bf b}{\bf c}}_{{\bf a}+{\bf b}+{\bf c}}=1$, as they obey the hexagon identity \eqref{hexagoneq} (see also figure~\ref{fig:hexagon1}).
Next, the $F$-symbols for threefold defects are determined by fixing the splitting states. The splitting of $\rho\to{\bf a}\times\rho$ and ${\bf a}\to\rho\times\overline\rho$ are defined by the Wilson string structures \begin{align}\left[\vcenter{\hbox{\includegraphics[width=0.3in]{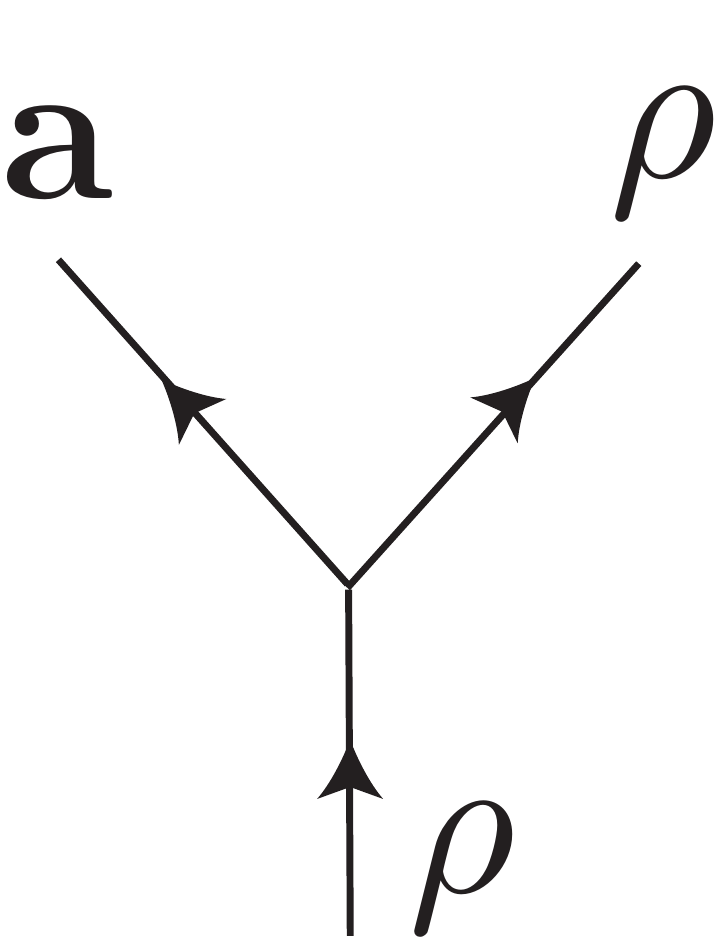}}}\right]=\vcenter{\hbox{\includegraphics[width=0.7in]{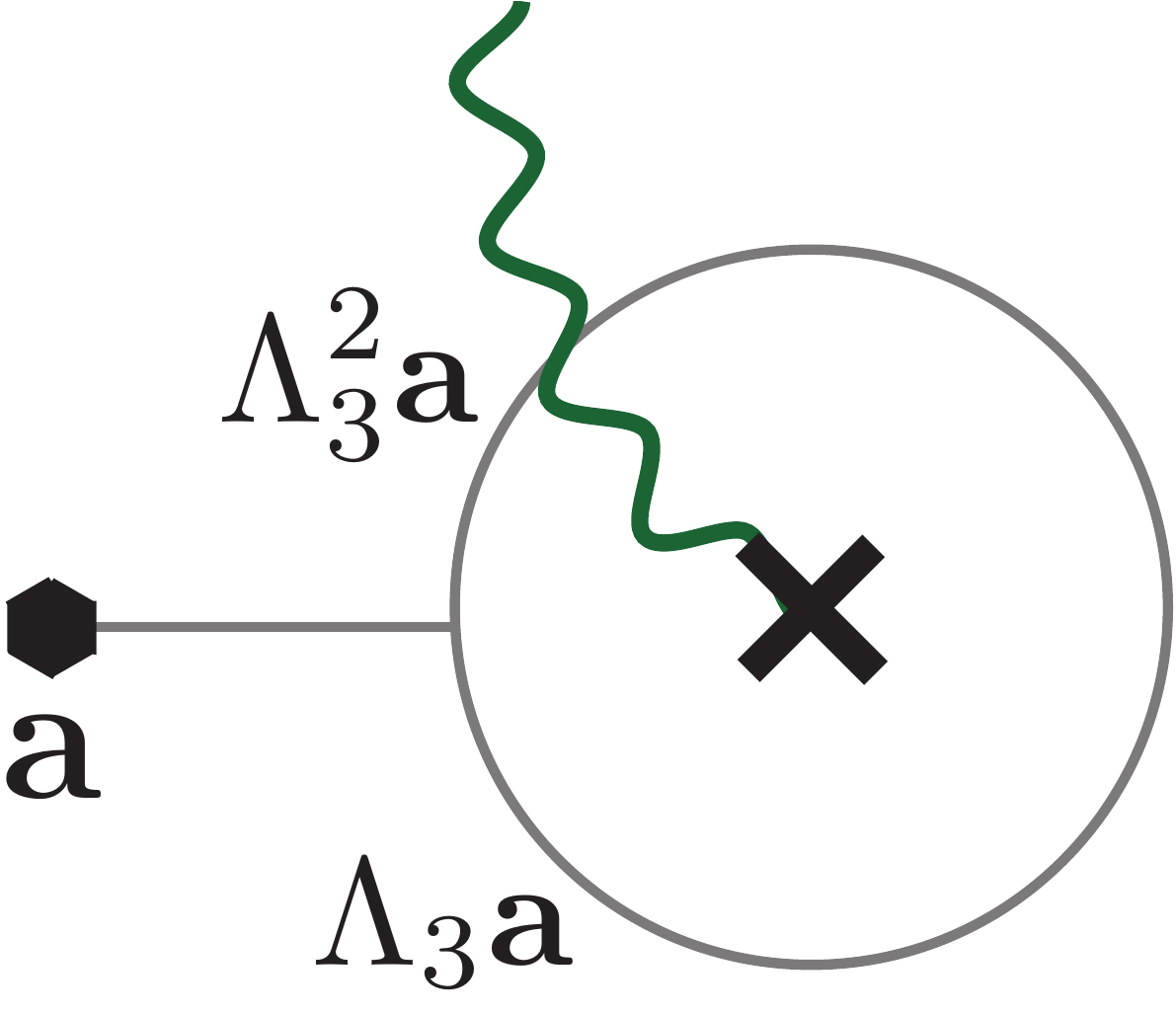}}},\quad\left[\vcenter{\hbox{\includegraphics[width=0.3in]{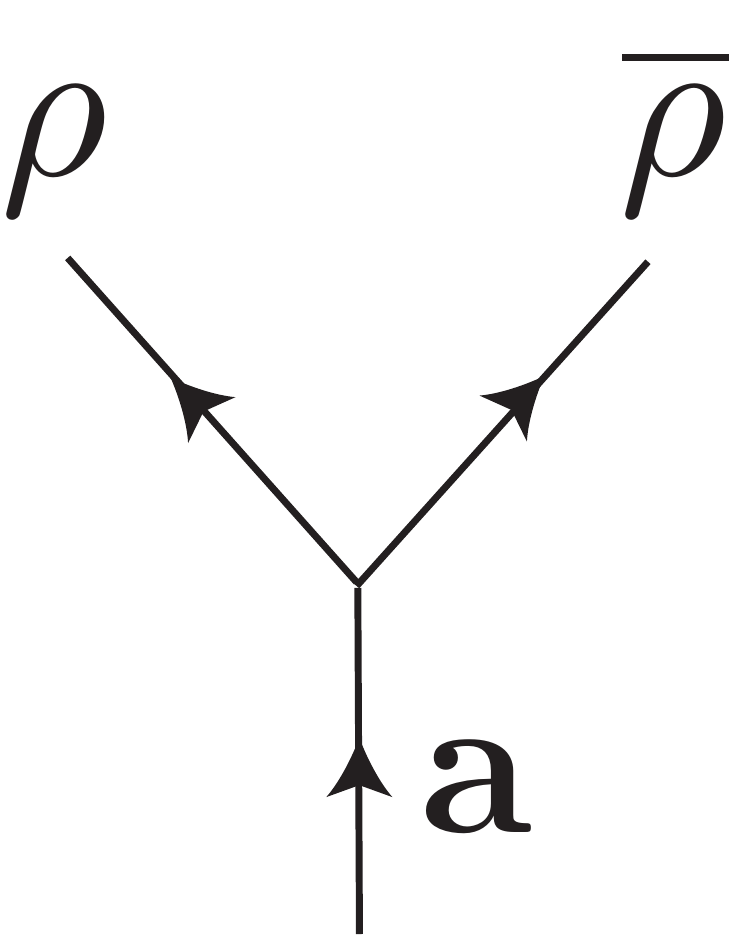}}}\right]=\vcenter{\hbox{\includegraphics[width=0.7in]{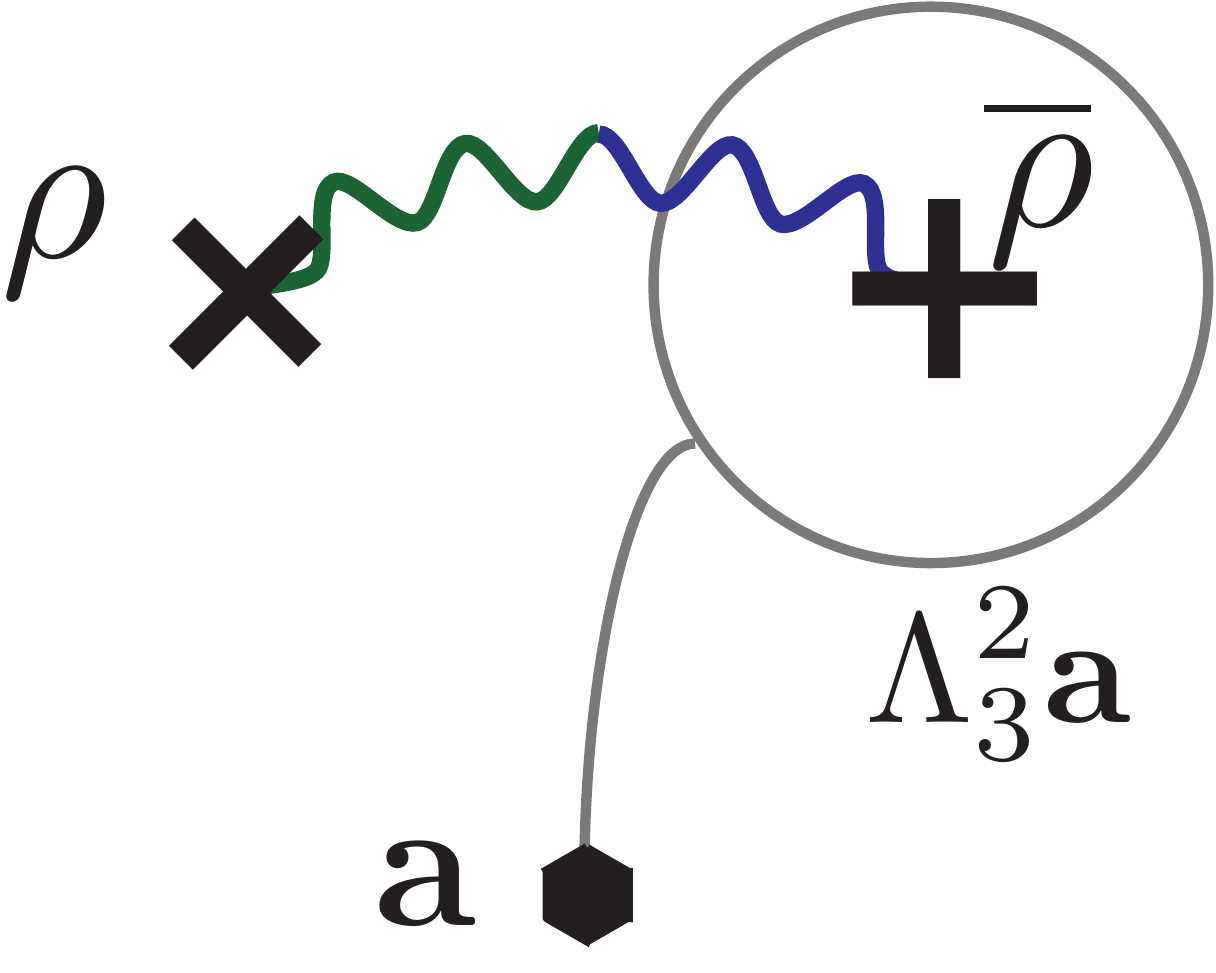}}}.\label{so(8)splittings}\end{align} Here a quasiparticle string changes type ${\bf a}\to\Lambda_3{\bf a}$ (or ${\bf a}\to\Lambda_3^{-1}{\bf a}$) as it goes from left to right across a $\rho$-branch cut represented by a green curvy line (resp.~a $\overline\rho$-branch cut represented by a blue curvy line).

Let $\mathcal{W}_{\bf b}$ be the ${\bf b}$-Wilson loop around the defect pair $\rho\times\overline\rho$. It commutes with the splitting state $1\to\rho\times\overline\rho$ in \eqref{so(8)splittings} because there are no Wilson strings intersecting $\mathcal{W}_{\bf b}$ and the overall vacuum state has unit eigenvalue with respect to $\mathcal{W}_{\bf b}$. On the ground state, this means \begin{align}\left|\vcenter{\hbox{\includegraphics[width=0.7in]{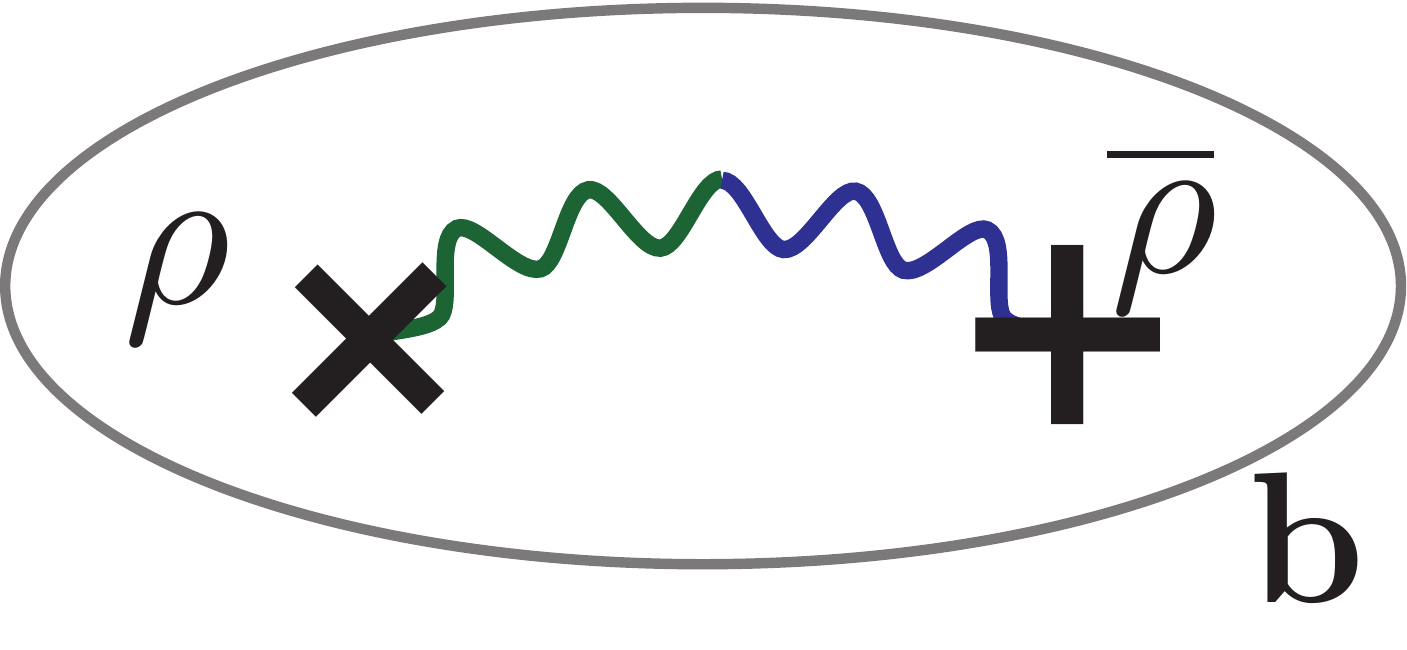}}}\right\rangle=\left|\vcenter{\hbox{\includegraphics[width=0.6in]{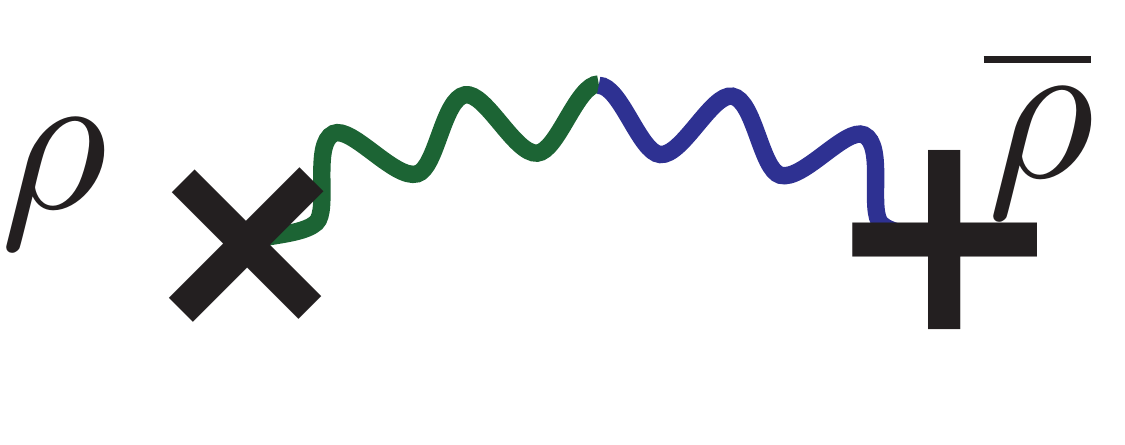}}}\right\rangle.\end{align} 

The two dimensional fusion degeneracy $N^{\overline\rho}_{\rho\rho}=2$ arises from an irreducible representation of the non-commuting algebra of Wilson operators \begin{align}\widehat{\mathcal{A}}_{\bf a}\left|\vcenter{\hbox{\includegraphics[width=0.3in]{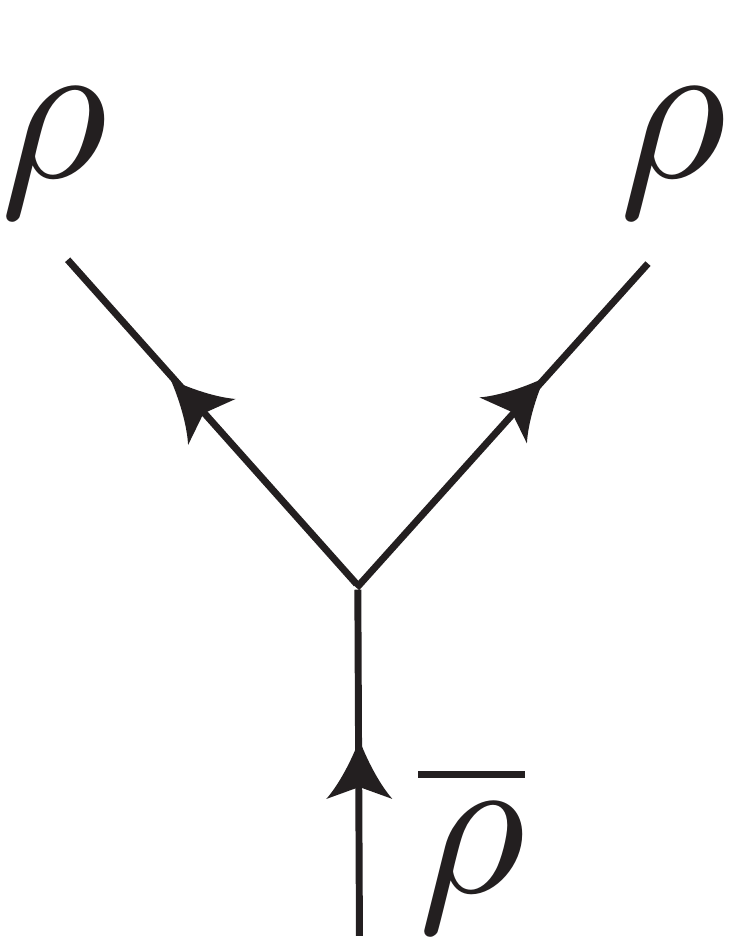}}}\right\rangle=\vcenter{\hbox{\includegraphics[width=0.8in]{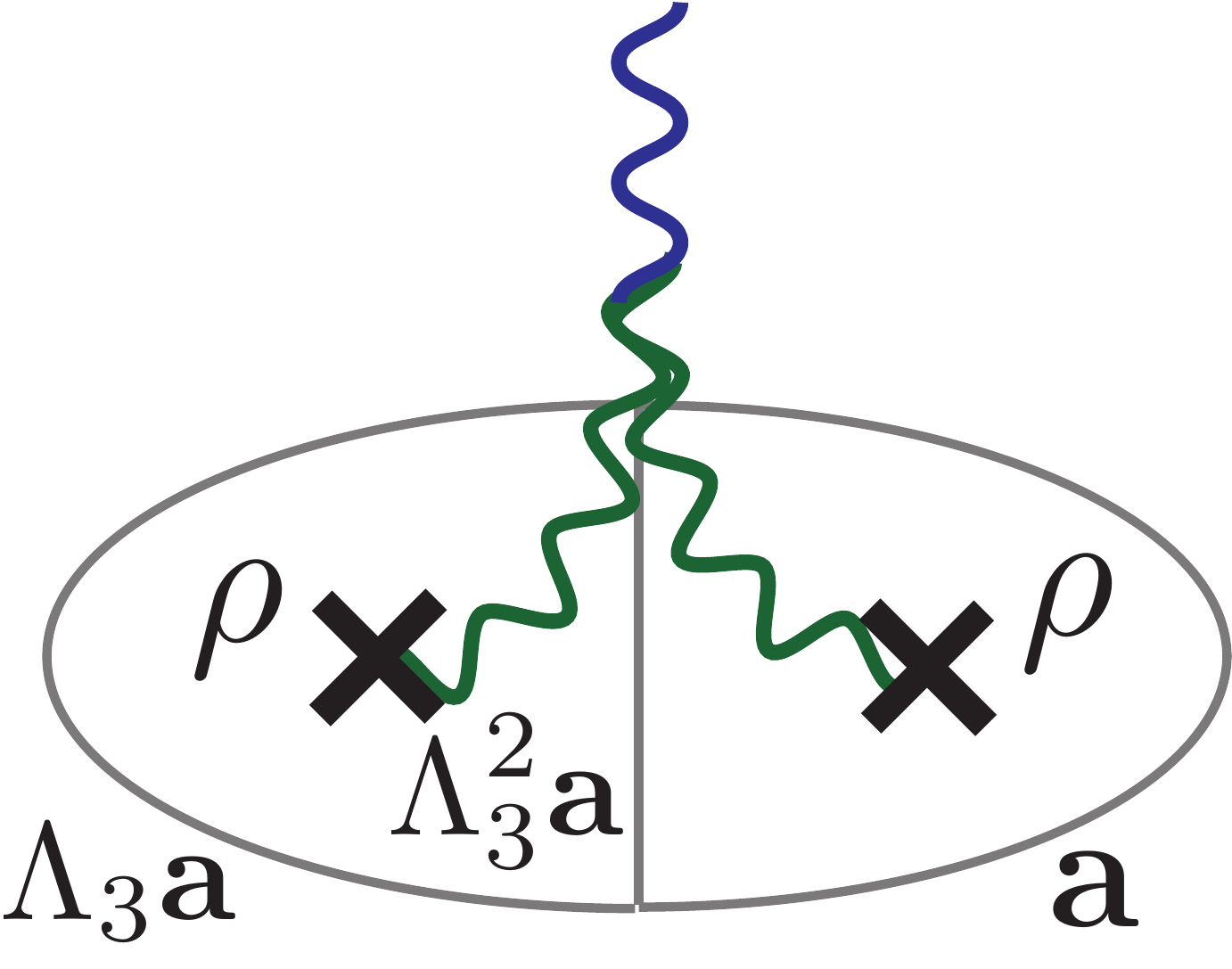}}}.\label{so(8)splittingrhorho}\end{align} By keeping track of the crossing phases of the Wilson strings, we find that the operators obey the algebraic relation \begin{align}\widehat{\mathcal{A}}_{\bf a}\widehat{\mathcal{A}}_{\bf b}=R^{{\bf a}{\bf b}}\widehat{\mathcal{A}}_{{\bf a}+{\bf b}}\label{so(8)AWilsonalgebra}\end{align} for $\widehat{\mathcal{A}}_a=1$. In particular, they satisfy the Clifford relation $\{\widehat{\mathcal{A}}_{\psi_i},\widehat{\mathcal{A}}_{\psi_j}\}=-2\delta_{ij},$ and exhibit the group structure of unit quaternions, $Q_8=\{\pm1,\pm\widehat{\mathcal{A}}_{\psi_1},\pm\widehat{\mathcal{A}}_{\psi_2},\pm\widehat{\mathcal{A}}_{\psi_3}\}$.  As such, they can be represented by $2\times2$ Pauli matrices \begin{align}\mathcal{A}_{\psi_1}=i\sigma_x,\quad\mathcal{A}_{\psi_2}=-i\sigma_y,\quad\mathcal{A}_{\psi_3}=i\sigma_z.\label{so(8)Arepapp}\end{align} In certain scenarios, it may be more appropriate to use a basis $|\pm\rangle$ which is symmetric under the threefold symmetry so that $\langle\pm|\widehat{\mathcal{A}}_{\psi_j}|\pm\rangle=\pm i/\sqrt{3}$ and $\langle+|\widehat{\mathcal{A}}_{\psi_j}|-\rangle=\sqrt{2/3}e^{2\pi ji/3}$. In addition to this algebraic structure, the splitting of $\rho\to\overline\rho\times\overline\rho$ has the same two-fold degeneracy. We thus have additional Wilson operators $\overline{\mathcal{A}}_{\bf a},$ which are defined similarly to \eqref{so(8)splittingrhorho}, except the threefold symmetry $\Lambda_3$ should be replaced by its inverse $\Lambda_3^{-1}=\Lambda_3^2$. The algebraic structure for $\overline{\mathcal{A}}_{\bf a}$ is identical to that of $\mathcal{A}_{\bf a}$ in \eqref{so(8)AWilsonalgebra}. Hence, we will not make the distinction between $\mathcal{A}$ and $\overline{\mathcal{A}}$ unless necessary. 


The $F$-symbols for threefold defects in $SO(8)_1$ are listed in Table~\ref{tab:so(8)Fsymbols}. They are evaluated by matching the Wilson strings of the splitting states $x\times(y\times z)$ and $(x\times y)\times z$. The $F$-transformations can be understood as a representation of the double cover of $A_4,$ which is the group of even permutations of 4 elements, or equivalently, the rotation group $T$ of a tetrahedron. The three $\mathcal{A}_{\psi_i}$ represent $\pi$-rotations about the $x$, $y,$ or $z$ axes and \begin{align}F^{\rho\rho\rho}_1=\exp\left[\frac{\pi}{3}\left(\frac{\mathcal{A}_{\psi_1}+\mathcal{A}_{\psi_2}+\mathcal{A}_{\psi_3}}{\sqrt{3}}\right)\right]\label{Frrrapp}\end{align} for instance, represents a $2\pi/3$ rotation about the $(111)$-axis. 

\subsubsection{\texorpdfstring{$S_3$}{S3} defects in the chiral ``4-Potts" state \texorpdfstring{$SU(2)_1/Dih_2$}{SU(2)/D2}}\label{sec:defect4Potts}
Next we move onto another example, the chiral ``4-Potts" state. It is a non-Abelian topological state presented in section~\ref{sec:trialitysymmetry}. Its gapless boundary is described by a CFT that matches with one of the chiral sectors of the 4-state Potts model~\cite{DijkgraafVafaVerlindeVerlinde99, CappelliAppollonio02}, or equivalently, the orbifold CFT~\cite{Ginsparg88, Harris88} $SU(2)_1/Dih_2=U(1)_4/\mathbb{Z}_2$. It supports 11 anyon types listes in table~\ref{tab:4statePottsanyons}. In particular there are three Abelian bosons $j_1,j_2,j_3$, three non-Abelian twist fields $\sigma_1,\sigma_2,\sigma_3$ with spin $h_\sigma=1/16$ and another three $\tau_1,\tau_2,\tau_3$ with spin $h_\tau=9/16$. The $S_3=\{1,\theta,\overline\theta,\alpha_1,\alpha_2,\alpha_3\}$ symmetry permutes the index $a=1,2,3$ for $(j_a,\sigma_a,\tau_a)$ (see eq.\eqref{S3operation4Potts}). This forms the defect fusion category \begin{align}\mathcal{C}=\mathcal{C}_1\oplus\mathcal{C}_\theta\oplus\mathcal{C}_{\overline\theta}\oplus\mathcal{C}_{\alpha_1}\oplus\mathcal{C}_{\alpha_2}\oplus\mathcal{C}_{\alpha_3}\end{align} where $\mathcal{C}_1$ is generated by the 11 anyons in the ``4-Potts" state, and the other sectors are generated by threefold ($\theta,\bar{\theta},\omega,\bar{\omega}$) and twofold ($\alpha,\mu$) twist defects: \begin{align}\mathcal{C}_\theta=\langle\theta,\omega\rangle,\quad\mathcal{C}_{\overline\theta}=\langle\overline\theta,\overline\omega\rangle,\quad\mathcal{C}_{\alpha_a}=\left\langle\alpha_a^0,\alpha_a^1,\alpha_a^2,\alpha_a^3,\boldsymbol\mu_a\right\rangle.\label{dfcategory4potts}\end{align} 
These defects can be understood as the fluxes of the octahedral point group $O\in SU(2)$. The defects $\theta$ and $\omega$ correspond to threefold rotations about diagonal axes like $(111)$, $\boldsymbol\mu_a$ corresponds to twofold rotations about axes such as $(110)$, and $\alpha_a$ correspond to fourfold rotations about the $x,y,z$-axes. We have chosen a bold symbol for $\boldsymbol\mu_a$ because, as we will see below, it will carry a larger quantum dimension than the $\alpha_a.$

We first explain the properties of the threefold defects in $\mathcal{C}_\theta$. The only non-trivial anyon in ``$SU(2)_1/Dih_2$" unaltered by a threefold symmetry is the semion super-sector $\Phi$. With this anyon we can  define a \emph{closed} Wilson loop \begin{align}\Theta_\Phi=\vcenter{\hbox{\includegraphics[width=0.05\textwidth]{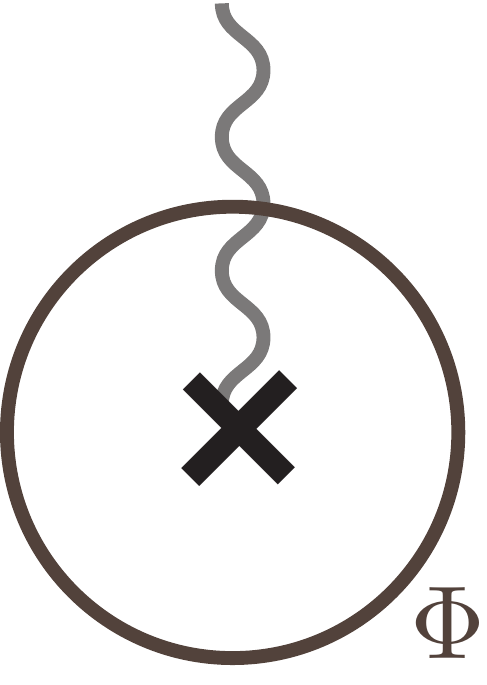}}}\end{align} around a threefold defect. From the $F$-symbols $F^{\Phi\Phi\Phi}_\Phi$ in \eqref{4PottsFsymbols}, it squares to unity. \begin{align}\Theta_\Phi^2=\vcenter{\hbox{\includegraphics[width=0.05\textwidth]{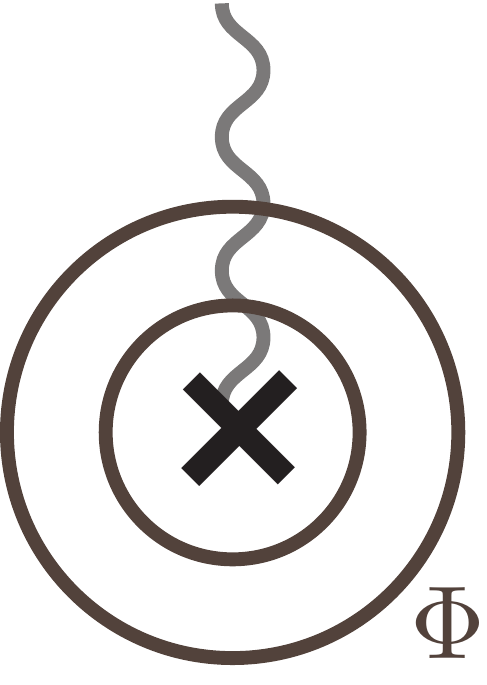}}}=\frac{1}{2}\left[\vcenter{\hbox{\includegraphics[width=0.06\textwidth]{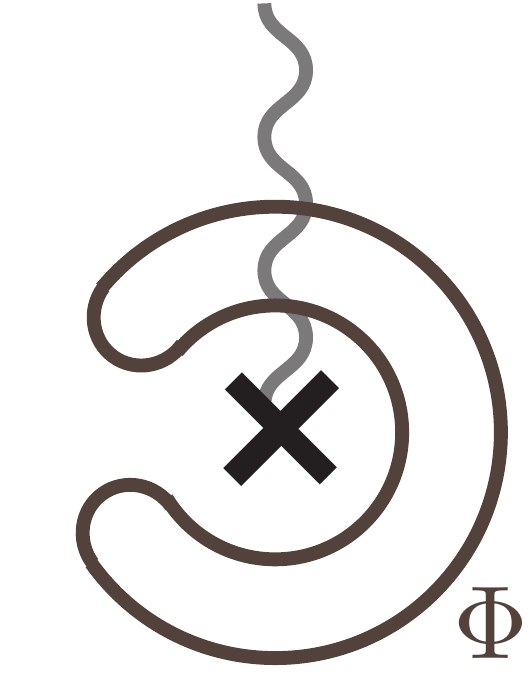}}}+\sum_{a=1}^3\vcenter{\hbox{\includegraphics[width=0.065\textwidth]{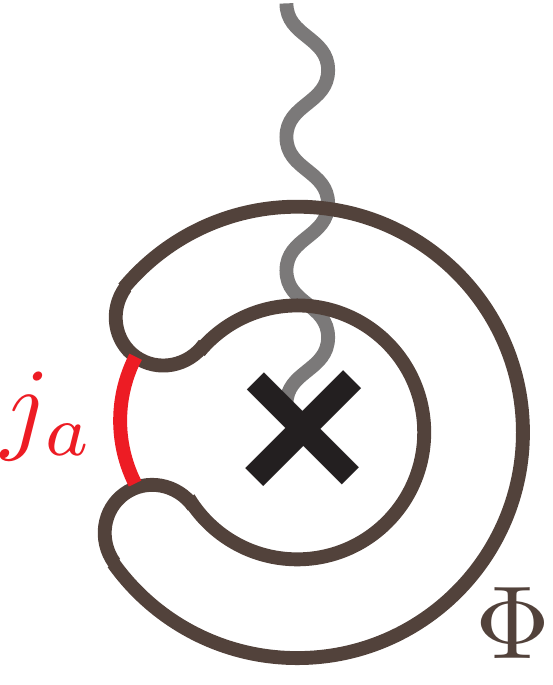}}}\right]=\frac{1}{2}\vcenter{\hbox{\includegraphics[width=0.05\textwidth]{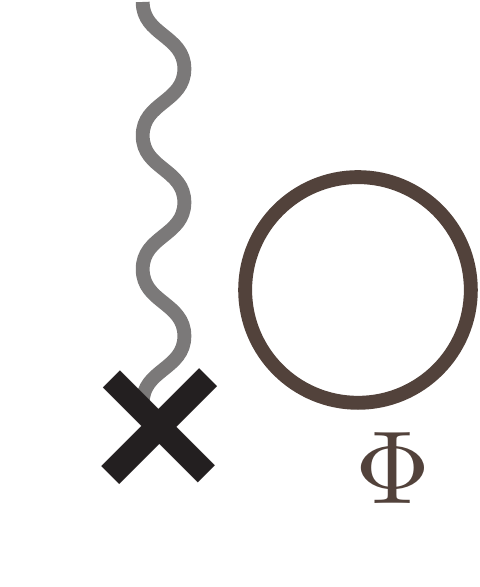}}}=\frac{d_\Phi}{2}\vcenter{\hbox{\includegraphics[width=0.033\textwidth]{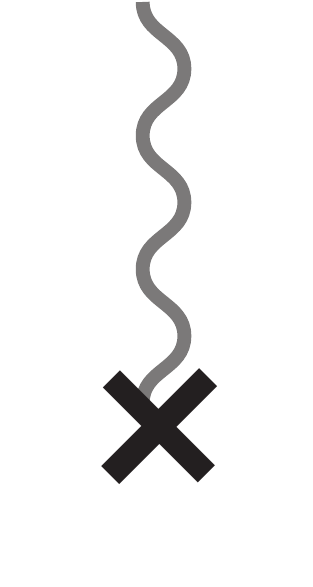}}}=1\label{ThetaPhisquare}\end{align} where each summand vanishes because the internal channel $j_a$ can be moved through the branch cut and changed so that the diagram vanishes: \begin{align}\vcenter{\hbox{\includegraphics[width=0.065\textwidth]{ThetaPhi21}}}=\vcenter{\hbox{\includegraphics[width=0.09\textwidth]{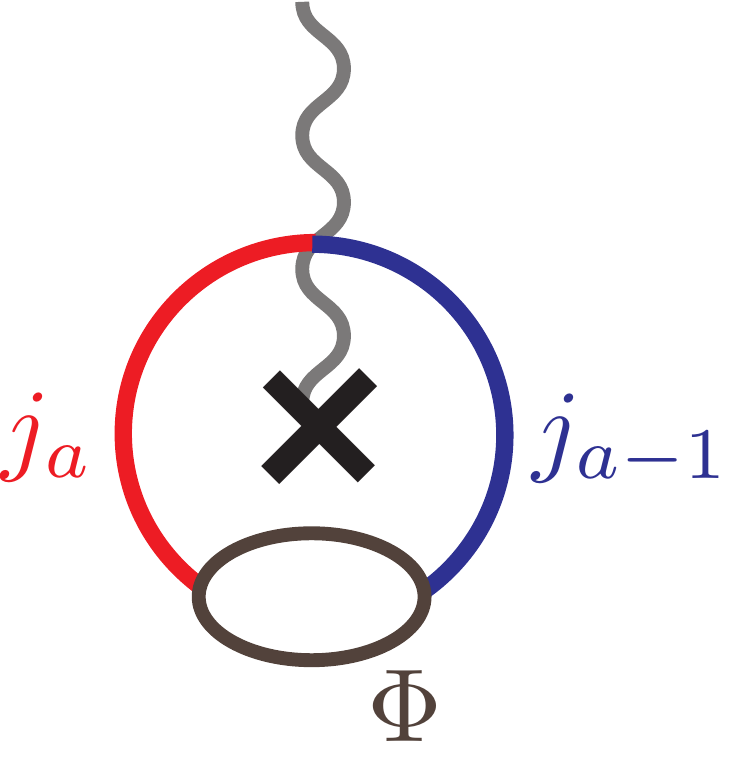}}}=0.\end{align} Eq.~\eqref{ThetaPhisquare} shows that the $\Phi$-loop has two possible eigenvalues $\Theta_\Phi=1$ and $-1$, which distinguish the two species of threefold defects $\theta$ and $\omega$ respectively: \begin{align}\vcenter{\hbox{\includegraphics[width=0.05\textwidth]{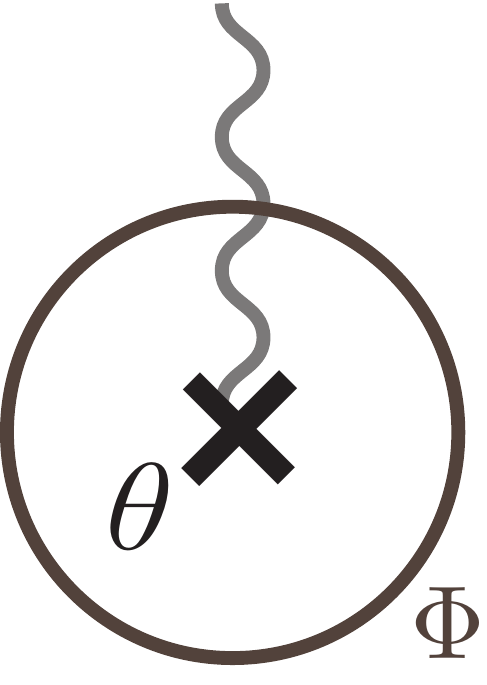}}}=+1,\quad\vcenter{\hbox{\includegraphics[width=0.05\textwidth]{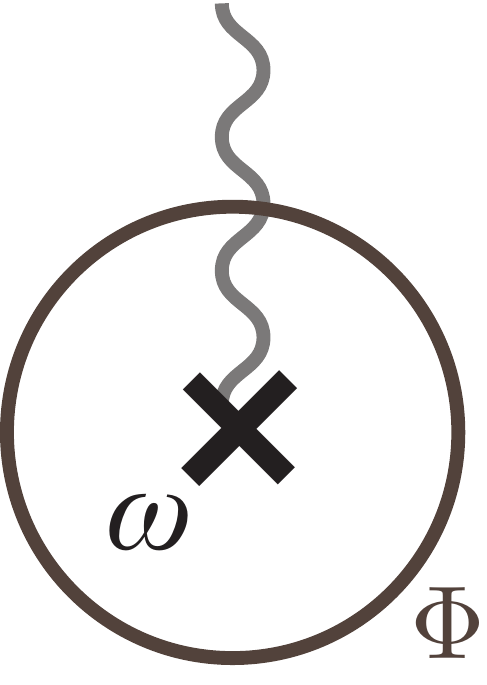}}}=-1.\label{Philoopapp}\end{align} The two species of threefold defects differ from each other by the fusion with the semion $\phi$ from the bosonic Laughlin state $SU(2)_1$ with $K=2$, i.e.~$\theta\times\phi=\omega$. In the chiral ``4-Potts" state, the semion super-sector $\Phi$ is two dimensional, and gives rise to a fusion degeneracy \begin{align}\theta\times\Phi=2\omega,\quad\omega\times\Phi=2\theta.\end{align} This means the two threefold defects $\theta$ and $\omega$ differ from each other by a semion string $\phi$. The additional $\phi$-string crosses the $\Phi$-loop around the defect in \eqref{Philoopapp}, and gives the extra minus sign for $\Theta_\Phi$. 

Fusing a threefold defect with the bosons $j_a$ will not alter $\Theta_\Phi$ as the braiding phase between $\Phi$ and $j_a$ is trivial, and therefore \begin{align}\theta\times j_a=\theta,\quad\omega\times j_a=\omega.\end{align} Fusion associativity -- $x\times(y\times z)=(x\times y)\times z$ -- requires \begin{align}\theta\times\sigma_a=\omega\times\sigma_a=\theta\times\tau_a=\omega\times\tau_a=\theta+\omega.\end{align}


From the $S_3$-group product relation $\theta^2=\overline\theta$, a pair of threefold defects fuses to its anti-partner, i.e.~$\theta\times\theta\to\overline\theta$ (or $\overline\omega$). The splitting states of $\theta\times\theta$ are labeled by the eigenvalues of the closed Wilson operators \begin{align}\hat{\mathcal{A}}_{j_a}=\vcenter{\hbox{\includegraphics[width=0.1\textwidth]{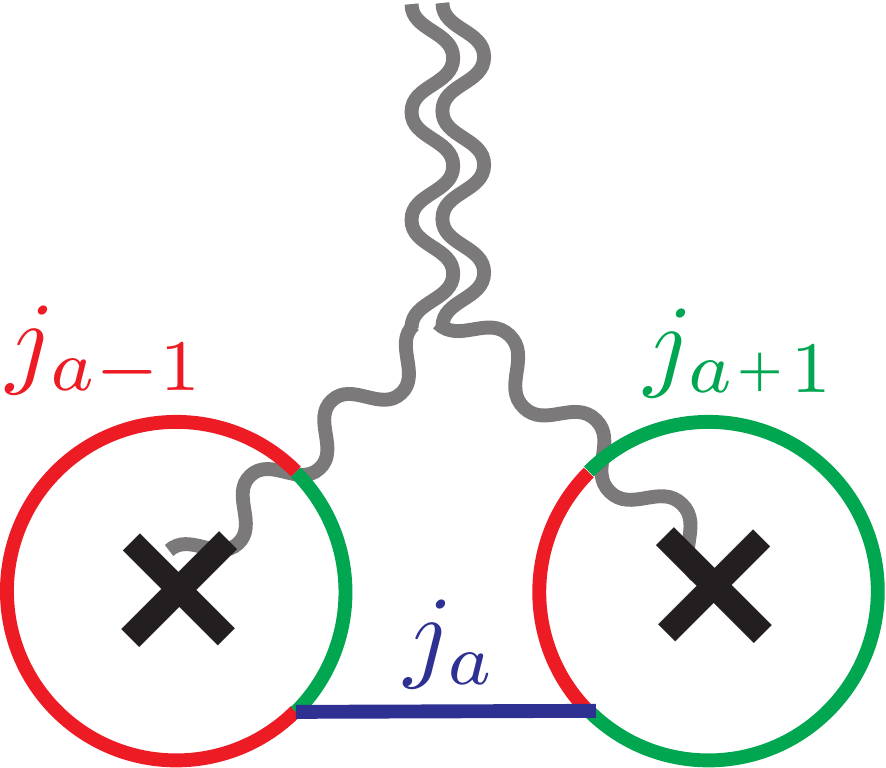}}}.\end{align} They mutually commute and satisfy the relations \begin{align}\hat{\mathcal{A}}_{j_1}\hat{\mathcal{A}}_{j_2}=\hat{\mathcal{A}}_{j_3},\quad\hat{\mathcal{A}}_{j_a}^2=1.\end{align} The four splitting states are specified by the simultaneous eigenvalues $\hat{\mathcal{A}}_{j_a}=(-1)^{s_a}$ for $(-1)^{s_1+s_2}=(-1)^{s_3}$. The fusion channels of $\theta\times\theta$ can be distinguished by a $\Phi$-loop, which, from the $F$-symbols in \eqref{4PottsFsymbols}, decomposes into \begin{gather}\vcenter{\hbox{\includegraphics[width=0.1\textwidth]{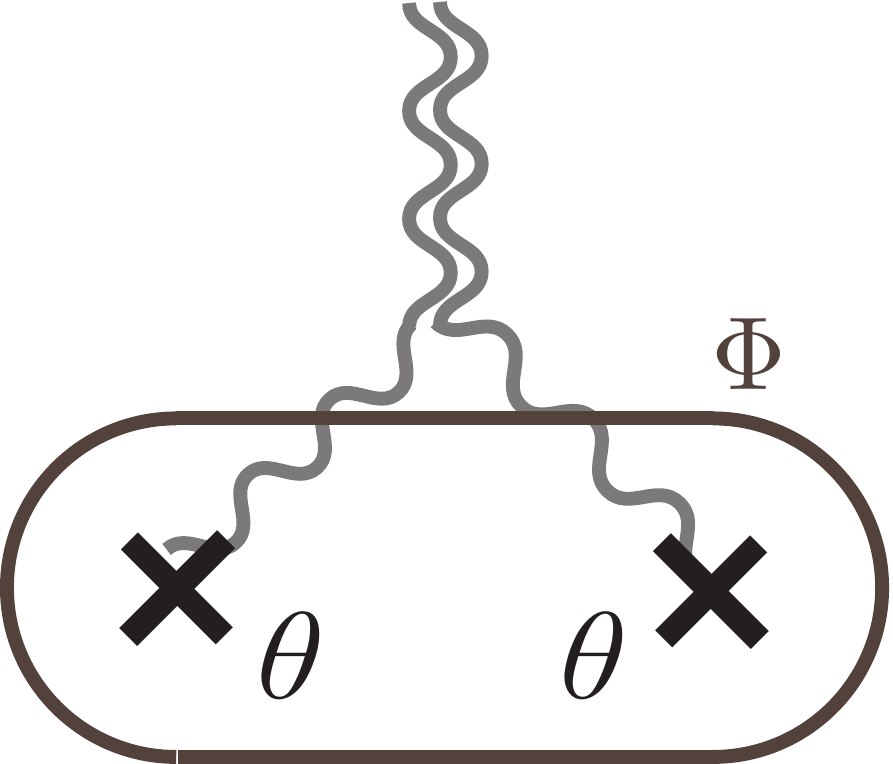}}}=\frac{1}{2}\left[\vcenter{\hbox{\includegraphics[width=0.1\textwidth]{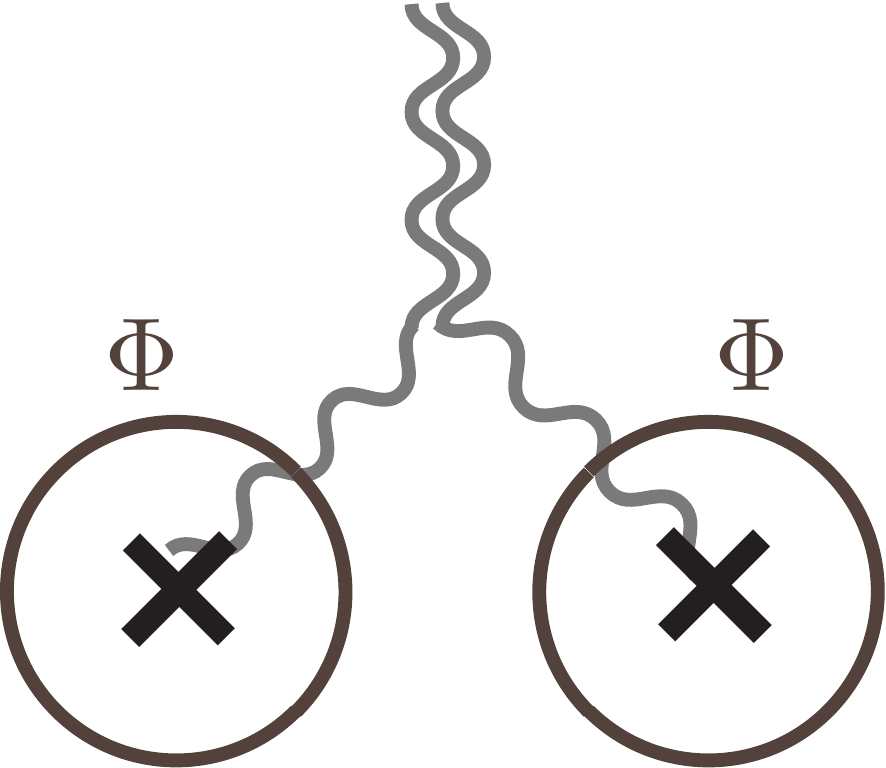}}}+\sum_{a=1}^3\vcenter{\hbox{\includegraphics[width=0.1\textwidth]{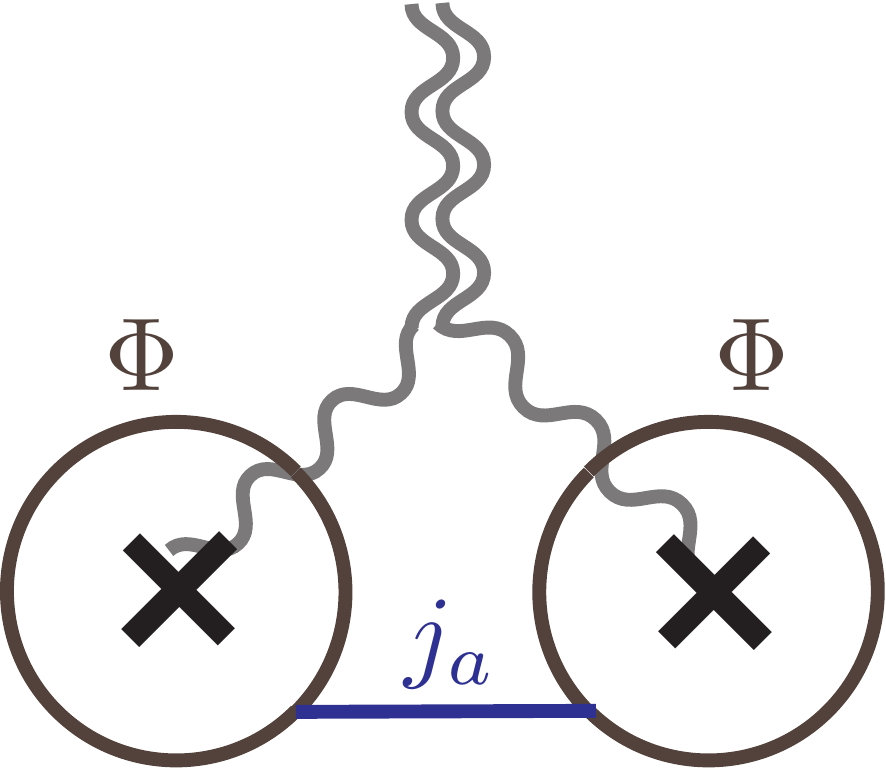}}}\right]=\frac{1}{2}\left[\vcenter{\hbox{\includegraphics[width=0.1\textwidth]{APhi1}}}-\sum_{a=1}^3\vcenter{\hbox{\includegraphics[width=0.11\textwidth]{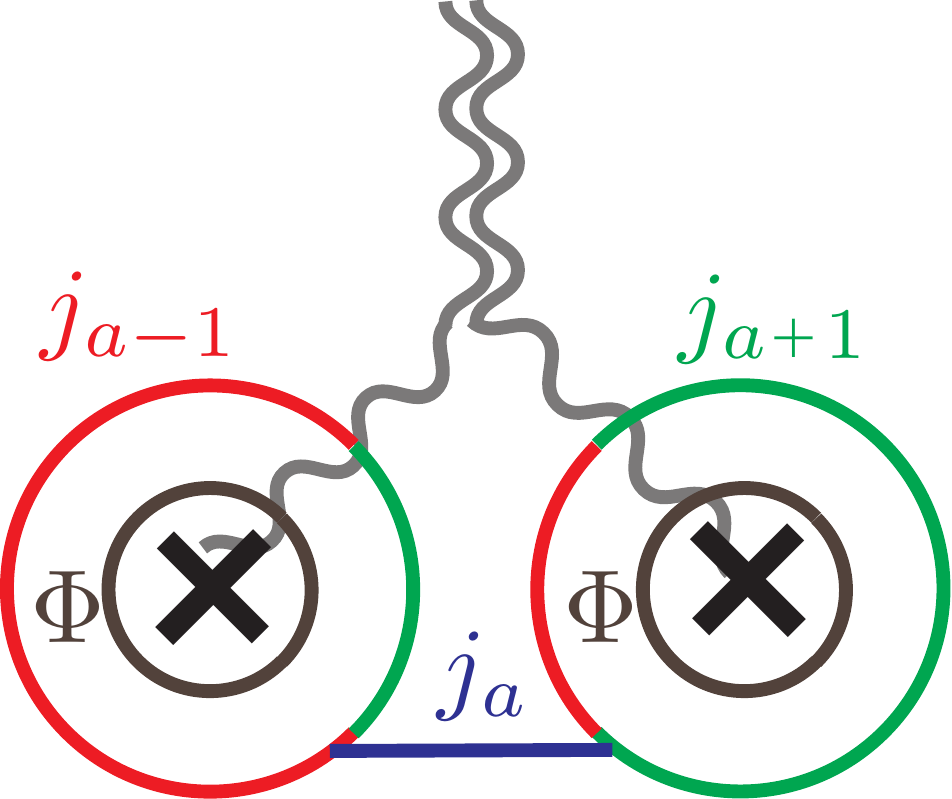}}}\right],\end{gather} where the small $\Phi$-loops around the $\theta$ defects are absorbed by the ground state. Hence \begin{align}\vcenter{\hbox{\includegraphics[width=0.1\textwidth]{APhi}}}=\frac{1}{2}\left[\vcenter{\hbox{\includegraphics[width=0.1\textwidth]{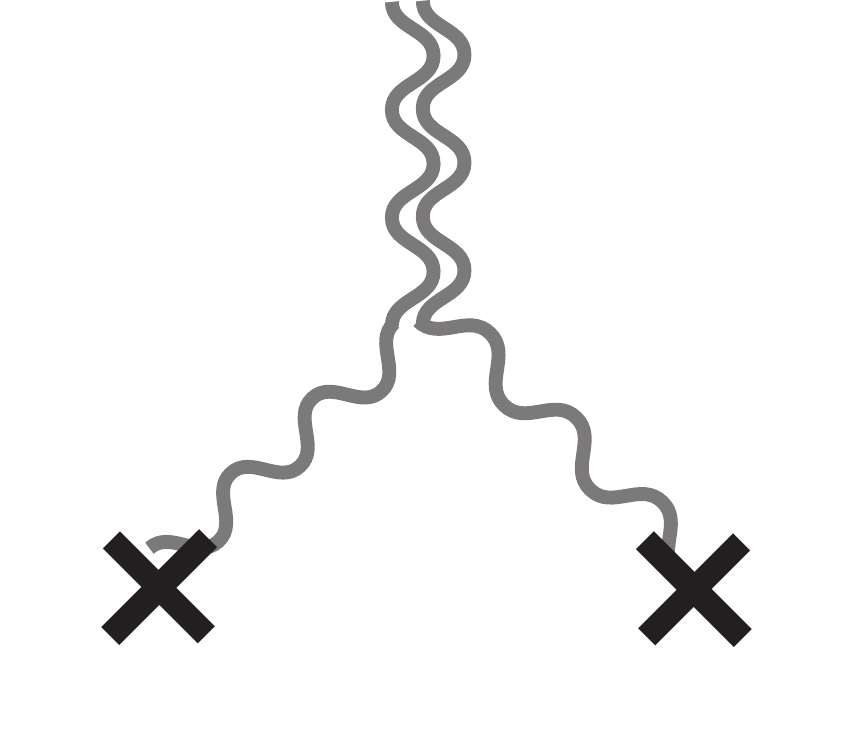}}}-\sum_{a=1}^3\vcenter{\hbox{\includegraphics[width=0.1\textwidth]{Aj}}}\right]=\frac{1}{2}\left(1-\sum_{a=1}^3\hat{\mathcal{A}}_{j_a}\right),\label{bigPhiloopthetabar}\end{align} which is $-1$ if $(-1)^{s_1}=(-1)^{s_2}=1$, or $+1$ if otherwise. This shows the fusion rules \begin{align}\theta\times\theta=\omega\times\omega=\overline\omega+3\overline\theta,\quad\theta\times\omega=\overline\theta+3\overline\omega\label{thetaxtheta}\end{align} where the fusion degeneracy, say in the first equation, comes from the fact that there are three distinct $(s_1,s_2)$ for \eqref{bigPhiloopthetabar} to be $+1$. The threefold fusion degeneracy is protected by the closed Wilson loop algebra \begin{align}\hat{\mathcal{A}}_{\sigma_a}=\vcenter{\hbox{\includegraphics[width=0.1\textwidth]{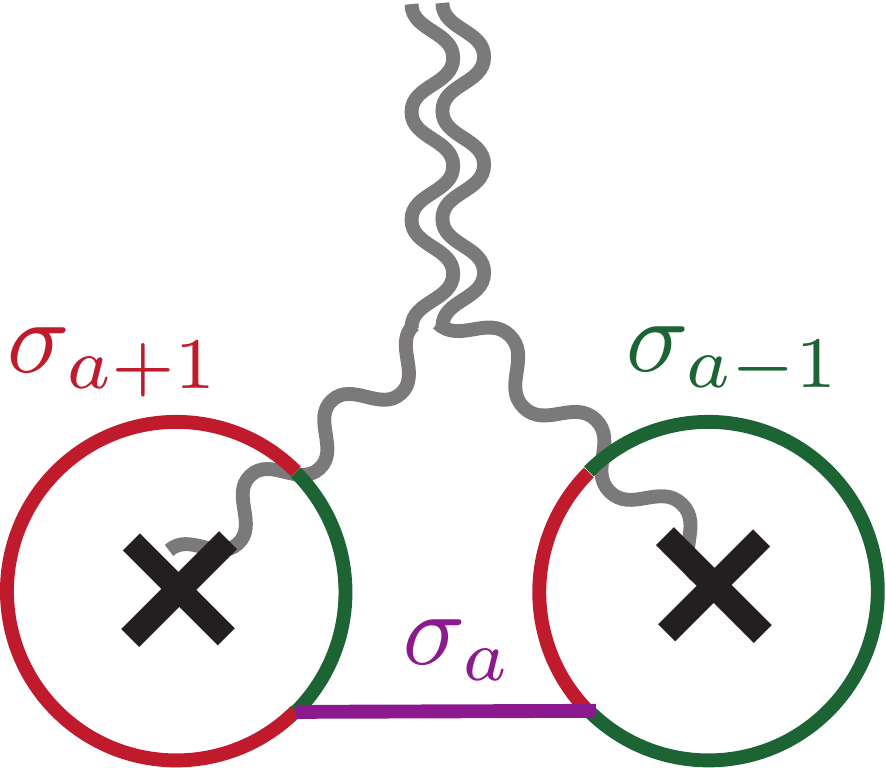}}},\end{align} which do not commute with the $\hat{\mathcal{A}}_{j_b}$'s. We also have \begin{align} \hat{\mathcal{A}}_{\sigma_a}\hat{\mathcal{A}}_{j_b}=-(-1)^{\delta_{ab}}\hat{\mathcal{A}}_{j_b}\hat{\mathcal{A}}_{\sigma_a}\end{align} due to the braiding phase $-(-1)^{\delta_{ab}}$ between $\sigma_a$ and $j_b$.

Eq.~\eqref{thetaxtheta} enforces that the quantum dimension of the threefold defects \begin{align}d_\theta=d_\omega=4.\end{align} This is consistent with the fusion rules between conjugate pairs \begin{align}\theta\times\overline\theta=\omega\times\overline\omega=1+\sum_{a=1}^3j_a+\sum_{a=1}^3\sigma_a+\sum_{a=1}^3\tau_a,\quad\theta\times\overline\omega=\omega\times\overline\theta=2\Phi+\sum_{a=1}^3\sigma_a+\sum_{a=1}^3\tau_a\end{align} where the non-trivial fusion channels of $\theta\times\overline\theta\to\chi_a$ can be generated by the Wilson structure \begin{align}\vcenter{\hbox{\includegraphics[width=0.13\textwidth]{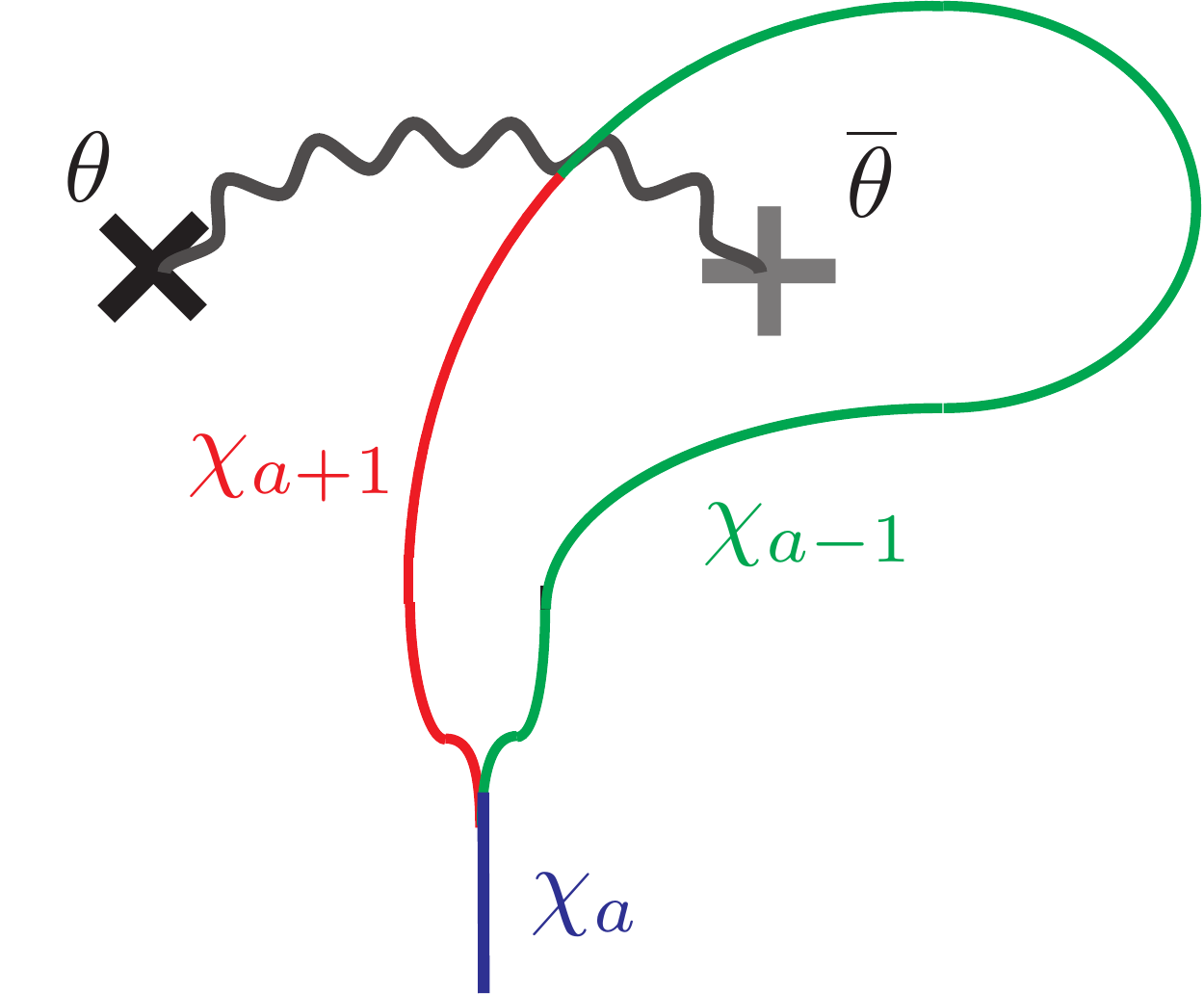}}}\end{align} for $\chi_a=j_a,\sigma_a,\tau_a$. The total quantum dimension for the defect sector $\mathcal{C}_\theta=\langle\theta,\omega\rangle$ is therefore \begin{align}\mathcal{D}_{\mathcal{C}_\theta}=\sqrt{d_\theta^2+d_\omega^2}=4\sqrt{2}=\mathcal{D}_0\end{align} which is also the total quantum dimension of the parent state.

Next, we explain the properties of the twofold defects in $\mathcal{C}_{\alpha_a}$ in \eqref{dfcategory4potts}. The non-trivial anyons in ``$SU(2)_1/Dih_2$" unaltered by the twofold symmetry $\alpha_a$ are $j_a$, $\Phi$, $\sigma_a,$ and $\tau_a$. These form the set of closed Wilson loops \begin{align}\Theta_{j_a}=\vcenter{\hbox{\includegraphics[width=0.05\textwidth]{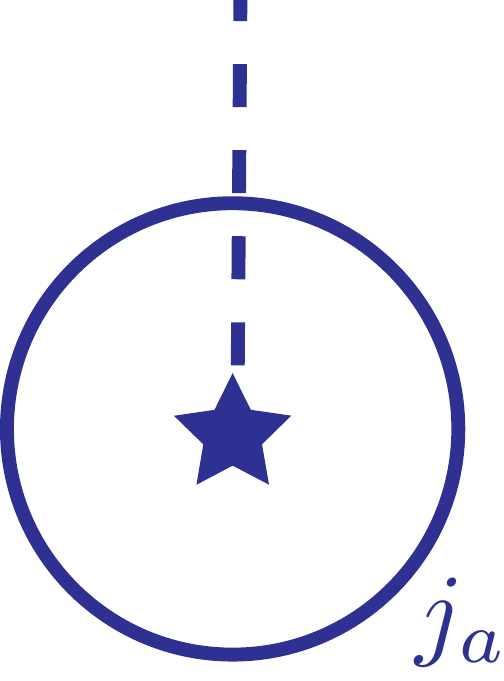}}},\quad\Theta_\Phi=\vcenter{\hbox{\includegraphics[width=0.05\textwidth]{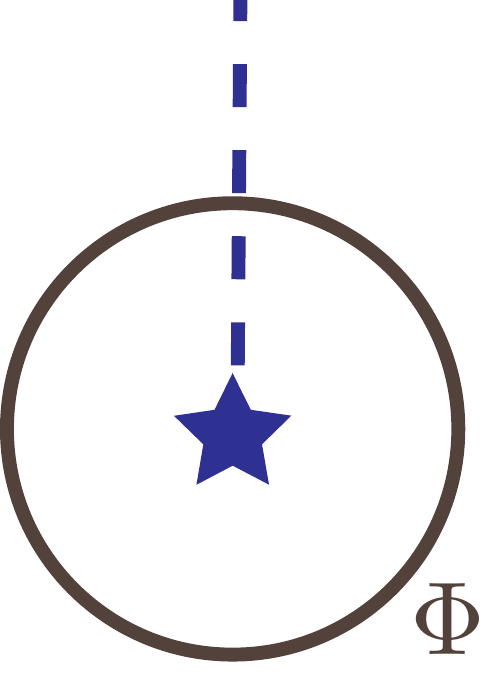}}},\quad\Theta_{\sigma_a}=\vcenter{\hbox{\includegraphics[width=0.05\textwidth]{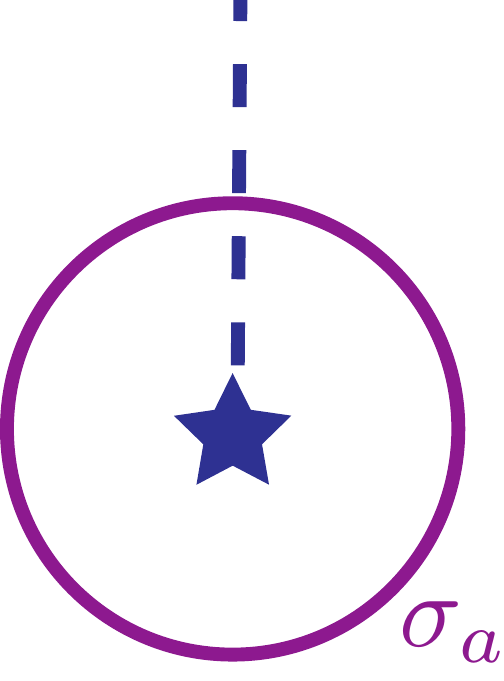}}},\quad\Theta_{\tau_a}=\vcenter{\hbox{\includegraphics[width=0.05\textwidth]{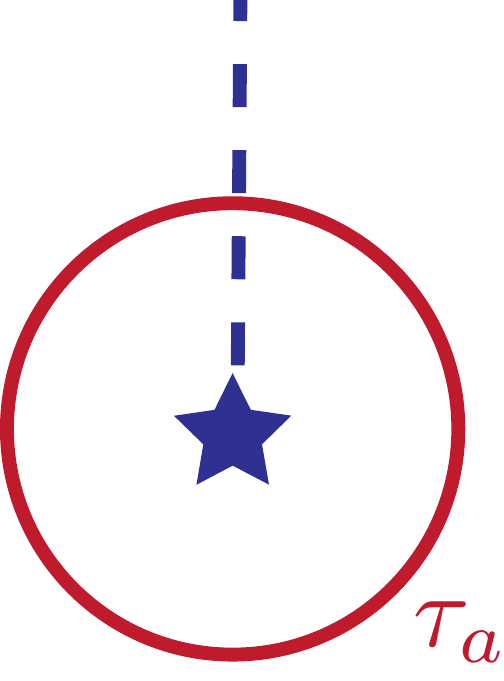}}}\end{align} around the twofold defect. They mutually commute as they can pass through one another without intersecting, but they are not independent. From the fusion rules \eqref{4statePottsfusion}, and the $F$-symbols in \eqref{4PottsFsymbols}, they satisfy \begin{gather}\Theta_{j_a}\Theta_\Phi=\Theta_\Phi,\quad\Theta_{j_a}\Theta_{\sigma_a}=\Theta_{\sigma_a},\quad\Theta_{j_a}\Theta_{\tau_a}=\Theta_{\tau_a}\nonumber\\\Theta_{j_a}^2=1,\quad\Theta_\Phi^2=1+\Theta_{j_a}\nonumber\\\Theta_{\sigma_a}\Theta_{\tau_a}=\Theta_\Phi\label{alphaloopalgebra}\\\Theta_{\sigma_a}^2=1+\Theta_{j_a}+\Theta_\Phi,\quad\Theta_{\tau_a}^2=1+\Theta_{j_a}-\Theta_\Phi\nonumber\\\Theta_\Phi\Theta_{\sigma_a}=\Theta_{\sigma_a}+\Theta_{\tau_a},\quad\Theta_\Phi\Theta_{\tau_a}=\Theta_{\sigma_a}-\Theta_{\tau_a}.\nonumber\end{gather} For example, $\Theta_\Phi^2$ can be evaluated by $F^{\Phi\Phi\Phi}_\Phi$ \begin{align}\vcenter{\hbox{\includegraphics[width=0.05\textwidth]{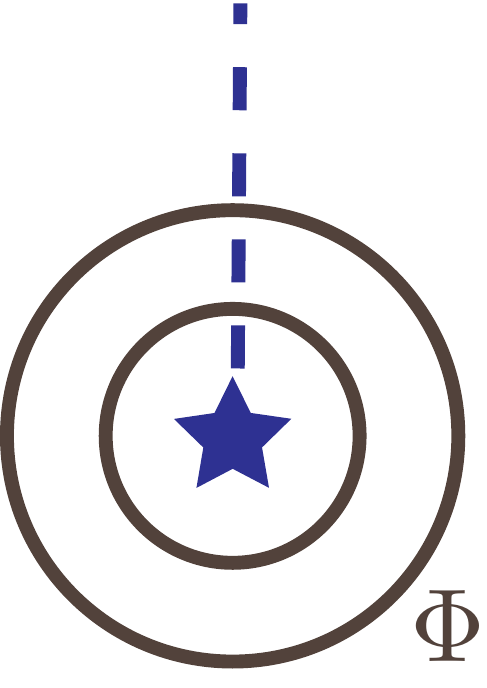}}}=\frac{1}{2}\left[\vcenter{\hbox{\includegraphics[width=0.06\textwidth]{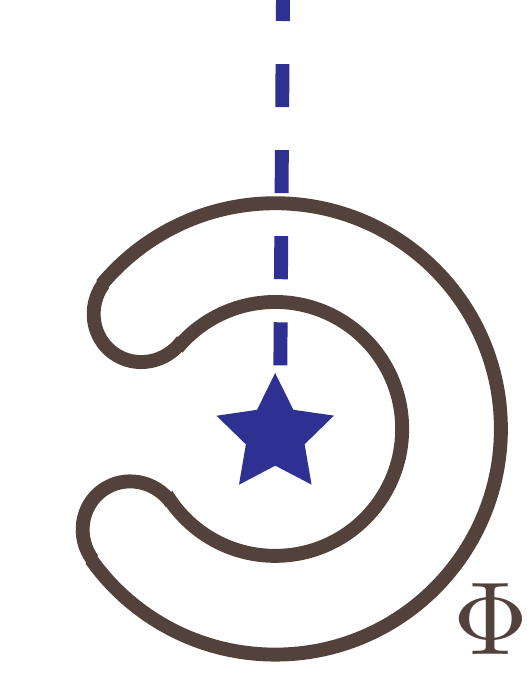}}}+\sum_{b=1}^3\vcenter{\hbox{\includegraphics[width=0.065\textwidth]{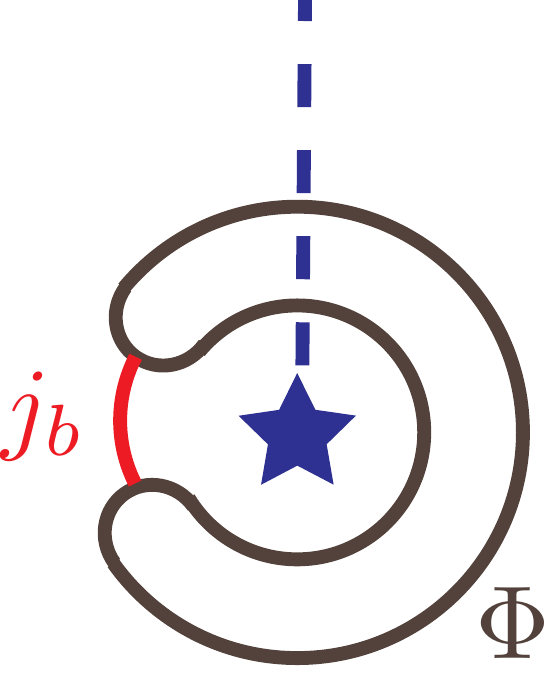}}}\right]=\frac{1}{2}\left[\vcenter{\hbox{\includegraphics[width=0.05\textwidth]{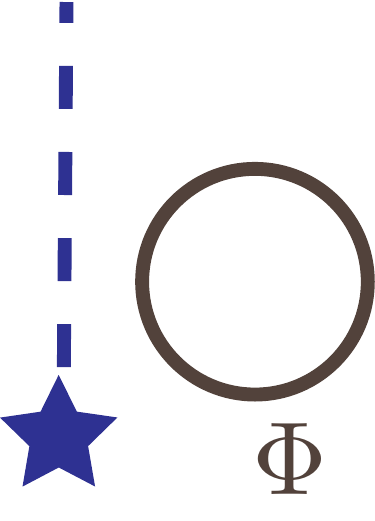}}}+\vcenter{\hbox{\includegraphics[width=0.065\textwidth]{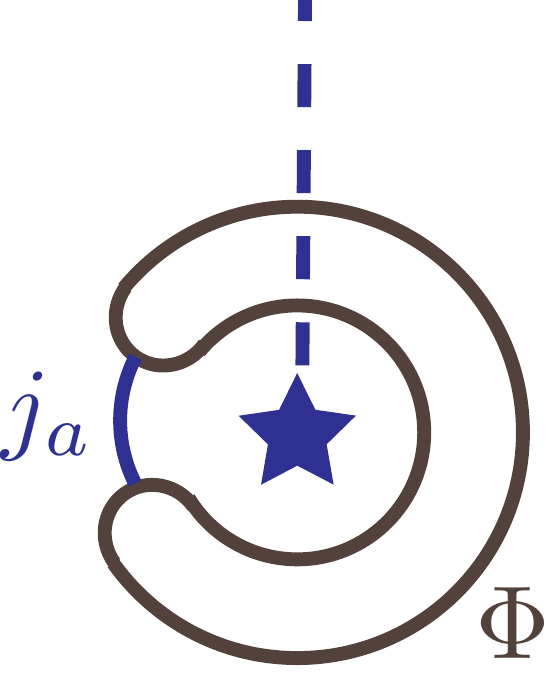}}}\right]=\vcenter{\hbox{\includegraphics[height=0.05\textwidth]{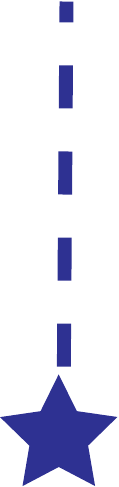}}}+\vcenter{\hbox{\includegraphics[width=0.05\textwidth]{alphaja}}}\end{align} where the other two summands involving $j_{a\pm1}$ vanish because their internal channels can be dragged past the branch cut and changed to lead to a vanishing diagram: \begin{align}\vcenter{\hbox{\includegraphics[width=0.065\textwidth]{alphaPhi21}}}=\vcenter{\hbox{\includegraphics[width=0.09\textwidth]{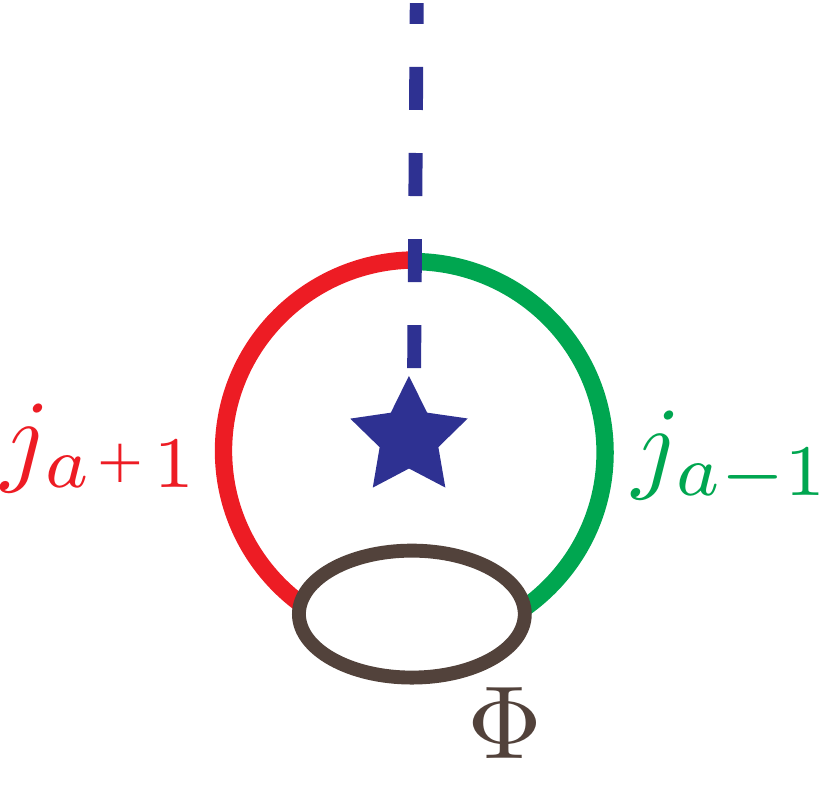}}}=0.\end{align} The Wilson loop $\Theta_{\sigma_a}^2$ can be evaluated by $F^{\sigma_a\sigma_a\sigma_a}_{\sigma_a}$ \begin{align}\vcenter{\hbox{\includegraphics[width=0.05\textwidth]{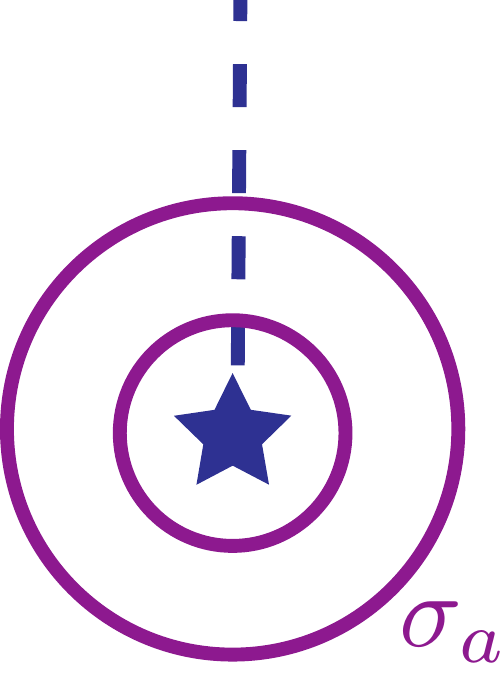}}}=\frac{1}{2}\left[\vcenter{\hbox{\includegraphics[width=0.06\textwidth]{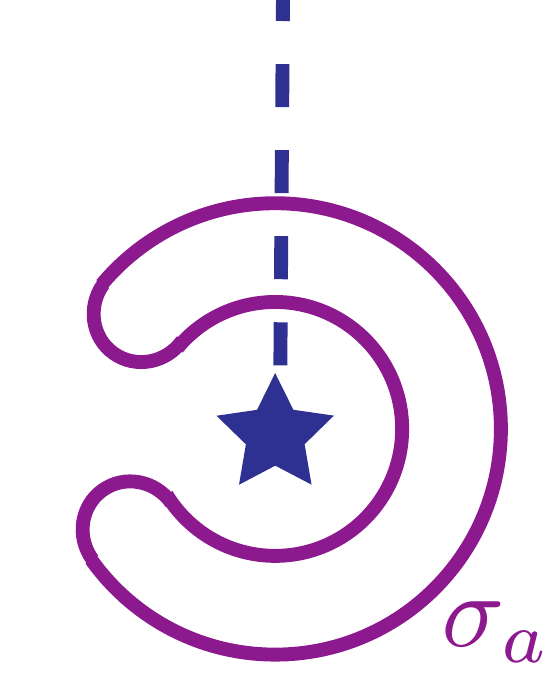}}}+\vcenter{\hbox{\includegraphics[width=0.065\textwidth]{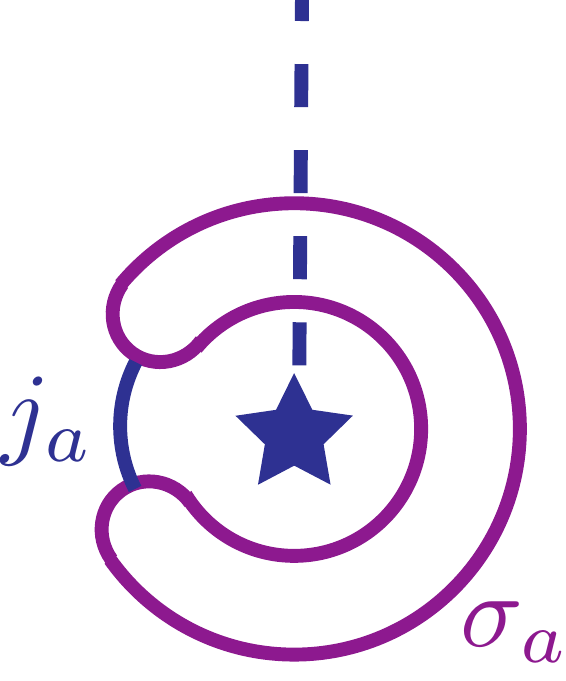}}}+\sqrt{2}\vcenter{\hbox{\includegraphics[width=0.065\textwidth]{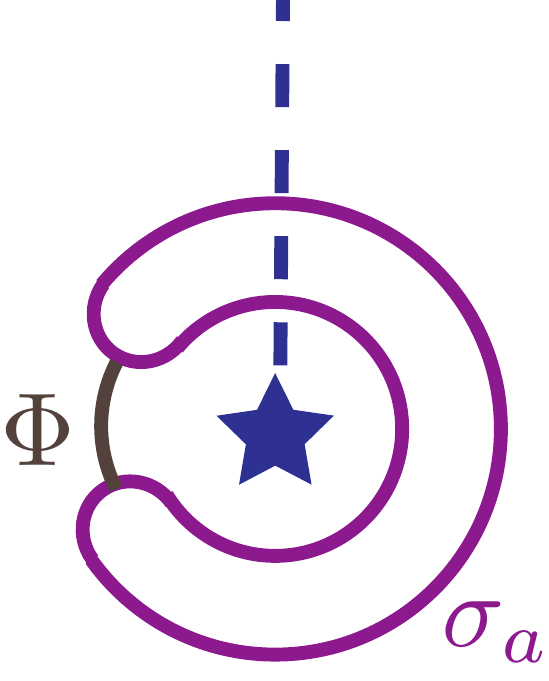}}}\right]=\vcenter{\hbox{\includegraphics[height=0.05\textwidth]{alphaPhi32}}}+\vcenter{\hbox{\includegraphics[width=0.05\textwidth]{alphaja}}}+\vcenter{\hbox{\includegraphics[width=0.05\textwidth]{alphaPhi}}}.\end{align} Other products of \eqref{alphaloopalgebra} are determined by associativity, i.e.~$\Theta_x(\Theta_y\Theta_z)=(\Theta_x\Theta_y)\Theta_z,$ and by a redefinition of signs in front of the Wilson operators $\Theta_x\to-\Theta_x$ if necessary. 

The Wilson loops $\Theta_{j_a},\Theta_{\Phi},\Theta_{\sigma_a},\Theta_{\tau_a}$ form an associative and commutative ring with identity. The species labels of twofold defects are representations of this ring. Let us see what this implies. First, $\Theta_{j_a}$ has eigenvalues $\pm1$ since it squares to unity. The product relations \eqref{alphaloopalgebra} then force the eigenvalues of the other Wilson operators to satisfy \begin{align}&\left\{\begin{array}{*{20}c}\Theta_\Phi=\sqrt{2}(-1)^s\hfill\\\Theta_{\sigma_a}=2\cos\left(\frac{\pi}{8}-\frac{s\pi}{2}\right)\\\Theta_{\tau_a}=2\sin\left(\frac{\pi}{8}-\frac{s\pi}{2}\right)\end{array}\right.,\quad\mbox{for }\Theta_{j_a}=+1\label{alphaloopvalues}\\&\Theta_\Phi=\Theta_{\sigma_a}=\Theta_{\tau_a}=0,\quad\mbox{for }\Theta_{j_a}=-1\label{muloopvalues}\end{align} for $2\cos(\pi/8)=\sqrt{2+\sqrt{2}}$, $2\sin(\pi/8)=\sqrt{2-\sqrt{2}}$.

The first possibility, which is listed in Eq.~\eqref{alphaloopvalues}, gives four twofold defects $\alpha^s_a,$ one for each value of $s\in\mathbb{Z}_4=\{0,1,2,3\}$. The $\Theta_\Phi$ and $\Theta_{\sigma_a}$ loops change signs upon fusing the defect with anyons that intersect non-trivially with $\Phi$ and $\sigma_a$. Fusion associativity requires \begin{gather}\alpha^s_a\times j_a=\alpha^s_a,\quad\alpha^s_a\times j_{a\pm1}=\alpha^{s+2}_a\nonumber\\\alpha^s_a\times\Phi=\alpha^{s+1}_a+\alpha^{s-1}_a\\\alpha^s_a\times\sigma_a=\alpha^{-s}_a+\alpha^{-s+1}_a.\nonumber\end{gather} The second possibility, which is listed in Eq.~\eqref{muloopvalues}, describes another twofold defect $\boldsymbol\mu_a$, which is related to the others by \begin{align}&\boldsymbol\mu_a=\alpha_a^s\times\sigma_{a\pm1}=\alpha_a^s\times\tau_{a\pm1}\label{musigmarelation}\\&\boldsymbol\mu_a\times\sigma_{a\pm1}=\boldsymbol\mu_a\times\tau_{a\pm1}=\sum_{s=0}^3\alpha_a^s.\end{align} 

The fusion channels of the a defect pair $\alpha^{s_1}_a\times\alpha^{s_2}_a$ are distinguished by the mutually commuting Wilson loops \begin{align}\mathcal{W}_x=\vcenter{\hbox{\includegraphics[width=0.08\textwidth]{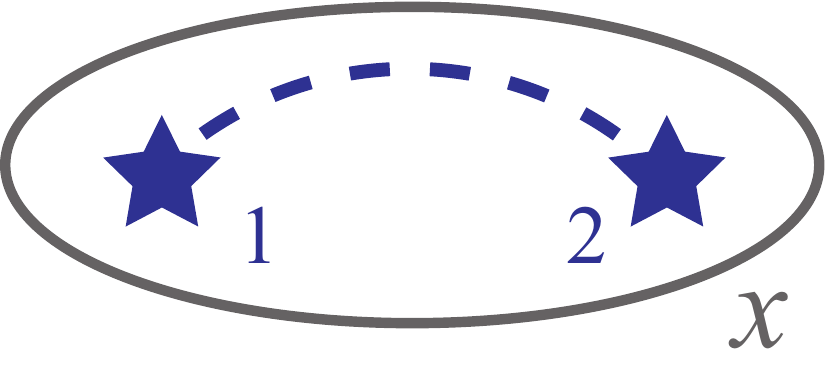}}}\end{align} where $x$ are anyons in the ``4-Potts" state. This Wilson algebra depends only on $\mathcal{W}_{j_{a-1}}$ and $\mathcal{W}_{\sigma_{a-1}}$ as the others are generated by the product relations: \begin{gather}\mathcal{W}_{j_a}=\Theta^1_{j_a}\Theta^2_{j_a}=1,\quad\mathcal{W}_{j_{a+1}}=\Theta^1_{j_a}\Theta^2_{j_a}\mathcal{W}_{j_{a-1}}=\mathcal{W}_{j_{a-1}}\nonumber\\\mathcal{W}_\Phi=\frac{1}{2}\Theta^1_\Phi\Theta^2_\Phi\left(1+\mathcal{W}_{j_{a-1}}\right)\nonumber\\\mathcal{W}_{\sigma_a}=\frac{1}{2}\left(\Theta^1_{\sigma_a}\Theta^2_{\sigma_a}+\Theta^1_{\tau_a}\Theta^2_{\tau_a}\mathcal{W}_{j_{a-1}}\right)\\\mathcal{W}_{\sigma_{a+1}}\mathcal{W}_{\sigma_{a-1}}=\frac{1}{\sqrt{2}}\left(\mathcal{W}_{\sigma_a}+\mathcal{W}_{\tau_a}\right)\nonumber\\\mathcal{W}_{\tau_a}=\mathcal{W}_{\sigma_a}\mathcal{W}_{j_{a-1}},\quad\mathcal{W}_{\tau_{a\pm1}}=\mathcal{W}_{\sigma_{a\pm1}}\mathcal{W}_{j_a}=\mathcal{W}_{\sigma_{a\pm1}}\nonumber\end{gather} where $\Theta^1$ (or $\Theta^2$) is the Wilson loop around only the first (resp.~second) defect. These product relations can be evaluated by using the $F$-symbols in \eqref{4PottsFsymbols}. For example, the $\Phi$-loop can be resolved into \begin{align}\vcenter{\hbox{\includegraphics[width=0.08\textwidth]{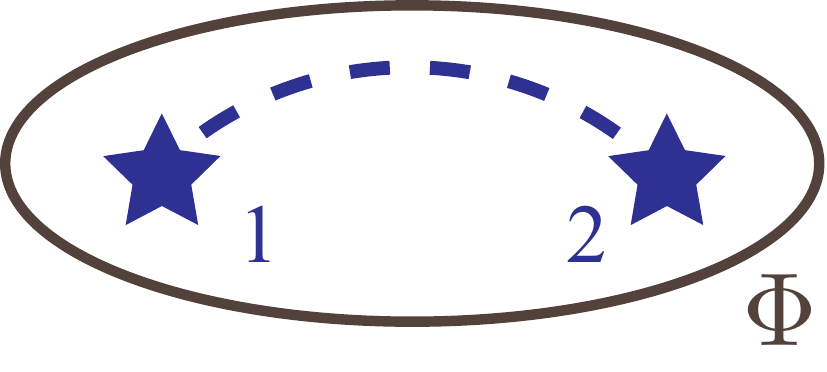}}}&=\frac{1}{2}\left[\vcenter{\hbox{\includegraphics[width=0.08\textwidth]{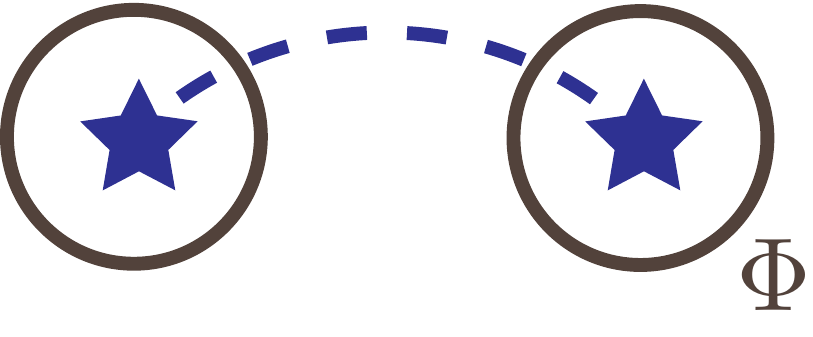}}}+\sum_{b=1}^3\vcenter{\hbox{\includegraphics[width=0.08\textwidth]{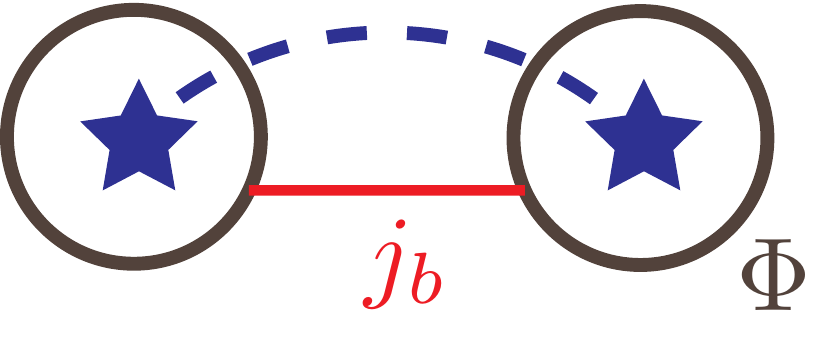}}}\right]\nonumber\\&=\frac{1}{2}\left[\vcenter{\hbox{\includegraphics[width=0.08\textwidth]{WPhi1}}}+\vcenter{\hbox{\includegraphics[width=0.09\textwidth]{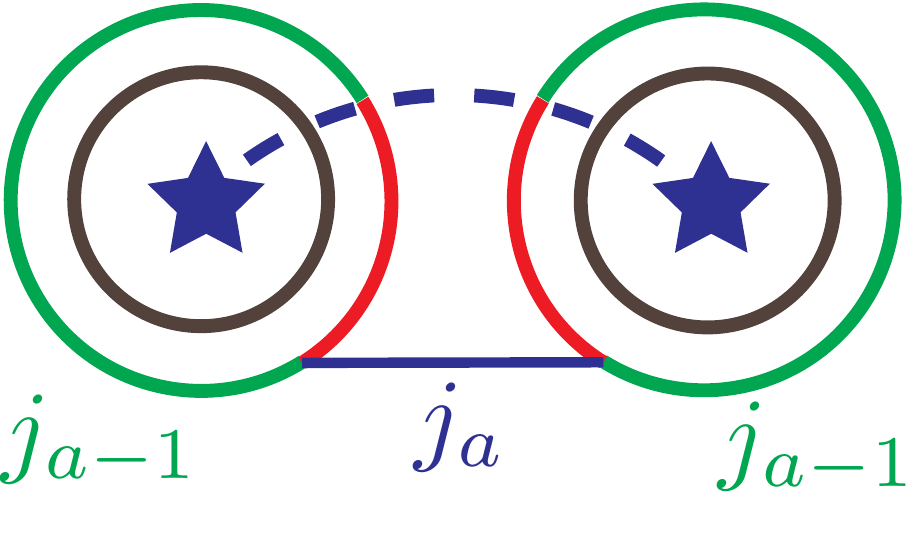}}}\right]=\frac{1}{2}\left[\vcenter{\hbox{\includegraphics[width=0.08\textwidth]{WPhi1}}}+\vcenter{\hbox{\includegraphics[width=0.09\textwidth]{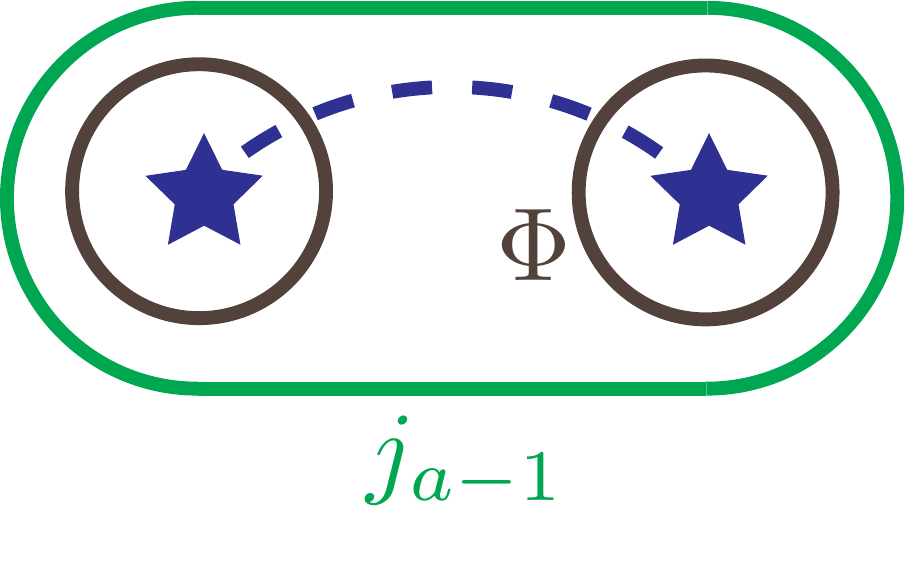}}}\right]\end{align} where the $j_b=j_{a\pm1}$ terms in the summand vanish. And the $\sigma_a$-loop is \begin{align}\vcenter{\hbox{\includegraphics[width=0.08\textwidth]{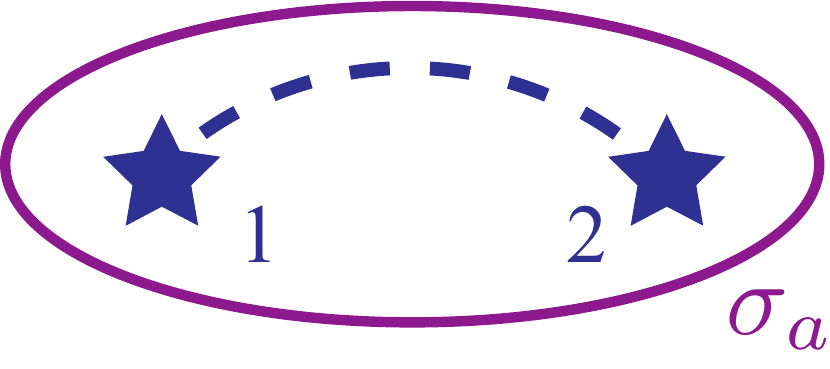}}}&=\frac{1}{2}\left[\vcenter{\hbox{\includegraphics[width=0.08\textwidth]{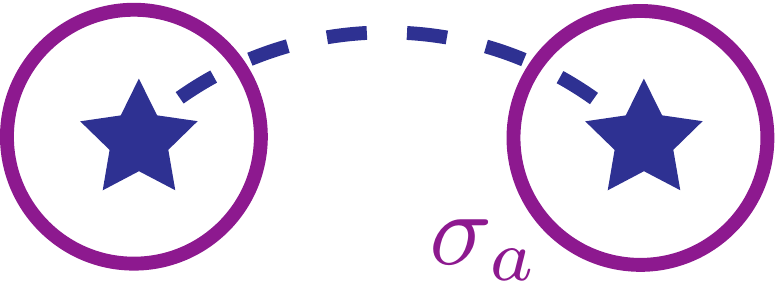}}}+\vcenter{\hbox{\includegraphics[width=0.08\textwidth]{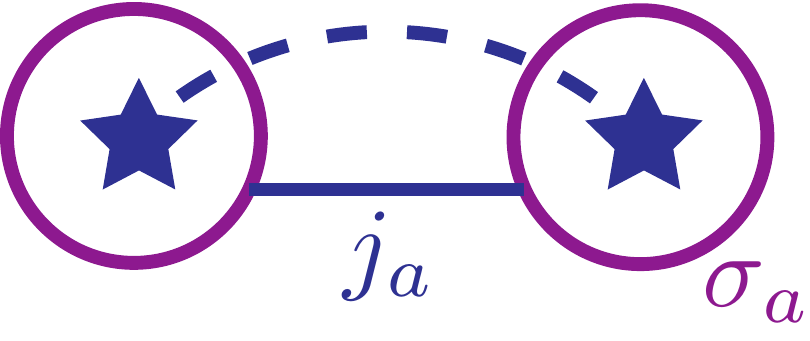}}}+\sqrt{2}\vcenter{\hbox{\includegraphics[width=0.08\textwidth]{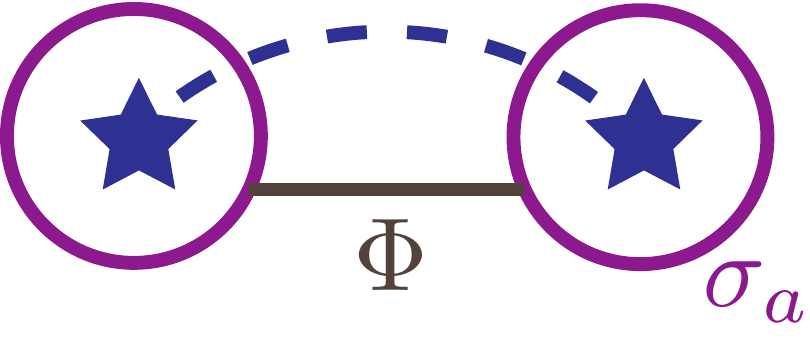}}}\right]\nonumber\\&=\frac{1}{2}\left[\vcenter{\hbox{\includegraphics[width=0.08\textwidth]{Wsigma1}}}+\vcenter{\hbox{\includegraphics[width=0.09\textwidth]{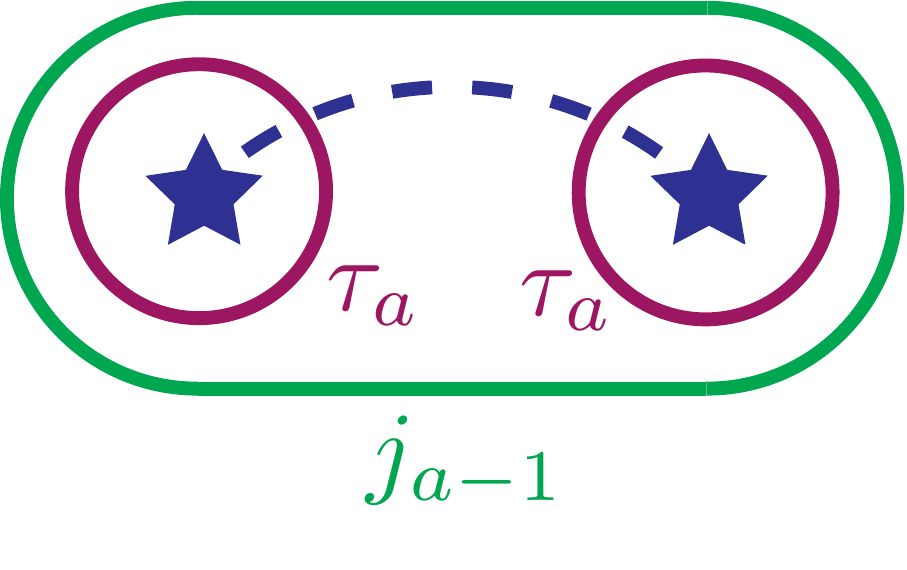}}}\right]\end{align} where the last term involving the intermediate channel $\Phi$ vanishes.

The fusion channels of $\alpha^{s_1}_a\times\alpha^{s_2}_a\to y$ are determined by the eigenvalues of the Wilson loop \begin{align}\mathcal{W}_x=\frac{1}{d_y}\mathcal{D}_0S_{xy}\end{align} where the $S$-matrix is given in \eqref{4statePottsSmatrix}, and $d_y$ is the quantum dimension of the fusion channel $y$. The fusion rules are summarized by \begin{align}\alpha_a^s\times\alpha_a^s&=\left\{\begin{array}{*{20}c}1+j_a+\sigma_a,&\mbox{for $s$ even}\\1+j_a+\tau_a,&\mbox{for $s$ odd}\end{array}\right.\nonumber\\\alpha_a^s\times\alpha_a^{s+2}&=\left\{\begin{array}{*{20}c}j_{a-1}+j_{a+1}+\tau_a,&\mbox{for $s$ even}\\j_{a-1}+j_{a+1}+\sigma_a,&\mbox{for $s$ odd}\end{array}\right.\label{alphaalpha}\\\alpha_a^s\times\alpha_a^{s+1}&=\left\{\begin{array}{*{20}c}\Phi+\sigma_a,&\mbox{for $s$ even}\\\Phi+\tau_a,&\mbox{for $s$ odd}\end{array}\right. .\nonumber\end{align} 

The fusion rule for a pair of $\boldsymbol\mu_a$'s can be deduced using \eqref{musigmarelation} \begin{align}\boldsymbol\mu_a\times\boldsymbol\mu_a=1+\sum_{b=1}^3j_b+2\Phi+2\sigma_a+2\tau_a.\label{mumu}\end{align} 

Eqs.~\eqref{alphaalpha} and \eqref{mumu} imply the defect quantum dimensions are \begin{align}d_{\alpha_a^s}=2,\quad d_{\boldsymbol\mu_a}=4.\end{align} The total quantum dimension for the defect sector $\mathcal{C}_{\alpha_a}=\langle\alpha^{0}_a,\alpha^{1}_a,\alpha^{2}_a,\alpha^{3}_a,\boldsymbol\mu_a\rangle$ is therefore \begin{align}\mathcal{D}_{\mathcal{C}_{\alpha_a}}=\sqrt{4d_{\alpha_a^{s}}^2+d_{\boldsymbol\mu_a}^2}=4\sqrt{2}=\mathcal{D}_0,\end{align} which again is the total quantum dimension of the parent state. 

Finally, we describe the fusion rules between twofold and threefold defects. Similar to the $SO(8)_1$ state, the non-abelian $S_3$-symmetry in the ``4-Potts" state implies non-commutative fusion rules: \begin{gather}\alpha_a^{s_1}\times\alpha_{a+1}^{s_2}=\left\{\begin{array}{*{20}c}\theta,&\mbox{for $s_1+s_2$ even}\\\omega,&\mbox{for $s_1+s_2$ odd}\end{array}\right.\nonumber\\\alpha_a^{s_1}\times\alpha_{a-1}^{s_2}=\left\{\begin{array}{*{20}c}\overline\theta,&\mbox{for $s_1+s_2$ even}\\\overline\omega,&\mbox{for $s_1+s_2$ odd}\end{array}\right.\nonumber\\\alpha_a^s\times\theta=\overline\theta\times\alpha_a^s=\alpha_{a+1}^s+\alpha_{a+1}^{s+2}+\boldsymbol\mu_{a+1}\nonumber\\\alpha_a^s\times\omega=\overline\omega\times\alpha_a^s=\alpha_{a+1}^{s-1}+\alpha_{a+1}^{s+1}+\boldsymbol\mu_{a+1}\nonumber\\\boldsymbol\mu_a\times\theta=\boldsymbol\mu_a\times\omega=2\boldsymbol\mu_{a+1}+\sum_{s=0}^3\alpha_{a+1}^s\nonumber\\\theta\times\boldsymbol\mu_a=\omega\times\boldsymbol\mu_a\times=2\boldsymbol\mu_{a-1}+\sum_{s=0}^3\alpha_{a-1}^s.\nonumber\end{gather} The first and second identities come from the the product relation $\alpha_a\alpha_{a+1}=\theta$ and $\alpha_a\alpha_{a-1}=\theta^{-1}$ of the symmetry group $S_3$, where the $\theta$ or $\omega$ channel is distinguished by the $\Phi$-loop \begin{align}\vcenter{\hbox{\includegraphics[width=0.09\textwidth]{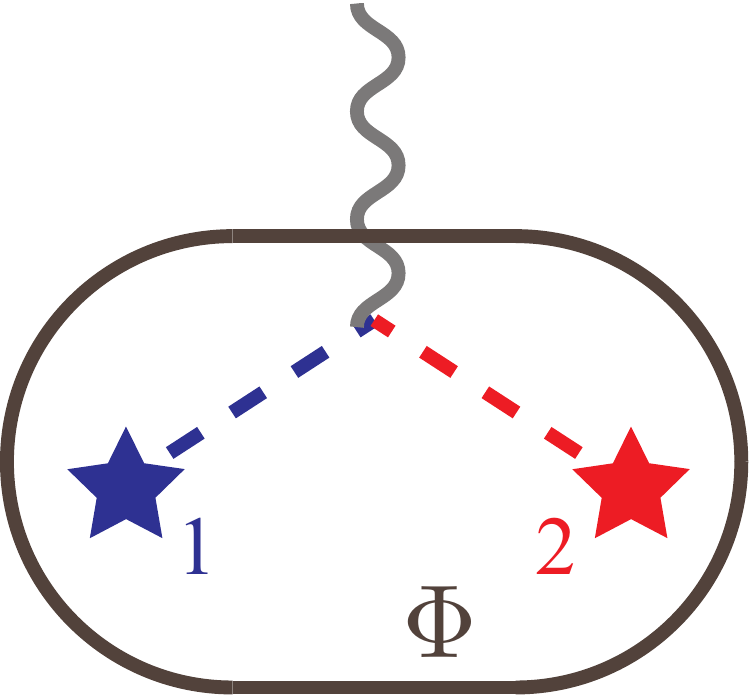}}}&=\frac{1}{2}\left[\vcenter{\hbox{\includegraphics[width=0.09\textwidth]{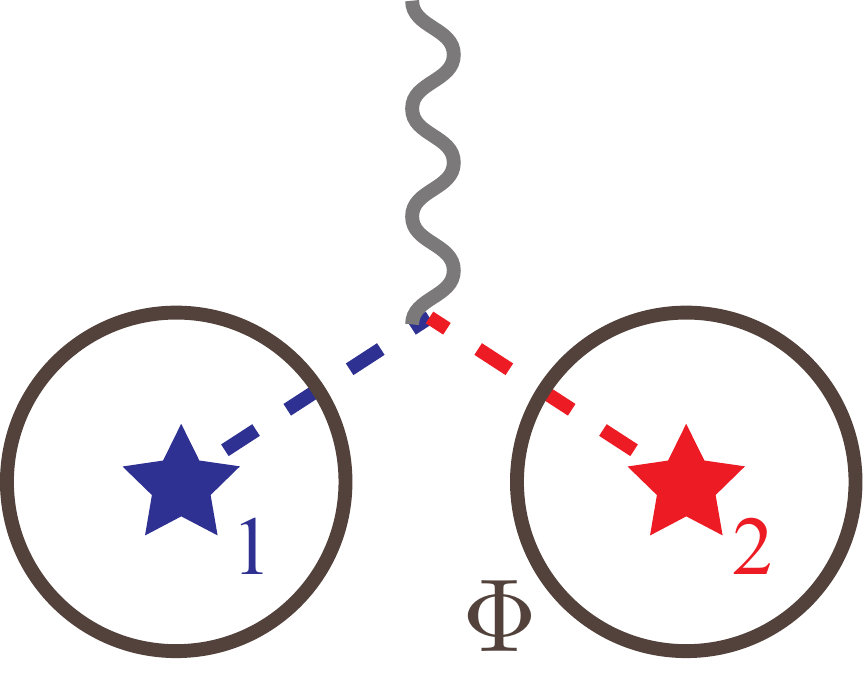}}}+\sum_{b=1}^3\vcenter{\hbox{\includegraphics[width=0.09\textwidth]{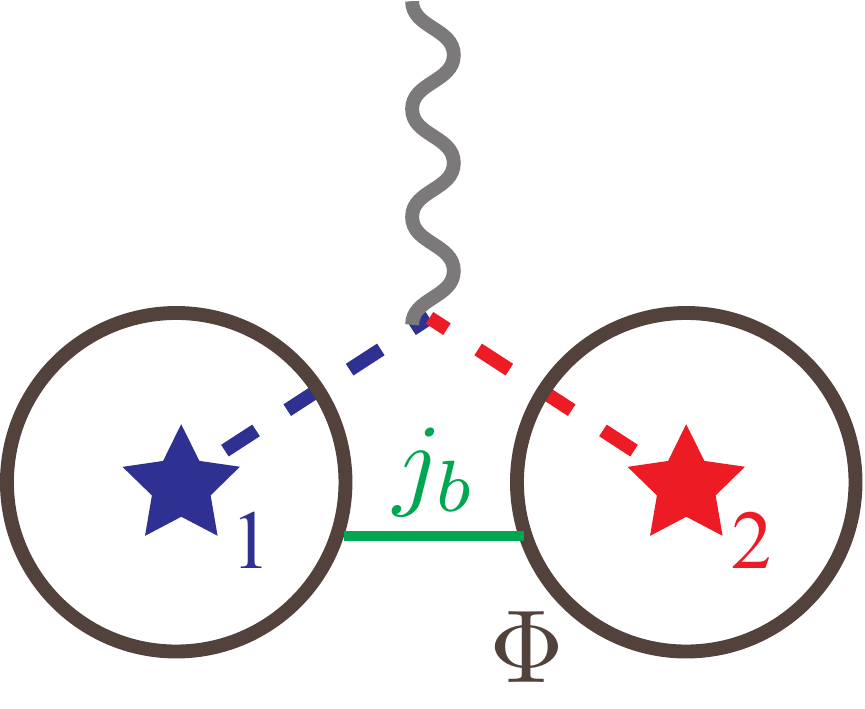}}}\right]\nonumber\\&=\frac{1}{2}\Theta^1_\Phi\Theta^2_\Phi=(-1)^{s_1+s_2}.\end{align} The other identities are consequences of fusion associativity.

\subsubsection{Braiding with a global texture}\label{sec:defectbraiding}
Topological defects are static objects with a global classical texture. Their exchange and braiding properties are fundamentally different from ordinary anyonic quantum excitations of a true topological state. Braiding operations are only {\em projective}~\cite{YouWen, YouJianWen, LindnerBergRefaelStern, ClarkeAliceaKirill, BarkeshliJianQi, TeoRoyXiao13long, teo2013braiding}. This is because when they move, their textures change accordingly. This costs energy and associates a non-universal dynamical phase. In a lot of cases, defects are stationary but they can effectively braid or hop around each other by adiabatically tunning inter-defects coupling parameters~\cite{LindnerBergRefaelStern, ClarkeAliceaKirill, BarkeshliJianQi}. For example, in a system with four defect at $x_i$ for $i=1,2,3,4$, inter-defect coupling can be generated by quasiparticle tunneling $\mathcal{W}_{ij}$, which is essentially a Wilson loop surrounding the $i,j$ defect pair. The adiabatic cyclic process \begin{align}H(t)=-J\left\{[1+\cos(2\pi t/T)]\mathcal{W}_{34}+[1+\cos(2\pi(t/T-1/3))]\mathcal{W}_{23}+[1+\cos(2\pi(t/T+1/3))]\mathcal{W}_{13}\right\}+h.c.\end{align} for $t\in[0,T]$, effectively exchanges the defects at $x_1$ and $x_2$ without physically moving them. 

The unitary braiding operation can be computed by solving the evolution of $H(t)$. It can also be equivalently derived by imagining that defects are physically dragged so that the {\em splitting operators} (see the previous subsection) between defects are deformed. We will demonstrate this explicitly in the case for the $\mathbb{Z}_2$ symmetric $\mathbb{Z}_n$ gauge theory, or equivalently the $(00n)$ FQH state (see section~\ref{sec:bilayerstates}), and will discuss the results for more general cases. Braiding can be systematically described by a mathematical framework known as a $G$-crossed category~\cite{BarkeshliBondersonChengWang14}. However we will focus more in the computation aspects and will omit the categorical approach.

The $\mathbb{Z}_n$ gauge theory, or quantum doube $D(\mathbb{Z}_n)$, generalizes the Kitaev's toric code. It is described by an effective Chern-Simons action \eqref{CSaction} with $K=n\sigma_x$. Quasiparticles are represented by lattice vector ${\bf a}=a_1{\bf e}+a_2{\bf m}=e^{a_1}\times m^{a_2}$ for $a_1,a_2=0,1,\ldots,n-1$ mod n, where ${\bf e}=(1,0)$ and ${\bf m}=(0,1)$ are the primitive charge and flux of the gauge theory with mutual monodromy $\mathcal{D}S_{{\bf e},{\bf m}}=e^{2\pi i/n}$. The exchange phase \eqref{Rsymboldefapp} between anyon ${\bf a}$ and ${\bf b}$ is $R^{\bf ab}=e^{\frac{2\pi i}{n}a_2b_1}$, which is gauge dependent when ${\bf a}\neq{\bf b}$. Under this choice the $F$-symbols can be chosen to be trivial, $F^{\bf abc}=1$, and satisfy the hexagon identities \eqref{hexagoneq}.

\begin{figure}[htbp]
\centering\includegraphics[width=0.55\textwidth]{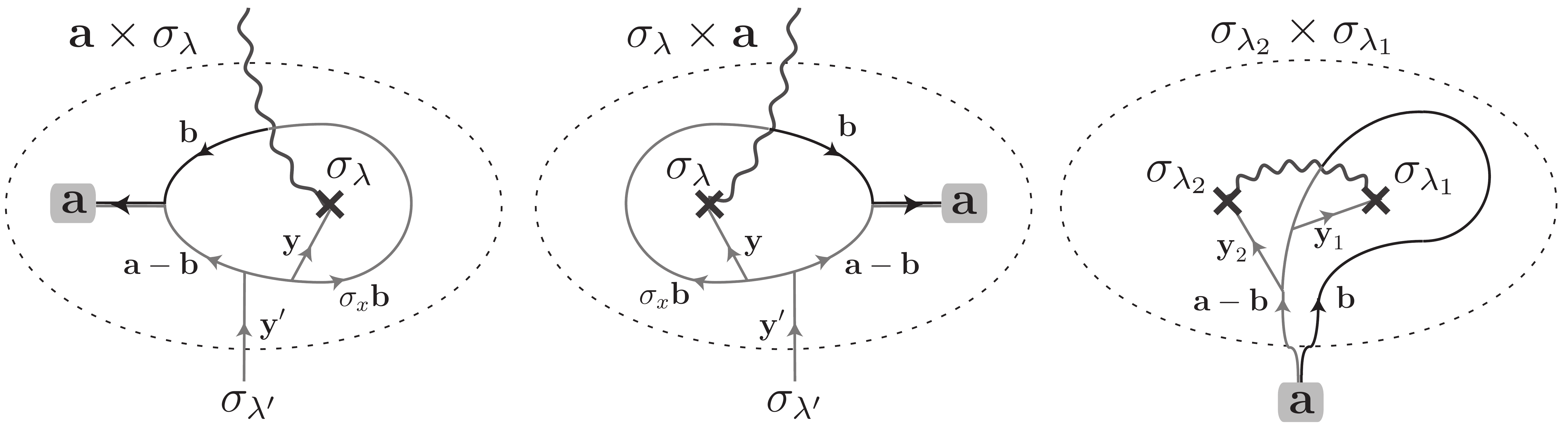}
\caption{Quasiparticle string configurations for splitting operators $V^{\psi_2\psi_1}_{\psi_3}$. Branch cuts (wavy lines) switch labels ${\bf b}\to\sigma_x{\bf b}$ of passing quasiparticles and end at twofold defects (crosses). Quasiparticle ${\bf a}=a_1{\bf e}+a_2{\bf m}$ is split into ${\bf b}=a_2{\bf m}$ and ${\bf a}-{\bf b}=a_1{\bf e}$. Twofold defect $\sigma_\lambda$ is attached with string ${\bf y}=\lambda{\bf e}$.}\label{fig:splittingstates}
\end{figure}

The $D(\mathbb{Z}_n)$ state has a twofold electric-magnetic (or bilayer) symmetry $\sigma_x$, and corresponds to twofold defects $\sigma_\lambda$, $\lambda=0,1,\ldots,n-1$ mod n. The species label characterizes defect bound anyons so that $\sigma_\lambda\times{\bf a}=\sigma_{\lambda+a_1+a_2}$. Defect pair fusion is given by \begin{align}\sigma_{\lambda_1}\times\sigma_{\lambda_2}=e^{\lambda_1+\lambda_2}\times\left(1+\Psi^2+\ldots+\Psi^{n-1}\right)\end{align} where $\Psi^j={\bf e}^j\times{\bf m}^j$ is the dyon with spin $h_{\Psi^j}=j^2/n$. Thus a twofold defect has quantum dimension $d_\sigma=\sqrt{n}$, and matches that of the parafermion or fractional Majorana in a FQSH-SC-FM heterostructure (c.f.~\eqref{parafermionfusion} in section~\ref{sec:parafermion}). The splitting operators are defined in figure~\ref{fig:splittingstates}. The $F$-symbols \eqref{Fsymboldef} for basis transformation can be computed similar to defects in the Kitaev's toric code in section~\ref{sec:Fbasistransformation} and are listed in table~\ref{tab:Fsymbols}.

\begin{table}[ht]
\centering
\begin{tabular}{ll}
\multicolumn{2}{c}{Defect $F$-symbols of $\mathbb{Z}_n$ gauge theory}\\\hline
$F^{{\bf a}{\bf b}{\bf c}}_{\bf d}$, $F^{{\bf a}{\bf b}\sigma}_{\sigma}$, $F^{\sigma{\bf a}{\bf b}}_{\sigma}$, $F^{{\bf a}\sigma\sigma}_{\bf b}$, $F^{\sigma\sigma{\bf a}}_{\bf b}$ & $1$ \\
$F^{{\bf a}\sigma{\bf b}}_{\sigma}$, $F^{\sigma{\bf a}\sigma}_{\bf b}$ & $e^{-\frac{2\pi i}{n}a_2b_2}$ \\
$\left[F^{\sigma\sigma\sigma}_{\sigma}\right]_{\bf a}^{\bf b}$ & $\frac{1}{\sqrt{n}}e^{\frac{2\pi i}{n}a_2b_2}$
\end{tabular}
\caption{Admissible $F$-symbols for twofold defects $\sigma$ in $D(\mathbb{Z}_n)$ with quasiparticles ${\bf a}=a_1{\bf e}+a_2{\bf m}$, ${\bf b}=b_1{\bf e}+b_2{\bf m}$.}\label{tab:Fsymbols}
\end{table}

\begin{figure}[ht]
	\centering\includegraphics[width=0.3\textwidth]{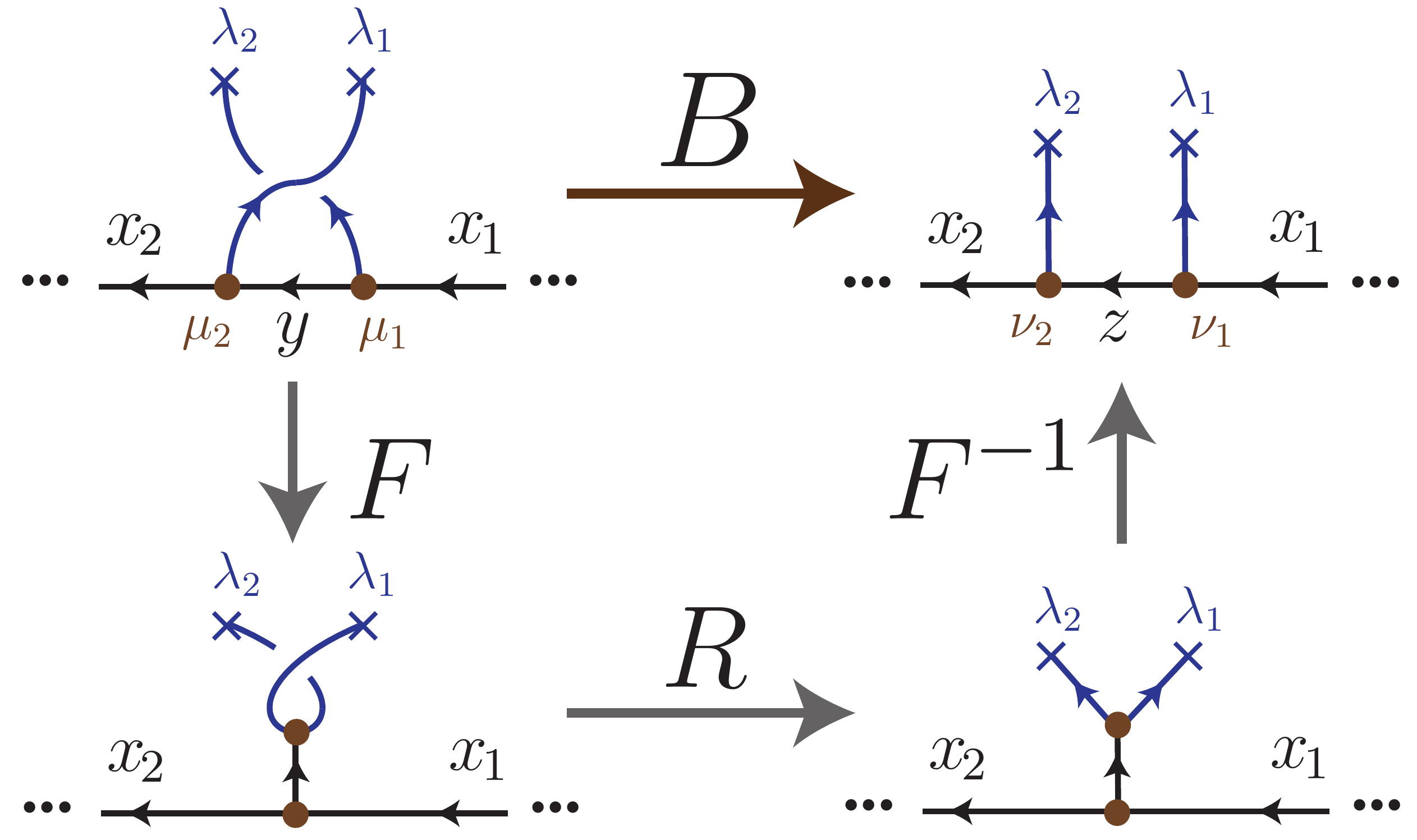}
	\caption{Braiding operation $B$ generated as a sequence of $F$ and $R$-moves. $\lambda_i$ are defects of the same $S_3$-type. Ground states are labeled by the intermediate channels $x_i,y,z$ and possible vertex degeneracies $\mu_i,\nu_i$.}\label{fig:Bmove}
\end{figure}

Non-abelian defect braiding operations are generated by fundamental exchange permutations of adjacent defects known as $B$-moves. Each $B$-move represents a counter-clockwise permutation of a pair of adjacent defects, and is a transformation between ground states labeled on the same fusion tree, i.e.~eigenstates of the same maximal set of commuting observables. It can be generated by a sequence of $F$ and $R$-moves as shown in figure~\ref{fig:Bmove} so that \begin{align}\left|\vcenter{\hbox{\includegraphics[width=0.5in]{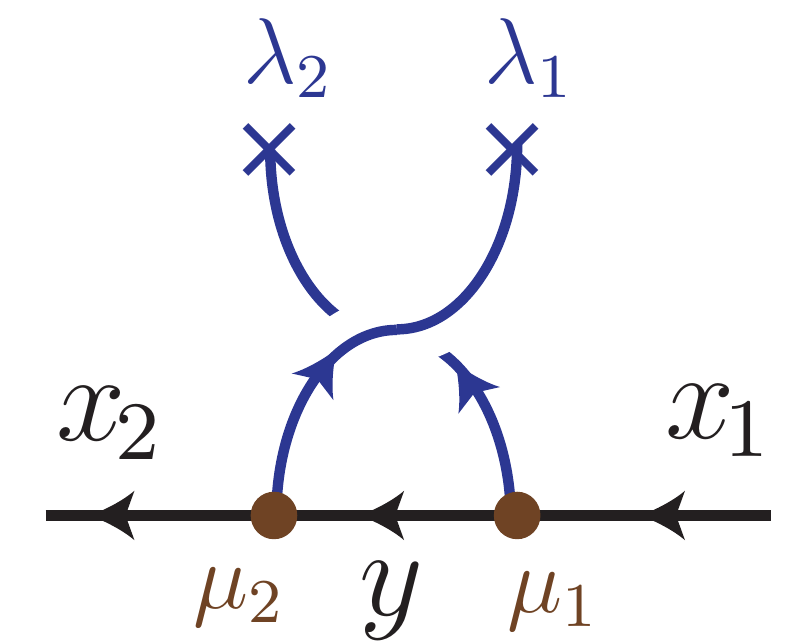}}}\right\rangle=\sum_{z,\nu_1,\nu_2}\left[B^{\lambda_2\lambda_1}_{x_2x_1}\right]_{y,\mu_2,\mu_1}^{z,\nu_1,\nu_2}\left|\vcenter{\hbox{\includegraphics[width=0.5in]{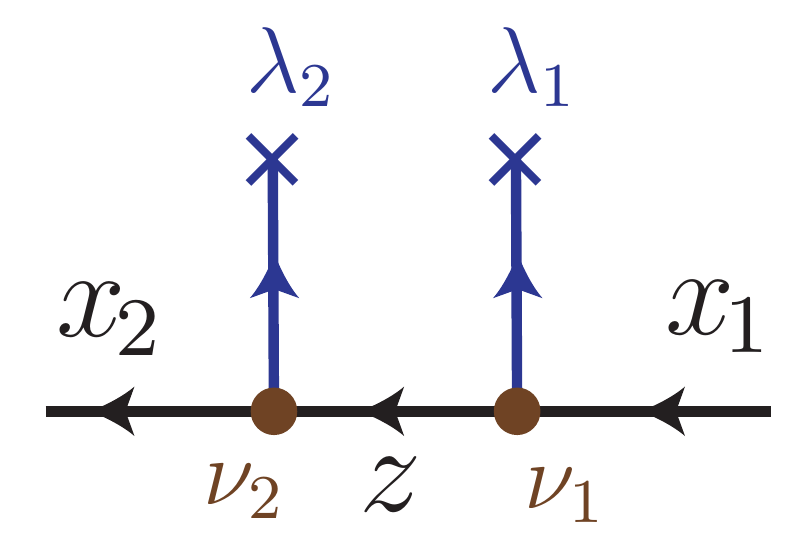}}}\right\rangle\label{Bmatrixdef1}\end{align} where the $B$-matrix is defined by \begin{align}\left[B^{\lambda_2\lambda_1}_{x_2x_1}\right]_{y,\mu_2,\mu_1}^{z,\nu_1,\nu_2}=\sum_{w\gamma_1\gamma_2\gamma_3}\left(\left[F^{x_2\lambda_2\lambda_1}_{x_1}\right]_{z,\nu_1\nu_2}^{w,\gamma_1,\gamma_3}\right)^\ast\left[R^{\lambda_1\lambda_2}_w\right]_{\gamma_2}^{\gamma_3}\left[F^{x_2\lambda_1\lambda_2}_{x_1}\right]_{y,\mu_2,\mu_1}^{w,\gamma_1,\gamma_2}.\label{Bmatrixdef2}\end{align} Here we define the braiding $B$-matrix in the most general setting, which may involve vertex degeneracies labeled by $\mu_i,\nu_i,\gamma_i$. The exchange operations $B^{\lambda_{i+1}\lambda_i}$ form the building blocks of the braid group of an ordered series of defects $\lambda_N,\ldots,\lambda_1$. 
\begin{figure}[ht]
	\centering\includegraphics[width=0.3\textwidth]{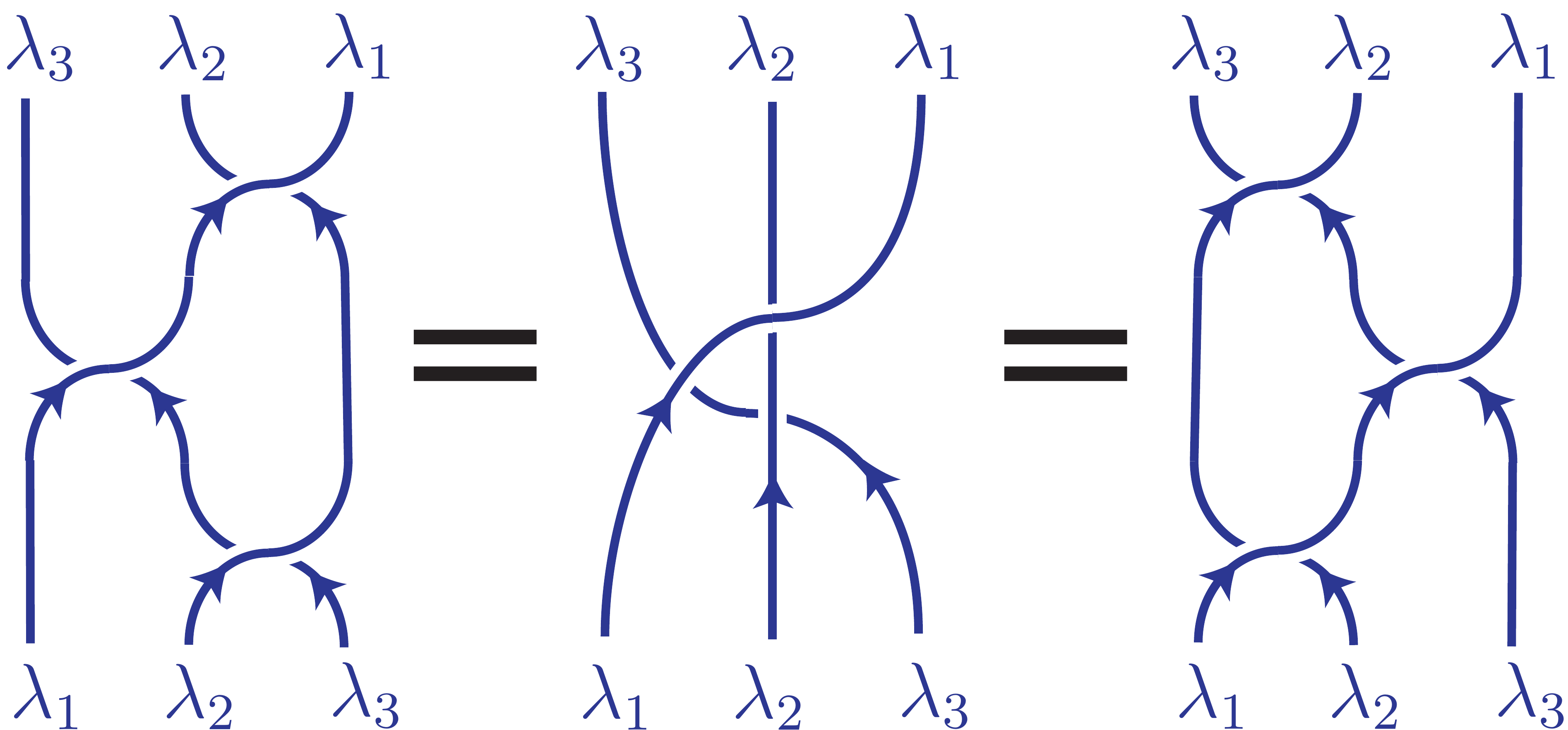}
	\caption{Yang-Baxter identity}\label{fig:YangBaxter}
\end{figure}
They obey the Yang-Baxter equation (see figure~\ref{fig:YangBaxter}) that characterizes braids. \begin{align}B^{\lambda_2\lambda_1}B^{\lambda_3\lambda_1}B^{\lambda_3\lambda_2}=B^{\lambda_3\lambda_2}B^{\lambda_3\lambda_1}B^{\lambda_2\lambda_1}\label{YangBaxtereq}\end{align} where the summation over intermediate channels and vertex degeneracies are suppressed. Defects $\lambda_i,\lambda_j$ are distinguishable when they have distinct species labels. A braiding operation is robustly represented only when the initial and final species labels configuration are identical.

\begin{figure}[htbp]
\centering\includegraphics[width=1\textwidth]{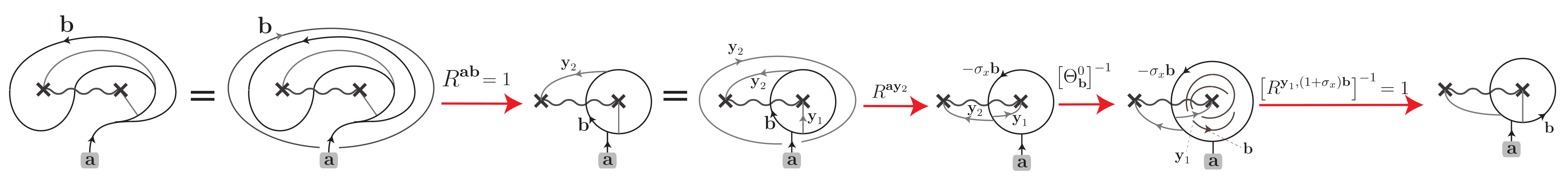}
\caption{Derivation of the defect $R$-symbol $R^{\sigma_{\lambda_2}\sigma_{\lambda_1}}_{\bf a}$ in a $\mathbb{Z}_n$ gauge theory. ${\bf a}=e^{a_1}m^{a_2}$, ${\bf b}=m^{a_2}$, ${\bf y}_j=e^{\lambda_j}$.}\label{fig:Rss}
\end{figure}
For defects in the $\mathbb{Z}_n$ gauge theory, the $R$-symbol $R^{\sigma_{\lambda_2}\sigma_{\lambda_1}}_{\bf a}$ relates the state before and after exchanging a pair of defects $\sigma_{\lambda_2}\times\sigma_{\lambda_1}$ with fixed fusion outcome ${\bf a}=a_1{\bf e}+a_2{\bf m}=e^{a_1}\times m^{a_2}$. It can be derived by twisting the quasiparticle strings of the splitting operator $V^{\sigma_2\sigma_1}_{\bf a}$ in figure~\ref{fig:splittingstates} by $180^\circ$ and deforming back into the untwisted configuration. \begin{align}R^{\sigma_{\lambda_2}\sigma_{\lambda_1}}_{\bf a}&=\left\langle\vcenter{\hbox{\includegraphics[width=0.5in]{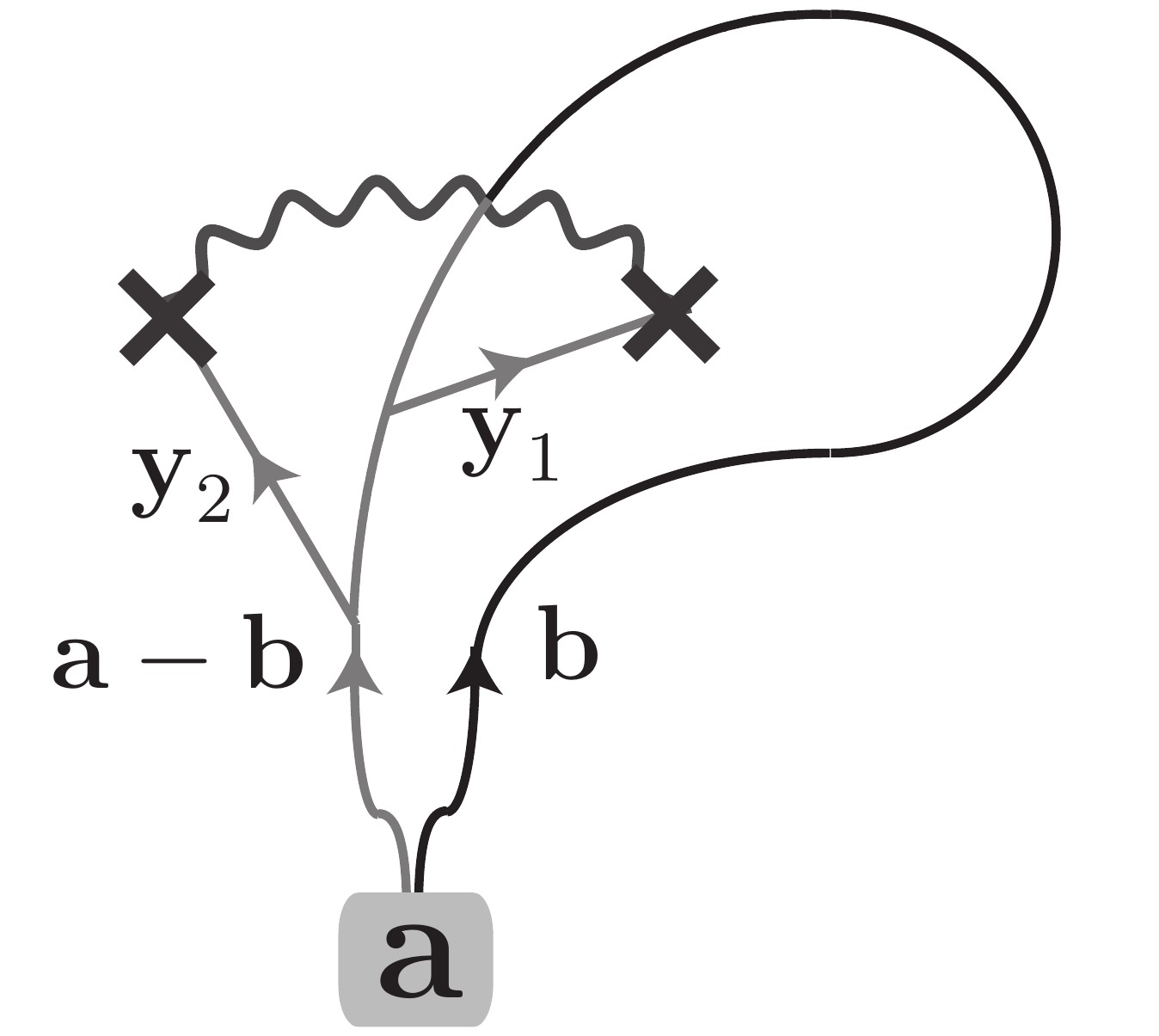}}}\left|\vcenter{\hbox{\includegraphics[width=0.7in]{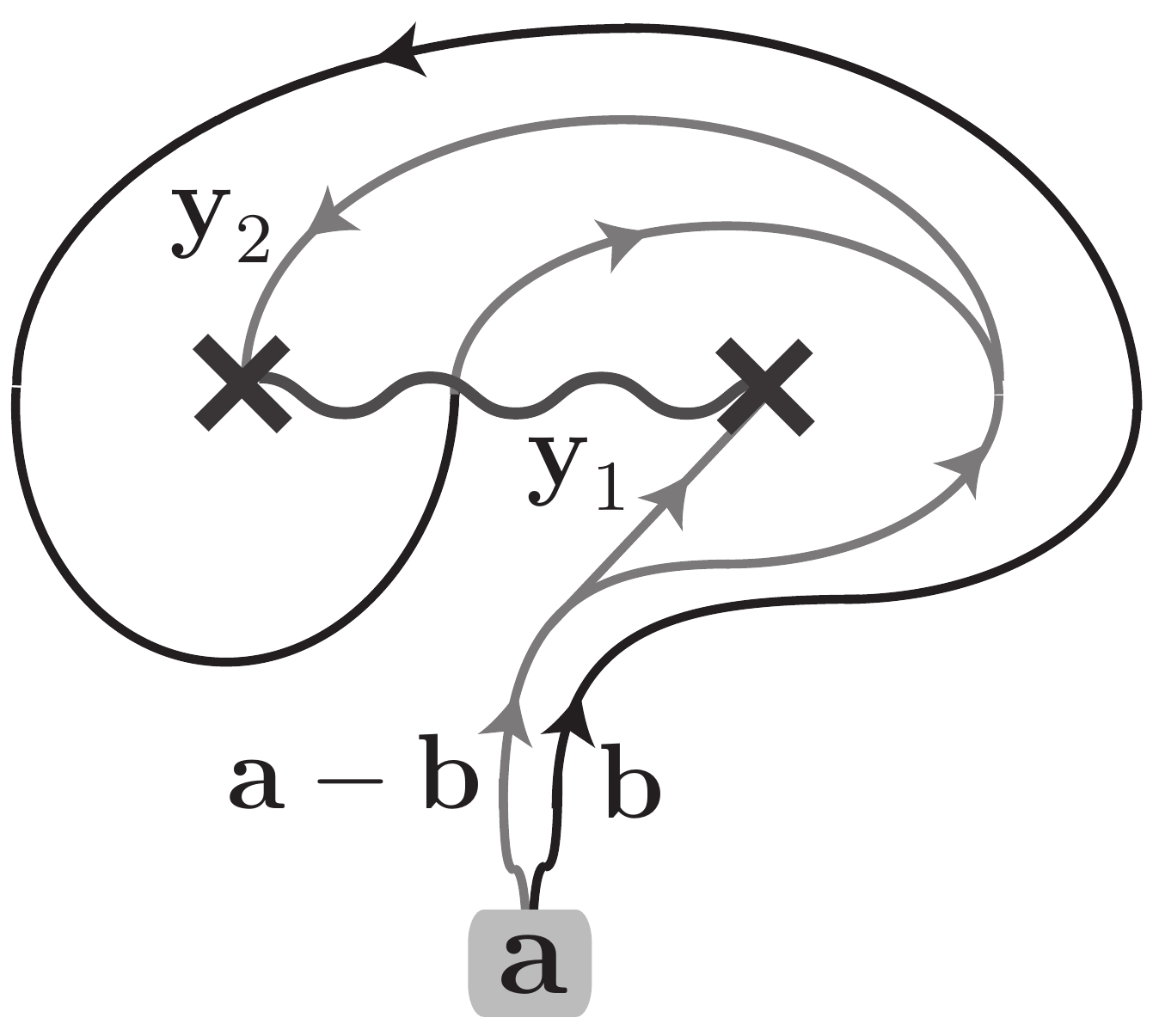}}}\right.\right\rangle=R^{{\bf ay}_2}\left(\Theta^0_{\bf b}\right)^{-1}=e^{\frac{2\pi i}{n}a_2\left(\lambda_2+\frac{n}{2}-\frac{1}{2}a_2\right)}\label{ZnRsymbol}\end{align} This can be computed systematically by a sequence of string crossings (see figure~\ref{fig:Rss}), and introducing a Wilson operator $\hat{\Theta}_{\bf b}$ that drags ${\bf b}=a_2{\bf m}=m^{a_2}$ twice around a defect. Since this operator is lying beneath the ${\bf y}_1$ string that gives the defect its species $\lambda_1$ (see the second last diagram in figure~\ref{fig:Rss}), its eigenvalue \begin{align}\Theta^0_{\bf b}=e^{\pi i\left(a_2^2/n+a_2\right)}\label{Theta0b}\end{align} corresponds to that of a bare defect $\sigma_0$. This eigenvalue is uniquely determined by self-intersection $\hat{\Theta}_{{\bf a}_1+{\bf a}_2}=e^{2\pi i{\bf a}_1K^{-1}\sigma_x{\bf a}_2}\hat{\Theta}_{{\bf a}_1}\hat{\Theta}_{{\bf a}_2}$ which implies $\hat\Theta_{\bf b}^n=(-1)^{(n-1)a_2^2}=(-1)^{(n-1)a_2}$, invariance under twofold symmetry $\hat{\Theta}_{\bf a}=\hat{\Theta}_{\sigma_x{\bf a}}$, and a linking phase shown in figure~\ref{fig:defect1}(c) that results in the phase $(\Theta^0_{\bf a})^2=e^{2\pi i{\bf a}^TK^{-1}\sigma_x{\bf a}}$ around a pair of self-conjugate twofold defects $\sigma_0$.

Applying \eqref{ZnRsymbol} in \eqref{Bmatrixdef2}, we get the $180^\circ$ exchange $B$-matrices \begin{align}B^{\sigma_{\lambda_2}\sigma_{\lambda_1}}_{{\bf a}{\bf b}}&=e^{\frac{2\pi i}{n}(b_2-a_2)\left[\lambda_2+\frac{n}{2}-\frac{1}{2}(b_2-a_2)\right]},\label{eqB1}\\\left[B^{\sigma_{\lambda_2}\sigma_{\lambda_1}}_{\sigma\sigma}\right]_{\bf a}^{\bf b}&=\frac{e^{\frac{i\pi}{4}(n\pm1)}}{\sqrt{|n|}}e^{\frac{2\pi i}{n}(\lambda_2+b_2-a_2)\left[\frac{1}{2}(\lambda_2+b_2-a_2)+\frac{n}{2}\right]}.\label{eqB2}\end{align} When acting on the quantum state of a sequence of defects \begin{align}|{\bf a}_i\rangle=\left|\vcenter{\hbox{\includegraphics[width=1.5in]{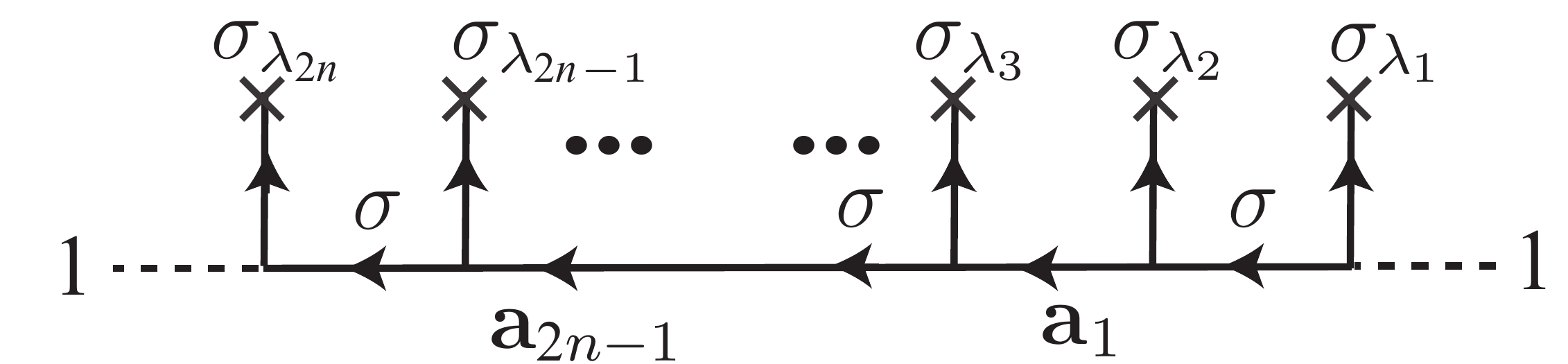}}}\right\rangle,\end{align} neighboring braiding operations \eqref{eqB1} and \eqref{eqB2} do not commute. These braiding matrices honesty obey the Yang-Baxter equations \eqref{YangBaxtereq} (see figure~\ref{fig:YangBaxter}) \begin{align}B^{\sigma_{\lambda_i}\sigma_{\lambda_{i-1}}}B^{\sigma_{\lambda_{i+1}}\sigma_{\lambda_i}}B^{\sigma_{\lambda_i}\sigma_{\lambda_{i-1}}}=B^{\sigma_{\lambda_{i+1}}\sigma_{\lambda_i}}B^{\sigma_{\lambda_i}\sigma_{\lambda_{i-1}}}B^{\sigma_{\lambda_{i+1}}\sigma_{\lambda_i}}\end{align} but form a projective representation of the sphere braid group when the system is compactified on a closed surface~\cite{TeoRoyXiao13long, teo2013braiding}. 

The exchange $R$-symbol \eqref{ZnRsymbol} defines the spin of a twofold defect by averaging over admissible quasiparticle fusion channel \begin{align}\theta_{\sigma_\lambda}=\frac{1}{d_\sigma}\sum_{\bf a}R^{\sigma_\lambda\sigma_\lambda}_{\bf a}=e^{\frac{i\pi}{4}(n-1)}e^{\pi i\left(\lambda^2/n+\lambda\right)}.\label{exchangespin}\end{align} A defect can be represented as an open end of a twofold branch cut which switches the bilayer, and therefore a $360^\circ$ rotation leaves behind an uncancelable branch cut surrounding the defect. The exchange spin \eqref{exchangespin} however identifies with $720^\circ$ rotation up to the constant $\mathbb{Z}_8$ phase $e^{\frac{i\pi}{4}(n-1)}$. The defect $\sigma_\lambda$ is equipped with a quasiparticle string ${\bf y}=\lambda{\bf e}$ that gives the species label to the defect. Upon rotation, the string wraps around the defect twice, and can be untwisted (through a trivial crossing $\theta_{\bf y}=R^{{\bf y}{\bf y}}=1$) by absorbing a double Wilson operator $\hat{\Theta}_{\bf y}$ into the condensate. \begin{align}\vcenter{\hbox{\includegraphics[width=0.35in]{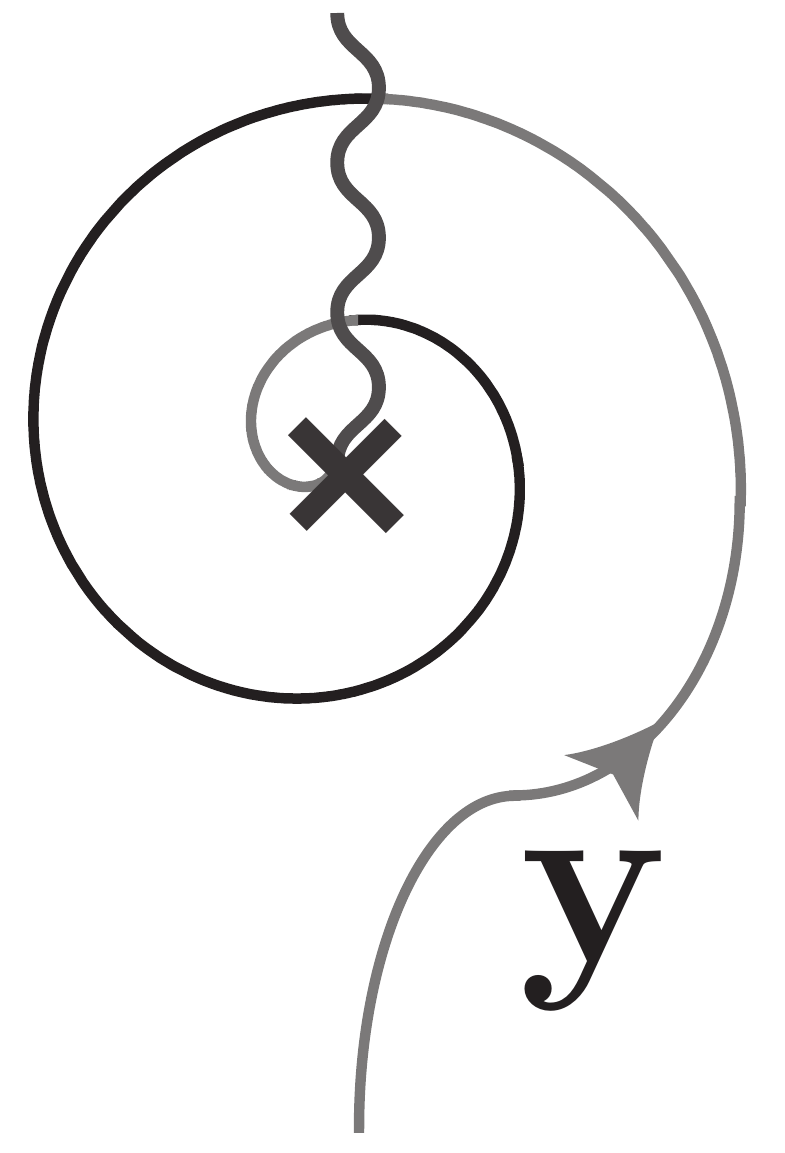}}}&=\theta_{\bf y}\vcenter{\hbox{\includegraphics[width=0.32in]{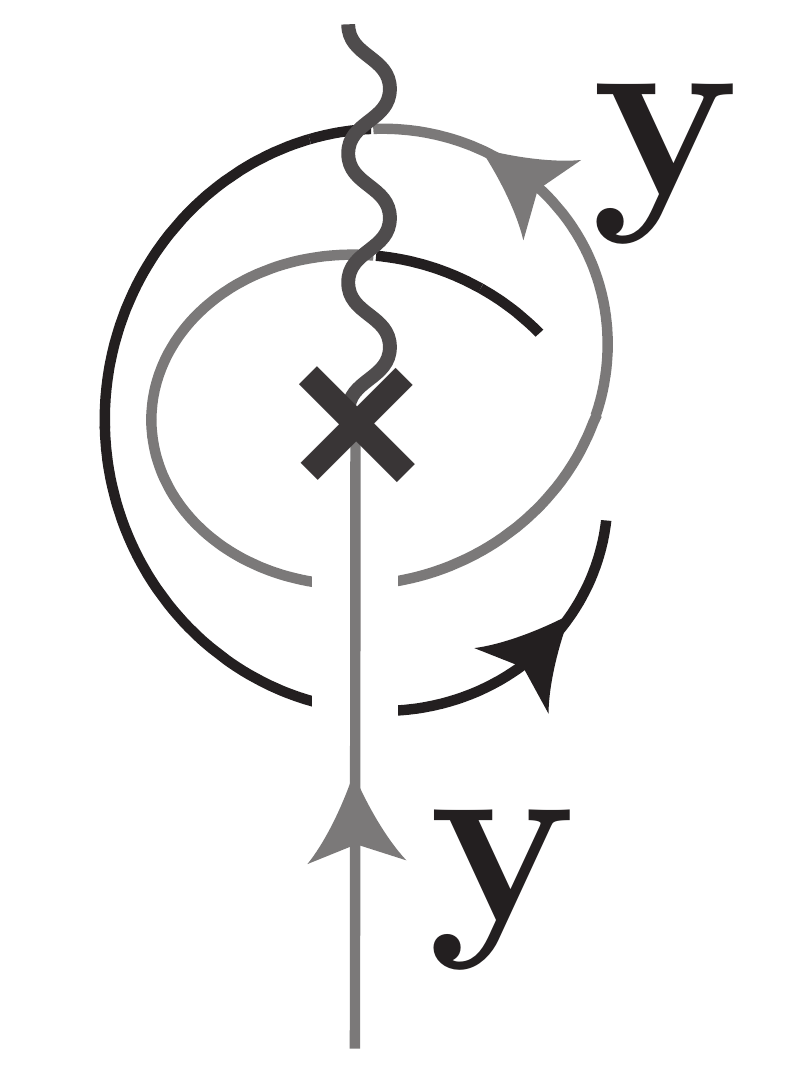}}}=\theta_{\bf y}\Theta^0_{\bf y}\vcenter{\hbox{\includegraphics[width=0.17in]{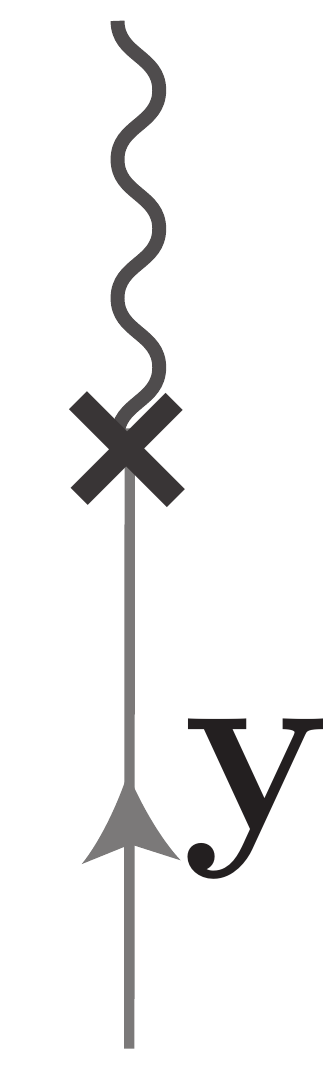}}}=e^{\pi i\left(\lambda^2/n+\lambda\right)}\vcenter{\hbox{\includegraphics[width=0.17in]{Z2spin3}}}.\label{720rotation}\end{align} 

Defect monodromy is encoded by the following average \begin{align}S_{\sigma_{\lambda_1}\sigma_{\lambda_2}}=\frac{1}{\mathcal{D}_{\sigma}}\vcenter{\hbox{\includegraphics[width=0.5in]{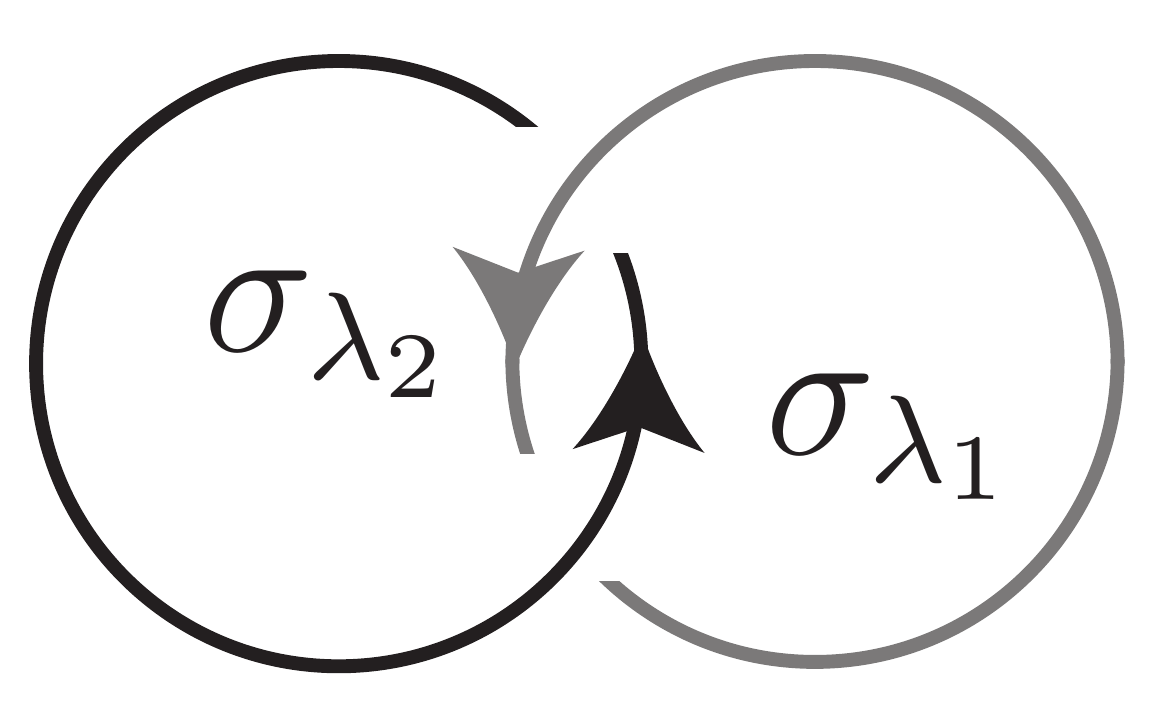}}}=\frac{1}{\mathcal{D}_{\sigma}}\sum_{\bf a}R^{\overline{\sigma_{\lambda_2}}\sigma_{\lambda_1}}_{\bf a}R^{\sigma_{\lambda_1}\overline{\sigma_{\lambda_2}}}_{\bf a}\label{defectSmatrix}\end{align} over admissible fusion channels $\sigma_{\lambda_1}\times\sigma_{\lambda_2}\to{\bf a}$ and mimics the modular $S$-matrix \eqref{braidingSapp}. For defects in the $\mathbb{Z}_n$ gauge theory, the total quantum dimension of the defect sector $\mathcal{C}_\sigma$ is $\mathcal{D}_\sigma=\sqrt{\sum_\lambda d_{\sigma_\lambda}^2}=n$, and the anti-particle of $\sigma_\lambda$ is $\overline{\sigma_\lambda}=\sigma_{-\lambda}$. The Gaussian sum has closed form solution \begin{align}S_{\sigma_{\lambda_1}\sigma_{\lambda_2}}=\frac{1}{n}\sum_{a=1}^{n-1}e^{\frac{2\pi i}{n}[-a^2+(\lambda_1-\lambda_2)a]}=\left\{\begin{array}{*{20}l}\frac{1}{\sqrt{n}}e^{-\frac{i\pi}{2}\left(\frac{n-1}{2}\right)^2}e^{\frac{2\pi i}{n}\left(\frac{n+1}{2}\right)^2(\lambda_1-\lambda_2)^2}&\mbox{for $n$ odd}\\\sqrt{\frac{2}{n}}e^{-\frac{i\pi}{4}}e^{\frac{2\pi i}{n}\frac{(\lambda_1-\lambda_2)^2}{4}}\left(\frac{1+(-1)^{\lambda_1-\lambda_2+n/2}}{2}\right)&\mbox{for $n$ even}\end{array}\right..\label{braidingSmatrix}\end{align}

\begin{figure}[ht]
\centering\includegraphics[width=0.7\textwidth]{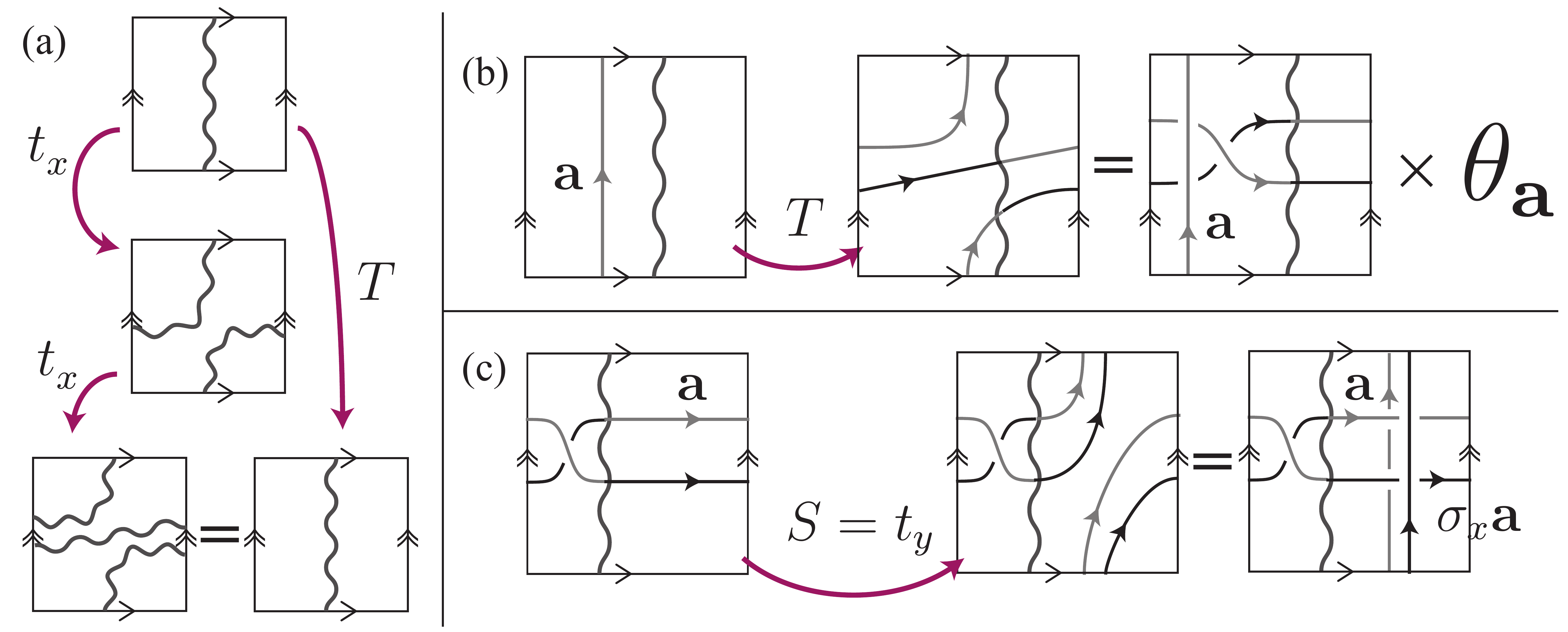}
\caption{Dehn twists. (a) Double Dehn twist $T=t_x^2$ along the horizontal direction that leaves the branch cut (curvy line) invariant. (b) $T$-action on a vertical Wilson loop $\hat{W}_{\bf a}$. (c) Dehn twist $S=t_y$ along the vertical (longitudinal) direction and its action on a double Wilson loop $\hat{\Theta}_{\bf a}$.}\label{fig:dehntwist}
\end{figure}
Defect braiding $S$ and exchange $T$ matrices can alternatively be evaluated as geometric {\em congruent} transformation of ground states on a torus with a twofold branch cut, which occupies a particular cycle, say the $y$-direction, on the torus (see figure~\ref{fig:dehntwist}). The subgroup of modular transformations that leaves the branch cut invariant [i.e. fixes the $y$-cycle with $\mathbb{Z}_2$ coefficient in $H_1(T^2;\mathbb{Z}_2)=\mathbb{Z}_2^2$] is known as a congruent subgroup \begin{align}\Gamma_0(2)=\left\langle\begin{array}{*{20}c}S=t_y\\T=t_x^2\end{array}\left|\begin{array}{*{20}c}(ST^{-1})^2=C,C^2=1\\SCS^{-1}=TCT^{-1}=C\end{array}\right.\right\rangle\end{align} where $t_x,t_y$ are Dehn twists on the torus along the $x,y$ direction. The Wilson operator algebra is generated by quasiparticle strings $\hat{W}_{\bf a}$ of moving ${\bf a}=e^{a_1}m^{a_2}$ along the vertical cycle and $\hat{\Theta}_{\bf a}$ twice along the horizontal cycle (see figure~\ref{fig:dehntwist}). $\hat{W}_{\bf a}$ corresponds the quasiparticle string attaching to a twofold defect and gives rise to its species $\lambda=a_1+a_2$, and $\hat{\Theta}_{\bf a}$ is equivalent to the Wilson operator that drags ${\bf a}$ twice around a defect. As these operator can pass across the branch cut and change its label, they obey the twofold symmetry $\hat{W}_{\bf a}=\hat{W}_{\sigma_x{\bf a}}$, $\hat{\Theta}_{\bf a}=\hat{\Theta}_{\sigma_x\bf a}$ and satisfy the fusion relations $\hat{W}_{{\bf a}+{\bf b}}=\hat{W}_{\bf a}\hat{W}_{\bf b}$, $\hat{\Theta}_{{\bf a}+{\bf b}}=e^{2\pi i{\bf a}^TK^{-1}\sigma_x{\bf b}}\hat{\Theta}_{\bf a}\hat{\Theta}_{\bf b}$.
They commute up to the braiding phase of intersection $\hat{\Theta}_{\bf a}\hat{W}_{\bf b}=e^{\frac{2\pi i}{n}(a_1+a_2)(b_1+b_2)}\hat{W}_{\bf b}\hat{\Theta}_{\bf a}$ and transform under the congruent generators $S$ and $T$ according to \begin{align}\hat{T}\hat{W}_{\bf a}\hat{T}^\dagger=\theta_{\bf a}\hat{W}_{\bf a}\hat{\Theta}_{\bf a},\quad\hat{S}\hat{\Theta}_{\bf a}\hat{S}^\dagger=\hat{W}_{\bf a}\hat{\Theta}_{\bf a}\hat{W}_{\sigma_x\bf a}\end{align} (see figure~\ref{fig:dehntwist}) while keeping $\hat{T}\hat{\Theta}_{\bf a}\hat{T}^\dagger=\hat{\Theta}_{\bf a}$ and $\hat{S}\hat{W}_{\bf a}\hat{S}^\dagger=\hat{W}_{\bf a}$.

The $n$-fold degenerate ground states are simultaneous eigenstates of $\hat{\Theta}_{\bf a}$ and $\hat{T}$ that corresponds defect species, \begin{align}\hat{\Theta}_{\bf a}|\lambda\rangle=\Theta^\lambda_{\bf a}|\lambda\rangle,\quad\hat{W}_{\bf a}|\lambda\rangle=|\lambda+{\bf a}\cdot{\bf t}\rangle\end{align} The eigenvalue $\Theta^\lambda_{\bf a}$ of $\hat\Theta_{\bf a}$ around defect $\sigma_\lambda$ can be evaluated by crossing the Wilson operator $\hat\Theta_{\bf a}$ with the ${\bf y}=\lambda{\bf e}=e^\lambda$ string into the defect before absorbing it to the ground state. \begin{gather}\vcenter{\hbox{\includegraphics[width=0.4in]{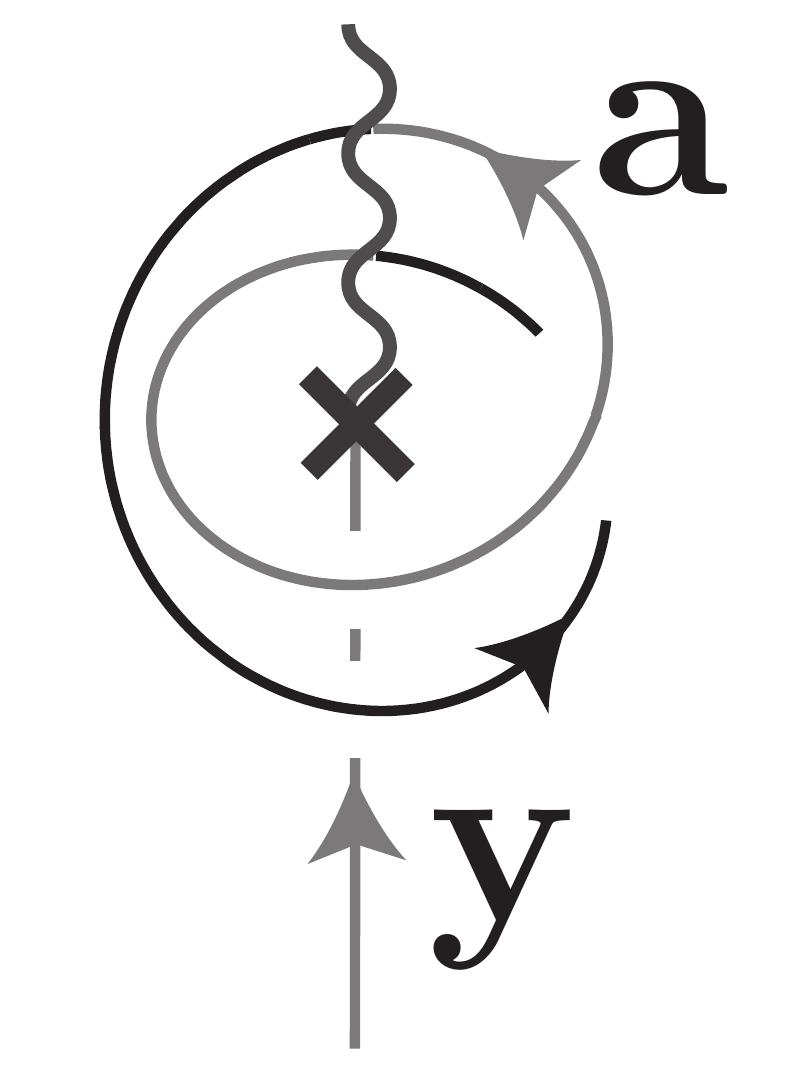}}}=e^{\frac{2\pi i}{n}(a_1+a_2)\lambda}\vcenter{\hbox{\includegraphics[width=0.4in]{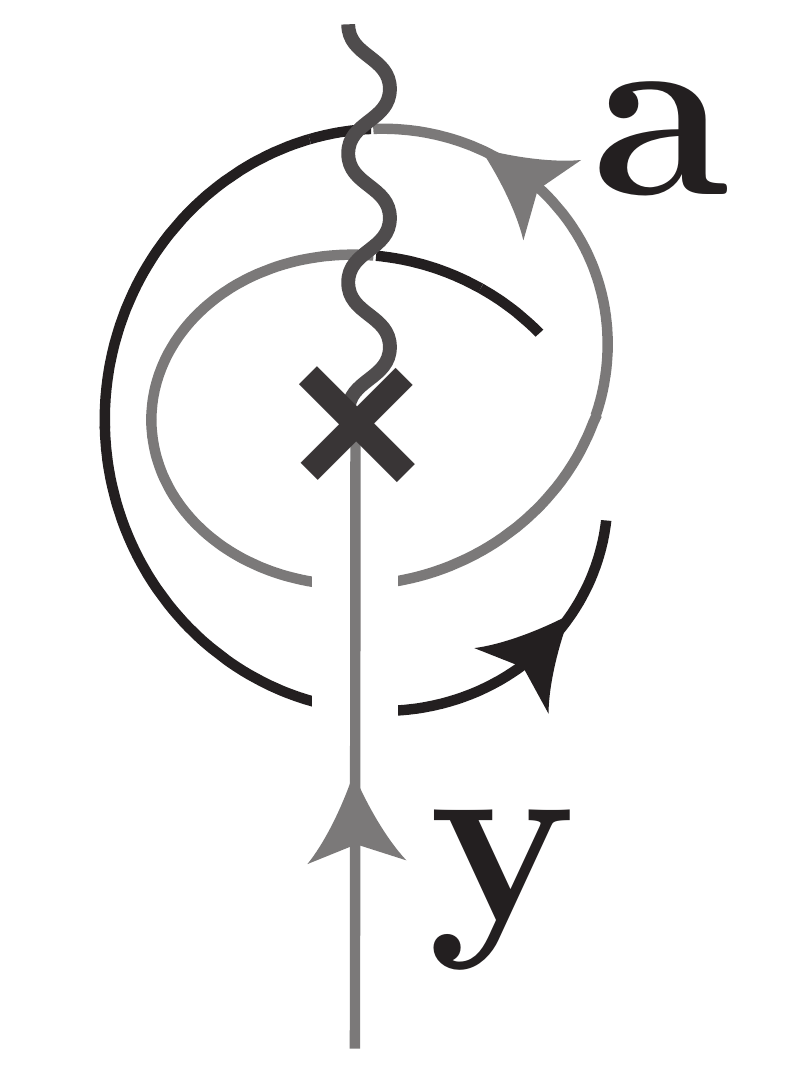}}}=e^{\frac{2\pi i}{n}(a_1+a_2)\lambda}\Theta^0_{\bf a}\vcenter{\hbox{\includegraphics[width=0.17in]{Z2spin3}}}.\nonumber\\\Theta^\lambda_{\bf a}=e^{\frac{2\pi i}{n}(a_1+a_2)\lambda}\Theta^0_{\bf a}=e^{\frac{2\pi i}{n}(a_1+a_2)\lambda}e^{\pi i({\bf a}\cdot{\bf a}/n+(a_1+a_2)\lambda)}\end{gather} where $\Theta^0_{\bf a}=e^{\pi i({\bf a}\cdot{\bf a}/n+(a_1+a_2)\lambda)}$ and can be evaluated in a manner similar to \eqref{Theta0b}.

The $T$-matrix that represents double Dehn twist $t_x^2$ is diagonal in this basis. \begin{align}\hat{T}|\lambda\rangle&=\hat{T}\hat{W}_{\bf a}|0\rangle=\theta_{\bf a}\hat{W}_{\bf a}\hat{\Theta}_{\bf a}\hat{T}|0\rangle=T_0\theta_{\bf a}\Theta^0_{\bf a}|\lambda\rangle=T_0\theta_{\lambda}|\lambda\rangle\end{align} where ${\bf a}$ is any quasiparticle that corresponds to the species $\lambda=a_1+a_2$, and $T_0$ is the eigenvalue $\langle0|\hat{T}|0\rangle$. And therefore up to a constant phase, the congruent $T$-matrix has entries $T_{\lambda_1\lambda_2}\propto\delta_{\lambda_1\lambda_2}\theta_{\lambda_1}$. The commuting $\hat{W}_{\bf a}$ and $\hat{S}$ share the simultaneous eigenstate $|\emptyset\rangle=\frac{1}{\sqrt{n}}\sum_\lambda|\lambda\rangle$, which generates all ground states by $|\lambda\rangle=\frac{1}{\sqrt{n}}\hat{W}_{\bf a}\sum_{{\bf y}=b{\bf e}}(\Theta^0_{\bf y})^{-1}\hat{\Theta}_{\bf y}|\emptyset\rangle$, where $\lambda=a_1+a_2$. The $S$-action on $|\lambda\rangle$ is digramatically represented in figure~\ref{fig:dehntwist}(c). The congruent $S$-matrix that represents Dehn twist $t_y$ thus has entries \begin{align}\langle\lambda_1|\hat{S}|\lambda_2\rangle=\frac{S_\emptyset}{n}\sum_{{\bf y}=b{\bf e}}\langle\emptyset|\hat{\Theta}_{\bf y}^\dagger\hat{W}_{{\bf y}_2-{\bf y}_1+{\bf y}}\hat{\Theta}_{\bf y}|\emptyset\rangle=\frac{S_\emptyset}{n}\sum_{b=0}^{n-1}e^{\frac{2\pi i}{n}b(-b+\lambda_1-\lambda_2)}\label{Sgaussiansum}\end{align} where $S_\emptyset$ is the eigenvalue $\langle\emptyset|\hat{S}|\emptyset\rangle$. The Gaussian sum \eqref{Sgaussiansum} equates the braiding $S$-matrix \eqref{braidingSmatrix} up to a constant phase. The braiding $S$ and exchange spin $T$ matrices in \eqref{braidingSmatrix} and \eqref{exchangespin} generates a unitary representation for the congruent subgroup $\Gamma_0(2)$ where \begin{align}C_{\lambda_1\lambda_2}\equiv\left[(ST^\dagger)^2\right]_{\lambda_1\lambda_2}=\delta_{\lambda_1,-\lambda_2}\end{align} serves as a charge conjugation operator to defect species.

\section{Conclusion and Outlook}\label{sec:conclusion}
In this review article, we focused on the theory of global symmetries, known as anyonic symmetries, of topological phases in two dimensions, and the twist defects supported by these systems. The anyonic excitations of a topological state are characterized by their fusion, exchange and braiding properties. For example, two anyons can fuse together when viewed in the long length scale into new anyons with totally different topological charges. The exchange phase $\theta_{\bf x}=e^{2\pi ih_{\bf x}}$ of a pair of identical quasiparticles ${\bf x}$ is equivalent to a $2\pi$ twist of a single one. Each anyon therefore carries a specific spin $h_{\bf x}$. The monodromy of orbiting an anyon around another is dictated by the fusion rules and quasiparticle spins through the ribbon identity \eqref{ribbonapp}. On the other hand, the quasiparticle fusion rules can be determined from monodromy through the Verlinde formula \eqref{Verlindeformulaapp}. There are also requirements, in particular the Gauss-Milgram formula \eqref{GaussMilgram} from modular symmetry, for quasiparticle spin and monodromy to satisfy. 

The topological order of a topological phase refers to a set of consistent anyon properties mentioned above, and anyonic symmetry is a symmetry of a topological order. The anyons are relabeled by the symmetry while all the fusion, exchange and braiding information remain invariant. The simplest example was given by the Kitaev's toric code (or a $\mathbb{Z}_2$ gauge theory) in section~\ref{sec:toriccode}, where the electric-magnetic symmetry permutes the two bosonic charge and flux. The symmetry was realized as a lattice translation of the exactly solvable lattice spin model. This concept was generalized to other exactly solvable lattice rotor models, including the Wen's $\mathbb{Z}_k$ plaquette mode and the tricolor model, in section~\ref{sec:symmlatticemodels}. In general, given an effective Chern-Simons description \eqref{CSaction} of an Abelian topological state, (a subset of) anyonic symmetries can be represented as matrices $M$, known as outer automorphisms, that leave the $K$-matrix invariant, $MKM^T=K$. This was applied to bilayer and conjugation symmetric Abelian topological phases presented in section~\ref{sec:bilayerstates}. Moreover the global anyonic symmetry group can also be non-Abelian. For example, the $S_3$-symmetric $SO(8)_1$ state was described in detail in section~\ref{sec:symmSO(8)}. The Abelian state supports three mutually semionic fermions $\psi_1,\psi_2,\psi_3$, and are permuted by the $S_3$ symmetry. This symmetries can be explicitly realized as lattice rotations and translations of the tricolor model, which is equivalent to the time reversal symmetric double $SO(8)_1\otimes\overline{SO(8)_1}$. Anyonic symmetry can also arise in non-Abelian topological phases. In section~\ref{sec:globalsymmetrynab}, we demonstrate this using the $S_3$ symmetric chiral ``4-Potts" state. It supports 11 anyon types within which there is a triplet of Abelian bosons $j_1,j_2,j_3$ and two triplets of non-Abelian quasiparticles $\sigma_1,\sigma_2,\sigma_3$ and $\tau_1,\tau_2,\tau_3$ with spins $h_\sigma=1/16$ and $h_\tau=9/16$. The $S_3$ symmetry simultaneously permutes these triplets while keeping the fusion rules, exchange and braiding information unchanged. 

One of the most promising consequence of anyonic symmetries in topological phases is the appearance of twist defects. They are static fluxes of the global symmetries that permute the anyon labels of orbiting quasiparticles. In other words, a quasiparticle string around a defect does not necessarily close back onto itself. The presence of multiple twist defects thus give rise to non-contractible Wilson loops and a non-commutative Wilson algebra. This was illustrated explicitly in the context of the Kitaev's toric code, bilayer topological states, and superconductor - ferromagnet - fractional quantum spin Hall insulator (SC-FM-FQSHI) heterostructure in section~\ref{sec:twistdefectA} (c.f.~eq.\eqref{Wilsonalgebratoriccode}, \eqref{Wilsonalgebrabilayer} and \eqref{Wilsonalgebraparafermion}). These Wilson operators then lead to a degenerate ground state Hilbert space, which is irreducibly acted upon by the non-commutative Wilson algebra. The degeneracy is topological protected and any splitting are exponentially suppressed by quasiparticle tunneling between defects. Twist defects therefore carry non-trivial quantum dimensions and obey non-Abelian fusion rules even in an Abelian medium (c.f.~\eqref{parafermionfusion}). The eigen-spectrum of Wilson loops around a defect pair can be summarized by the fusion rules of the pair. They were shown explicitly in the Kitaev's toric code, the $\mathbb{Z}_k$ gauge theory, the tricolor model, the bilayer topological states, and the SC-FM-QSHI heterostructure in section~\ref{sec:twistdefectA}.

Defect fusion rules can differ from conventional fusion rules of anyonic excitations. For instance they can contain degeneracies and can even be non-commutative. The reason is the fusion rules are related to the multiplication of elements in the symmetry group, which can be non-Abelian. This was demonstrated in the the $S_3$ symmetric tricolor model in section~\ref{sec:defectlatticemodel} as well as the $SO(8)_1$ state and the ``4-Potts" state in section~\ref{sec:defectSO(8)} and \ref{sec:defect4Potts} respectively. Moreover defects can also be combined with a normal anyonic excitation in the underlying topological state. This give rise to distinct {\em species} among twist defects associated to the same anyonic symmetry. The topological charge that characterizes defect species can be statistically measured by local Wilson operators, which involve closing a quasiparticle strings by wrapping it multiple times around the defect. This was demonstrated in the Abelian examples in section~\ref{sec:twistdefectA} and can be generalized to all Abelian scenarios. Defect species in a non-Abelian medium requires a more elobarate treatment. For instance, different defects associating to the same symmetry can have unidentical quantum dimensions. This was illustrated in the non-Abelian chiral ``4-Potts" state in section~\ref{sec:defect4Potts}, where the local Wilson operators, which distinguished defect species, themselves formed a ring structure (see eq.\eqref{alphaloopalgebra}) rather than a group structure like the previous Abelian examples.

The degenerate quantum states of a multi-defect system can be labeled by the simultaneous eigenvalue of a maximal set of mutually commuting Wilson operators. Equivalently, they can be characterized by the internal channels of a fusion/splitting tree (see figure~\ref{fig:Fmoves} in section~\ref{sec:MTC}). This is because there is a one-to-one correspondence between the admissible (possibly degenerate) fusion channels between two defect objects and the eigenvalues of the Wilson loops around the defect pair. And different internal fusion channels of a fusion tree correspond to eigenvalues of non-intersecting and therefore commuting Wilson operators (see figure~\ref{toriccodefusiontree} for the toric code case for example). There are multiple choices of maximal sets of mutually commuting Wilson operators. For example the Wilson operators in an commuting set $\{\mathcal{W}^I\}$ may intersect with those in another set $\{\mathcal{W}^{II}\}$. Diagramatically, they corresponds to distinct fusion trees $\mathcal{T}^I$ and $\mathcal{T}^{II}$. As the two sets do not share simultaneous eigenstates, quantum states labeled by the two fusion trees are transformed into each other by non-diagonal basis transformations (see figure~\ref{fig:Fmoves}). These basis transformations are generated by primitive ones called the $F$-symbols, which relates the splittings ${\bf d}\to{\bf x}\times{\bf c}\to({\bf a}\times{\bf b})\times{\bf c}$ and ${\bf d}\to{\bf a}\times{\bf y}\to{\bf a}\times({\bf b}\times{\bf c})$. They were defined in eq.\eqref{Fsymboldef} and were computed in detail for defects in the Kitaev's toric code in section~\ref{sec:Fbasistransformation}. These $F$-symbols are gauge dependent matrices that depends on certain gauge fixing. In particular they can be dervied by defining a set of {\em splitting operators} $V_{\bf c}^{\bf ab}$ that encode the Wilson strings involved in an admissible splitting ${\bf c}\to{\bf a}\times{\bf b}$. Examples were demonstrated for defects in the Kitaev's toric code (figure~\ref{fig:splittingstatestoriccode}), the $SO(8)_1$ state (eq.\eqref{so(8)splittings} and \eqref{so(8)splittingrhorho}) as well as the $\mathbb{Z}_n$ gauge theory (figure~\ref{fig:splittingstates}). Their defect $F$-transformations were summarized in table~\ref{tab:Fsymbolstoriccode}, \ref{tab:so(8)Fsymbols} and \ref{tab:Fsymbols} respectively.

We discussed the projective and non-Abelian defects braiding operation in section~\ref{sec:defectbraiding}. We focused on the $\mathbb{Z}_n$ gauge theory, where electric-magnetic twist defects behaves like parafermions or fractional Majorana bound states in SC-FM-FQSHI heterostructures. The quantum exchange braiding operations $B^{\sigma_{\lambda_2}\sigma_{\lambda_1}}$ were computed using a sequence of $R$ and $F$-moves (see figure~\ref{fig:Bmove}). The defect exchange $R$-symbols were defined and computed in eq.\eqref{ZnRsymbol} and figure~\ref{fig:Rss}. These braiding operations $B$ obey the Yang-Baxter equation \eqref{YangBaxtereq} (see also figure~\ref{fig:YangBaxter}) and formed a projective representation of the sphere braid group. However, unlike conventional $R$-symbols which satisfy the hexagon identities \eqref{hexagoneq} (see figure~\ref{fig:hexagon1}), the defects $R$-symbols follow a set of modified consistency relations, known as the $G$-crossed heptagon equations, that take the anyon relabeling into account. These are out of the scope of this review article, but these concepts has been recently defined by Barkeshil {\em et.al.} in Ref.~\cite{BarkeshliBondersonChengWang14}.

We conclude this review by remarking a couple of prospects beyond twist defects. Firstly, twist defects are static objects rely on a global symmetry in a topological state. Until now, the most part of this review article as well as most of the work on twist defects has focused on their exciting semiclassical properties. However, it would be interesting to understanding the consequences of their {\em quantum} dynamics. Given a parent (topological) phase with a global (anyonic) symmetry we can ask what new phase emerges when defects associated to the non-trivial elements of the symmetry group become deconfined quantum excitations, i.e.~when the symmetry is converted from a global symmetry to a local one. Perhaps it is appropriate to draw the relation to discrete gauge theories (see section~\ref{sec:fluxcharge}) where the locally symmetric topological state is promoted by a {\em gauging} phase transition (see eq.\eqref{D(G)gauging}) from a globally symmetric trivial boson condensate. In general, a globally symmetric topological phase can also be promoted and gauged into a locally symmetric one. \begin{align}\begin{diagram}\stackrel{\mbox{Topological Phase with}}{\mbox{Global Anyonic Symmetry}}&\pile{\rTo^{\mbox{\small Gauging}}\\\lTo_{\mbox{\small Condensation}}}&\mbox{Twist Liquid}\end{diagram}.\label{gaugingtransition}\end{align} The resulting new topological phase, the {\em twist liquid}, now supports quantum anyonic excitations which were previously the semiclassical twist defects. There have already been some descriptions of a quantum dynamical theory of twist defects in the special cases of a bilayer FQH state~\cite{BarkeshliWen11}, and an Abelian $\mathbb{Z}_N$ theory~\cite{BarkeshliWen10, BarkeshliWen12}. A more general gauging framework has recently been proposed in Ref.~\cite{BarkeshliBondersonChengWang14, TeoHughesFradkin15}. Through bulk-boundary correspondence, the gauging transition can also be understood as an orbifold construction~\cite{bigyellowbook,Ginsparg88,DijkgraafVafaVerlindeVerlinde99,CappelliAppollonio02,chenroyteoinprogress} of the conformal field theory that lives along the $(1+1)$D boundary. However a more realistic gauging {\em mechanism} out of the exactly solved string-net regime that can be applied to known globally symmetric topological phases is still missing.  

Secondly, new topological systems can arise from coupled arrays of twist defects. A tremendous effort has already been expended on quasi-1D topological systems such as the AKLT chain~\cite{Haldanespinchain, AKLT}, anyonic quantum spin chains~\cite{FeiguinTrebstLudwigTroyerKitaevWangFreedman07, Fendley12, GilsArdonneTrebstHuseLudwigTroyerWang13}, classification of symmetry protected phases~\cite{ChenGuLiuWen11, GuWen12, ChenGuLiuWen12}, gapped edges of fractional topological insulators~\cite{MotrukTurnerBergPollmann13}, as well as parafermion chains~\cite{BondesanQuella13, JermynMongAliceaFendley14, Fendley14, LiYangTuCheng15, Yuchenroyteoinprogress}. Defects can also be coupled in a two dimensional array~\cite{mongg2, Vaeziparafermion14, VaeziKim13, VaeziKitaevspin14, BarkeshliJiangThomaleQi15, StoudenmireClarkeMongAlicea15, AliceaFendley15} and give rise to a new non-Abelian topological state that support Fibonacci anyons, which are powerful enough for universal topological quantum computing. Twist defects therefore provide very promising routes in the construction of new topological phases in all dimensions.

{}


%

\end{document}